\documentclass[11pt]{article}

\usepackage[a4paper, margin=1in]{geometry}
\usepackage{amsmath, amssymb, amsthm}
\usepackage{mathtools}
\usepackage{booktabs}
\usepackage{longtable}
\usepackage{multirow}
\usepackage{array}
\usepackage{graphicx}
\usepackage{xcolor}
\usepackage[hidelinks]{hyperref}
\usepackage{caption}
\usepackage{subcaption}
\usepackage{enumitem}
\usepackage{siunitx}
\usepackage{microtype}
\usepackage{authblk}
\usepackage{cite}
\usepackage{float}
\usepackage[normalem]{ulem}

\newcommand{\R}{\mathbb{R}}
\newcommand{\bx}{\mathbf{x}}

\newcommand{\bs}{\mathbf{s}}
\newcommand{\bA}{\mathbf{A}}
\newcommand{\bI}{\mathbf{I}}
\newcommand{\bX}{\mathbf{X}}

\newcommand{\bZ}{\mathbf{Z}}
\newcommand{\bU}{\mathbf{U}}
\newcommand{\bV}{\mathbf{V}}
\newcommand{\bone}{\mathbf{1}}

\definecolor{forestgreen}{rgb}{0.33,0.61,0.34}

\title{Transferable reconstruction of equilibria in network dynamics from sentinel nodes}
\author[1, 2, 3, *]{Naoki Masuda}
\author[2]{Bisna Mary Eldo}
\author[1]{Tharusha Bandara}
\affil[1]{Gilbert S.\ Omenn Department of Computational Medicine and
          Bioinformatics, University of Michigan, Ann Arbor, MI, USA}
\affil[2]{Department of Mathematics, University of Michigan, Ann Arbor,
          MI, USA}
\affil[3]{Center for Computational Social Science, Kobe University,
          Kobe, Japan}
\affil[*]{Corresponding author. E-mail address: \texttt{naokimas@umich.edu}}
\date{\today}

\newcommand{\ba}{\mathbf{a}}
\newcommand{\bmu}{\boldsymbol{\mu}}
\newcommand{\bW}{\mathbf{W}}
\newcommand{\bJ}{\mathbf{J}}
\newcommand{\bP}{\mathbf{P}}

\begin{document}

\maketitle

\begin{abstract}
Reconstructing the state of a large networked system from measurements at only a few nodes is a challenge for monitoring, prediction, and intervention. Here we ask whether such reconstruction is possible for equilibrium states and whether it can transfer across different nonlinear dynamics. We study this question across various networks and across $17$ nonlinear dynamics on networks from different domains. For each network, we observe only a vanishingly small fraction of sentinel nodes and train either a neural network or linear decoder. We find that accurate reconstruction is often possible, and that transferability is structured rather than universal. Eleven dynamics with adjacency-matrix-type coupling form a robust transferable class. By contrast, six diffusively coupled dynamics form isolated transfer components. Non-random sentinel selection is consistently important, and linear decoders often approach neural-network performance. These results show that sparse node observations can encode enough information to reconstruct full network equilibria across broad families of nonlinear dynamics, while also revealing sharp limits to universal transfer.
\end{abstract}

\section{Introduction}

Many nonlinear dynamics of practical interest occur on networks,
including epidemic spreading in populations, ecological interactions
among species, gene regulation, neural activity, and transport in
metapopulation systems. In such systems, the dynamical state of
interest may be difficult to measure at all nodes because of cost,
accessibility, or invasiveness, even though node-level state
information is often needed for monitoring, prediction, and
intervention. For example, estimating infection prevalence in every
patch of a metapopulation network or measuring biomass across many
species or sites in an ecosystem can be impractical. These examples motivate a practical question:
can the full node-wise state of a networked dynamical system be
reconstructed from observations at only a small fraction of nodes?
We address this problem, focusing on reconstruction of equilibrium (or final) states of nonlinear dynamics on networks.

Sparse reconstruction has a long history in signal processing and
data-driven modeling. Compressed sensing shows that high-dimensional
signals can be recovered from few measurements when they are sparse or
compressible in a known basis~\cite{Donoho2006IeeeTransInfoTheory, Candes2008IeeeSigProcMag}. In
many such settings, the measurements are linear projections of the
full state; for a signal on network nodes, these projections take the
form $R_\ell=\sum_{j=1}^{N}a_{\ell j}x_j$, where $x_j$ is the signal
at the $j$th node. Dimension reduction theory for network dynamics
also often uses such collective variables; a broad class of nonlinear
dynamics on networks can be represented accurately by a small number
of variables $R_1,R_2,\ldots$~\cite{GaoBarzelBarabasi2016Nature, Tu2017PhysRevE, laurence2019,
Thibeault2020PhysRevResearch, Tu2021Iscience, Zhang2022NatEcolEvol,
KunduKoriMasuda2022PhysRevE, Masuda2022PhysRevResearch,
Vegue2023PnasNexus}, particularly when the adjacency matrix has low
effective rank~\cite{Valdano2019PhysRevX, Thibeault2024NatPhys}.
These approaches reveal low-dimensional structure, but they do not
directly solve the monitoring problem considered here because each
$R_\ell$ is typically a non-sparse combination of many node states.
Here we ask instead whether the full network state,
$(x_1,\ldots,x_N)$, generated by nonlinear network dynamics can be
reconstructed from the observed values of $x_i$ at only a small number
of sentinel nodes.

Sparse sensor placement methods~\cite{Willcox2006CompFluids,
Drmac2016SIAM, manohar2018} are closer to our measurement setting
because they select a small number of sensor locations and reconstruct
the full state from those sensor values. The sparse sensor placement
methods most relevant here, however, usually assume that the target
signals are well represented in a prescribed or previously learned
low-dimensional basis. For nonlinear dynamics on heterogeneous
networks, an appropriate low-dimensional basis for $(x_1,\ldots,x_N)$
is not known a priori. Graph-signal sampling also reconstructs signals
on all nodes from values observed at selected sensor nodes
~\cite{Chen2015IEEE, Anis2016IEEE}, but it typically relies on
bandlimitedness or related assumptions and requires knowledge of the
network structure. The states of heterogeneous nonlinear dynamics on
networks need not satisfy these assumptions, and the network itself
may not be fully known in practical monitoring settings.

Network observability frameworks also consider a small number of
sensor nodes with the aim of observing the entirety or a large part of
the network in some sense. In control theory, observability is dual to
controllability and concerns whether the state of a dynamical system
can be inferred from limited measurements. This perspective has led to
work on observability of complex networks, often with the goal of
estimating the initial condition of networked dynamics
~\cite{LiuSlotineBarabasi2013PNAS, Yan2015NatPhys, Haber2018TCNS},
as well as distinguishability of initial conditions
~\cite{FornasiniValcher2013TAC} and functional observability of target
$R_\ell$-like variables in linear systems
~\cite{Fernando2010IEEE, Montanari2022PNAS}. Other network
observability studies ask whether sensor nodes and their neighbors
cover a large fraction of the network
~\cite{YangWangMotter2012PRL, Hasegawa2013PRE}. However, these
frameworks do not directly address the data-driven task of
reconstructing $x_i$ at every node of a potentially large nonlinear
networked system from observations at a small set of sentinel nodes.
Existing combinatorial and algebraic approaches to nonlinear
observability also appear difficult to scale beyond networks with tens
of nodes~\cite{Letellier2018SciRep}.

\begin{figure}
\centering
\includegraphics[width=\textwidth]{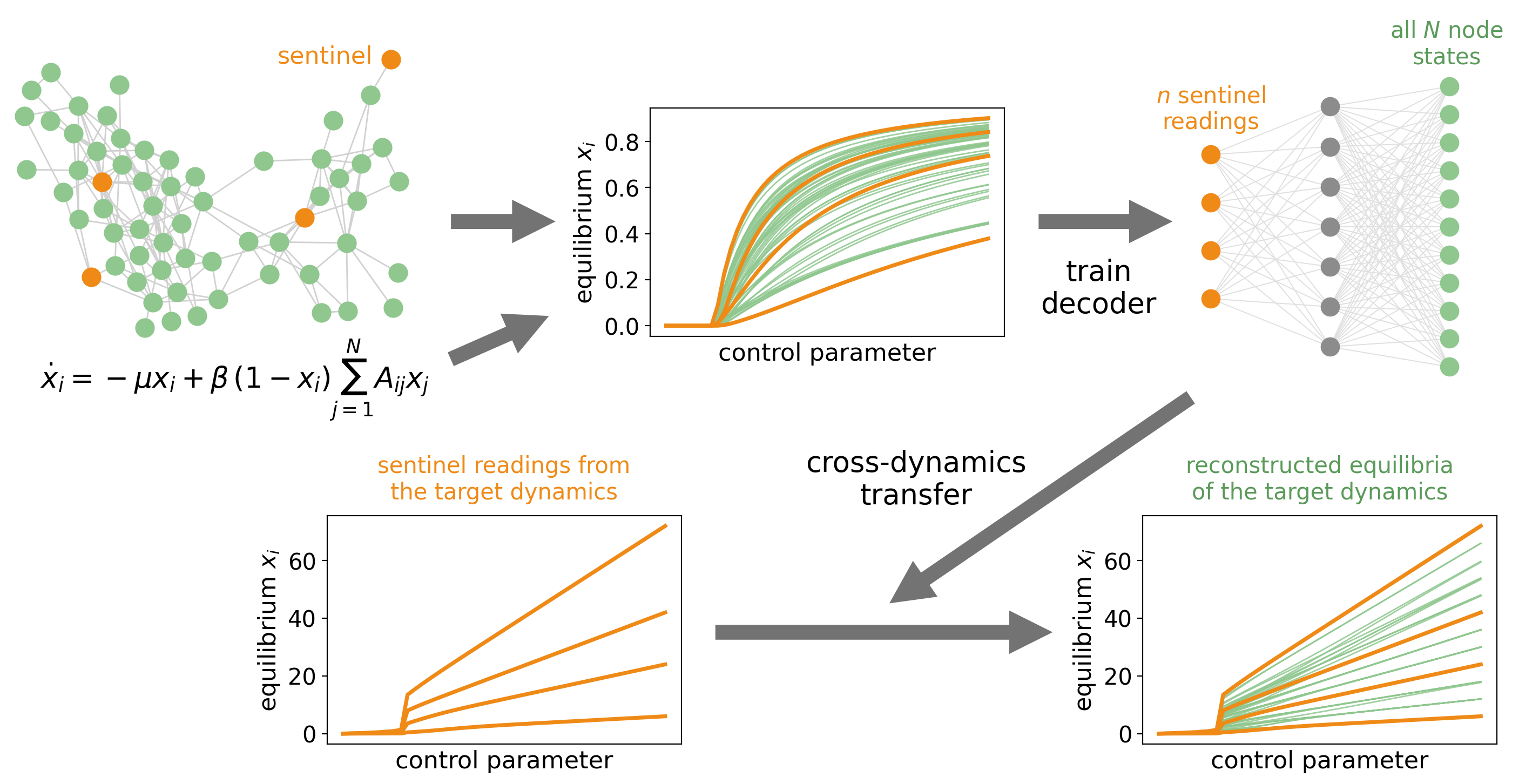}
\caption{Schematic of equilibrium reconstruction from sentinel nodes
and cross-dynamics transfer. A nonlinear dynamics on a network produces an
equilibrium state $(x_1,\ldots,x_N)$. We observe only $n\ll N$ sentinel
nodes, shown in orange, and use a decoder to reconstruct the states of the
remaining nodes, shown in green. The decoder, either a shallow neural
network or linear ridge regression, is trained to map the $n$ sentinel
readings to the full $N$-dimensional equilibrium state. In cross-dynamics
transfer, a decoder trained on a source dynamics is applied to sentinel
readings generated by a different target dynamics and returns an estimate
of the target equilibrium $(x_1,\ldots,x_N)$.}
\label{fig:schem}
\end{figure}

In the present study, we investigate this practical notion of network
dynamics reconstructability. We fit either a shallow neural-network
decoder, motivated by recent studies showing that such decoders can
reconstruct nonlinear spatiotemporal fluid dynamics from sparse
sensors on spatial grids~\cite{Erichson2020PRSA, wzk2024}, or a
linear decoder to reconstruct the equilibrium state
$(x_1,\ldots,x_N)$ from sentinel observations; related reservoir computing
approaches have also been used to reconstruct spatiotemporal chaotic
dynamics~\cite{Lu2017Chaos}. We focus on equilibria and final-size
observables because many applications of network dynamics, including
endemic infection levels, species abundances, and resilience analysis,
are naturally formulated in terms of such states. Beyond
within-dynamics reconstruction, we ask whether reconstructability
transfers across different dynamical systems.
%
%
See Fig.~\ref{fig:schem} for a schematic workflow.
We find that the $17$ nonlinear dynamics considered in this
study are organized into robust transfer classes.
%


\section{Results}
\label{sec:results}

\subsection{Reconstructing network equilibria from sentinel nodes}
\label{sec:framework}

We consider a fixed network with $N$ nodes and an equilibrium state
$\bx=(x_1,\ldots,x_N)^\top\in\R^N$ generated by a nonlinear dynamics
on that network. The reconstruction task is to infer the full node
state $\bx$ from the readings of only
$n=\lfloor \ln N \rfloor$ sentinel nodes
(as has been considered in compressed sensing studies~\cite{Donoho2006IeeeTransInfoTheory, Candes2008IeeeSigProcMag}),
implying $n\ll N$. For each
network and dynamics, we generate equilibrium snapshots by sweeping
one control parameter over $M=51$ values. A decoder is trained to map
the $n$ sentinel readings at a given control-parameter value to the
full $N$-dimensional equilibrium state, $\bx$. See the top half of Fig.~\ref{fig:schem} for a schematic.

We compare two decoder classes. The first is a shallow neural network (NN) decoder with
one hidden layer and ReLU activation~\cite{Erichson2020PRSA, wzk2024}. The second is a linear ridge
decoder, which provides a simpler benchmark and tests how much of the
reconstruction can be explained by linear relationships between
sentinel and non-sentinel nodes. In both cases, the trained decoders are evaluated on held-out
control-parameter values. Unless otherwise stated, we use a bias-free
NN decoder so that a zero sentinel input is mapped to zero output, a
natural constraint for many of the dynamics considered in this study. Full
architectural and optimization details are given in Methods.

The choice of sentinel nodes is part of the reconstruction problem. We
use $n=\lfloor \ln N \rfloor$ sentinels by default and compare three
selection methods: a greedy minimum-correlation method (MinCorr), a
QR-pivoted sensor method (QR) \cite{manohar2018, wzk2024}, and uniformly random selection. The
first two methods choose sentinels from the training data and are
designed to capture diverse directions of variation in the equilibrium
snapshots.

A central question in this study is whether reconstruction transfers
across dynamics. We call the dynamics used to train a decoder the
source dynamics, denoted by $\mathrm{S}$, and the dynamics on which
the trained decoder is evaluated the target dynamics, denoted by
$\mathrm{T}$. Because different dynamics can have different physical
scales, we calibrate cross-dynamics deployment using only the target
readings at the sentinel nodes. (See the bottom half of Fig.~\ref{fig:schem}.) Specifically, target sentinel readings
are first affinely mapped into the source decoder's sentinel-input
range, and the decoder output is then affinely mapped back to the
target dynamics' scale using a two-parameter fit on the same sentinel nodes.
Thus, no non-sentinel data in the target dynamics are used for cross-dynamics
calibration. Furthermore, we test cross-dynamics reconstruction
at conditions of the target dynamics that are not used for cross-dynamics calibration, ensuring out-of-sample nature of testing.
See Methods for the precise definitions.

We quantify reconstruction performance using two complementary
measures. The first is the normalized root-mean-squared error
(n-RMSE), obtained by dividing the RMSE by the global range of the
target dynamics. The second is the ratio
$r(\mathrm{S}\to\mathrm{T})$ between the decoder's n-RMSE and the
n-RMSE of a trivial baseline that predicts every node by the mean of
the sentinel readings (see Methods for details). A successful reconstruction should have both
small n-RMSE and $r(\mathrm{S}\to\mathrm{T})<1$.

\subsection{Source--target reconstruction: an illustrative example}
\label{sec:cross_matrices}

\begin{figure}
\centering
\includegraphics[width=\textwidth]{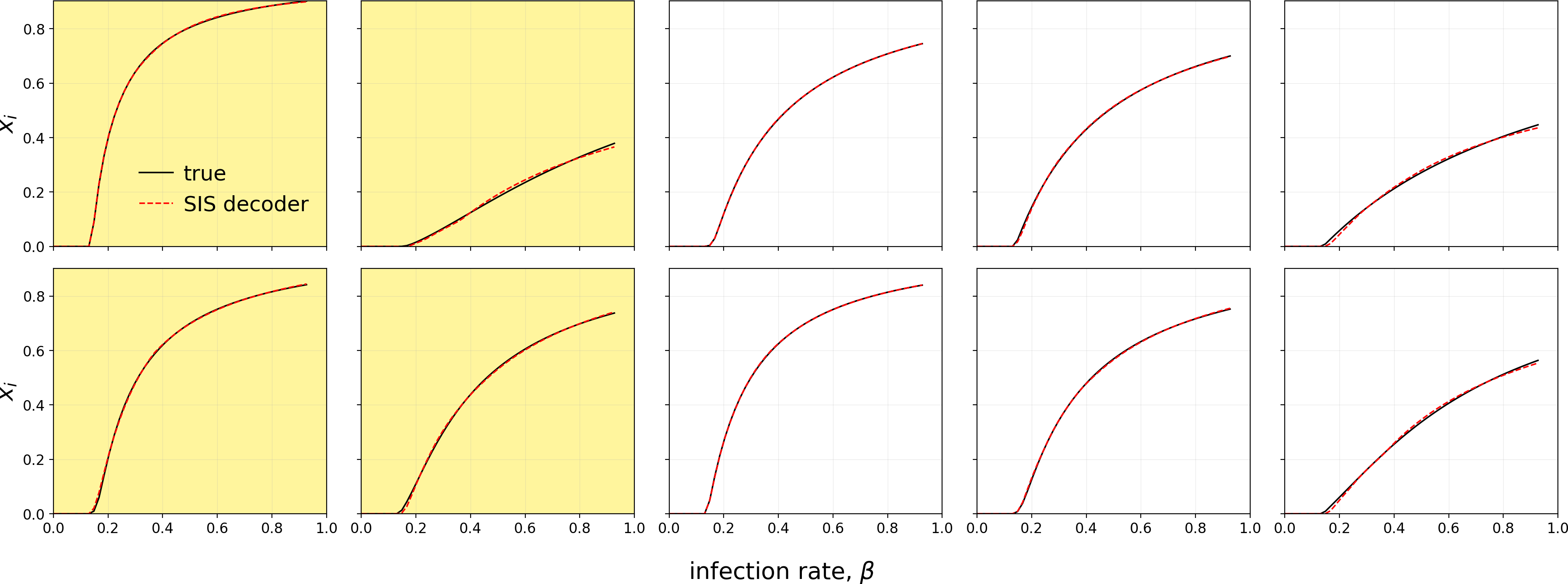}
\caption{Within-dynamics reconstruction on the dolphin network
($N=62$). We trained a shallow NN decoder on SIS equilibria using
MinCorr sentinels ($n=\lfloor\ln N\rfloor=4$) and evaluated it on
held-out SIS equilibria from the same dynamics. Each panel
corresponds to one node; we show $10$ of the $62$ nodes. The first
two columns show the four sentinel nodes, highlighted in yellow. Among the non-sentinel
nodes, the third, fourth, and fifth columns show, respectively, the two nodes
with the smallest node-wise n-RMSE, the two nodes whose node-wise n-RMSE is
closest to the median across all $62$ nodes, and the two nodes with the
largest node-wise n-RMSE. The solid line shows the true equilibrium $x_i$ as a
function of the infection rate $\beta$. The dashed line shows the decoder
prediction, which almost completely overlaps the true $x_i$ for all nodes.}
\label{fig:sis-sis-dolphin}
\end{figure}

We first illustrate the source--target reconstruction procedure using
the smallest network in our data set, the dolphin social network
($N = 62$ nodes). Figure~\ref{fig:sis-sis-dolphin} shows a
within-dynamics test for the susceptible--infectious--susceptible
(SIS) model. We trained a shallow NN decoder on SIS equilibria using
MinCorr sentinels and evaluated it on held-out SIS equilibria from
the same network. The horizontal axis shows the infection rate
$\beta$, which is the control parameter for the SIS model. Each panel
corresponds to one node, and the panels corresponding to the
$n = \lfloor \ln N \rfloor = 4$ sentinel nodes are highlighted in
yellow. To keep the figure readable, we show $10$ representative nodes
rather than all the $62$ nodes; the caption explains how we select them.
The decoder accurately reconstructs the equilibrium values
$x_i$ for both sentinel and non-sentinel nodes across the control
parameter range.

\begin{figure}
\centering
\includegraphics[width=\textwidth]{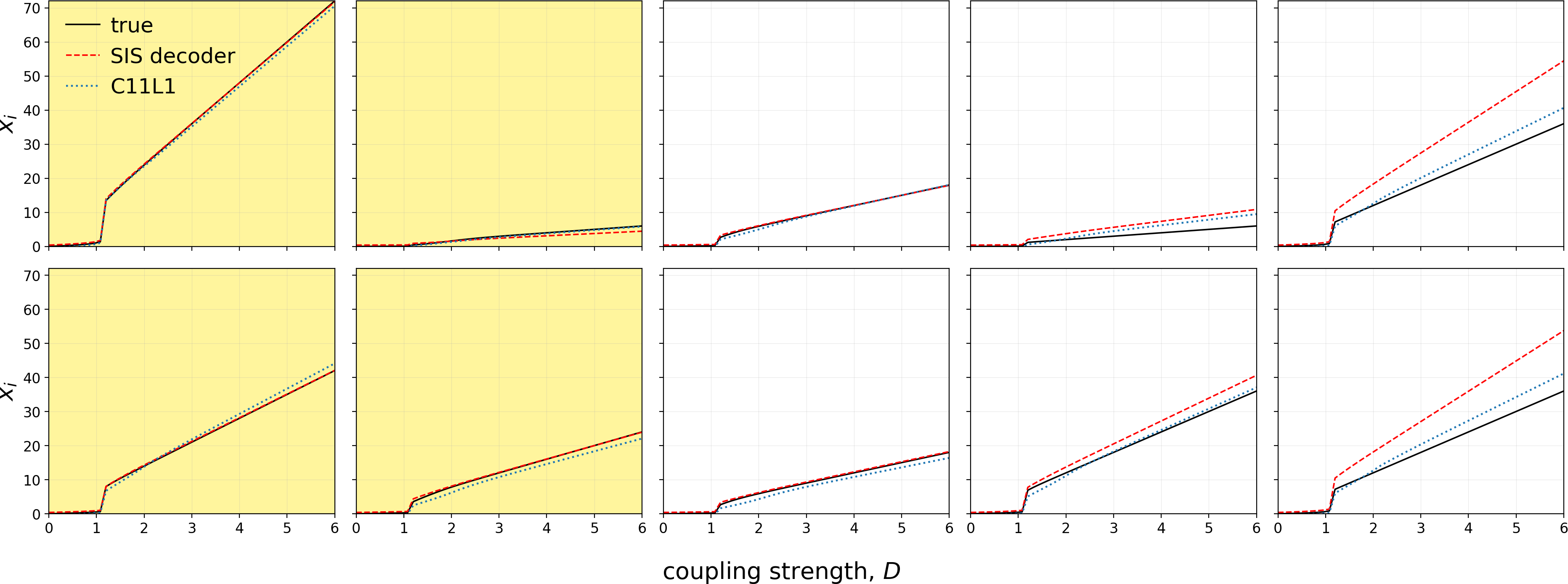}
\caption{Cross-dynamics reconstruction on the dolphin network
($N=62$). The dashed lines show the predictions of the SIS-trained NN
decoder used in Fig.~\ref{fig:sis-sis-dolphin}, now evaluated on WC
equilibria. The dotted lines show the predictions of a C11L1 NN decoder,
i.e., a pooled leave-one-dynamics-out decoder trained on the $10$
transferable dynamics other than WC and evaluated on WC. Each panel
corresponds to one node; we show $10$ of
the $62$ nodes using the same layout as in
Fig.~\ref{fig:sis-sis-dolphin}. The first two columns show the four sentinel
nodes of the SIS-trained decoder, highlighted in yellow; these are the same
nodes as in Fig.~\ref{fig:sis-sis-dolphin}. The nodes in the third, fourth,
and fifth columns are selected according to the node-wise n-RMSE of the
SIS-trained decoder, using the same criteria as in
Fig.~\ref{fig:sis-sis-dolphin}. Except for one node, the
non-sentinel nodes shown here differ from those in
Fig.~\ref{fig:sis-sis-dolphin}. The solid line shows the true WC
equilibrium.}
\label{fig:sis-wc-dolphin}
\end{figure}

The dashed lines of Fig.~\ref{fig:sis-wc-dolphin} show a cross-dynamics deployment of
the same SIS-trained decoder on the Wilson--Cowan (WC) dynamics on
the same network. Although the reconstruction is less accurate for
some non-sentinel nodes, the decoder captures the equilibrium curves
reasonably well for most nodes and across most values of the
control parameter (which is the coupling strength). This transfer is notable because the SIS and WC
models differ substantially in equation form, in how $x_i$ depends on
the control parameter, and in the degree of node-to-node heterogeneity
in the equilibrium states. In particular, SIS equilibria are relatively
homogeneous across nodes, whereas the WC dynamics produces a wider
range of node-dependent responses.

The reconstruction error varies substantially across nodes. To
quantify this variability, we compute the node-wise n-RMSE, defined as the
n-RMSE without averaging over nodes (see Supplementary
Section~\ref{si:nodewise} for the methods and full results). For the
examples shown in Figs.~\ref{fig:sis-sis-dolphin} and
\ref{fig:sis-wc-dolphin}, the node-wise n-RMSE varies across nodes by
factors of $16$ and $44$, respectively, and its distribution is skewed
toward large values. The sentinel nodes are not systematically the most
accurately reconstructed nodes. Furthermore, the association between
node-wise n-RMSE and degree is weak and inconsistent in sign across the
cases examined. However, when each node's error is normalized by that
node's own equilibrium range, thereby measuring the error relative to the
scale of variation of each node, high-degree nodes are consistently
reconstructed more accurately than low-degree nodes.

\subsection{Source--target reconstruction across dynamical systems}

To examine the generality of the illustrative results in
Figs.~\ref{fig:sis-sis-dolphin} and~\ref{fig:sis-wc-dolphin}, we next
consider $12$ model families of continuous-time dynamics on networks,
which yield $17$ one-parameter dynamics in total. These dynamics span
five broad physical settings: epidemic dynamics (SIS,
susceptible--infectious--recovered (SIR), metapopulation SIS (MSIS),
and metapopulation SIR (MSIR)); neural dynamics (WC); gene-regulatory
dynamics (GEN); ecological dynamics (mutualistic interaction (MUT),
generalized Lotka--Volterra (gLV), Noy--Meir grazing (NM),
eutrophication (EUT), and vegetation--turbidity feedback (VEG)); and
archetypal bistable dynamics (coupled double well (DW) and its Laplacian-coupling variant (LDW)). For MSIS,
MSIR, EUT, and VEG, we use two parameter variants, denoted by MSIS1
and MSIS2, for example. We treat these variants as distinct dynamics,
because the parameter change substantially alters the equilibrium
response. While SIR-type models (i.e., SIR and MSIR) show transient dynamics and we measure final epidemic sizes for them, we call them equilibria for simplicity. For each of the resulting $17$ dynamics, we vary one control
parameter, typically a bifurcation or forcing parameter. See Methods
for the equations, parameter values, and sweep ranges.

Before systematically evaluating reconstruction, we ask whether the
equilibria visited during a parameter sweep are intrinsically
low-dimensional, because low-dimensional equilibrium data would support the
possibility that observations from a small fraction of nodes are
sufficient. For each dynamics and network, we use the equilibrium matrix
$\bX\in\R^{N\times M}$ defined above and compute its singular values
$\sigma_1\ge\cdots\ge\sigma_M$. There are $M=51$ singular values because
$M<N$ for every network considered. We quantify effective rank using four
measures derived from these singular values. For most dynamics--network
pairs, all four effective ranks are much smaller than $M$ (and hence than
$N$), and are typically no greater than $3$; see Supplementary
Section~\ref{si:rank} for their definitions and the results. This
finding motivates our use of both linear and nonlinear sentinel-based
reconstruction across a variety of dynamics and networks.

\begin{figure}
\centering
\includegraphics[width=\textwidth]{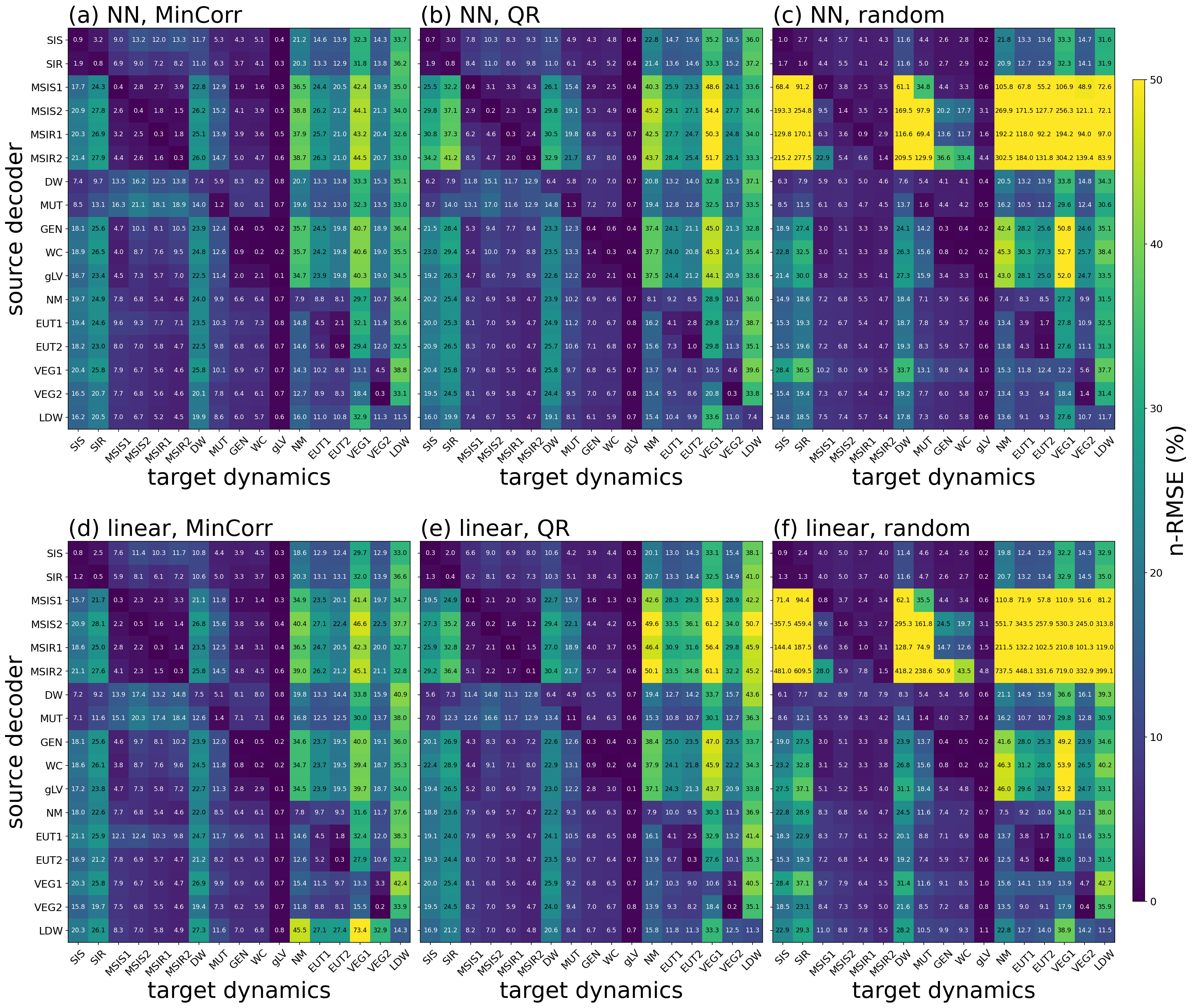}
\caption{Normalized root-mean-squared error (n-RMSE) for each source--target pair, decoder architecture,
and sentinel-selection method. (a)~NN decoder with MinCorr
sentinels. (b)~NN decoder with QR-pivoted sentinels.
(c)~NN decoder with uniformly random sentinels. (d)~Linear
decoder with MinCorr sentinels. (e)~Linear decoder with
QR-pivoted sentinels. (f)~Linear decoder with uniformly random
sentinels. Each cell gives the n-RMSE, averaged over the $14$
networks, for a decoder trained on the source dynamics and evaluated
on the target dynamics. Values are reported in percent. The colormap
is clipped at $50\%$.}
\label{fig:heat-combined}
\label{fig:heat-nn-minc}\label{fig:heat-nn-qr}\label{fig:heat-nn-random}%
\label{fig:heat-lin-minc}\label{fig:heat-lin-qr}\label{fig:heat-lin-random}
\end{figure}

For each of the $14$ networks (see Methods), we trained each decoder
type on each of the $17$ dynamics. We then evaluated each trained
decoder on both its source dynamics and the other target dynamics on
the same network. Figure~\ref{fig:heat-combined} shows the normalized
root-mean-squared error (n-RMSE) for all source--target pairs and for
each combination of decoder architecture and sentinel-selection
method. Each panel corresponds to a pair of a decoder architecture, either the
shallow NN or the linear ridge decoder, and a sentinel-selection
method, either MinCorr, QR, or random. Each cell gives the n-RMSE of a
source decoder evaluated on a target dynamics, averaged over the
$14$ networks.

Figure~\ref{fig:heat-combined} reveals several patterns. First,
within-dynamics reconstruction is usually accurate; most diagonal
entries have small n-RMSE, consistent with the example in
Fig.~\ref{fig:sis-sis-dolphin}. The main exceptions are VEG1 and LDW and, to a
lesser extent, DW, NM, and EUT1. Second, many off-diagonal entries are also small,
showing that reconstruction can transfer across different dynamical
systems. For example, with the shallow NN decoder and MinCorr sentinels
(Fig.~\ref{fig:heat-combined}(a)), the n-RMSE is at most $3\%$ for $12\%$ of
the $272$ off-diagonal entries, at most $6\%$ for $25\%$ of them, and at most
$10\%$ for $44\%$ of them. As a scale for interpretation, the within-dynamics SIS
example in Fig.~\ref{fig:sis-sis-dolphin} gives
$\mathrm{n\text{-}RMSE}=0.43\%$, and the cross-dynamics
SIS$\to$WC example in Fig.~\ref{fig:sis-wc-dolphin} gives
$\mathrm{n\text{-}RMSE}=5.54\%$. Third, some target dynamics, such as WC and gLV, are comparatively easy to
reconstruct from many different source decoders, whereas others, such as VEG1 and LDW, are difficult to reconstruct from most non-self sources. Fourth, some source decoders transfer well to many target dynamics, but
none transfers well to all targets. Among the $17$ per-dynamics
sources, every decoder has at least one target dynamics with
n-RMSE of approximately $28\%$ or larger. 
Fifth, related to the third and fourth observations, cross-dynamics reconstruction performance is not symmetric between the source and target dynamics. If a decoder trained with dynamics
$\mathrm{S}$ transfers well to reconstruct equilibria of dynamics $\mathrm{T}$, the same type of decoder trained with $\mathrm{T}$ may poorly reconstruct equilibria of 
$\mathrm{S}$. Sixth, random sentinel
selection can lead to catastrophic failures, especially for source
decoders trained on MSIS or MSIR, and to a lesser extent for GEN, WC,
and gLV. Seventh, the NN and linear decoders show similar performance. For
example, the median n-RMSE over the off-diagonal cells of
Fig.~\ref{fig:heat-combined} is $12.54\%$ for the NN decoder and $12.09\%$
for the linear decoder with MinCorr sentinels, and $11.73\%$ and $11.52\%$,
respectively, with QR-pivoted sentinels. This similarity suggests that
successful reconstruction primarily reflects the low dimensionality of the
equilibrium data revealed by our effective-rank analysis, rather than the
use of a particular decoder architecture. We therefore continue to compare
the two decoder architectures below.
Finally, as expected, a decoder trained on one parameter
variant of a model tends to reconstruct the other variant accurately
(e.g., MSIS1$\to$MSIS2 and MSIS2$\to$MSIS1).

\begin{figure}
\centering
\includegraphics[width=\textwidth]{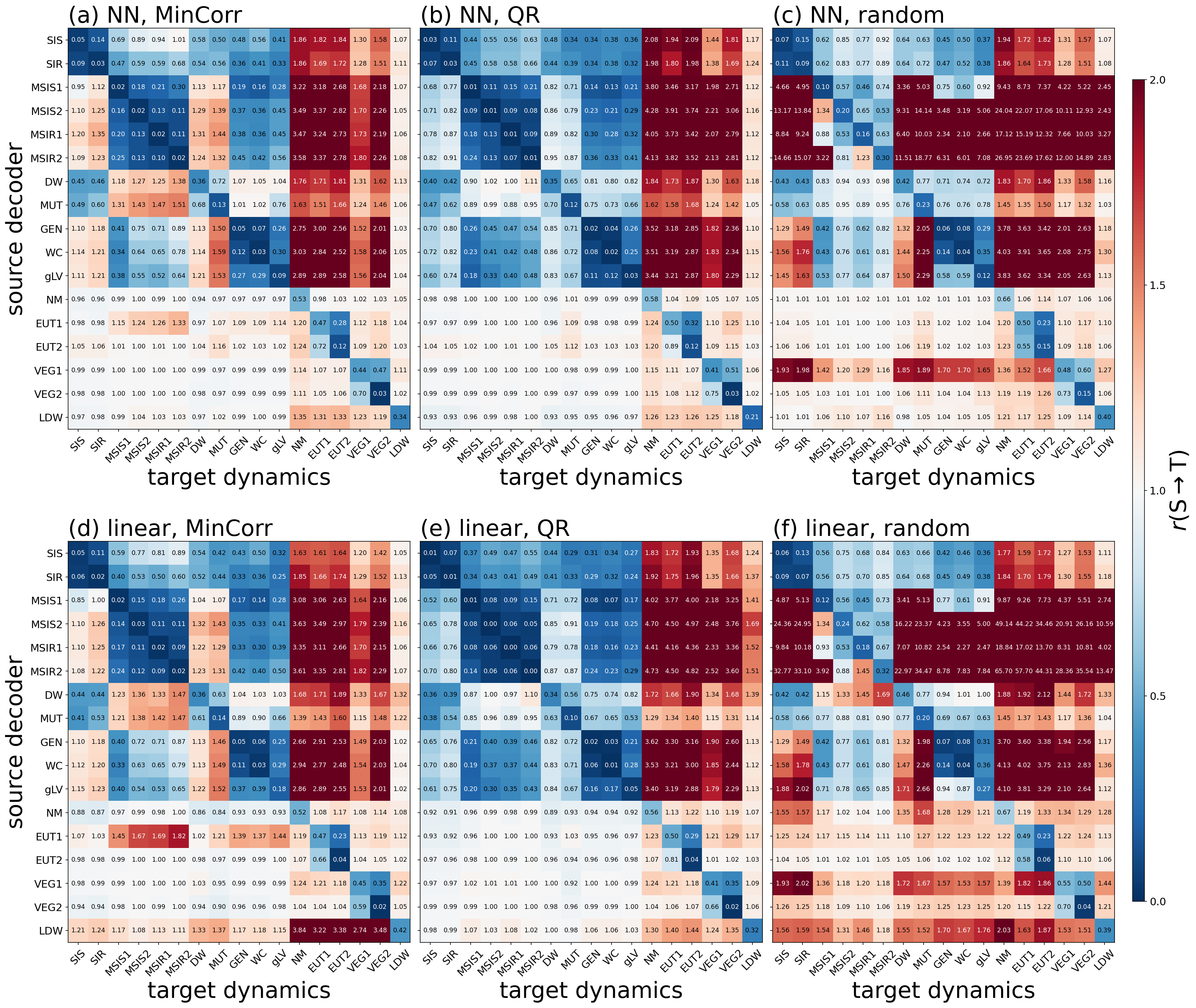}
\caption{Ratio $r(\mathrm{S}\to\mathrm{T})$ of the decoder's n-RMSE
to the n-RMSE of the mean-prediction baseline. The panel definitions
are the same as in Fig.~\ref{fig:heat-combined}. Each cell gives the
n-RMSE averaged over the $14$ networks divided by the mean-baseline n-RMSE
averaged over the $14$ networks. Blue cells
($r<1$) indicate source--target pairs for which the decoder
outperforms the baseline; red cells ($r>1$) indicate pairs for which
the baseline is more accurate.}
\label{fig:ratio}
\end{figure}

As a further test, we compared each trained decoder with a trivial
baseline that predicts the same value for every node, namely the mean
of the observed sentinel values at the same control-parameter value.
Figure~\ref{fig:ratio} shows the ratio
$r(\mathrm{S}\to\mathrm{T})$ of the decoder's n-RMSE to the
n-RMSE of this mean-prediction baseline. A value
$r(\mathrm{S}\to\mathrm{T})<1$ indicates that the decoder outperforms
this trivial mean-prediction baseline, whereas $r(\mathrm{S}\to\mathrm{T})>1$ indicates that
the baseline is more accurate. As reference values, the SIS
within-dynamics example in Fig.~\ref{fig:sis-sis-dolphin} gives
$r(\mathrm{SIS}\to\mathrm{SIS})=0.026$, and the SIS$\to$WC
cross-dynamics example in Fig.~\ref{fig:sis-wc-dolphin} gives
$r(\mathrm{SIS}\to\mathrm{WC})=0.431$.

The ratio in Fig.~\ref{fig:ratio} provides information that is
complementary to the n-RMSE in Fig.~\ref{fig:heat-combined}. A
decoder should ideally have both a small n-RMSE and $r(\mathrm{S}\to\mathrm{T})$
substantially below $1$. We make three observations. First, beating
the mean-prediction baseline is not automatic; many source--target
pairs have $r(\mathrm{S}\to\mathrm{T})>1$. This situation can occur even when
the n-RMSE in Fig.~\ref{fig:heat-combined} is small, because some
target dynamics have relatively homogeneous node states for which the
mean of the sentinels is already a competitive predictor. Second,
despite this difference between the n-RMSE and $r(\mathrm{S}\to\mathrm{T})$, the broad patterns in
$r(\mathrm{S}\to\mathrm{T})$ resemble those in the n-RMSE heatmaps
across decoder architectures and sentinel-selection methods. Third,
consistent with Fig.~\ref{fig:heat-combined}, no single per-dynamics
source decoder outperforms the mean-prediction baseline across all
target dynamics.

\subsection{Robustness of decoding}
\label{sec:robust}

We carry out several robustness tests in this section.

First, the mean-prediction baseline is deliberately simple. We therefore also
compare our decoders with gappy POD, a classical linear method for
reconstructing a full state from a small number of observed
entries~\cite{EversonSirovich1995, manohar2018, Brunton2022book,
Erichson2020PRSA}. 
Gappy POD uses the equilibrium data matrix of the source dynamics
to learn its
low-dimensional linear basis. It then
reconstructs the full state from the sentinel readings by fitting their
coefficients in this basis; see Supplementary Section~\ref{si:gappy} for
the precise formulation.
We substitute gappy POD
for the trained decoder. With QR-pivoted sentinels, gappy POD is approximately
as accurate as our decoders. In particular, when the mean-prediction
baseline in $r(\mathrm{S}\to\mathrm{T})$ is replaced by gappy POD, the
resulting ratio is, on average, close to one. Thus, the NN and linear
decoders do not systematically outperform gappy POD. This finding is not
surprising because gappy POD is itself a linear map from the sentinel state
to the full network state and is therefore closely related to our linear
decoder. It further supports our conclusion that successful reconstruction
does not depend strongly on a particular decoder architecture. Nevertheless,
our decoding task remains nontrivial because some dynamics produce substantially
higher-dimensional equilibrium data than others
(Supplementary Section~\ref{si:rank}), and our decoders use only the
sentinel readings rather than projections of the full state onto the leading
left singular vectors.

Second, the results reported above use a uniformly random split of the $M=51$
control-parameter values into $\lfloor0.8M\rfloor=41$ training values and
$10$ test values. Thus, the held-out values are generally interspersed among
the training values, and the decoder primarily interpolates along the
equilibrium branch. As more challenging tests, we repeat the
within-dynamics analysis using either the first $41$ or the last $41$
control-parameter values for training, such that the $10$ test values lie
outside the training range and require extrapolation. Averaged over the
dynamics and networks, the within-dynamics n-RMSE increases from
$2.5$--$3.1\%$ under the interspersed split to $3.0$--$3.8\%$ under the
first-$41$ split and $3.1$--$4.5\%$ under the last-$41$ split, where each
range covers the four combinations of decoder architecture and non-random
sentinel-selection method. We omit random sentinels because they perform
worse than MinCorr and QR-pivoted sentinels
(Figs.~\ref{fig:heat-combined} and~\ref{fig:ratio}). The NN and linear
decoders nevertheless outperform the mean-prediction baseline for most
dynamics under the first-$41$ and last-$41$ split. See Supplementary Section~\ref{si:extrap} for the methods
and results. We therefore conclude that successful
within-dynamics reconstruction is not merely an artifact of interpolation.

Cross-dynamics transfer is also largely insensitive to whether the
source decoder is trained on interspersed values, the first $41$ values, or
the last $41$ values of the source sweep
(Supplementary Section~\ref{si:extrap}). By contrast, the choice of
calibration values on the target side has a larger effect. In the analyses
above, we use $41$ randomly interspersed target control-parameter values to
calibrate the coordinate maps and evaluate reconstruction on the remaining
$10$ values. We additionally examine contiguous target-side calibration
sets consisting of the first or last $41$ values. Both contiguous choices
degrade performance relative to the baseline ratio. Across distinct
source--target pairs, the fraction satisfying
$r(\mathrm{S}\to\mathrm{T})<1$ decreases from $41$--$59\%$ with
interspersed calibration to $35$--$53\%$ with first-$41$ calibration and
$11$--$26\%$ with last-$41$ calibration, where each range covers the four
combinations of decoder architecture and non-random sentinel-selection
method. Under first-$41$ calibration, the $10$ test values lie
above the entire calibration range. The coordinate maps defined by
Eqs.~\eqref{eq:tgt-sent-min}--\eqref{eq:v22_input_map} must therefore be
extrapolated to target-state values not observed during calibration, which
is often difficult. Under last-$41$ calibration, the test values are the
first $10$ values of the sweep, where node states are often nearly
indistinguishable, as in the disease-free equilibria of epidemic models. In
this regime, the mean-prediction baseline is nearly exact, so
$r(\mathrm{S}\to\mathrm{T})$ can exceed one even when the decoder's n-RMSE
is small. See
Supplementary Section~\ref{si:extrap} for the full methods and results. Thus, contiguous target-side calibration confounds transfer
performance with the intrinsic difficulty of the test window. We therefore
use interspersed target-side calibration in the subsequent analyses.

Third, the coupled double-well dynamics (DW) is multistable because the
single-node dynamics is bistable. The results reported above use a lower
initial condition at each node, so the system settles onto the lower
equilibrium branch whenever possible. We repeat the simulations using an
upper initial condition at every node and refer to this case as DW-up. We
verify that DW and DW-up converge to different equilibria for many
control-parameter values. DW-up performs comparably to the other dynamics
as both a source and a target and, overall, performs better than DW. These
results suggest that successful source--target reconstruction is not
restricted to a particular equilibrium branch. See Supplementary Section~\ref{si:dw} for
the methods and results.

Fourth, we ask whether the framework extends to dynamics governed by two
control parameters rather than one. We construct two two-parameter families
from models already considered here. MSIS2P varies both the infection rate
and the diffusion rate of infectious individuals. MUT2P varies both
the coupling strength and the immigration rate. Within-dynamics
reconstruction of MSIS2P and MUT2P, their reconstruction using
source decoders trained on one-parameter dynamics, and
reconstruction of other target dynamics using source decoders
trained on MSIS2P and MUT2P are approximately as
accurate as the one-parameter results in
Figs.~\ref{fig:heat-combined} and~\ref{fig:ratio}. These results suggest that our framework extends to dynamics for
which more than one control parameter is varied. See Supplementary
Section~\ref{si:2d} for the methods and results.

Fifth, because each cell in Figs.~\ref{fig:heat-combined} and
\ref{fig:ratio} is an average over $14$ networks, we examine variability
across networks and across four additional random factors: NN
initialization, the split of control-parameter values into training and
held-out sets, the choice of uniformly random sentinels, and the
heterogeneous node parameters in NM, EUT1, EUT2, VEG1, and VEG2. We have found that most of
the variability in n-RMSE and $r(\mathrm{S}\to\mathrm{T})$ arises from
differences among networks. The main exceptions are the random-sentinel
method, for which the choice of sentinel nodes and heterogeneous node
parameters also contributes appreciably. Because random sentinels perform
worse than MinCorr and QR-pivoted sentinels
(Figs.~\ref{fig:heat-combined} and~\ref{fig:ratio}), these exceptions do
not compromise our main conclusions. Although performance varies
substantially across networks even with MinCorr and QR-pivoted sentinels,
the strongest source dynamics, such as SIS and SIR, outperform the
mean-prediction baseline for a large majority of networks. See
Supplementary Section~\ref{si:var} for the methods and results.

Sixth, we examine robustness of the results to measurement noise
(Supplementary Section~\ref{si:noise}). We add independent Gaussian noise to every recorded equilibrium value and repeat the four cases
examined in Supplementary Section~\ref{si:nodewise}. The noise standard
deviation is the same for every node and equals a fraction $\epsilon$ of the
median node-wise dynamic range. Reconstruction
degrades gradually rather than failing abruptly as $\epsilon$ increases. At
$\epsilon=0.1$, the n-RMSE remains below $8\%$ in all four cases, and the
decoder remains more accurate than the mean-prediction baseline in three of the four cases.
These results indicate that the framework is reasonably robust to
measurement noise.

Finally, we examine how reconstruction performance depends on the
number of sentinel nodes, $n$, and the network size, $N$
(Supplementary Section~\ref{sec:c-sweep}). Our default choice is
$n=\lfloor\ln N\rfloor$. First, across the $14$ networks, the effective
rank of the equilibrium data, which indicates how many independent
directions must be captured, increases with $N$ no faster than $\ln N$.
Second, we generate Barab\'asi--Albert
networks~\cite{BarabasiAlbert1999Science} of varying size but approximately
constant mean degree and evaluate reconstruction for $n$ ranging from $2$
to $10$. The number of sentinels required to attain a given n-RMSE 
shows no evidence of increasing faster than $\ln N$ over the tested
range. Although these analyses are
not exhaustive, they support $n=\lfloor\ln N\rfloor$ as a sufficient and
practically reasonable default.

\subsection{Transfer classes of network dynamics}
\label{sec:wccnet}

The results above show that no per-dynamics source decoder transfers
accurately to all target dynamics. We therefore asked whether the
dynamics instead form smaller classes within which reconstruction is
transferable. To this end, for each combination of decoder architecture
and sentinel-selection method, we constructed a directed
cross-transfer network whose nodes are the $17$ dynamics. For each
ordered source--target pair $(\mathrm{S}, \mathrm{T})$ with
$\mathrm{S}\neq \mathrm{T}$, we drew a directed edge from
$\mathrm{S}$ to $\mathrm{T}$ if and only if
$\mathrm{n\text{-}RMSE}(\mathrm{S}\to\mathrm{T}) < 0.15$ and
$r(\mathrm{S}\to\mathrm{T}) < 0.8$.
Thus, an edge represents cross-dynamics reconstruction with both
reasonably small error and at least a $20\%$ improvement over
the mean-prediction baseline.

\begin{figure}
\centering
\includegraphics[width=0.62\textwidth]{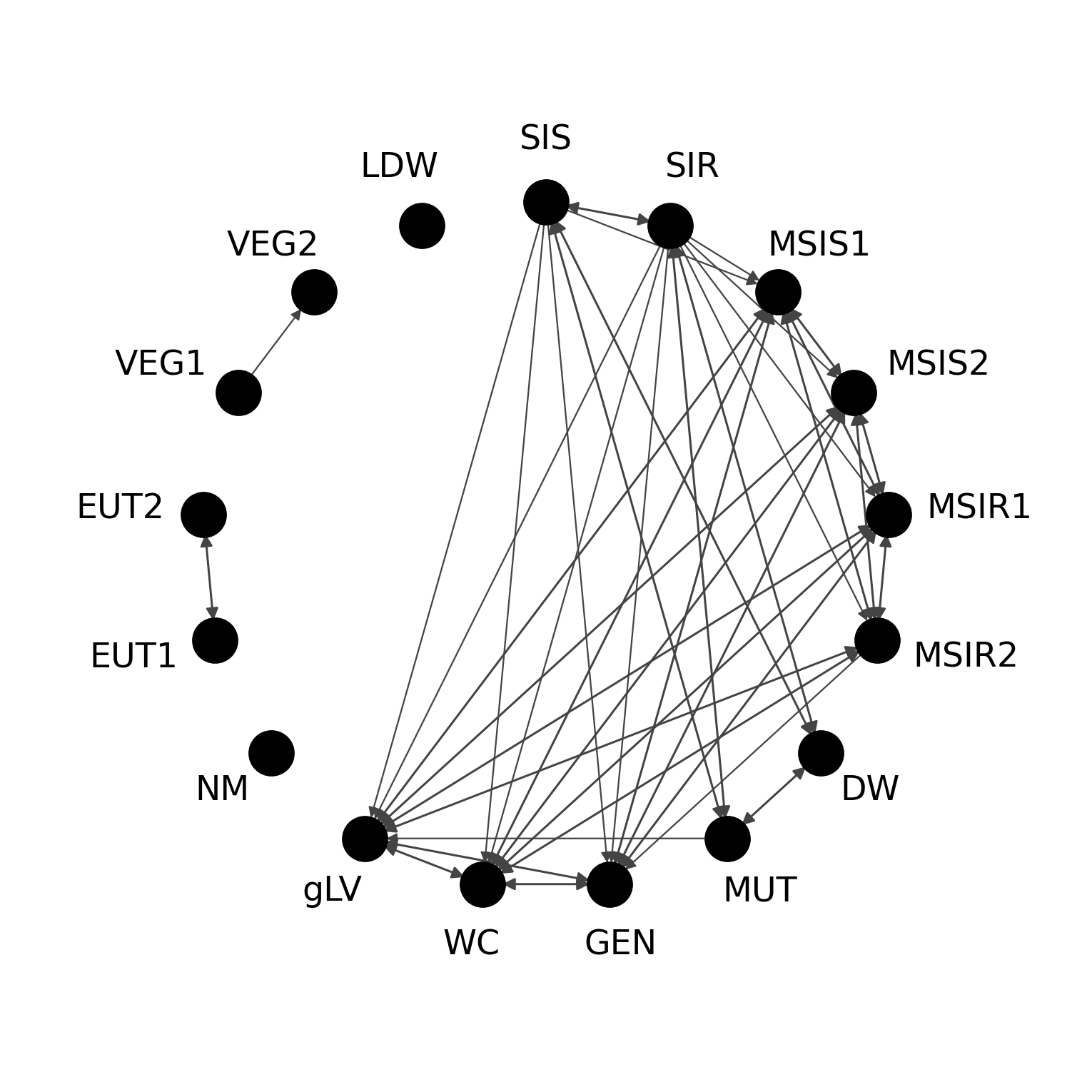}
\caption{Cross-transfer network for the NN decoder with MinCorr
sentinels. Each node is one of the $17$ dynamics. We draw a directed
edge from source dynamics $\mathrm{S}$ to target dynamics
$\mathrm{T}$ if
$\mathrm{n\text{-}RMSE}(\mathrm{S}\to\mathrm{T}) < 0.15$ and
$r(\mathrm{S}\to\mathrm{T}) < 0.8$.
}
\label{fig:wccnet}
\end{figure}

Figure~\ref{fig:wccnet} shows the resulting cross-transfer network for
the NN decoder with MinCorr sentinels. The network separates into five
weakly connected components: one dense component containing $11$
dynamics (SIS, SIR, MSIS1, MSIS2, MSIR1, MSIR2, DW, MUT, GEN, WC, and
gLV), one component containing EUT1 and EUT2, one component containing
VEG1 and VEG2, and the two singletons NM and LDW.
We use weakly connected
components, rather than strongly connected components, to define the
transfer classes, because our
goal here is to identify groups of dynamics linked by reliable
transfer in at least one direction. 

Weak connectivity is nevertheless a permissive criterion because a
single successful transfer in either direction is sufficient to connect two
dynamics. We therefore examine the largest component in more detail; see
Supplementary Section~\ref{app:wcc-robust} for the methods and full results.
Its edge density ranges from $0.46$ to $0.69$ across the six combinations
of decoder architecture and sentinel-selection method. Moreover, when
uniformly random sentinels are excluded, the directed edges are highly
reciprocal. With QR-pivoted sentinels, the largest weakly connected
component is itself strongly connected. With MinCorr sentinels, it consists
of only two strongly connected components. Thus, although the largest
component is not always strongly connected, it is dense and tightly
connected when MinCorr or QR-pivoted sentinels are used. Because these two
methods outperform uniformly random sentinel selection
(Figs.~\ref{fig:heat-combined} and~\ref{fig:ratio}), excluding the
random-sentinel results from this conclusion does not weaken our main
claims.

The same five-component structure, with the same membership of each
component, is obtained for all the other five combinations of the decoder
architecture and sentinel-selection method (see
Supplementary Section~\ref{app:wcc-robust}). The component structure is also robust
to substantial changes in the two edge-definition thresholds (see Supplementary
Section~\ref{app:wcc-robust}).
A standard hierarchical clustering method that avoids thresholding the n-RMSE values
also recovers exactly the same five groups for all six
combinations of the decoder architecture and sentinel-selection method.
Therefore, the partition is not an artifact
of a particular decoder, sentinel-selection method, or threshold
choice.

The component structure is also not an artifact of averaging over
networks. To demonstrate this, we apply the default edge rule, i.e.,
$\mathrm{n\text{-}RMSE}(\mathrm{S}\to\mathrm{T})<0.15$ and
$r(\mathrm{S}\to\mathrm{T})<0.8$, separately to each network using its own
performance values. For each of the four combinations of decoder
architecture and non-random sentinel-selection method (MinCorr or QR), all
$11$ dynamics in the original largest component belong to a single weakly
connected component on every one of the $14$ networks. In $42$ of the $56$
combinations of network, decoder architecture, and non-random
sentinel-selection method ($14\times2\times2$), the largest component
consists of exactly these $11$ dynamics. In the other $14$ combinations,
the largest component is larger, not smaller; in $10$ of these cases, it
contains only one additional dynamics. Thus, the coupling-based
cross-transfer component is not produced by averaging over networks but is
also present in individual networks when a non-random sentinel-selection
method is used. See Supplementary Section~\ref{app:wcc-robust} for the
detailed results.

The robust partition into five weakly connected components suggests
that the six dynamics outside the largest component, namely NM,
EUT1, EUT2, VEG1,  LDW, and VEG2, are difficult to reconstruct using
decoders that transfer well among the other $11$ dynamics. It is also
natural that the two parameter variants of EUT form one component and
that the two parameter variants of VEG form another. A notable common
feature of these six isolated dynamics is that they are the only
scalar node-state models in our collection whose
variable is coupled directly through linear diffusion, i.e., by the unnormalized graph Laplacian.
By contrast, the $11$ dynamics in the largest component use neighbor
states as adjacency-matrix-mediated inputs or interactions, with the
metapopulation models adding degree-normalized compartmental mobility;
none directly smooths the node state variable by the
unnormalized graph Laplacian.

The $17$ dynamical systems considered above differ in both their
single-node dynamics and their coupling type, except for DW and LDW, which
differ only in whether their coupling is adjacency- or Laplacian-mediated.
To reduce this potential confounding, we repeat the analysis using an
extended set of $21$ dynamics containing five matched pairs, including the
DW--LDW pair. Specifically, we add a Laplacian-coupled variant of MUT and
adjacency-coupled variants of NM, EUT1, and VEG1; see Supplementary
Section~\ref{si:fourdyn} for the equations. We introduce these four variants as benchmarks rather than as realistic models. Among the $21$
dynamics, $14$ are adjacency coupled and seven are Laplacian coupled. With
QR-pivoted or uniformly random sentinels, the largest weakly connected
component consists of exactly the $14$ adjacency-coupled dynamics for both
decoder architectures. With MinCorr sentinels, it consists of these $14$
dynamics together with VEG1 and VEG2. In particular, the adjacency-coupled
variants of NM, EUT1, and VEG1 join the largest component. These results
provide further evidence that the largest transferable component is
characterized by adjacency-mediated coupling rather than by particular
single-node dynamics. See Supplementary Section~\ref{si:fourdyn} for details.

The preceding results identify coupling type as an organizing feature
of the transfer classes but do not explain why it matters. We hypothesize
that adjacency-coupled dynamics often transfer successfully because their
equilibrium state vectors $(x_1^*,\ldots,x_N^*)$ share a similar
degree-dependent organization, with $x_i^*$ tending to increase with node
degree. Consistent with this hypothesis, the equilibrium state vectors of
adjacency-coupled dynamics are strongly correlated both with one another and
with the degree and log-degree vectors. The Laplacian-coupled dynamics lack
this organization, and their equilibrium state vectors are not correlated
with those of the adjacency-coupled dynamics. Furthermore, among the
adjacency-coupled dynamics, greater similarity between the source and target
equilibrium state vectors is strongly associated with smaller
$\mathrm{n\text{-}RMSE}$ and $r(\mathrm{S}\to\mathrm{T})$, particularly
with MinCorr and QR-pivoted sentinels. These results suggest that
adjacency-mediated coupling induces a common, approximately
degree-dependent organization of equilibrium states that enables a decoder
trained on one dynamics to transfer to another. See Supplementary
Section~\ref{si:why} for the methods and full results.

These results suggest a transferable class of equilibrium states
among the dynamics in the largest component. Although these
models come from different application domains, their equilibria
appear to share enough structure across networks, driven by adjacency-matrix coupling, and control
parameters that a decoder trained on one member dynamics can reconstruct node states for other members of the class.
By contrast, the six dynamics with direct Laplacian diffusion
form separate transfer components and are not well captured by the
same decoders. In the following sections, we therefore focus on two
settings: reconstruction across the $11$ transferable dynamics and
reconstruction across all $17$ dynamics.

\subsection{Pooled training improves cross-dynamics reconstruction}
\label{sec:winning}

The cross-transfer network suggests that the $11$ dynamics in the
largest component form a transferable class. We therefore asked which
source decoders best reconstruct this class. For each of the six
combinations of decoder architecture and sentinel-selection method, we
ranked the source decoders by the n-RMSE and $r(\mathrm{S}\to\mathrm{T})$, each
averaged over the $11$ transferable target dynamics.

In addition to the $11$ per-dynamics source decoders, we considered
two pooled source decoders. The C17 decoder is trained on the pooled
equilibrium data from all $17$ dynamics. The C11 decoder is trained on
the pooled equilibrium data from only the $11$ transferable dynamics
identified in the last section (i.e., SIS, SIR, MSIS1,
MSIS2, MSIR1, MSIR2, DW, MUT, GEN, WC, and gLV). The motivation for
C11 is that including the six isolated dynamics may degrade
performance if their equilibrium structure is not compatible with that
of the transferable class.

The resulting rankings are shown in
Supplementary Section~\ref{sec:decoder-ranking}. The pooled
decoders consistently perform best. C11 ranks first for both performance
measures, both decoder architectures, and all three sentinel-selection
methods. C17 performs worse than C11 in every ranking, indicating that
including the six dynamics outside the transferable class in the training
pool reduces performance. Nevertheless, C17 outperforms every per-dynamics
source decoder in $10$ of the $12$ rankings. The two exceptions both involve
the linear decoder.
Among the per-dynamics source decoders, SIS and SIR are the most
competitive: they are the two best per-dynamics sources in
eight of the $12$ rankings, with MSIS1
displacing one of them in the remaining four. We conclude that pooling
across multiple dynamics produces a decoder that is substantially more
transferable than any single-dynamics source, particularly when the
training pool is restricted to the $11$ transferable dynamics.

\begin{figure}[H]
\centering
\includegraphics[width=0.85\textwidth]{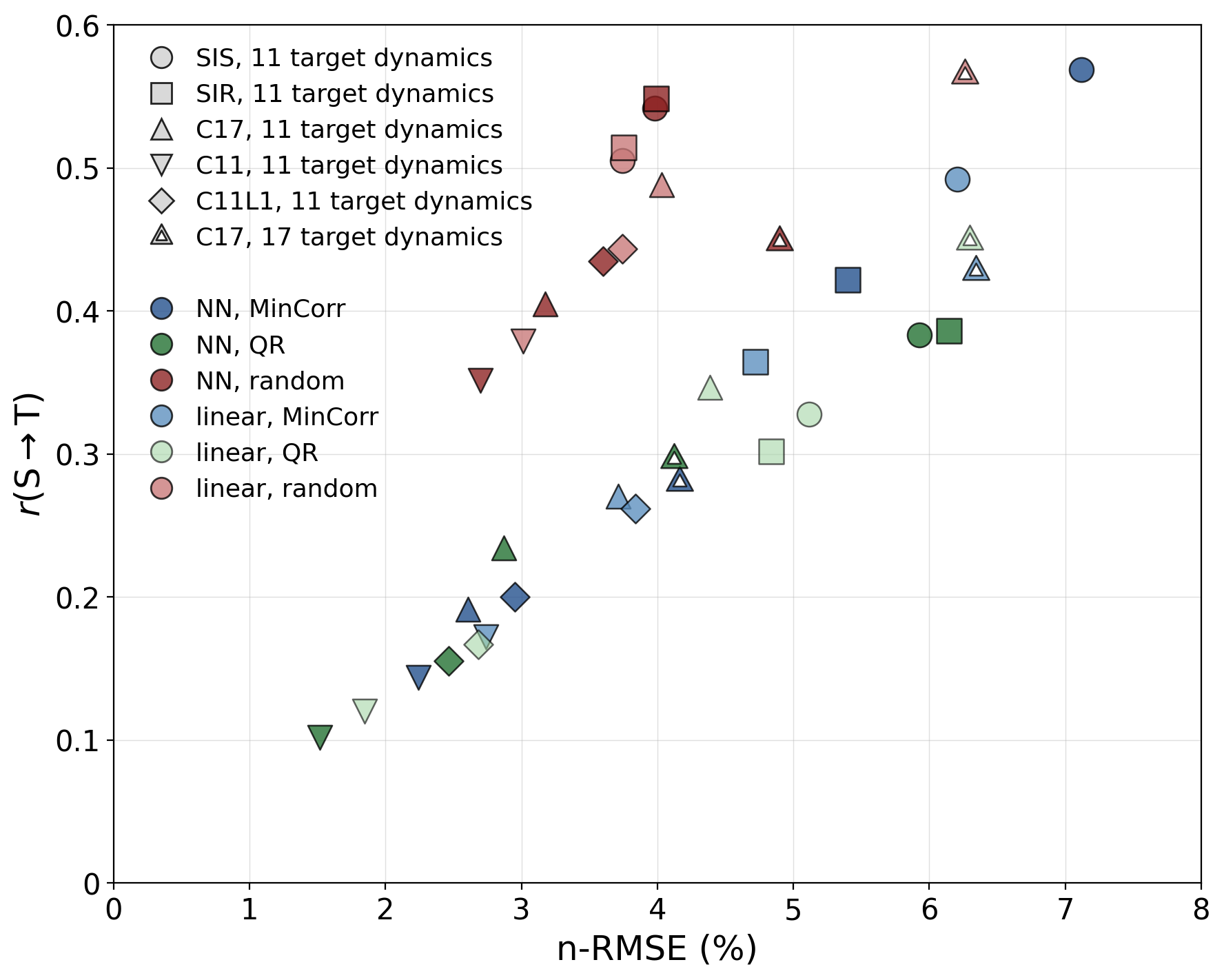}
\caption{Mean n-RMSE (horizontal axis) and mean $r(\mathrm{S}\to\mathrm{T})$ (vertical axis), each averaged over
the $11$ transferable target dynamics. Points show the leading source
decoders SIS, SIR, C17, and C11 across the six combinations of decoder
architecture and sentinel-selection method. The diamonds
show C11L1, the leave-one-dynamics-out pooled decoder trained on the
other $10$ transferable dynamics and evaluated on the held-out
transferable target. The double triangles
%
%
show the C17 source decoder, for which the n-RMSE and $r(\mathrm{S}\to\mathrm{T})$ are averaged over
all $17$ target dynamics rather than only the $11$ transferable
targets. Symbol shape codes the source decoder; fill color codes the
combination of decoder architecture and sentinel-selection method,
with darker hues for the NN decoder and paler hues for the linear
decoder.}
\label{fig:scatter-topfour}
\end{figure}

We therefore focus on C11, C17, SIS, and SIR as representative
high-performing source decoders for the $11$ transferable dynamics.
Figure~\ref{fig:scatter-topfour} shows, for these source decoders, the
n-RMSE and $r(\mathrm{S}\to\mathrm{T})$ averaged over the
$11$ transferable target dynamics. Each point corresponds to one combination
of source decoder, decoder architecture, and sentinel-selection
method. Three patterns are apparent. First, random sentinels perform
substantially worse than the two non-random sentinel methods. This
result supports that sentinel selection is important for cross-dynamics
reconstruction. Second, among the non-random sentinel methods, C11
substantially outperforms the best
per-dynamics sources, SIS and SIR, in terms of both n-RMSE and
$r(\mathrm{S}\to\mathrm{T})$. C17 also substantially outperforms SIS
and SIR with the NN decoder, and to a lesser extent with the linear
decoder and MinCorr sentinels. Third, within the pooled decoders, C11
outperforms C17, NN decoders outperform linear
decoders, and QR sentinels usually outperform MinCorr sentinels.
These differences are modest compared with the performance gap between
pooled and per-dynamics training, and with the degradation caused by
random sentinel selection.

Although C11 is trained and tested using disjoint control-parameter
values for each target dynamics, its strong performance could still
reflect the fact that it has seen training data from every one of the
$11$ dynamics. To test genuine transfer to an unseen dynamics within
the transferable class, we constructed a leave-one-dynamics-out pooled
decoder, which we call C11L1. For each held-out target among the
$11$ transferable dynamics, C11L1 is trained on the pooled equilibrium
data from the other $10$ transferable dynamics and then evaluated on
the held-out dynamics. We then average the n-RMSE and
$r(\mathrm{S}\to\mathrm{T})$ over the $11$ possible held-out target dynamics
and over all networks.

As expected, C11L1 performs worse than C11 because the target
dynamics is excluded from its training pool. C11L1 nevertheless outperforms
C17 in half of the rankings, even though C17 was trained on the target
dynamics. This result suggests shows that a compatible training pool (i.e., only using the adjacency-coupled dynamics) is worth more than the
inclusion of the target dynamics. For
non-random sentinel methods, C11L1 remains substantially better than
the best per-dynamics sources. This result indicates that pooled training across several dynamics
mediated by adjacency-matrix coupling can learn shared node-state
relationships that transfer to a new dynamics of the same broad class.

The pooled decoders use more training data than a per-dynamics decoder:
C11 and C11L1 are trained on $11\times41=451$ and $10\times41=410$
control-parameter values, respectively, whereas a per-dynamics decoder is
trained on $41$ values. To separate the benefit of pooling multiple
dynamics from that of simply using more training data, we repeat the
comparison using budget-matched variants of C11 and C11L1, denoted by C11B
and C11L1B. Each budget-matched decoder is trained on exactly $41$
control-parameter values in total, drawn from its member dynamics, and
therefore uses the same amount of training data as a per-dynamics decoder.
Both C11B and C11L1B continue to outperform every per-dynamics decoder.
Moreover, their n-RMSE and $r(\mathrm{S}\to\mathrm{T})$ remain close to
those of C11 and C11L1, respectively; see Supplementary
Section~\ref{sec:decoder-ranking} for details. Thus, the advantage of pooled
training arises not merely from a larger training set but from exposing the
decoder to several different adjacency-coupled dynamics.

The example in Fig.~\ref{fig:sis-wc-dolphin} illustrates this point
at the node level. The dotted lines show a C11L1 decoder trained
on the $10$ transferable dynamics other than WC and evaluated on WC
equilibria on the dolphin network. This leave-one-dynamics-out pooled
decoder reconstructs the WC equilibrium curves more accurately than
the SIS-trained decoder for nearly every node. In this example,
C11L1 yields $\mathrm{n\text{-}RMSE}=2.02\%$ and
$r(\mathrm{S}\to\mathrm{T})=0.163$, compared with
$\mathrm{n\text{-}RMSE}=5.54\%$ and
$r(\mathrm{S}\to\mathrm{T})=0.431$ for the SIS-trained decoder.
Together, these results show that training on multiple dynamics from
the transferable class enables accurate reconstruction even for an
unseen target dynamics in that class.

Reconstruction performance degrades when we try to learn across all
dynamics, including the six dynamics outside the largest transfer
component. To quantify this effect, we ranked $19$ candidate source
decoders by performance averaged over all $17$ target dynamics: the
$17$ per-dynamics decoders, C17, and C17L1. Here C17L1 is the
leave-one-dynamics-out analogue of C17: for each held-out target
dynamics, it is trained on the pooled equilibrium data from the other
$16$ dynamics and evaluated on the held-out target dynamics.

The rankings, shown in Supplementary Section~\ref{sec:decoder-ranking}, indicate
that C17 is the best source decoder across all $17$ targets, and
C17L1 is second, for every combination of decoder architecture,
sentinel-selection method, and performance measure. However, their
errors are substantially larger than those of C11 and C11L1 when the
evaluation of C11 and C11L1 is restricted to the $11$ transferable dynamics as in Fig.~\ref{fig:scatter-topfour}. Thus,
broadening the training and target set to include the six isolated
dynamics comes at a clear cost. We conclude that accurate
cross-dynamics reconstruction is best achieved by restricting the
training pool to a structurally compatible class of dynamics, rather
than by seeking a single decoder that is universal across all dynamics
considered here.

\subsection{Robustness of the transfer structure to sentinel count and decoder specification}

We performed several robustness checks to assess whether the main
 claims on the transfer structure of dynamics decoding depend on the number of sentinel nodes, the NN decoder
architecture, or the ridge regularization strength. 
Whereas Section~\ref{sec:robust} examined how the number of sentinel
nodes affects reconstruction accuracy, here we examine its effects on the
transfer structure and on the relative ranking of source decoders,
including the pooled decoders.
First, we varied
the number of sentinels around the default choice
$n=\lfloor \ln N \rfloor$ by setting
 $n \in [0.25  \lfloor \ln N \rfloor, 2 \lfloor \ln N \rfloor]$ (see Supplementary Section~\ref{sec:c-sweep}). The main results
are stable over this range. In particular, the relative ranking of the
main source decoders is preserved, pooled decoders remain
substantially better than the best per-dynamics decoders, and the
performance gain from using more than
$n=\lfloor \ln N \rfloor$ sentinels is modest.

Second, we replaced the bias-free NN decoder by a bias-restored NN
decoder, matching the architecture used in previous shallow-decoder
studies for fluid dynamics~\cite{Erichson2020PRSA, wzk2024} (Supplementary Section~\ref{app:baseline}).
Third, for the linear ridge decoder,
we increased the ridge strength from $\alpha=10^{-2}$ to
$\alpha=10^{-1}$. These two types of changes
generally increase reconstruction error, especially for some
lower-ranked source decoders (see Supplementary Section~\ref{app:baseline}). However, they do not alter the qualitative
picture. In particular, the same broad transfer structure remains:
the $11$ dynamics mediated by adjacency-matrix coupling form the dominant transferable
class, the six scalar dynamics with direct Laplacian diffusion remain
comparatively isolated, non-random sentinel selection outperforms
random sentinel selection, and pooled training yields the most
transferable decoders.

\section{Discussion}
\label{sec:discussion}

We have shown that equilibrium states of nonlinear dynamics on
networks can often be reconstructed from observations at only
$n=\lfloor \ln N \rfloor$ sentinel nodes, and that this
reconstructability can transfer across dynamical systems. Across
$14$ networks and $17$ nonlinear dynamics, accurate reconstruction was
not universal, but it was structured. Specifically, $11$ dynamics with
adjacency-mediated coupling out of the $17$ dynamics formed a robust transferable class. Pooled
decoders trained on this class consistently outperformed all
single-dynamics decoders and retained strong performance even when the
target dynamics was held out from training. By contrast, six scalar
dynamics with direct Laplacian diffusion formed isolated transfer
components, revealing a limit to cross-dynamics generalization.
We also found that non-random sentinel selection is consistently
important and that linear ridge decoders often approached the
performance of shallow NN decoders. Together, these
results suggest that sparse node observations can encode enough
information to reconstruct full network equilibria across adjacency-coupled
families of nonlinear dynamics, while also identifying the coupling
architecture as a potential key determinant of when such transfer is possible.
Because the present study considers only $17$ dynamical systems
(together with four additional benchmark variants in Supplementary
Section~\ref{si:fourdyn}), extending the analysis to a broader range of
dynamics and networks is an important direction for future work.

We used a shallow NN decoder inspired by previous work on
reconstruction of fluid dynamics in spatial domains
~\cite{Erichson2020PRSA, wzk2024}. For nonlinear dynamics on
networks, the NN decoder slightly but consistently outperformed the
linear decoder across comparisons. Because the NN decoder has many
trainable parameters, overfitting is a concern. We mitigated
this concern in three ways. First, all within-dynamics evaluations use
held-out control-parameter values that were not used for training.
Second, our stronger transfer test, C11L1, holds out an entire target
dynamics from the training pool and evaluates the decoder on that
unseen dynamics. Third, the main qualitative conclusions are reproduced
by the linear ridge decoder, which has a much smaller
capacity. Specifically, our NN decoder, with a hidden layer of
$H=64$ units, has $H(n+N)=64(n+N)$ trainable parameters. With
$n=\lfloor \ln N\rfloor$, this number ranges from
$64\times(4+62)=4{,}224$ on the smallest network (dolphin,
$N=62$) to $64\times(9+11{,}558)=740{,}288$ on the largest network
(Spanish, $N=11{,}558$). In contrast, the linear decoder fits each
node's $x_{i, k}$ using only $n$ coefficients, namely, by a linear combination of the
sentinel readings $x_{j,k}$ with $j\in\mathcal{S}$, from $M=51$
control-parameter samples, with regularization. Therefore, the
central conclusion is not that a high-capacity NN is required, but
rather that the equilibrium states contain low-dimensional information
accessible from a small number of sentinels. In particular, accurate
cross-dynamics reconstruction remains possible even with a linear
decoder, provided that one uses pooled training, such as C11 or C11L1,
together with a non-random sentinel-selection method.

Although the pooled decoders are the strongest overall, the
per-dynamics decoders suggest what types of
single dynamics can serve as transferable sources. A 
pooled decoder is attractive when equilibrium data are
available for only one dynamical process. Among the $11$ transferable
dynamics, the SIS and SIR decoders were consistently the strongest
single-dynamics sources. This is notable because SIS and SIR are among
the simplest dynamical systems on networks in our collection. In particular, they have no free
parameters beyond the control parameter swept in our experiments.
Thus, reasonably transferable reconstruction does not require a highly
specialized or highly parameterized source dynamics. Rather, even a
simple adjacency-matrix-mediated epidemic process can expose node-to-node
variation that is useful for reconstructing equilibria of many other
dynamics in the same transfer class. At the same time, the
per-dynamics decoders remain clearly inferior to C11 and C11L1, even if we give C11 and C11L1 only the same amount of training data as that used by a per-dynamics decoder (see Sections~\ref{sec:decoder-ranking} and~\ref{app:baseline}).
We therefore view the strong performance of SIS and SIR as evidence
for a useful low-data baseline, rather than as a substitute for pooled
training when multi-dynamics data or simulations are available.

The present study focuses on reconstructing equilibria, or final-size
observables, rather than dynamic trajectories. This choice distinguishes
our work from a large body of machine-learning approaches for
forecasting trajectories of nonlinear dynamical systems
~\cite{Kong2021PRR, Patel2023Chaos, Pathak2018PRL}, including shallow-decoder approaches for
spatiotemporal fluid dynamics from sparse sensors
~\cite{Erichson2020PRSA, wzk2024}. We have focused on equilibria for two
reasons. First, equilibria (and final sizes) are often the main objects of interest in
network dynamics, including endemic infection levels, species
abundances, gene-expression levels, and final epidemic sizes. They
are also often analyzed in dimension-reduction and resilience theory for
network dynamics
~\cite{GaoBarzelBarabasi2016Nature, Tu2017PhysRevE, laurence2019,
Thibeault2020PhysRevResearch, Tu2021Iscience, Zhang2022NatEcolEvol,
KunduKoriMasuda2022PhysRevE, Liu2022PhysRep,
Masuda2022PhysRevResearch, Vegue2023PnasNexus}, as well as to early
warning signals for tipping transitions in networks
~\cite{dakos2010, Dakos2018EcolIndic, Aparicio2021PNAS, MacLaren2023JRSI, masuda2024natcommun, Maclaren2025JRSocInterface}. Second, in many applications, it may be
more realistic to observe equilibrium or quasi-equilibrium states
across environmental or control-parameter conditions than to observe
densely sampled node-level time series. Learning transient trajectories
for the dynamics considered here would require substantially richer
data, because many of these systems relax quickly to equilibrium and
short transients provide limited coverage of the high-dimensional
state space. In contrast, oscillatory, synchronous, or chaotic network
dynamics can generate sustained trajectories from a small number of
initial conditions~\cite{Arenas2008PhysRev, Rodrigues2016PhysRep}.
%
%
Extending sentinel-based reconstructability from
equilibria to such time-dependent dynamics is an important
direction for future work.

Sentinel selection affected reconstruction
performance. The two data-driven methods, MinCorr and QR, consistently
outperformed uniformly random sentinel selection, showing that the
identity of the observed nodes matters. This finding is consistent with related
work on selecting sensor nodes for early warning signals in network
dynamics~\cite{Aparicio2021PNAS, MacLaren2023JRSI,
masuda2024natcommun} and for estimating the population mean of
equilibrium states from sparse sentinel observations~\cite{maclaren2025}.
However, both MinCorr and QR are heuristic choices. A natural next
step is to optimize sentinel placement directly for reconstruction
accuracy, transferability across dynamics, or robustness to noise and
missing observations. Jointly optimizing the sentinel set and decoder
architecture may further improve performance.
%
%
A limitation of the present framework is that it requires
network-specific equilibrium data for decoder training and sentinel
selection. Whether decoders or sentinel sets can transfer across networks is
an important question for future work.

We also ask whether network structure alone is sufficient to select
sentinel nodes, which would eliminate the need for equilibrium data during
sentinel selection; see Supplementary Section~\ref{si:sentsel} for the
methods and full results. The sentinel sets selected by MinCorr and QR are
far from random. Compared with uniformly random node sets of the same size,
they have larger mean degree and betweenness centrality, greater dispersion
in degree, and substantially greater overlap with each other. However, this
structural signature does not yield an effective selection rule. Selecting
the $n$ nodes with the largest degree or betweenness centrality performs
comparably to MinCorr and QR for within-dynamics reconstruction but
substantially worse for cross-dynamics transfer. A likely explanation is redundancy
among the selected nodes: structurally similar hubs provide overlapping
information, whereas MinCorr penalizes redundancy by construction and QR
reduces it through pivoting. Thus, network structure helps characterize the nodes selected by
MinCorr and QR but is insufficient for selecting effective sentinels
without dynamical data. This finding supports data-driven sentinel
selection.

Finally, we restricted our analysis to dynamics on undirected and
unweighted networks. This choice provides a controlled setting in
which to compare many dynamics and networks. However, the decoder framework
itself does not require knowledge of the network structure or of the
equations governing the dynamics. It only requires paired observations
of sentinel-node states and full network states for training. Therefore, our framework
is directly testable on directed or weighted
networks, signed networks, hypergraphs, multilayer networks, and other
structured populations, provided that suitable training data are
available. Together, these results suggest that sentinel-based reconstruction can
serve as a practical route to monitoring networked systems when full
observation is unavailable.
%
%
We expect this perspective to motivate further
work on transferable decoders, principled sentinel placement, and
reconstruction beyond equilibria.


\section{Methods}

\subsection{Decoder architectures, training, and sentinel selection}
\label{sec:decoder-methods}

We denote by $x_i$ the equilibrium or final-size observable at the
$i$th node, where $i\in\{1,\ldots,N\}$. For a fixed network and
dynamics, we assemble the recorded states over the swept
control-parameter values into a data matrix
$\bX\in\R^{N\times M}$, whose $(i,k)$ entry is $x_{i,k}$, i.e., the
state of the $i$th node at the $k$th control-parameter value. We use
$M=51$ control-parameter values.

We split the $M$ columns of $\bX$ into a training set and a held-out
test set. Specifically, we select
$\lfloor 0.2M\rfloor=10$ columns uniformly at random and hold them out
for testing. The remaining
$M_{\rm train}=M-\lfloor 0.2M\rfloor=41$ columns are used for
training. We denote by
$\mathcal{K}_{\rm train}$ the set of training column indices. We also denote by
$\bX_{\rm train}\in\R^{N\times M_{\rm train}}$ the submatrix of
$\bX$ consisting of the training columns. The same train--test split
is used across decoder architectures and sentinel-selection methods
for a given network and dynamics.

All decoders are trained in a normalized state space. For each
per-dynamics decoder, we min--max scale the data by
$\tilde{\bX}
=
(\bX-x_{\min})/(x_{\max}-x_{\min})$,
where
$x_{\min}
=
\min_{i\in\{1,\ldots,N\},\, k\in \mathcal{K}_{\rm train}} x_{i,k}$,
and $x_{\max}
=
\max_{i\in\{1,\ldots,N\},\, k\in \mathcal{K}_{\rm train}} x_{i,k}$.
The corresponding normalized training matrix is
$\tilde{\bX}_{\rm train}$, the submatrix of $\tilde{\bX}$ restricted
to the columns in $\mathcal{K}_{\rm train}$. For dynamics whose physical minimum is
$x_i=0$, this normalization maps the physical zero state to zero.

Let $\mathcal{S}\subseteq\{1,\ldots,N\}$ denote the set of sentinel
nodes, with $|\mathcal{S}|=n$. For a single control-parameter value,
we write $\bs\in\R^n$ for the vector of sentinel readings
$\{x_i:i\in\mathcal{S}\}$, and $\tilde{\bs}$ for the corresponding
normalized vector. The decoder is trained to map $\tilde{\bs}$ to the
full normalized state vector $\tilde{\bx}\in\R^N$. For the pooled
decoders described below, min--max scaling is applied separately to
each dynamics-specific training block before pooling.

\subsubsection{Shallow NN decoder}
\label{sec:nn}

The NN decoder is a single-hidden-layer feedforward network. Let
$\bs\in\R^n$ denote the vector of sentinel readings at one
control-parameter value, where $n$ is the number of sentinels. We use
the bias-free decoder
\begin{equation}
\hat{\tilde{\bx}}(\tilde{\bs})
=
\mathbf{W}_2
\mathrm{ReLU}\!\left(\mathbf{W}_1\tilde{\bs}\right),
\label{eq:nn-decoder-nobias}
\end{equation}
where $\hat{\tilde{\bx}}\in\R^N$ is the reconstructed normalized
state, $\mathbf{W}_1\in\R^{H\times n}$,
$\mathbf{W}_2\in\R^{N\times H}$, and $H=64$ is the hidden-layer
width. The ReLU function acts elementwise. The final layer is linear,
following the shallow decoder architecture used in previous work on
fluid-flow reconstruction~\cite{Erichson2020PRSA, wzk2024}.

We use the bias-free architecture because many of the dynamics
considered here admit a physical disease-free, extinct, or quiescent
state with $x_i=0$ for all nodes. For such dynamics, zero sentinel
input should ideally map to zero output, especially when transferring
a decoder across dynamics. The bias-free decoder satisfies this
constraint by construction. We assess the corresponding bias-restored
NN decoder as a robustness check in Supplementary Section~\ref{app:baseline}.

We train the NN decoder with the Adam optimizer using learning rate
$10^{-2}$ and the mean-squared-error loss between the predicted
normalized state and the corresponding normalized training column.
Training proceeds for $200$ epochs for per-dynamics NN decoders.

\subsubsection{Linear ridge decoder}
\label{sec:linear}

For comparison, we also fit a linear decoder in the normalized space.
To remain consistent with the bias-free NN decoder, we omit the
intercept and fit
\begin{equation}
\hat{\tilde{\bx}}_{\rm lin}(\tilde{\bs})
=
\mathbf{W}_{\rm lin}\tilde{\bs}.
\end{equation}
Let $\bZ\in\R^{n\times M_{\rm train}}$ denote the matrix of
normalized sentinel readings in the training set, i.e., the
sentinel-row submatrix of $\tilde{\bX}_{\rm train}$. We solve
$(\bZ\bZ^\top+\alpha\bI)\mathbf{W}_{\rm lin}^\top
=
\bZ\tilde{\bX}_{\rm train}^\top$
with $\alpha=10^{-2}$. The ridge penalty controls the Frobenius norm
of $\mathbf{W}_{\rm lin}$ and prevents unstable weights when sentinel
rows are nearly collinear. We assess the effect of increasing the
ridge strength to $\alpha=10^{-1}$ in Supplementary Section~\ref{app:baseline}.

\subsubsection{Sentinel selection}
\label{sec:sentinels}

We observe only
$n=\lfloor \ln N \rfloor$ sentinel nodes and denote their index set by
$\mathcal{S}\subseteq\{1,\ldots,N\}$, with $|\mathcal{S}|=n$. We
verified that the main results are robust to varying $n$ around this
default value (Supplementary Section~\ref{sec:c-sweep}). We compare
three sentinel-selection methods.

In the greedy minimum-correlation method, which we call MinCorr, we
initialize $\mathcal{S}$ with the row of the training matrix that has
the largest variance across training columns. We then iteratively add
the row whose largest absolute Pearson correlation with the
already-selected rows is smallest.

In the QR-pivoted sensor method \cite{manohar2018, wzk2024}, which we call QR, we compute the
rank-$n$ truncated SVD of the training matrix,
\[
\bX_{\rm train}\approx \bU_n\boldsymbol{\Sigma}_n\bV_n^\top,
\]
apply row-pivoted QR factorization to $\bU_n$, and take the first
$n$ pivoted row indices as the sentinel set
$\mathcal{S}$~\cite{manohar2018,wzk2024}. This method selects nodes
whose rows best span the leading left singular subspace of the
training data.

In the random method, we draw $\mathcal{S}$ uniformly at random from
$\{1,\ldots,N\}$ without replacement.

\subsection{Pooled C17 and C11 decoders}
\label{sec:combined}

In addition to the $17$ per-dynamics decoders, we train a pooled
``C17'' decoder for each combination of decoder architecture (NN or
linear) and sentinel-selection method. For a fixed network, the C17
decoder is trained on equilibrium data pooled across all $17$
dynamics and across the swept control-parameter values. Within each
dynamics, $20\%$ of the columns are held out for testing, leaving
$M_{\rm train}=M-\lfloor 0.2M\rfloor=41$ training columns. Thus, the
pooled C17 training matrix has
$17M_{\rm train}=17\times 41=697$ columns and $N$ rows.

Because the range of $x_i$ can differ substantially across dynamics,
even on the same network, we apply min--max scaling separately to the
training block of each dynamics before pooling. In other words, each
dynamics-specific block is rescaled to $[0,1]$ using its own
$x_{\min}$ and $x_{\max}$. For each pooled decoder, the sentinel set
is selected from the corresponding pooled training matrix using the
specified sentinel-selection method, and the same set of $n$ sentinel
nodes is then used for all dynamics contributing to that pooled
training set. We train
the pooled NN decoder with the same learning rate and number of epochs as
the per-dynamics NN decoder.

We also train a C11 decoder for each combination of decoder
architecture and sentinel-selection method. The C11 decoder is defined
in the same way as C17, except that its training pool is restricted to
the $11$ transferable dynamics, i.e., SIS, SIR, MSIS1, MSIS2, MSIR1, MSIR2, DW,
MUT, GEN, WC, and gLV. Thus, for each network, the pooled C11
training matrix has $11M_{\rm train}=11\times 41=451$ columns and
$N$ rows.

To test whether pooled decoders generalize to a dynamics that is
absent from the training pool, we evaluate leave-one-dynamics-out
variants, denoted by C11L1 and C17L1. For C11L1, we hold out one of
the $11$ transferable dynamics, train a pooled decoder on the
remaining $10$ transferable dynamics, and evaluate the trained decoder
on the held-out target dynamics. Repeating this procedure for each of
the $11$ possible held-out targets yields $11$ leave-one-out decoders.
We report the n-RMSE and $r(\mathrm{S}\to\mathrm{T})$ averaged over
these $11$ held-out target dynamics and over networks.

C17L1 is the analogous leave-one-dynamics-out experiment over all
$17$ dynamics. For each held-out target dynamics, we train a pooled
decoder on the remaining $16$ dynamics and evaluate it on the held-out
target dynamics. Repeating this procedure for all $17$ choices of the held-out
dynamics yields $17$ leave-one-out decoders. We report the n-RMSE and
$r(\mathrm{S}\to\mathrm{T})$ averaged over the $17$ held-out target dynamics
and over all the networks. Unlike C11L1, C17L1 includes the six held-out
cases in which the target is one of the isolated dynamics, NM, EUT1,
EUT2, VEG1, VEG2, or LDW.

\subsection{Dynamical systems on networks}
\label{sec:dynamics}

We use $12$ model families of continuous-time dynamics on networks,
yielding $17$ one-parameter dynamics in total after including the
parameter variants described below. Four of these dynamics (SIS, DW,
MUT, and GEN) were also used in our previous study~\cite{maclaren2025}.
Throughout, $\bA \in \{0,1\}^{N \times N}$ denotes the adjacency
matrix, which is symmetric and has zero diagonal, and
$k_i = \sum_{j=1}^{N} A_{ij}$ denotes the degree of the $i$th node.
Each dynamics has a single scalar control parameter $p$, which we
sweep over $M=51$ equally spaced values. At each value of $p$, we
integrate the ODE from a specified initial condition to time
$t_{\max}$ and record the final state $\bx(p) \in \R^N$, which we use
as the equilibrium or final-size observable. The only exception is
gLV, for which we use the analytical coexistence equilibrium.

\textbf{SIS:} The deterministic approximation to the
susceptible--infectious--susceptible (SIS) dynamics on the contact
network, with infection rate $\beta$ and recovery rate $\mu$, is
given by
\begin{equation}
\dot{x}_i = -\mu x_i + \beta (1 - x_i) \sum_{j=1}^{N} A_{ij} x_j,
\qquad x_i \in [0, 1],
\label{eq:SIS}
\end{equation}
where $x_i$ represents the probability that the $i$th node is infectious.
We fix $\mu=1$, so that $\beta$ is the control parameter. We sweep
$\beta$ over $[0,\beta_{\max}]$. The epidemic threshold of
Eq.~\eqref{eq:SIS} is
$\beta_{\mathrm{c}} = 1/\alpha_{\max}$, where $\alpha_{\max}$ is the
largest eigenvalue of $\bA$. Because $\alpha_{\max}$ varies
substantially across networks, using the same fixed value of
$\beta_{\max}$ would place the threshold at different relative
positions in the sweep for different networks. To standardize the
threshold location, we set
\[
\beta_{\max} = \frac{\beta_{\mathrm{c}}}{0.15}
             = \frac{1}{0.15\alpha_{\max}}
\]
for each network, so that $\beta_{\mathrm{c}}$ lies at the $15\%$
position of the interval $[0,\beta_{\max}]$. Equivalently, the upper
$85\%$ of the swept $\beta$ values are above the epidemic threshold.

\textbf{SIR:} The deterministic approximation to the
susceptible--infectious--recovered (SIR) dynamics is given by
\begin{align}
\dot{S}_i &= -\beta S_i \sum_{j=1}^{N} A_{ij} I_j, \\
\dot{I}_i &= \beta S_i \sum_{j=1}^{N} A_{ij} I_j - \mu I_i,
\end{align}
where $S_i$ and $I_i$ are the probabilities that the $i$th node is
susceptible and infectious, respectively. The probability of the recovered state is
$R_i = 1 - S_i - I_i$. We set $\mu=1$ and use $\beta$ as the control
parameter. The observable is the final expected fraction of recovered nodes given by
$x_i \equiv R_i(t_{\max}) = 1 - S_i(t_{\max}) - I_i(t_{\max})$.
We sweep $\beta$ over the same network-dependent range as for the SIS
model.

\textbf{MSIS1/2:} We consider a metapopulation SIS model on a network
of $N$ patches~\cite{Colizza2007NatPhys}. The state of the $i$th node
is given by the densities of susceptible and infectious individuals in
the corresponding patch, denoted by $\rho_{S,i}$ and $\rho_{I,i}$,
respectively. Infection and recovery occur within each patch, whereas
individuals move between neighboring patches. The dynamics are
\begin{align}
\dot{\rho}_{S,i} &= -\beta \rho_{S,i} \rho_{I,i} + \mu \rho_{I,i}
        + D_S\!\left(\sum_{j=1}^{N} \frac{A_{ij}}{k_j} \rho_{S,j}
                     - \rho_{S,i}\right), \\
\dot{\rho}_{I,i} &= \beta \rho_{S,i} \rho_{I,i} - \mu \rho_{I,i}
        + D_I\!\left(\sum_{j=1}^{N} \frac{A_{ij}}{k_j} \rho_{I,j}
                     - \rho_{I,i}\right),
\end{align}
where $D_S$ and $D_I$ are the diffusion rates of susceptible and
infectious individuals, respectively. We set $\mu=1$ and $D_S=1$.
We use two values of $D_I$, i.e., $D_I=1$ and
$D_I=0.2$, which define MSIS1 and MSIS2, respectively. A small
$D_I$ tends to concentrate infectious individuals at equilibrium in
high-degree patches. The observable is the density of infectious individuals,
$x_i \equiv \rho_{I,i}(t_{\max})$.

The epidemic threshold of the MSIS model is generally larger than the
SIS threshold $1/\alpha_{\max}$~\cite{Masuda2010NJP, GomezGardenes2018NatPhys, SorianoPanos2020JSTAT}. Therefore, the
SIS-matched range
$\beta \in [0,1/(0.15\alpha_{\max})]$ can lie entirely below the
metapopulation epidemic threshold. We therefore set $\beta_{\max}$
adaptively for MSIS and MSIR, as we describe in
section~\ref{sec:autotrim}.

\textbf{MSIR1/2:} We use the metapopulation SIR analogue of the MSIS
model, in which susceptible and infectious individuals diffuse between
patches~\cite{Colizza2007PRL, Colizza2008JTB}. Recovered individuals
do not diffuse because their mobility does not affect epidemic
spreading in this model. The dynamics are given by
\begin{align}
\dot{\rho}_{S,i} &= -\beta \rho_{S,i} \rho_{I,i}
        + D_S\!\left(\sum_{j=1}^{N} \frac{A_{ij}}{k_j} \rho_{S,j}
                     - \rho_{S,i}\right), \\
\dot{\rho}_{I,i} &= \beta \rho_{S,i} \rho_{I,i} - \mu \rho_{I,i}
        + D_I\!\left(\sum_{j=1}^{N} \frac{A_{ij}}{k_j} \rho_{I,j}
                     - \rho_{I,i}\right), \\
\dot{\rho}_{R,i} &= \mu \rho_{I,i},
\end{align}
where $\rho_{S,i}$, $\rho_{I,i}$, and $\rho_{R,i}$ are the densities
of susceptible, infectious, and recovered individuals, respectively,
in the $i$th patch. We set $\mu=1$ and
$D_S=1$. As for MSIS, we use two values of the diffusion
rate for infectious individuals, $D_I=1$ and $D_I=0.2$, which define MSIR1 and MSIR2,
respectively. The observable is $x_i \equiv \rho_{R,i}(t_{\max})$. We use the same sweep range and
adaptive widening rule as for the MSIS model.

\textbf{DW:} A coupled double-well (DW) dynamics on networks is given by~\cite{Wunderling2020NJP, MacLaren2023JRSI, masuda2024natcommun, maclaren2025}
\begin{equation}
\dot{x}_i = -(x_i - r_1) (x_i - r_2) (x_i - r_3)
            + D \sum_{j=1}^{N} A_{ij} x_j,
\label{eq:dw}
\end{equation}
with $r_1=1$, $r_2=3$, and $r_3=5$~\cite{masuda2024natcommun, maclaren2025}. In the absence of network coupling, i.e., when $D=0$, each node is bistable,
with stable equilibria at $x_i=r_1$ and $x_i=r_3$ and an unstable
equilibrium at $x_i=r_2$ separating their basins of attraction.
We sweep $D$ over $[0,D_{\max}]$. Using the same $D_{\max}$ for all
networks would produce different effective coupling strengths because
the average total input to a node is approximately proportional to
$D\langle k\rangle$, where $\langle k\rangle$ is the mean degree. To
standardize this input scale across networks, we keep
$D_{\max}\langle k\rangle$ constant. We chose this constant using the
dolphin network, for which $\langle k\rangle=5.13$ and the first
saddle-node bifurcation encountered as $D$ increases from zero occurs
at approximately $D=0.19$. We set $D_{\max}=0.5$ for the dolphin
network, placing this saddle node at roughly the $40\%$ position of
the sweep. For every other network, we set
$D_{\max}$ by
$D_{\max}\langle k\rangle = 0.5 \times 5.13 = 2.56$.
%
%

\textbf{MUT:} The mutualistic interaction (MUT) dynamics on networks,
with logistic growth, an Allee effect, and saturating mutualistic
coupling, is given by
\begin{equation}
\dot{x}_i = B + x_i \left(1 - \frac{x_i}{K}\right)
                 \left(\frac{x_i}{C} - 1\right)
        + D \sum_{j=1}^{N} A_{ij}
            \frac{x_i  x_j}{\tilde D + E  x_i + H  x_j},
\end{equation}
where $x_i$ represents the abundance of the $i$th species. We set $B=0.1$ (immigration rate), $K=5$ (carrying
capacity), $C=1$ (Allee threshold), and
$\tilde D=5$, $E=0.9$, and $H=0.1$ (saturation constants)
~\cite{GaoBarzelBarabasi2016Nature}.
We sweep $D$ over $[0,D_{\max}]$. As for DW, the saddle-node
bifurcation in MUT is controlled by the average total input from
neighboring nodes, which is approximately proportional to
$D\langle k\rangle$. We therefore use the same mean-degree-based
rescaling as for DW, keeping $D_{\max}\langle k\rangle$ constant
across networks. On the dolphin network, the first saddle node
encountered as $D$ increases occurs at approximately $D=0.78$. We
therefore set $D_{\max}=2$ for the dolphin network and set
$D_{\max}$ for every other network by
$D_{\max}\langle k\rangle = 2 \times 5.13 = 10.26$.

\textbf{GEN:} The gene-regulatory (GEN) dynamics on networks, with
Hill-saturating production and linear degradation, is given by
\begin{equation}
\dot{x}_i = -B x_i^F + D \sum_{j=1}^{N} A_{ij} \frac{x_j^H}{1 + x_j^H},
\end{equation}
where $x_i$ represents the expression level of the $i$th gene. We set $B=1$, $F=1$, and $H=2$
~\cite{GaoBarzelBarabasi2016Nature}.
We sweep $D$ over $[0,D_{\max}]$. As in the SIS model, the all-zero
state is always an equilibrium, and a nonzero branch appears as $D$
increases past a threshold. On the dolphin network, this threshold is
located at approximately $D_{\mathrm{c}}=0.28$. We set
$D_{\max}=1.6$ for the dolphin network, placing the threshold at
approximately the $17.5\%$ position of the sweep. Because the
threshold coupling strength scales approximately as
$1/\alpha_{\max}$, we keep $D_{\max}\alpha_{\max}$ constant across
networks by imposing
$D_{\max}\alpha_{\max} = 1.6 \times 7.19 = 11.51$; the dolphin network yields
$\alpha_{\max}\approx 7.19$.

\textbf{WC:} We use a modified Wilson--Cowan (WC) model for firing
rates of neuronal populations on networks given by
\begin{equation}
\dot{x}_i = -x_i + D \sum_{j=1}^{N}
\frac{A_{ij}}{1 + e^{-\tau(x_j - \overline{\mu})}},
\end{equation}
with $\tau=1$ and $\overline{\mu}=3$~\cite{laurence2019}.
We sweep $D$ over $[0,D_{\max}]$. At $\bx=0$, the network input to the $i$th node is
$D\sum_{j=1}^{N} A_{ij}/(1+e^{\tau\overline{\mu}})
= D k_i/(1+e^{\tau\overline{\mu}})$, so the average input scale is
proportional to $D\langle k\rangle$. We therefore choose
$D_{\max}$ so that $D_{\max}\langle k\rangle$ is constant across
networks. We calibrated this constant using the dolphin network
(having $\langle k\rangle=5.13$), for which the equilibrium response shows a
marked increase around $D\approx 1$. Setting $D_{\max}=6$ for this
network places this rapid-response region at approximately the
$17\%$ position of the sweep. Therefore, we impose
$D_{\max}\langle k\rangle = 6 \times 5.13 = 30.77$ for all networks.

\textbf{gLV:} The generalized Lotka--Volterra (gLV) model on networks
is given by
\begin{equation}
\dot{x}_i = x_i\left(\overline{\lambda} + \sum_{j=1}^{N} W_{ij} x_j\right),
\label{eq:glv-componentwise}
\end{equation}
where $x_i$ represents the abundance of the $i$th species, and $\overline{\lambda}>0$ is the intrinsic growth
rate~\cite{Tu2017PhysRevE, laurence2019}. We set
$\overline{\lambda}=1$. The interaction matrix has self-regulation
$W_{ii}=-(\alpha_{\max}+1)$~\cite{KunduKoriMasuda2022PhysRevE,
Masuda2022PhysRevResearch} and off-diagonal interactions
$W_{ij}=D A_{ij}$ for $i\neq j$. With
$c=\alpha_{\max}+1$, the coexistence equilibrium is
\begin{equation}
\bx = -\overline{\lambda} \left( D \bA - c \bI \right)^{-1}\bone,
\label{eq:glv-equilibrium}
\end{equation}
where $\bone \in \R^N$ is the column vector of all ones.
We sweep $D$ over
$D \in [0.05D_{\rm c},\; D_{\rm c} - 0.5/\alpha_{\max}]$, where
$D_{\rm c} = (\alpha_{\max}+1)/\alpha_{\max}$.
In this range of $D$, this coexistence
equilibrium is positive at each node, and the equilibrium is locally
asymptotically stable; see
Section~\ref{si:glv} for proofs.
At $D=D_{\rm c}$, the matrix $D\bA-c\bI$ becomes singular. At the upper
endpoint of the sweep, the leading eigenvalue of
$D\bA-c\bI$ equals $-0.5$ for every network.

\textbf{NM:} The Noy--Meir (NM) grazing model~\cite{Noymeir1975JEcol,
May1977Nature} with diffusive coupling~\cite{dakos2010} is given by
\begin{equation}
\dot{x}_i = r_i x_i \left(1 - \frac{x_i}{K}\right)
            - c \frac{x_i^2}{x_i^2 + 1}
            + D \sum_{j=1}^{N} A_{ij} (x_j - x_i),
\end{equation}
where $x_i$ represents the vegetation biomass at the $i$th node. We
set $K=10$ and $D=0.05$. The per-node growth rates $r_i$ are drawn
independently from the uniform distribution on $[0.6,1]$, once for
each network, and then held fixed across all values of the grazing
pressure $c$ in the sweep, following~\cite{dakos2010}. We use $c$ as
the control parameter and sweep $c \in [1,3]$.

\textbf{EUT1/2:} The eutrophication (EUT) model~\cite{Carpenter1999EcolAppl}
with diffusive coupling~\cite{dakos2010} is given by
\begin{equation}
\dot{x}_i = \alpha - b_i x_i + r \frac{x_i^p}{x_i^p + 1}
            + D \sum_{j=1}^{N} A_{ij} (x_j - x_i),
\label{eq:EUT}
\end{equation}
where $x_i$ represents the nutrient concentration, or more specifically
phosphorus concentration, at the $i$th node. The first, second, third, and fourth terms on the right-hand side of Eq.~\eqref{eq:EUT} represent
external nutrient loading, linear loss, nonlinear internal recycling,
and diffusive exchange between neighboring nodes, respectively. We set
$r=1$, $D=0.05$, and use two Hill exponents. The variant with $p=8$~\cite{dakos2010}
is denoted by EUT1, and the variant with $p=2$ is denoted by EUT2.
For each network, the per-node loss rates $b_i$ are drawn
independently from the uniform distribution on $[0.8,1.2]$ and then
held fixed across all values of $\alpha$ and across both EUT1 and
EUT2. We use $\alpha$ as the control parameter and sweep
$\alpha \in [0.1,0.9]$.

\textbf{VEG1/2:} The vegetation--turbidity feedback (VEG)
model~\cite{Scheffer1998book} with diffusive coupling~\cite{dakos2010}
is given by
\begin{equation}
\dot{x}_i = r_V x_i
            \left[1 - \frac{x_i (h_{E,i}^p + E_i^p)}{h_{E,i}^p}\right]
            + D \sum_{j=1}^{N} A_{ij} (x_j - x_i),
\end{equation}
where $x_i$ represents the vegetation biomass at the $i$th node. The
local turbidity is
$E_i = \frac{E_{\rm o} h_V}{h_V + x_i}$, where
$E_{\rm o}$ represents the background turbidity.
We set $h_V=0.2$, $r_V=0.5$, and $D=0.05$.
The variant with $p=4$~\cite{dakos2010} is denoted by VEG1, and the variant with
$p=1$ is denoted by VEG2. The per-node
half-saturation parameters $h_{E,i}$ are drawn independently from the
uniform distribution on $[1,3]$, once for each network, and then held
fixed across all values of $E_{\rm o}$ and across both VEG variants.
We use $E_{\rm o}$ as the control parameter and
sweep $E_{\rm o} \in [2,12]$.

Lowering the Hill exponent reduces the switch-like character of the
eutrophication and vegetation--turbidity feedback models. Thus, EUT2
and VEG2 are smooth and monostable along the entire sweep, in contrast
to the bistable bifurcation diagrams of EUT1 and VEG1.
%
%
We include EUT2 and VEG2 as robustness checks on the role of bistable
saturation in cross-decoder transfer.

\textbf{LDW:} The double-well dynamics defined above couples nodes
through the adjacency matrix. Replacing this coupling with diffusive
coupling gives the Laplacian-coupled double-well dynamics (LDW):
\begin{equation}
\dot{x}_i = -(x_i-r_1)(x_i-r_2)(x_i-r_3)
            +D\sum_{j=1}^{N}A_{ij}(x_j-x_i).
\label{eq:ldw}
\end{equation}
As for DW, we set $r_1=1$, $r_2=3$, and $r_3=5$, and use $D$ as the
control parameter. The spatially uniform states $x_i=r_1$ and $x_i=r_3$ for
all $i$ are stable equilibria of the $N$-dimensional system, whereas the
uniform state $x_i=r_2$ is unstable. Because the single-node dynamics is
bistable, Laplacian coupling allows nonuniform equilibria to
persist over a range of $D$. The LDW coupling term has the same form as
those of NM, EUT1/2, and VEG1/2. We include LDW because it isolates the
effect of the coupling operator and is related to previously studied
diffusively coupled bistable systems~\cite{Keener1987SIAM, Zinner1992JDE,
Eom2018PRE}. LDW and DW have identical single-node dynamics and differ only
in whether the coupling is mediated by the Laplacian or adjacency matrix.

We sweep $D$ over $[0,D_{\max}]$ and determine the range separately
for each network. The rescaling rules used for adjacency-coupled dynamics,
which hold either $D_{\max}\langle k\rangle$ or
$D_{\max}\alpha_{\max}$ constant across networks, do not adequately align
the transition in LDW. The value $D_{\rm c}$ at which the spatially
nonuniform equilibrium disappears depends on the competition between
diffusion and the bistable single-node dynamics, and we find numerically
that neither product captures $D_{\rm c}$ consistently across networks. We
therefore determine $D_{\rm c}$ numerically for each network, defining it
as the smallest value of $D$ for which the equilibrium is spatially uniform,
i.e., $\max_i x_i-\min_i x_i<10^{-6}$, and sweep
$D\in[0,D_{\rm c}/0.9]$. This choice places the transition to the spatially
uniform equilibrium within the sweep for every network.

\subsection{Numerical methods}
\label{sec:numerics}

\subsubsection{Integration horizon}

We set the integration horizon to $t_{\max}=200$ for SIS, DW, MUT,
GEN, WC, NM, EUT, and VEG. We use $t_{\max}=400$ for SIR and
$t_{\max}=100$ for MSIS and MSIR. The only network-specific exception
is DW on the Reactome network, for which we use $t_{\max}=50$ because
the network is dense ($\langle k\rangle \approx 49$) and relaxation is
fast despite the large number of nodes. For a representative control
parameter value, we verified that the final $x_i$ values for this
case are virtually unchanged when we instead use $t_{\max}=200$.
For LDW, we set $t_{\max}=200$ and then, if
$\max_i |\dot{x}_i| \ge 10^{-6}$ at $t=t_{\max}$, repeat the integration with
$t_{\max}$ multiplied by ten, until $\max_i |\dot{x}_i| < 10^{-6}$. We do so because, for LDW, relaxation is slow
near the value of $D$ at which spatially frozen patterns disappear.
%
%

\subsubsection{Initial condition and sweep direction}
\label{sec:icpolicy}

For all dynamics except gLV, we obtain the recorded state by direct
time integration of the ODE system. For gLV, we instead use the
analytical coexistence equilibrium given in
Eq.~\eqref{eq:glv-componentwise}. For the time-integrated dynamics,
we use one of two sweep policies, depending on the dynamical system model.

First, we use a cold-start policy for all dynamical systems but MSIS. Under this policy, we use the
same initial condition at every value of the control parameter. The
initial conditions for all nodes are $x_i(0)=0.9$ for SIS;
$S_i(0)=0.999$ and $I_i(0)=0.001$ for SIR;
$\rho_{S,i}(0)=0.999$, $\rho_{I,i}(0)=0.001$, and
$\rho_{R,i}(0)=0$ for MSIR; $x_i(0)=1$ for DW, corresponding to the
lower branch; $x_i(0)=0.001$ for MUT, corresponding to the lower
branch; $x_i(0)=2$ for GEN, corresponding to the high-expression
branch; $x_i(0)=0.1$ for WC; $x_i(0)=8$ for NM, corresponding to high
vegetation biomass; $x_i(0)=0.5$ for EUT; and $x_i(0)=0.6$ for VEG.
For LDW, we draw each $x_i(0)$ independently from the uniform
distribution on $[r_1,r_3]$, once for each network, and use the resulting
initial condition at every value of $D$. We use a spatially heterogeneous
rather than uniform initial condition because the uniform states
$x_i=r_1$ and $x_i=r_3$ are stable equilibria of
Eq.~\eqref{eq:ldw}.

Second, we use a reverse warm-start policy for MSIS. We use a reverse
rather than a cold start because our preliminary numerical simulations indicated that the MSIS dynamics
are multistable over part of the swept range in a complicated manner. Under this
policy, we sweep the control parameter in descending order and
initialize each sweep point from the final state obtained at the
previous, larger control parameter value. The initial
condition at $\beta=\beta_{\max}$ is
$\rho_{S,i}(0)=0.9$ and $\rho_{I,i}(0)=0.1$; changing it to
$\rho_{S,i}(0)=0.1$ and $\rho_{I,i}(0)=0.9$ yielded virtually the same equilibria.

\subsubsection{Clipping of $x_i$}

For GEN, the numerical integrator can produce small transient
excursions with $x_j<0$ at intermediate steps. Because $x_j$
represents a gene-expression level, we clip these values to the
physically meaningful range by replacing $x_j$ with $\max(x_j,0)$ at
each time step. We do not apply analogous clipping in DW, MUT, WC, NM,
EUT, LDW, or VEG, because the integrator does not produce negative values
of $x_j$ in our simulations.

\subsubsection{Auto-trimming of the control-parameter range}
\label{sec:autotrim}

We apply an auto-trimming procedure to the control-parameter range for
all dynamics except gLV, for which we use the analytical equilibrium
given by Eq.~\eqref{eq:glv-componentwise}, and except LDW, whose range
is determined separately for each network as described above.
For most
pairs of network and dynamics, the original control-parameter range
$[p_{\min},p_{\max}]$ does not require trimming. For a small number of
cases, however, the original range contains a long flat subrange in
which the node-averaged recorded state,
$\overline{x}(p) = \sum_{i=1}^{N} x_i(p)/N$,
is close to zero. Such a subrange contributes little dynamical signal
for decoder training. We detect and trim these flat regions as
follows.

First, for MSIS and MSIR, the SIS-matched infection-rate range
$[0,1/(0.15\alpha_{\max})]$ can lie entirely below the endemic
threshold. We therefore use an adaptive widening rule for
$\beta_{\max}$ before running the full sweep. We initialize
$\beta_{\max}=1/(0.15\alpha_{\max})$. We then run an ODE simulation
at $\beta=\beta_{\max}/2$ until $t=t_{\max}$ and check whether
$\sum_{i=1}^{N}\rho_{I,i}(t_{\max}) / N \geq 0.1$
for MSIS, or
$\sum_{i=1}^{N}\rho_{R,i}(t_{\max}) / N \geq 0.1$
for MSIR. If this condition is not satisfied, we double
$\beta_{\max}$ and repeat the test. Once the condition is satisfied,
we accept the current value of $\beta_{\max}$ and run the full sweep.

Second, we trim flat regions at the low end of the range. We set a
probe point at the $20\%$ position of the current interval,
$p_q = p_{\min} + 0.2(p_{\max}-p_{\min})$.
We integrate the ODE at $p=p_q$ and compute $\overline{x}(p_q)$. If
$\overline{x}(p_q)<0.02$, we replace $p_{\min}$ by $p_q$ and repeat
the procedure. Otherwise, we accept the current value of $p_{\min}$.
Under the assumption that $x_i$ monotonically increases over the trimmed control-parameter region, the
remaining flat portion of the final interval occupies at most the
lowest $20\%$ of the sweep. Across all $14$ networks and the $15$
 dynamics to which auto-trimming applies, this low-end trimming was triggered only for
the metapopulation epidemic dynamics, MSIS1, MSIS2, MSIR1, and MSIR2,
on many of the networks.

Third, we analogously trim flat regions at the high end of the range.
We set a probe point at the $80\%$ position of the current interval,
$p_q = p_{\min} + 0.8(p_{\max}-p_{\min})$.
We integrate the ODE at $p=p_q$ and compute $\overline{x}(p_q)$. If
$\overline{x}(p_q)<0.02$, indicating a flat high-end region, we
replace $p_{\max}$ by $p_q$ and repeat the procedure. Otherwise, we
accept the current value of $p_{\max}$. This high-end trimming was
triggered only for VEG on the Spanish-language word-association
network, where $p_{\max}$ was reduced from $10.75$ to $10.00$.

\subsection{Networks}
\label{sec:networks}

We use $14$ undirected and unweighted networks drawn from the data set
used in our previous study~\cite{maclaren2025}. For each network, we
extract the largest connected component and use this component in all
analyses. The $14$
networks and their basic statistics are given in
Supplementary Section~\ref{si:networks}.

\subsection{Cross-transfer}
\label{sec:fair}

We assess cross-transfer by training a decoder on one source dynamics
$\mathrm{S}$ and deploying it on another target dynamics
$\mathrm{T}$ on the same network. We denote the target data by
$x^{(\mathrm{T})}_{i,k}$, where $i \in \{1,\ldots,N\}$ indexes nodes
and $k \in \{1,\ldots,M\}$ indexes the swept control-parameter values.
The decoder observes only the $x^{(\mathrm{T})}_{i,k}$ values at the sentinel nodes,
$i \in \mathcal{S}$, and aims to reconstruct
$x^{(\mathrm{T})}_{i,k}$ for all nodes and all control-parameter
values.

A decoder trained on the source dynamics expects inputs on the source
dynamics' numerical scale and produces outputs on the same scale.
Deployment on a target dynamics therefore requires two affine maps. The
input map rescales the target sentinel readings to the interval spanned by
the source sentinel inputs during training. Without this map, differences
in scale between the source and target dynamics can place the target
inputs far outside the decoder's training range. The output map then
transforms the decoder output back to the scale of the target dynamics.
Each affine map is specified by two parameters.

We construct the input map using extrema. On the source side, we
compute the extrema of the decoder's training data over all $N$ nodes and
the training control-parameter indices $\mathcal{K}_{\rm train}$. On the
target side, only sentinel readings are observable, so we compute the
extrema over $i\in\mathcal{S}$ alone. We fit the output map by least squares,
again using only target values at the sentinel nodes. Thus, no non-sentinel
target value is used at any stage. We also separate calibration from
evaluation on the target side. Specifically, we use the source-side
train--test split on the target side as well. In other words, the
calibration set is $\mathcal{K}_{\rm cal}=\mathcal{K}_{\rm train}$, where we
recall that $\mathcal{K}_{\rm train}$ is the set of the $M_{\rm train}=41$
control-parameter indices used for training the decoder, defined in
Section~\ref{sec:decoder-methods}, and the evaluation set
$\mathcal{K}_{\rm eval}=\{1,\ldots,M\}\setminus\mathcal{K}_{\rm cal}$
contains the remaining $10$ indices. Only the values indexed by
$\mathcal{K}_{\rm cal}$ are used to fit the input and output maps, whereas
reconstruction accuracy is evaluated only on
$\mathcal{K}_{\rm eval}$. Because the two sides share the same split, no
control-parameter index at which accuracy is evaluated is used for either
training the decoder or calibrating the two maps. This property matters in
particular when the source and target dynamics are the same.

To construct the input map, for a per-dynamics source decoder, we define
\begin{align}
x_{\min}^{(\mathrm{S})}
   &= \min_{i \in \{1,\ldots,N\},\, k \in \mathcal{K}_{\rm train}}
      x_{i,k}^{(\mathrm{S})},\\
x_{\max}^{(\mathrm{S})}
   &= \max_{i \in \{1,\ldots,N\},\, k \in \mathcal{K}_{\rm train}}
      x_{i,k}^{(\mathrm{S})}.
\end{align}
Similarly, we define the sentinel-only extrema of the source dynamics by
\begin{align}
x_{\min,\mathrm{sent}}^{(\mathrm{S})}
   &= \min_{i \in \mathcal{S},\, k \in \mathcal{K}_{\rm train}}
      x_{i,k}^{(\mathrm{S})},\\
x_{\max,\mathrm{sent}}^{(\mathrm{S})}
   &= \max_{i \in \mathcal{S},\, k \in \mathcal{K}_{\rm train}}
      x_{i,k}^{(\mathrm{S})}.
\end{align}
Because
$x_{\min,\mathrm{sent}}^{(\mathrm{S})} \geq x_{\min}^{(\mathrm{S})}$
and
$x_{\max,\mathrm{sent}}^{(\mathrm{S})} \leq x_{\max}^{(\mathrm{S})}$,
the normalized sentinel inputs used to train the source decoder lie in
the interval
\begin{equation}
\left[
   \frac{x_{\min,\mathrm{sent}}^{(\mathrm{S})} - x_{\min}^{(\mathrm{S})}}
        {x_{\max}^{(\mathrm{S})} - x_{\min}^{(\mathrm{S})}},
   \quad
   \frac{x_{\max,\mathrm{sent}}^{(\mathrm{S})} - x_{\min}^{(\mathrm{S})}}
        {x_{\max}^{(\mathrm{S})} - x_{\min}^{(\mathrm{S})}}
\right].
\label{eq:source-sentinel-interval}
\end{equation}
For pooled decoders, there is no single physical source dynamics from
which to define source-side extrema. Instead, we define the
source-side sentinel-input interval directly in the normalized pooled
training data. Let $\mathcal{D}_{\mathrm{S}}$ be the set of dynamics
used for training the pooled source decoder. For example,
$\mathcal{D}_{\mathrm{S}}$ contains all $17$ dynamics for C17 and the
$11$ transferable dynamics for C11. For each
$\mathrm{S}'\in\mathcal{D}_{\mathrm{S}}$, let
$\tilde{x}^{(\mathrm{S}')}_{i,k}$ denote the value of
$x^{(\mathrm{S}')}_{i,k}$ after min--max scaling that dynamics'
training data to $[0,1]$. Then, the pooled source has
$x_{\min}^{(\mathrm{S})}=0$ and $x_{\max}^{(\mathrm{S})}=1$ by
construction, and its sentinel-only extrema are
\begin{align}
x_{\min,\mathrm{sent}}^{(\mathrm{S})}
&=
\min_{\mathrm{S}'\in\mathcal{D}_{\mathrm{S}}}
\min_{i\in\mathcal{S},\, k\in\mathcal{K}_{\rm train}}
\tilde{x}^{(\mathrm{S}')}_{i,k},\\
x_{\max,\mathrm{sent}}^{(\mathrm{S})}
&=
\max_{\mathrm{S}'\in\mathcal{D}_{\mathrm{S}}}
\max_{i\in\mathcal{S},\, k\in\mathcal{K}_{\rm train}}
\tilde{x}^{(\mathrm{S}')}_{i,k}.
\end{align}
Thus, the source-side interval in
Eq.~\eqref{eq:source-sentinel-interval} is the range of normalized
sentinel inputs that the pooled decoder actually saw during training.

We next define the sentinel-only extrema of the target dynamics:
\begin{align}
x_{\min,\mathrm{sent}}^{(\mathrm{T})}
   &= \min_{i \in \mathcal{S},\, k \in \mathcal{K}_{\rm cal}}
      x_{i,k}^{(\mathrm{T})},\label{eq:tgt-sent-min}\\
x_{\max,\mathrm{sent}}^{(\mathrm{T})}
   &= \max_{i \in \mathcal{S},\, k \in \mathcal{K}_{\rm cal}}
      x_{i,k}^{(\mathrm{T})}.\label{eq:tgt-sent-max}
\end{align}
We map each target sentinel reading $x^{(\mathrm{T})}_{i,k}$, with
$i \in \mathcal{S}$, to the source sentinel-input interval by
\begin{equation}
z^{(\mathrm{T})}_{i,k} =
  \frac{1}{x_{\max}^{(\mathrm{S})} - x_{\min}^{(\mathrm{S})}}
  \left[
    x_{\min,\mathrm{sent}}^{(\mathrm{S})} - x_{\min}^{(\mathrm{S})}
    + \left(x^{(\mathrm{T})}_{i,k} - x_{\min,\mathrm{sent}}^{(\mathrm{T})}\right)
      \frac{x_{\max,\mathrm{sent}}^{(\mathrm{S})} - x_{\min,\mathrm{sent}}^{(\mathrm{S})}}
           {x_{\max,\mathrm{sent}}^{(\mathrm{T})} - x_{\min,\mathrm{sent}}^{(\mathrm{T})}}
  \right].
\label{eq:v22_input_map}
\end{equation}
This is the input map.
By construction, the scaled target sentinel values
$z^{(\mathrm{T})}_{i,k}$ lie in the same normalized interval as the
source sentinel inputs used during training. For each
control-parameter index $k$, we feed the vector
$\{z^{(\mathrm{T})}_{i,k}: i\in\mathcal{S}\}$ into the source decoder
and obtain predicted normalized values
$\hat{z}^{(\mathrm{T})}_{i,k}$ for all nodes
$i \in \{1,\ldots,N\}$.

The decoder output is still in the source decoder's normalized output
coordinates. To convert it to the physical scale of the target
dynamics, we fit an affine calibration using only the target values at
the sentinel nodes. Specifically, we choose $(a,b)\in\R^2$ to minimize
\[
\sum_{i \in \mathcal{S}}\sum_{k \in \mathcal{K}_{\rm cal}}^{}
\left(a\hat{z}^{(\mathrm{T})}_{i,k} + b
      - x^{(\mathrm{T})}_{i,k}\right)^2 .
\]
We then apply this same affine map to every node:
\begin{equation}
\hat{x}^{(\mathrm{T})}_{i,k}
   = a\hat{z}^{(\mathrm{T})}_{i,k} + b,
\qquad
i \in \{1,\ldots,N\},\quad k \in \{1,\ldots,M\}.
\end{equation}
This is the output map.
Thus, cross-transfer uses information about the target dynamics only
through the sentinel readings $x^{(\mathrm{T})}_{i,k}$ with
$i\in\mathcal{S}$: first to scale the decoder inputs through
Eqs.~\eqref{eq:tgt-sent-min}--\eqref{eq:v22_input_map}, and then to fit the
two calibration constants $(a, b)$.

\subsection{Performance measures}
\label{sec:performance}

We quantify reconstruction performance for each source decoder and
target dynamics using two measures. The source decoder may be a
per-dynamics decoder or a pooled decoder such as C11 or C17. We remind that
$x^{(\mathrm{T})}_{i,k}$ represents the recorded state of
the target dynamics, $\mathrm{T}$, where
$i \in \{1,\ldots,N\}$ indexes nodes and
$k \in \{1,\ldots,M\}$ indexes control-parameter values.

The first measure is the root-mean-squared error (RMSE) between the
decoder reconstruction and the ground truth. We define
\begin{equation}
\mathrm{RMSE}(\mathrm{S}\to\mathrm{T}) =
\sqrt{
   \frac{1}{|\mathcal{K}_{\rm eval}|N}
   \sum_{k \in \mathcal{K}_{\rm eval}}
   \sum_{i=1}^{N}
   \left(\hat{x}^{(\mathrm{T})}_{i,k}
         - x^{(\mathrm{T})}_{i,k}\right)^2
     },
\label{eq:RMSE_v15}
\end{equation}
where $\mathrm{S}$ denotes the source decoder. We recall that 
$\mathcal{K}_{\rm eval}$ is the set of
control-parameter indices over which the error is evaluated.

Because the physical range of $x_i$ differs across dynamics and
networks, we normalize the RMSE by the global range of the target
dynamics:
\begin{equation}
\mathrm{n\text{-}RMSE}(\mathrm{S}\to\mathrm{T})
   =
   \frac{\mathrm{RMSE}(\mathrm{S}\to\mathrm{T})}
        {
        \max_{i \in \{1,\ldots,N\},\, k \in \{1,\ldots,M\}}
        x^{(\mathrm{T})}_{i,k}
        -
        \min_{i \in \{1,\ldots,N\},\, k \in \{1,\ldots,M\}}
        x^{(\mathrm{T})}_{i,k}
        } .
\label{eq:relRMSE_v15}
\end{equation}

The second measure compares the decoder against a simple
mean-prediction baseline. Consider a source decoder with sentinel set
$\mathcal{S}$. The baseline predicts the same value for
all nodes at a given control-parameter value, by the mean of the
target readings over the source decoder's sentinel nodes:
\begin{equation}
\hat{x}^{(\mathrm{T}),{\rm mean}}_{i,k} =
   \frac{1}{n}\sum_{i' \in \mathcal{S}}
   x^{(\mathrm{T})}_{i',k},
\qquad
i \in \{1,\ldots,N\},\quad k \in \mathcal{K}_{\rm eval}.
\label{eq:meanbase}
\end{equation}
This baseline uses the same sentinel set as the source decoder but
does not use any trained mapping; it simply broadcasts the sentinel
mean to all nodes. Substituting
$\hat{x}^{(\mathrm{T}),{\rm mean}}_{i,k}$ for
$\hat{x}^{(\mathrm{T})}_{i,k}$ in Eq.~\eqref{eq:RMSE_v15}, using the
same index set $\mathcal{K}_{\rm eval}$, and then applying
the normalization in Eq.~\eqref{eq:relRMSE_v15} gives the
mean-baseline n-RMSE, denoted by
$\mathrm{n\text{-}RMSE}^{\rm mean}(\mathrm{S}\to\mathrm{T})$.

We define the baseline ratio
\begin{equation}
r(\mathrm{S}\to\mathrm{T})
   =
   \frac{\mathrm{n\text{-}RMSE}(\mathrm{S}\to\mathrm{T})}
        {\mathrm{n\text{-}RMSE}^{\rm mean}(\mathrm{S}\to\mathrm{T})}.
\label{eq:ratio}
\end{equation}
A value $r(\mathrm{S}\to\mathrm{T})<1$ indicates that the source
decoder reconstructs the target dynamics more accurately than the
mean-prediction baseline, whereas
$r(\mathrm{S}\to\mathrm{T})>1$ indicates that the baseline is more
accurate.

\section*{Acknowledgments}

N.M. acknowledges financial support by the Japan Science and Technology Agency (JST) Moonshot R\&D (under Grant No. JPMJMS2021), the National Science Foundation (under grant no.\,2200173), and JSPS KAKENHI (under grant nos.\,JP 23H03414, 24K14840, and 24K03013).
During the preparation of this manuscript, the authors used Claude Opus 4.7 and 5.5 for manuscript editing and code development assistance. All scientific interpretations, conclusions, code, and manuscript text were verified by the authors.

\section*{Author contributions}

N.M. conceived the study, developed the methods, wrote the code, performed numerical simulations, and drafted the manuscript. T.B. performed preliminary numerical simulations. N.M., B.M.E., and T.B. discussed the results. N.M. and B.M.E. revised and checked the manuscript and code. 

\section*{Competing interests}

The authors declare no competing interests.

\section*{Data availability}

Empirical networks used in this study are available from \href{https://networks.skewed.de}{https://networks.skewed.de} or the original sources as described in the SI. All other data was produced by simulation and is available at 
Zenodo upon publication of this article.
%

\section*{Code availability}

Optimization, simulation, and analysis code used in this study is available at \newline \href{https://github.com/naokimas/decoder}{https://github.com/naokimas/decoder}.

\bibliographystyle{plain}

\clearpage
\setcounter{page}{1}
\makeatletter\Hy@pageanchorfalse\makeatother 
\renewcommand{\theHsection}{SI.\arabic{section}}
\renewcommand{\theHfigure}{SI.\arabic{figure}}
\renewcommand{\theHtable}{SI.\arabic{table}}
\renewcommand{\theHequation}{SI.\arabic{equation}}

\clearpage
\suppressfloats[t]  
\begin{center}
{\Large\bfseries Supplementary Materials for:\\[0.3em]
Transferable reconstruction of nonlinear network dynamics from sentinel nodes}\\[0.6em]
{\large Naoki Masuda, Bisna Mary Eldo, Tharusha Bandara}
\end{center}
\vspace{1em}
\setcounter{section}{0}
\setcounter{figure}{0}
\setcounter{table}{0}
\setcounter{equation}{0}
\renewcommand{\thesection}{S\arabic{section}}
\renewcommand{\thefigure}{S\arabic{figure}}
\renewcommand{\thetable}{S\arabic{table}}
\renewcommand{\theequation}{S\arabic{equation}}

\section{Node-wise reconstruction error}\label{si:nodewise}

\subsection{Methods}

The n-RMSE defined in the main text averages the squared reconstruction
error over all $N$ nodes. To examine how the error is distributed across
individual nodes, we omit this average while retaining the average over the
$10$ control-parameter values used for performance evaluation. This defines
the node-wise n-RMSE for node $i$:
\begin{equation}
\mathrm{n\text{-}RMSE}_i(\mathrm{S}\to\mathrm{T})
 = \frac{
   \sqrt{\dfrac{1}{|\mathcal{K}_{\rm eval}|}
         \sum_{k \in \mathcal{K}_{\rm eval}}
         \left(\hat{x}^{(\mathrm{T})}_{i,k}-x^{(\mathrm{T})}_{i,k}\right)^2}}
   {\max_{i',k} x^{(\mathrm{T})}_{i',k} - \min_{i',k} x^{(\mathrm{T})}_{i',k}}.
\label{eq:nodewise}
\end{equation}
The denominator is the global range of the target dynamics, consistent
with the normalization used in the main text. With this choice, the
node-wise values are directly comparable across nodes. We note that
$\mathrm{n\text{-}RMSE} = \sqrt{\sum_{i=1}^{N}
\mathrm{n\text{-}RMSE}_i^2 / N}$.

We use the same decoder and sentinel-selection method as in
Figs.~\ref{fig:sis-sis-dolphin} and~\ref{fig:sis-wc-dolphin} of the main
text, namely, the shallow NN decoder without bias terms and the MinCorr
sentinels. We consider four cases. The first two use the network and dynamics
examined in Figs.~\ref{fig:sis-sis-dolphin} and~\ref{fig:sis-wc-dolphin} of
the main text: the dolphin network ($N=62$) with
$\mathrm{S}=\mathrm{T}=\mathrm{SIS}$ and with
$\mathrm{S}=\mathrm{SIS}$ and $\mathrm{T}=\mathrm{WC}$, respectively, where
$\mathrm{S}$ and $\mathrm{T}$ denote the source and target dynamics. The
other two cases use a larger network, the email network with $N=1133$ nodes,
with $\mathrm{S}=\mathrm{MSIR1}$ and either
$\mathrm{T}=\mathrm{MSIR1}$ or $\mathrm{T}=\mathrm{GEN}$.

\subsection{Results}

\begin{figure}[htbp]
\centering
\includegraphics[width=\textwidth]{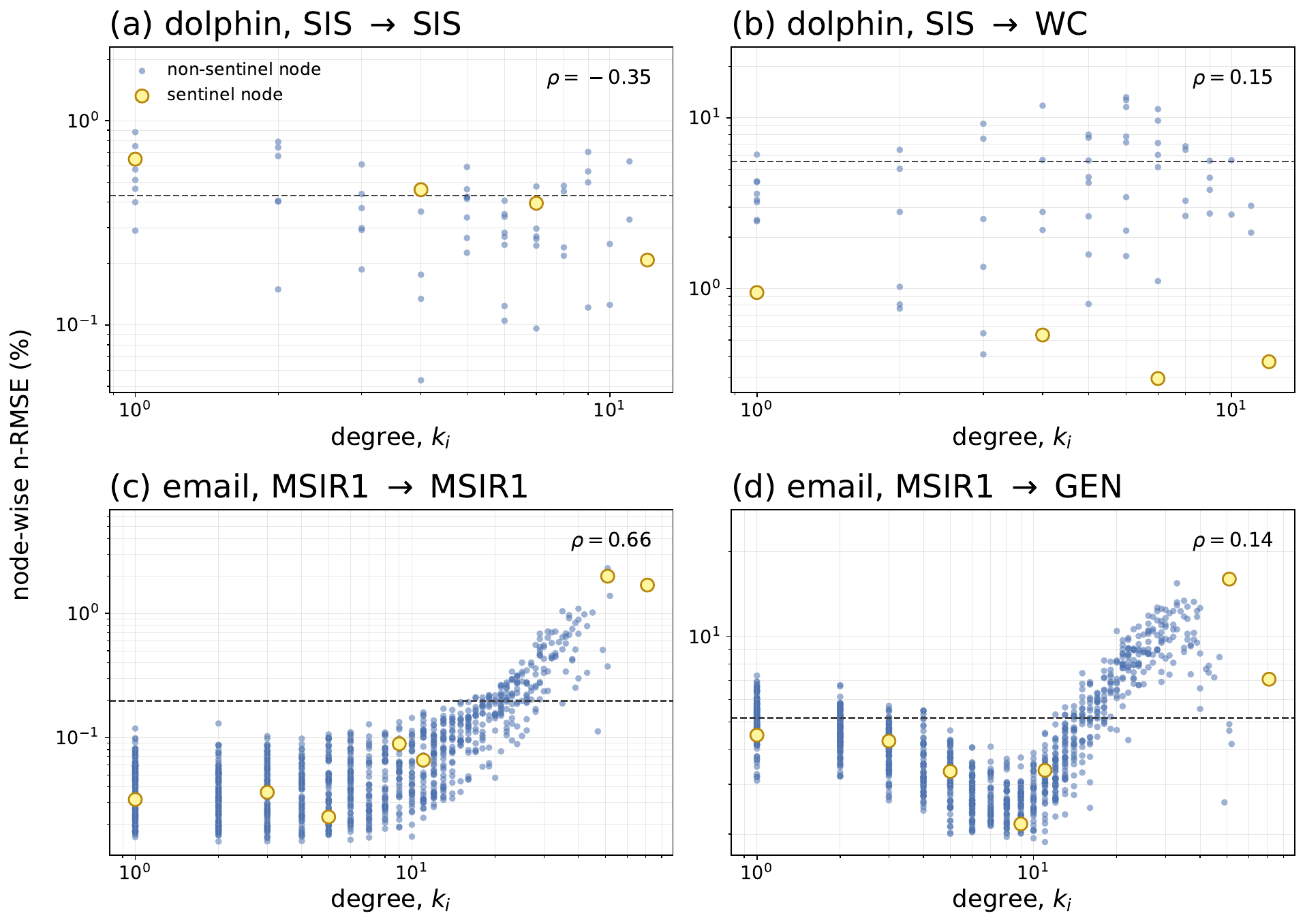}
\caption{Node-wise n-RMSE, defined by Eq.~\eqref{eq:nodewise}, as a
function of the node's degree. (a) Dolphin network, with the SIS decoder
evaluated on SIS, corresponding to Fig.~\ref{fig:sis-sis-dolphin} of the
main text. (b) Dolphin network, with the SIS decoder evaluated on WC,
corresponding to Fig.~\ref{fig:sis-wc-dolphin} of the main text.
(c) Email network, with the MSIR1 decoder evaluated on MSIR1. (d) Email
network, with the MSIR1 decoder evaluated on GEN. The dashed line represents
the n-RMSE aggregated over all $N$ nodes, i.e., the n-RMSE defined in the
main text. Its value is $0.43\%$, $5.54\%$, $0.20\%$, and $5.16\%$ in
(a), (b), (c), and (d), respectively. The value of $\rho$ shown in each
panel is the Spearman rank correlation coefficient between the degree and
node-wise n-RMSE.
}
\label{fig:nodewise}
\end{figure}

Figure~\ref{fig:nodewise} shows the node-wise n-RMSE as a function of
the node's degree for the four cases. The error is far from uniform across
nodes. The dependence of the
node-wise n-RMSE on the node's degree is weak and inconsistent in sign across
the four cases. The Spearman rank correlation coefficient between the degree and
node-wise n-RMSE is $-0.35$, $0.15$, $0.66$, and $0.14$ in
Fig.~\ref{fig:nodewise}(a), (b), (c), and (d), respectively. Therefore, hubs
are not systematically reconstructed either more or less accurately than
other nodes across the four cases.

\begin{figure}[htbp]
\centering
\includegraphics[width=\textwidth]{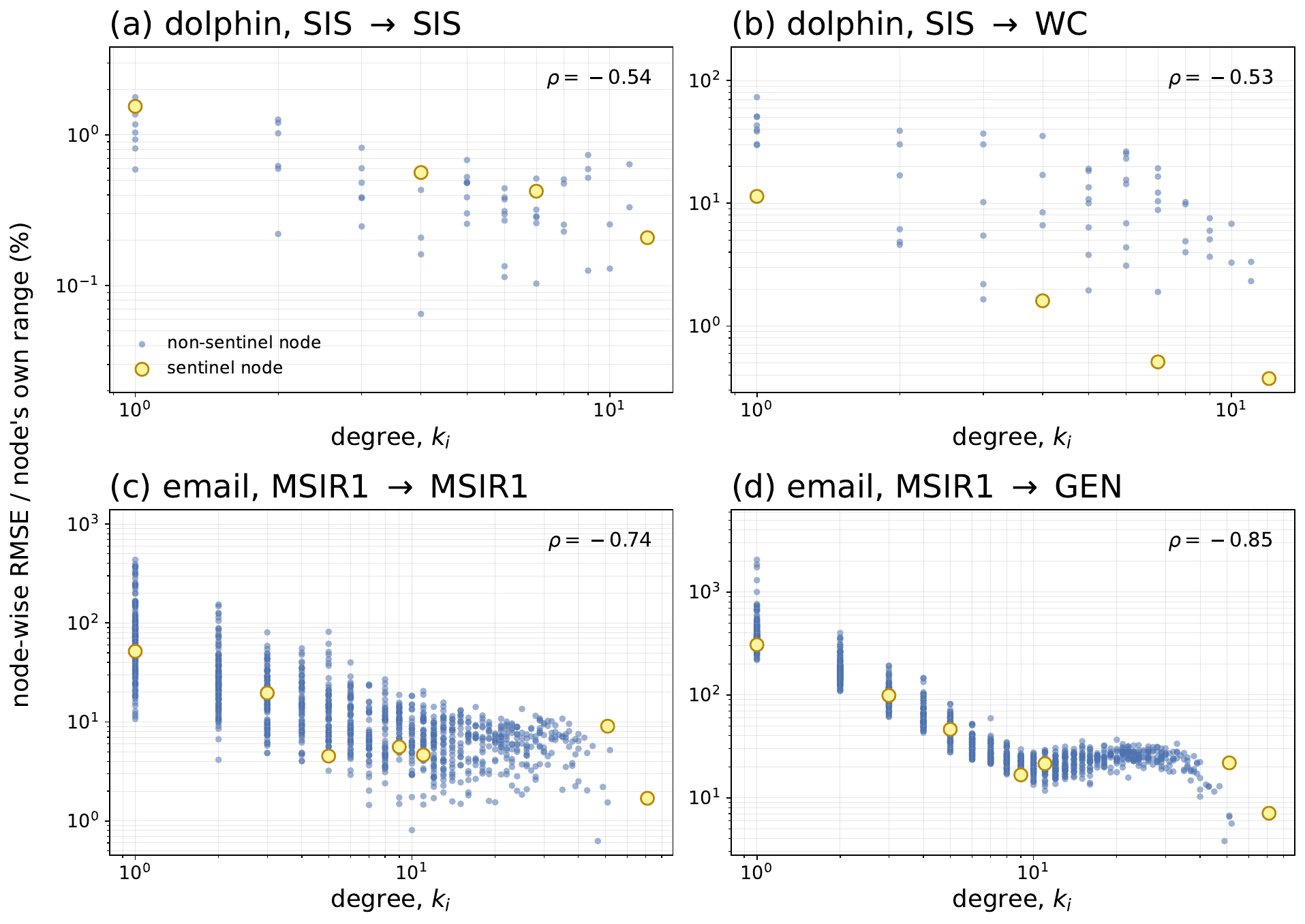}
\caption{Node-wise RMSE normalized by each node's own range,
$\max_k x^{(\mathrm{T})}_{i,k}-\min_k x^{(\mathrm{T})}_{i,k}$, rather than
by the global range of the target dynamics, as a function of the node's
degree. The four panels correspond to the same cases as those in
Fig.~\ref{fig:nodewise}. The value of $\rho$ shown in each panel is the
Spearman rank correlation coefficient between the degree and the normalized
node-wise RMSE.}
\label{fig:nodewise-own}
\end{figure}

Because Eq.~\eqref{eq:nodewise} normalizes every node's error by the
same global range, a node whose equilibrium value varies little over the
control-parameter sweep tends to have a small node-wise n-RMSE by
construction. Therefore, in Fig.~\ref{fig:nodewise-own}, we show the
node-wise RMSE normalized by each node's own range rather than by the global
range. With this normalization, the Spearman rank correlation coefficient
between degree and normalized node-wise RMSE is negative in all four cases:
$-0.54$, $-0.53$, $-0.74$, and $-0.85$ in
Fig.~\ref{fig:nodewise-own}(a), (b), (c), and (d), respectively. Relative to
the extent of their variation, high-degree nodes are reconstructed more
accurately than low-degree nodes.

\newpage
\clearpage

\section{Effective ranks of the equilibrium data}\label{si:rank}

\subsection{Definitions of effective ranks}

For each of the $17$ dynamics and each of the $14$ networks, we use
the equilibrium matrix $\bX\in\R^{N\times M}$ defined in the main text,
whose $(i,k)$ entry is the state of node $i$ at the $k$th control-parameter
value. We denote the singular values of $\bX$ by
$\sigma_1\ge\sigma_2\ge\cdots\ge\sigma_M\ge 0$. Because $M=51$ is smaller
than $N$ for every network considered ($N\ge 62$), there are $M=51$
singular values in every case. Therefore, the spectra are directly
comparable across networks.

Our primary measure is the energy-based effective rank~\cite{Thibeault2024NatPhys-SI}
\begin{equation}
r_\theta \;=\; \min\Bigl\{\, r \;:\;
\sum_{i=1}^{r}\sigma_i^{2} \;\ge\; \theta \sum_{i=1}^{M}\sigma_i^{2} \Bigr\},
\label{eq:eff-rank}
\end{equation}
i.e., the number of leading singular directions required to account for a
fraction $\theta$ of the total squared singular-value mass. We report
$r_\theta$ at $\theta=0.99$ and $\theta=0.999$.

Because any choice of the threshold $\theta$ is arbitrary, we also
report two threshold-free measures of effective rank. The first is the
entropy effective rank~\cite{RoyVetterli2007EUSIPCO-SI}:
\begin{equation}
r_{\mathrm{ent}} \;=\; \exp\Bigl(-\sum_{i=1}^{M} p_i \ln p_i\Bigr),
\label{eq:ent-rank}
\end{equation}
where
\begin{equation}
p_i \;=\; \frac{\sigma_i}{\sum_{j=1}^{M}\sigma_j}.
\end{equation}
The entropy effective rank is equal to $k$ when $k$ singular values are equal
and nonzero and all the remaining singular values vanish. The second is the
stable rank~\cite{RudelsonVershynin2007JACM-SI}:
\begin{equation}
r_{\mathrm{stab}} \;=\;
\frac{\sum_{i=1}^{M}\sigma_i^{2}}{\sigma_1^{2}}.
\label{eq:stable-rank}
\end{equation}

Because our equilibria are nonnegative and share a substantial common
level across the control-parameter sweep, $\sigma_1$ may be dominated by
this common level rather than by the variation of interest. We therefore
repeat the effective-rank calculations using the centered matrix obtained
by subtracting, for each node, its mean over the $M$ control-parameter
values.

\subsection{Results}

Table~\ref{tab:rank} reports the effective ranks for each of the $17$
dynamics as the median and range over the $14$ networks. The results indicate
that the equilibrium data are low dimensional for most combinations of
dynamics and network. Across the $238$ dynamics--network pairs, the median
$r_{99\%}$ is $2$, the interquartile range is $1$--$2$, and the maximum is
$7$; $89\%$ of the pairs satisfy $r_{99\%}\le 3$. At the stricter threshold,
the median $r_{99.9\%}$ is $2.5$, the interquartile range is $2$--$5$, and
the maximum is $30$. The different measures of effective rank broadly agree
on which dynamics are low dimensional. Across the $17$ dynamics, the
Spearman rank correlation between the per-dynamics medians of $r_{99\%}$ and
$r_{\mathrm{ent}}$ is $0.92$, and that between $r_{99\%}$ and
$r_{\mathrm{stab}}$ is $0.92$. The corresponding correlations for
$r_{99.9\%}$ are $0.95$ and $0.86$, respectively.

The dependence of the effective ranks on the dynamics is related to the
shape of the transition traversed by the control-parameter sweep. We group
the dynamics, at least phenomenologically, into those whose sweep involves a
single smooth global shift (SIS, SIR, MSIS1, MSIR1, MUT, GEN, WC, gLV, EUT2,
and VEG2) and those whose sweep passes through a node-by-node cascade or a
bistable switch (DW, MSIS2, MSIR2, NM, EUT1, VEG1, and LDW). The first group
tends to have lower effective ranks (median $r_{99\%}=1$,
$r_{99.9\%}=2$, $r_{\mathrm{ent}}=1.4$, and
$r_{\mathrm{stab}}=1.00$) than the second group (median
$r_{99\%}=2$, $r_{99.9\%}=7$, $r_{\mathrm{ent}}=3.2$, and
$r_{\mathrm{stab}}=1.03$). 
The mechanism is intuitive: a smooth global shift moves all nodes primarily
along a single direction, so one singular vector captures much of the
variation. By contrast, a cascade in which nodes switch at different points
requires additional directions for different groups of nodes that switch
together.

\begin{table}[t]
\centering
\caption{Effective ranks of the equilibrium matrix $\bX$ for each of
the $17$ dynamics. For each dynamics and each effective-rank measure, we
report the median and range over the $14$ networks.}
\label{tab:rank}
\small
\begin{tabular}{lcccccccc}
\toprule
dynamics & \multicolumn{2}{c}{$r_{99\%}$} & \multicolumn{2}{c}{$r_{99.9\%}$}
& \multicolumn{2}{c}{$r_{\mathrm{ent}}$}
& \multicolumn{2}{c}{$r_{\mathrm{stab}}$}\\
\cmidrule(lr){2-3}\cmidrule(lr){4-5}\cmidrule(lr){6-7}\cmidrule(lr){8-9}
 & median & range & median & range & median & range & median & range\\
\midrule
SIS & 1 & 1--2 & 2 & 1--3 & 1.5 & 1.1--2.0 & 1.01 & 1.00--1.02 \\
SIR & 2 & 1--2 & 2 & 1--3 & 1.7 & 1.1--2.2 & 1.01 & 1.00--1.03 \\
MSIS1 & 2 & 1--2 & 2 & 1--3 & 1.7 & 1.1--2.2 & 1.01 & 1.00--1.03 \\
MSIS2 & 2 & 1--3 & 3 & 2--5 & 2.1 & 1.5--3.1 & 1.03 & 1.01--1.07 \\
MSIR1 & 2 & 1--2 & 2 & 1--4 & 1.7 & 1.1--2.7 & 1.02 & 1.00--1.11 \\
MSIR2 & 2 & 1--3 & 3 & 2--5 & 2.1 & 1.4--3.3 & 1.03 & 1.01--1.12 \\
DW & 4 & 1--6 & 18 & 1--30 & 7.0 & 1.1--10.8 & 1.04 & 1.00--1.06 \\
MUT & 1 & 1--3 & 1 & 1--15 & 1.3 & 1.0--6.5 & 1.00 & 1.00--1.03 \\
GEN & 1 & 1--1 & 2 & 1--4 & 1.3 & 1.0--2.1 & 1.00 & 1.00--1.01 \\
WC & 1 & 1--1 & 1 & 1--2 & 1.2 & 1.0--1.5 & 1.00 & 1.00--1.00 \\
gLV & 2 & 1--2 & 2 & 1--3 & 1.6 & 1.1--2.0 & 1.02 & 1.00--1.07 \\
NM & 2 & 1--3 & 6 & 1--12 & 3.0 & 1.0--4.7 & 1.02 & 1.00--1.04 \\
EUT1 & 2 & 1--2 & 7 & 1--10 & 3.0 & 1.0--3.8 & 1.01 & 1.00--1.02 \\
EUT2 & 1 & 1--1 & 1 & 1--1 & 1.2 & 1.0--1.2 & 1.00 & 1.00--1.00 \\
VEG1 & 6 & 1--7 & 24 & 1--29 & 9.1 & 1.1--10.7 & 1.10 & 1.00--1.12 \\
VEG2 & 1 & 1--1 & 2 & 1--2 & 1.3 & 1.1--1.3 & 1.00 & 1.00--1.00 \\
LDW & 3 & 2--6 & 10 & 4--17 & 4.7 & 3.2--7.7 & 1.04 & 1.01--1.17 \\
\bottomrule
\end{tabular}
\end{table}

Table~\ref{tab:rank-centered} reports the same effective ranks for the
centered matrices. Centering increases the effective ranks somewhat, as
expected, because it removes the common level that would otherwise
contribute substantially to $\sigma_1$. Nevertheless, the
ordering of the dynamics is largely unchanged. Across the $17$ dynamics,
the Spearman rank correlation between the raw and centered medians is $0.85$
for $r_{99\%}$, $0.95$ for $r_{99.9\%}$, $0.93$ for
$r_{\mathrm{ent}}$, and $0.82$ for $r_{\mathrm{stab}}$. Therefore, the
conclusion that the equilibrium data are low dimensional for most of the
dynamics considered in this study remains valid after removal of the common
level.

\begin{table}[t]
\centering
\caption{Effective ranks of the centered equilibrium matrix for each of
the $17$ dynamics. As in
Table~\ref{tab:rank}, we report the median and range over the $14$ networks
for each dynamics and each effective-rank measure.}
\label{tab:rank-centered}
\small
\begin{tabular}{lcccccccc}
\toprule
dynamics & \multicolumn{2}{c}{$r_{99\%}$} & \multicolumn{2}{c}{$r_{99.9\%}$}
& \multicolumn{2}{c}{$r_{\mathrm{ent}}$}
& \multicolumn{2}{c}{$r_{\mathrm{stab}}$}\\
\cmidrule(lr){2-3}\cmidrule(lr){4-5}\cmidrule(lr){6-7}\cmidrule(lr){8-9}
 & median & range & median & range & median & range & median & range\\
\midrule
SIS & 2 & 1--2 & 2 & 1--4 & 1.7 & 1.1--2.4 & 1.02 & 1.00--1.05 \\
SIR & 2 & 1--3 & 3 & 2--4 & 2.0 & 1.2--2.8 & 1.03 & 1.00--1.09 \\
MSIS1 & 2 & 1--3 & 3 & 1--4 & 2.2 & 1.2--2.6 & 1.04 & 1.00--1.08 \\
MSIS2 & 3 & 2--4 & 5 & 3--7 & 3.0 & 1.7--4.5 & 1.11 & 1.01--1.29 \\
MSIR1 & 2 & 1--3 & 3 & 1--5 & 2.4 & 1.1--3.5 & 1.10 & 1.00--1.32 \\
MSIR2 & 3 & 2--4 & 4 & 2--6 & 2.9 & 1.7--4.7 & 1.20 & 1.03--1.81 \\
DW & 12 & 1--23 & 25 & 1--41 & 12.0 & 1.1--20.5 & 1.19 & 1.00--1.38 \\
MUT & 1 & 1--10 & 2 & 1--30 & 1.6 & 1.0--13.4 & 1.01 & 1.00--1.15 \\
GEN & 1 & 1--2 & 2 & 1--6 & 1.4 & 1.0--2.7 & 1.00 & 1.00--1.01 \\
WC & 1 & 1--1 & 1 & 1--2 & 1.3 & 1.0--1.8 & 1.00 & 1.00--1.00 \\
gLV & 1 & 1--2 & 2 & 1--2 & 1.2 & 1.0--1.7 & 1.00 & 1.00--1.04 \\
NM & 3 & 1--5 & 8 & 1--15 & 3.8 & 1.0--6.3 & 1.03 & 1.00--1.07 \\
EUT1 & 4 & 1--5 & 10 & 1--13 & 4.2 & 1.0--5.3 & 1.04 & 1.00--1.06 \\
EUT2 & 1 & 1--1 & 2 & 1--2 & 1.3 & 1.0--1.3 & 1.00 & 1.00--1.00 \\
VEG1 & 10 & 1--15 & 30 & 1--35 & 12.6 & 1.1--15.2 & 1.19 & 1.00--1.29 \\
VEG2 & 1 & 1--1 & 2 & 1--2 & 1.2 & 1.1--1.3 & 1.00 & 1.00--1.00 \\
LDW & 12 & 5--17 & 18 & 11--36 & 11.5 & 6.7--17.3 & 1.44 & 1.10--1.93 \\
\bottomrule
\end{tabular}
\end{table}

\newpage
\clearpage

\section{Gappy POD}\label{si:gappy}

In the main text, we compare the decoders with the deliberately simple
mean-prediction baseline defined by Eq.~\eqref{eq:meanbase}. In this
section, we compare them with gappy POD~\cite{EversonSirovich1995-SI,Willcox2006CompFluids-SI,manohar2018-SI,Brunton2022book-SI,Erichson2020PRSA-SI}, a classical linear method for
reconstructing a full state vector from a small number of observed entries. It is a substantially stronger baseline. All results in this section use the
$17$ dynamics and $14$ networks considered in the main text, enabling direct
comparison with Figs.~\ref{fig:heat-combined} and~\ref{fig:ratio} of the main
text.

\subsection{Methods}

We denote the source training-snapshot matrix by
$\bX_{\rm train} \in \R^{N\times M_{\rm train}}$, where
$M_{\rm train}=41$, and the sentinel set by $\mathcal{S}$, where
$|\mathcal{S}|=n=\lfloor\ln N\rfloor$, following the notation of
Section~\ref{sec:decoder-methods} in the main text. Gappy POD assumes that
each state lies close to the span of the leading $r$ left singular vectors,
$\bU_r$, of $\bX_{\rm train}$. The sentinel observations satisfy
$\bs=(\bU_r)_{\mathcal{S}}\ba$, where $(\bU_r)_{\mathcal{S}} \in \R^{n \times r}$ denotes the
restriction of $\bU_r$ to the rows indexed by $\mathcal{S}$ and $\ba$ is the
vector of expansion coefficients. Solving for $\ba$ in the least-squares
sense yields
\begin{equation}
\hat{\bx}=\bU_r\,(\bU_r)_{\mathcal{S}}^{+}\,\bs ,
\label{eq:gappy}
\end{equation}
where ${}^{+}$ denotes the Moore--Penrose pseudoinverse.

The centered variant of gappy POD first subtracts the training mean
$\bmu\in\R^{N}$. The $i$th entry of $\bmu$ is the mean state of node $i$
over the $M_{\rm train}$ training control-parameter values:
\begin{equation}
\mu_i=\frac{1}{M_{\rm train}}\sum_{k\in\mathcal{K}_{\rm train}}x_{i,k}.
\end{equation}
The reconstruction in the centered variant is
\begin{equation}
\hat{\bx}=\bmu+\bU_r\,(\bU_r)_{\mathcal{S}}^{+}\,(\bs-\bmu_{\mathcal{S}}).
\label{eq:gappy-centered}
\end{equation}
We report both variants because the uncentered variant,
Eq.~\eqref{eq:gappy}, is more directly comparable to our linear decoder,
whereas the centered variant, Eq.~\eqref{eq:gappy-centered}, is the
standard form of gappy POD. Gappy POD defines a linear map from $\bs$ to
$\hat{\bx}$ that is determined entirely by the training data through
$\bU_r$. If the normalized training matrix has rank exactly $n$ and $r=n$,
this map coincides with the zero-ridge limit of our linear decoder, defined
in Section~\ref{sec:linear} of the main text. The two methods therefore
differ only in truncation, ridge regularization, and centering.

Before applying either the uncentered or centered variant, we replace
$\bX_{\rm train}$ with the normalized training matrix
$\tilde{\bX}_{\rm train}$ defined in
Section~\ref{sec:decoder-methods} of the main text, whose entries are
\begin{equation}
\tilde{x}_{i,k}=\frac{x_{i,k}-x_{\min}}{x_{\max}-x_{\min}},
\label{eq:gappy-minmax}
\end{equation}
where $x_{\min}$ and $x_{\max}$ are the minimum and maximum, respectively,
over all $N$ nodes and all training control-parameter values in
$\mathcal{K}_{\rm train}$.

Because $n$ sentinel observations cannot identify more than $n$
coefficients, we require $r\le n$, where $n$ ranges from four to nine across
our networks. We use two rules to select $r$. 
The first sets $r=n$, making $(\bU_n)_{\mathcal{S}}$ square. With
QR-pivoted sentinels, Eq.~\eqref{eq:gappy} is then closely related to the
Q-DEIM reconstruction~\cite{ChaturantabutSorensen2010-SI,Drmac2016SIAM-SI}.
The second selects the smallest $r$ for which the singular values account for
$99.9\%$ of the squared singular-value mass of the normalized training
matrix $\tilde{\bX}_{\rm train}$, with $r$ capped at $n$.

We use gappy POD in place of the NN or linear decoder while leaving the
remainder of the cross-transfer procedure described in
Section~\ref{sec:fair} of the main text unchanged. Specifically, the map
from the sentinel observations to the full state is now given by
Eq.~\eqref{eq:gappy} or Eq.~\eqref{eq:gappy-centered}, whereas the input
map, given by Eqs.~\eqref{eq:tgt-sent-min}--\eqref{eq:v22_input_map} of the
main text, and the output map remain unchanged. As for the NN and linear
decoders, we fit both the input and output maps using the $41$ calibration
control-parameter values of the target dynamics, i.e., those in
$\mathcal{K}_{\rm cal}$, and evaluate the reconstruction on the remaining
ten control-parameter values, i.e., those in $\mathcal{K}_{\rm eval}$.

\subsection{Results}

\begin{figure}[t]
\centering
\includegraphics[width=\textwidth]{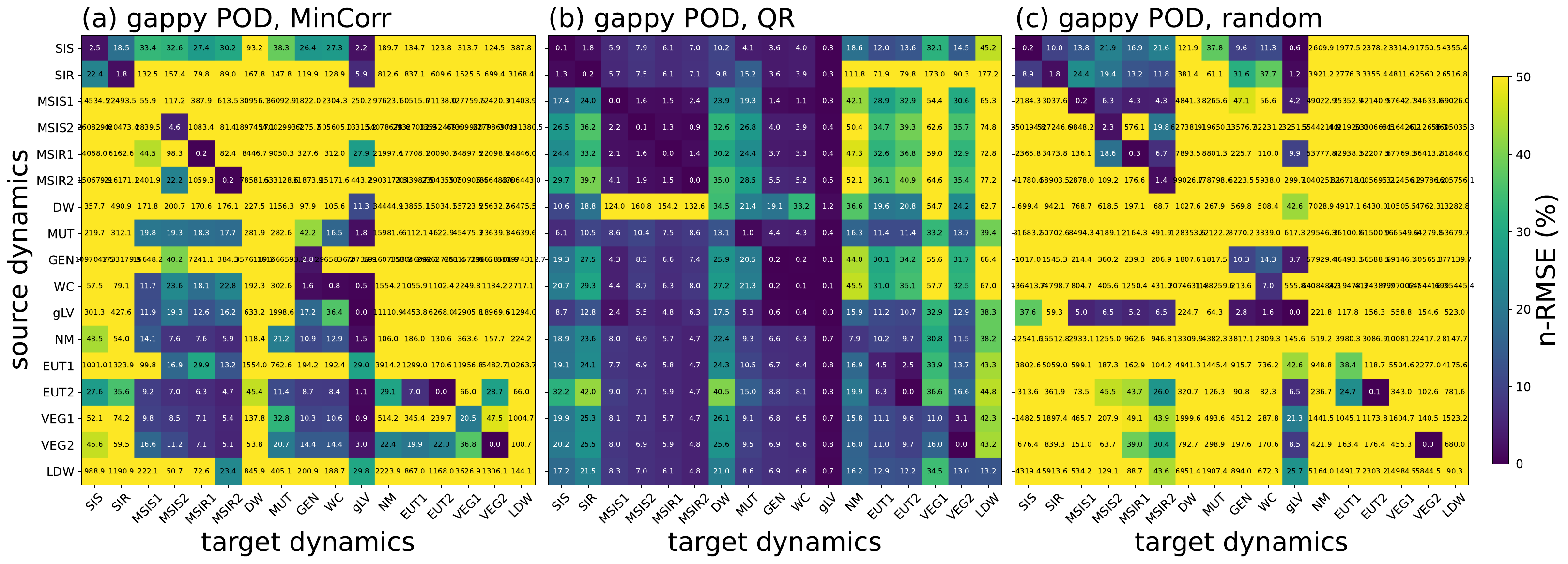}
\caption{n-RMSE of the uncentered gappy POD variant with $r=n$ under
the three sentinel-selection methods. (a)~MinCorr sentinels.
(b)~QR-pivoted sentinels. (c)~Uniformly random sentinels.}
\label{fig:gappy-nrmse}
\end{figure}

Figure~\ref{fig:gappy-nrmse} shows the n-RMSE of gappy POD for the three
sentinel-selection methods, using the same $17\times17$ layout as
Fig.~\ref{fig:heat-combined} of the main text.
Figure~\ref{fig:gappy-nrmse}(b), which uses QR-pivoted sentinels, resembles
the corresponding results for the NN and linear decoders in the main text,
i.e., Figs.~\ref{fig:heat-combined}(b) and (e). Most diagonal entries have small
n-RMSE values, the adjacency-to-adjacency off-diagonal entries tend to have
moderately small n-RMSE values, and the Laplacian-coupled target dynamics
are reconstructed poorly. By contrast, Figs.~\ref{fig:gappy-nrmse}(a) and
(c), which use MinCorr and random sentinels, respectively, show large
n-RMSE values for most source--target pairs. Gappy POD is therefore effective
with QR-pivoted sentinels but not with MinCorr or random sentinels. This
difference among the three sentinel-selection methods arises from the
conditioning of $(\bU_r)_{\mathcal{S}}$, which QR pivoting is designed to
control, whereas MinCorr and random sentinel selection are not. The median
condition number of $(\bU_r)_{\mathcal{S}}$ over the $17$
source dynamics and $14$ network pairs is $5.2$ for QR, compared with
$2.1\times10^{2}$ and $2.5\times10^{2}$ for MinCorr and random sentinels,
respectively.

Table~\ref{tab:gappy-summary} compares the gappy POD variants with the
NN and linear decoders used in the main text in terms of median n-RMSE. The
results are consistent with those in Fig.~\ref{fig:gappy-nrmse}. With
QR-pivoted sentinels, gappy POD performs similarly to the NN and linear
decoders. This result is not surprising: gappy POD is itself a linear map
from the sentinel observations to the full network state and is therefore
closely related to our linear decoder, which, as shown in the main text,
performs almost as well as the NN decoder.

\begin{table}[t]
\centering
\caption{Median n-RMSE (\%) over the $17\times17$ source--target
dynamics pairs, reported separately for within-dynamics reconstruction
(i.e., diagonal cells) and cross-dynamics transfer (i.e., off-diagonal
cells). We first average the n-RMSE for each source--target dynamics pair over the
$14$ networks, and then take the median over the diagonal or
off-diagonal cells. We report medians because some cells have extremely
large n-RMSE values.}
\label{tab:gappy-summary}
\small
\begin{tabular}{llcc}
\toprule
method & sentinels & diagonal & off-diagonal\\
\midrule
gappy POD, uncentered, $r=n$   & MinCorr & 2.84 & 231.96\\
                             & QR      & 0.10 & 12.10\\
                             & random  & 2.32 & 989.80\\
\addlinespace
gappy POD, uncentered, $99.9\%$ & MinCorr & 0.93 & 30.78\\
                              & QR      & 1.00 & 13.88\\
                              & random  & 1.41 & 82.73\\
\addlinespace
gappy POD, centered, $r=n$   & MinCorr & 1.56 & 619.83\\
                             & QR      & 0.10 & 21.56\\
                             & random  & 2.95 & 1696.45\\
\addlinespace
gappy POD, centered, $99.9\%$ & MinCorr & 0.45 & 40.43\\
                              & QR      & 0.49 & 19.59\\
                              & random  & 0.74 & 116.21\\
\addlinespace
shallow NN decoder           & MinCorr & 0.81 & 12.54\\
                             & QR      & 0.65 & 11.73\\
                             & random  & 1.40 & 11.67\\
\addlinespace
linear ridge decoder         & MinCorr & 0.49 & 12.09\\
                             & QR      & 0.32 & 11.52\\
                             & random  & 1.33 & 12.56\\
\bottomrule
\end{tabular}
\end{table}

\begin{figure}[t]
\centering
\includegraphics[width=\textwidth]{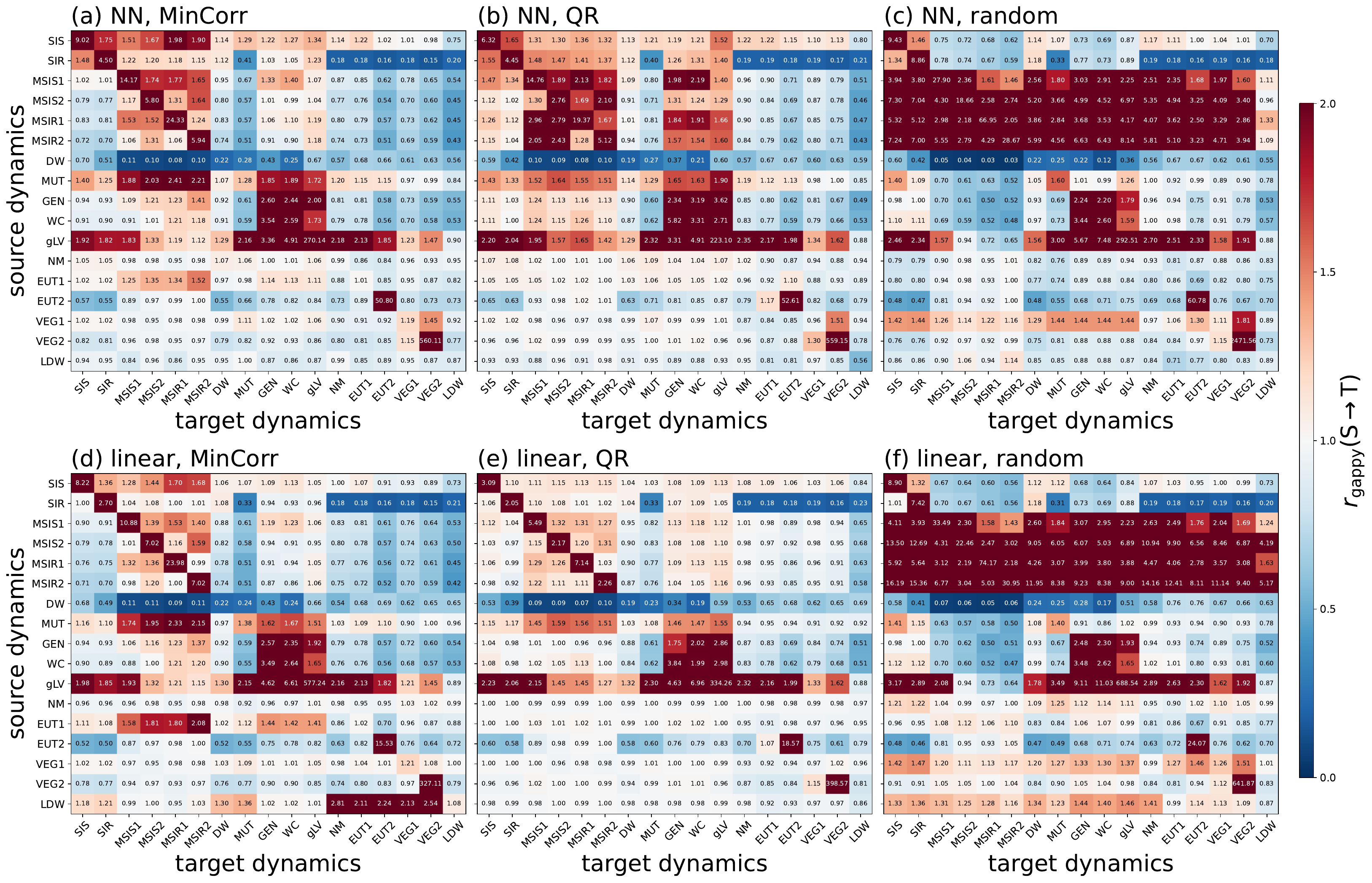}
\caption{Ratio $r_{\rm gappy}(\mathrm{S}\to\mathrm{T})$, defined as
the decoder's n-RMSE divided by that of the uncentered gappy POD variant with
$r=n$ and QR-pivoted sentinels. Each cell shows the decoder's n-RMSE
averaged over the $14$ networks divided by the n-RMSE of gappy POD averaged
over the $14$ networks. Each panel corresponds to a combination of
decoder architecture and sentinel-selection method. (a)~NN decoder with
MinCorr sentinels. (b)~NN decoder with QR-pivoted sentinels. (c)~NN decoder
with uniformly random sentinels. (d)~Linear decoder with MinCorr sentinels.
(e)~Linear decoder with QR-pivoted sentinels. (f)~Linear decoder with
uniformly random sentinels.}
\label{fig:gappy-ratio}
\end{figure}

We next replace the mean-prediction baseline with gappy POD in the
definition of the ratio $r(\mathrm{S}\to\mathrm{T})$ given by
Eq.~\eqref{eq:ratio} in the main text. To distinguish the resulting ratio
from $r(\mathrm{S}\to\mathrm{T})$, which uses the mean-prediction baseline,
we define
\begin{equation}
r_{\rm gappy}(\mathrm{S}\to\mathrm{T})
=\frac{\text{n-RMSE of the decoder}}{\text{n-RMSE of gappy POD}},
\label{eq:r-gappy}
\end{equation}
such that $r_{\rm gappy}(\mathrm{S}\to\mathrm{T})<1$ indicates that the
decoder is more accurate than gappy POD. Figure~\ref{fig:gappy-ratio} shows
$r_{\rm gappy}(\mathrm{S}\to\mathrm{T})$ for the NN and linear decoders
under the three sentinel-selection methods, paralleling
Fig.~\ref{fig:ratio} of the main text. In Fig.~\ref{fig:gappy-ratio}, we
use the uncentered gappy POD variant with $r=n$ and QR-pivoted sentinels in
the denominator. We use QR-pivoted sentinels because gappy POD performs
poorly with MinCorr and random sentinels
(Fig.~\ref{fig:gappy-nrmse} and Table~\ref{tab:gappy-summary}). We use the
uncentered variant with $r=n$ because it is the most accurate of the four
gappy POD configurations listed in Table~\ref{tab:gappy-summary}. Thus,
Fig.~\ref{fig:gappy-ratio} compares our decoders with the strongest gappy
POD configuration considered here. Table~\ref{tab:gappy-ratio} additionally
reports the medians of $r_{\rm gappy}(\mathrm{S}\to\mathrm{T})$ over the
off-diagonal cells, including the other three gappy POD
configurations, all with QR-pivoted sentinels.

Table~\ref{tab:gappy-ratio} indicates that, although the individual values of
$r_{\rm gappy}(\mathrm{S}\to\mathrm{T})$ depend on the source and target
dynamics (Fig.~\ref{fig:gappy-ratio}), their medians are close to $1$ for
both decoder architectures and all three sentinel-selection methods. Table~\ref{tab:gappy-summary} indicates that the uncentered gappy POD
variant is somewhat more accurate than the centered variant.
Table~\ref{tab:gappy-ratio} further shows that the performance of the
uncentered variant is generally closer to that of the NN and linear
decoders, as indicated by values of
$r_{\rm gappy}(\mathrm{S}\to\mathrm{T})$ closer to $1$. These results
reinforce our finding that gappy POD with QR-pivoted sentinels performs
similarly to the NN and linear decoders.

In summary, with QR-pivoted sentinels, gappy POD performs about as well
as our shallow NN and linear ridge decoders. This agreement is reassuring
because it shows that the reconstruction quality obtained with QR-pivoted
sentinels does not depend strongly on the particular decoder used: a
classical linear method based on the same sentinels achieves similar
accuracy. The main difference between the approaches lies not in their
accuracy but in their requirements on the sentinel set. Gappy POD requires
a sentinel set that keeps $(\bU_r)_{\mathcal{S}}$ well conditioned, a
requirement met only by QR pivoting among the three sentinel-selection
methods considered here. By contrast, the NN and linear ridge decoders
remain effective with MinCorr, as shown in the main text and various sections in this supplementary document.

\begin{table}[t]
\centering
\caption{Median $r_{\rm gappy}(\mathrm{S}\to\mathrm{T})$ over the
off-diagonal source--target dynamics pairs. The columns correspond to the
four gappy POD configurations used in the denominator, all with QR-pivoted
sentinels. The rows correspond to combinations of the decoder architecture
and the sentinel-selection method used in the numerator. A value below one
indicates that the decoder is more accurate than gappy POD.}
\label{tab:gappy-ratio}
\small
\begin{tabular}{llcccc}
\toprule
 & & \multicolumn{2}{c}{uncentered} & \multicolumn{2}{c}{centered}\\
\cmidrule(lr){3-4}\cmidrule(lr){5-6}
decoder & sentinels & $r=n$ & $99.9\%$ & $r=n$ & $99.9\%$\\
\midrule
NN & MinCorr & $0.953$ & $0.922$ & $0.769$ & $0.887$ \\
NN & QR & $1.001$ & $0.999$ & $0.883$ & $0.987$ \\
NN & random & $0.940$ & $0.940$ & $0.821$ & $0.908$ \\
\addlinespace
linear & MinCorr & $0.956$ & $0.905$ & $0.746$ & $0.840$ \\
linear & QR & $0.987$ & $0.981$ & $0.871$ & $0.963$ \\
linear & random & $1.052$ & $1.045$ & $0.873$ & $0.968$ \\
\bottomrule
\end{tabular}
\end{table}

\newpage
\clearpage

\section{Robustness to interpolation and extrapolation across control-parameter ranges}\label{si:extrap}

To train a decoder using equilibrium data from a single dynamical
system, we split the $M=51$ control-parameter values in each sweep into
$\lfloor0.8M\rfloor=41$ training values and $10$ held-out values. In
Figs.~\ref{fig:heat-combined} and~\ref{fig:ratio}, we selected the $41$
training values uniformly at random. The held-out values are therefore
generally interspersed among the training values, such that evaluation on
the held-out values requires interpolation along the relevant equilibrium
branch. This task may be easier than extrapolation, which is required when
the training control-parameter values form a contiguous block and the
held-out values lie outside it. The same distinction between interpolation
and extrapolation also arises on the target-dynamics side. In this section,
we report the results of extrapolation experiments.

\subsection{Within-dynamics reconstruction}

In addition to the interspersed protocol for splitting the $51$
control-parameter values, we consider two contiguous splits. In the
``first-$41$'' split, the first $41$ values of the sweep are used for
training, and the $10$ held-out values lie beyond the upper end of the
training range. In the ``last-$41$'' split, the last $41$ values are used
for training, and the $10$ held-out values lie beyond the lower end of the
training range. Thus, in both contiguous splits, the decoder must extrapolate
beyond the range of control-parameter values used for training, in contrast
to the interspersed split.

Throughout this section, we compute the n-RMSE using only the held-out
values and normalize it by the target dynamics' full range over all $M$
control-parameter values. This normalization places the errors from the
interspersed, first-$41$, and last-$41$ splits on a common scale. In the
remainder of this section, we report results for the two non-random
sentinel-selection methods, averaged over the $14$ networks, and omit the
random sentinel-selection method. We do so because random sentinels perform
worse than non-random sentinels with both the NN and linear decoders under
the interspersed protocol (Figs.~\ref{fig:heat-combined} and
\ref{fig:ratio} of the main text).

Table~\ref{tab:extrap-nn} reports the within-dynamics n-RMSE and $r(\mathrm{S}\to\mathrm{T})$ for the shallow NN decoder under each of
the three splits. Table~\ref{tab:extrap-lin} reports the corresponding
results for the linear decoder. Averaged over the dynamics and networks, the
n-RMSE is $2.5$--$3.1\%$ under the interspersed split, compared with
$3.0$--$3.8\%$ under the first-$41$ split and $3.1$--$4.5\%$ under the
last-$41$ split, where each range covers the four combinations of decoder
architecture and sentinel-selection method. The decoder outperforms the
mean-prediction baseline for all $17$ dynamics under the interspersed split,
for $15$--$16$ dynamics under the first-$41$ split, and for $12$--$17$
dynamics under the last-$41$ split.

The performance degradation under the first-$41$ and last-$41$ splits,
relative to the interspersed split, is concentrated in a subset of the
dynamics. Dynamics whose sweeps show a single smooth global shift (SIS,
SIR, MSIS1, MSIR1, MUT, GEN, WC, gLV, EUT2, and VEG2) are barely affected by
the change in splitting protocol. These dynamics also have the lowest
effective ranks: their median $r_{99.9\%}$ is $2$, compared with $7$ for the
remaining dynamics (Section~\ref{si:rank}). By contrast, dynamics whose
sweeps pass through a node-by-node cascade or a bistable switch (DW, MSIS2,
MSIR2, NM, EUT1, VEG1, and LDW), at least phenomenologically, account for
nearly all of the increase in error.

We conclude that the decoder performs worse under extrapolation (i.e.,
the first-$41$ and last-$41$ splits) than under interpolation, but the
degradation is not severe except for a few dynamics. Under the interspersed
split, all $17$ dynamics satisfy $r(\mathrm{S}\to\mathrm{T})<1$ for each of
the four combinations of decoder architecture and sentinel-selection
method. Under the extrapolation splits, six dynamics reach
$r(\mathrm{S}\to\mathrm{T})\ge 1$ in at least one of the eight combinations
of split, decoder architecture, and sentinel-selection method. These
dynamics are LDW and VEG1 under the first-$41$ split, and DW, MUT, EUT1,
EUT2, and VEG1 under the last-$41$ split. The most difficult dynamics is
LDW, which is Laplacian-coupled and, as shown in the main text, does not
belong to the transferable class. For the transferable dynamics, which are
characterized by adjacency-matrix coupling, the decoder generally
reconstructs its source dynamics accurately even when evaluated over a
control-parameter range not encountered during training. Therefore, the
extrapolation results do not substantially alter our main conclusions.

\begin{table}[t]
\centering
\caption{Within-dynamics reconstruction using the shallow NN decoder
under the three splitting protocols: interspersed (abbreviated as
``intersp.''), first-$41$, and last-$41$. We compute the n-RMSE (\%) and
$r(\mathrm{S}\to\mathrm{T})$ using the $10$ held-out control-parameter
values, following the evaluation procedure used in
Figs.~\ref{fig:heat-combined} and~\ref{fig:ratio} of the main text.}
\label{tab:extrap-nn}
\scriptsize\setlength{\tabcolsep}{3.4pt}
\begin{tabular}{l*{12}{c}}
\toprule
& \multicolumn{6}{c}{MinCorr sentinels} & \multicolumn{6}{c}{QR sentinels}\\
\cmidrule(lr){2-7}\cmidrule(lr){8-13}
& \multicolumn{3}{c}{n-RMSE (\%)} & \multicolumn{3}{c}{$r(\mathrm{S}\to\mathrm{T})$}
& \multicolumn{3}{c}{n-RMSE (\%)} & \multicolumn{3}{c}{$r(\mathrm{S}\to\mathrm{T})$}\\
\cmidrule(lr){2-4}\cmidrule(lr){5-7}\cmidrule(lr){8-10}\cmidrule(lr){11-13}
dynamics & intersp. & first-$41$ & last-$41$ & intersp. & first-$41$ & last-$41$
& intersp. & first-$41$ & last-$41$ & intersp. & first-$41$ & last-$41$\\
\midrule
SIS & 0.9 & 2.0 & 1.9 & 0.05 & 0.09 & 0.53 & 0.7 & 1.8 & 1.8 & 0.03 & 0.06 & 0.38 \\
SIR & 0.8 & 2.7 & 3.0 & 0.03 & 0.11 & 0.44 & 0.8 & 2.5 & 3.1 & 0.03 & 0.08 & 0.50 \\
MSIS1 & 0.4 & 0.7 & 1.2 & 0.02 & 0.04 & 0.28 & 0.4 & 0.7 & 1.2 & 0.01 & 0.02 & 0.19 \\
MSIS2 & 0.4 & 1.1 & 1.1 & 0.02 & 0.05 & 0.14 & 0.2 & 0.7 & 1.0 & 0.01 & 0.02 & 0.12 \\
MSIR1 & 0.3 & 0.7 & 0.5 & 0.02 & 0.03 & 0.08 & 0.3 & 0.6 & 0.6 & 0.01 & 0.02 & 0.09 \\
MSIR2 & 0.3 & 0.7 & 0.9 & 0.02 & 0.04 & 0.12 & 0.3 & 0.5 & 0.8 & 0.01 & 0.02 & 0.09 \\
DW & 7.4 & 5.8 & 3.3 & 0.36 & 0.36 & 0.93 & 6.4 & 6.3 & 5.1 & 0.35 & 0.34 & 1.08 \\
MUT & 1.2 & 0.9 & 6.0 & 0.13 & 0.08 & 1.43 & 1.3 & 0.9 & 5.9 & 0.12 & 0.06 & 1.48 \\
GEN & 0.4 & 0.4 & 0.6 & 0.05 & 0.03 & 0.58 & 0.4 & 0.4 & 0.6 & 0.02 & 0.01 & 0.37 \\
WC & 0.2 & 0.2 & 0.3 & 0.03 & 0.02 & 0.36 & 0.3 & 0.3 & 0.3 & 0.02 & 0.01 & 0.22 \\
gLV & 0.1 & 1.5 & 0.0 & 0.09 & 0.21 & 0.47 & 0.1 & 1.0 & 0.0 & 0.03 & 0.08 & 0.22 \\
NM & 7.9 & 0.4 & 1.7 & 0.53 & 0.72 & 0.59 & 8.1 & 0.4 & 2.0 & 0.58 & 0.69 & 0.61 \\
EUT1 & 4.5 & 2.3 & 4.4 & 0.47 & 0.27 & 5.71 & 4.1 & 3.2 & 3.0 & 0.50 & 0.34 & 3.52 \\
EUT2 & 0.9 & 1.5 & 5.0 & 0.12 & 0.16 & 2.26 & 1.0 & 1.4 & 4.7 & 0.12 & 0.14 & 2.03 \\
VEG1 & 13.1 & 15.4 & 1.9 & 0.44 & 1.41 & 1.39 & 10.5 & 17.4 & 5.6 & 0.41 & 1.34 & 5.69 \\
VEG2 & 0.3 & 1.3 & 2.7 & 0.03 & 0.14 & 0.36 & 0.3 & 1.3 & 2.7 & 0.03 & 0.13 & 0.36 \\
LDW & 11.5 & 27.6 & 35.0 & 0.34 & 2.65 & 0.73 & 7.4 & 18.2 & 39.0 & 0.21 & 1.67 & 0.70 \\
\bottomrule
\end{tabular}
\end{table}

\begin{table}[htbp]
\centering
\caption{Within-dynamics reconstruction using the linear decoder under
the three splitting protocols. All conventions are the same as in
Table~\ref{tab:extrap-nn}.}
\label{tab:extrap-lin}
\scriptsize\setlength{\tabcolsep}{3.4pt}
\begin{tabular}{l*{12}{c}}
\toprule
& \multicolumn{6}{c}{MinCorr sentinels} & \multicolumn{6}{c}{QR sentinels}\\
\cmidrule(lr){2-7}\cmidrule(lr){8-13}
& \multicolumn{3}{c}{n-RMSE (\%)} & \multicolumn{3}{c}{$r(\mathrm{S}\to\mathrm{T})$}
& \multicolumn{3}{c}{n-RMSE (\%)} & \multicolumn{3}{c}{$r(\mathrm{S}\to\mathrm{T})$}\\
\cmidrule(lr){2-4}\cmidrule(lr){5-7}\cmidrule(lr){8-10}\cmidrule(lr){11-13}
dynamics & intersp. & first-$41$ & last-$41$ & intersp. & first-$41$ & last-$41$
& intersp. & first-$41$ & last-$41$ & intersp. & first-$41$ & last-$41$\\
\midrule
SIS & 0.8 & 1.6 & 0.6 & 0.05 & 0.07 & 0.16 & 0.3 & 0.9 & 0.3 & 0.01 & 0.03 & 0.07 \\
SIR & 0.5 & 2.0 & 0.6 & 0.02 & 0.08 & 0.08 & 0.4 & 1.3 & 0.4 & 0.01 & 0.04 & 0.07 \\
MSIS1 & 0.3 & 0.7 & 0.7 & 0.02 & 0.04 & 0.17 & 0.1 & 0.5 & 0.4 & 0.01 & 0.01 & 0.06 \\
MSIS2 & 0.5 & 1.1 & 0.8 & 0.03 & 0.05 & 0.11 & 0.2 & 0.5 & 0.6 & 0.00 & 0.01 & 0.07 \\
MSIR1 & 0.3 & 0.6 & 0.3 & 0.02 & 0.03 & 0.06 & 0.1 & 0.3 & 0.1 & 0.00 & 0.01 & 0.02 \\
MSIR2 & 0.3 & 0.9 & 1.0 & 0.02 & 0.05 & 0.14 & 0.1 & 0.3 & 0.4 & 0.00 & 0.01 & 0.04 \\
DW & 7.5 & 5.8 & 3.4 & 0.36 & 0.35 & 0.95 & 6.4 & 6.3 & 5.0 & 0.34 & 0.34 & 1.05 \\
MUT & 1.4 & 1.1 & 2.7 & 0.14 & 0.09 & 0.64 & 1.1 & 0.9 & 2.2 & 0.10 & 0.06 & 0.55 \\
GEN & 0.4 & 0.4 & 0.3 & 0.05 & 0.03 & 0.26 & 0.3 & 0.2 & 0.2 & 0.02 & 0.01 & 0.15 \\
WC & 0.2 & 0.2 & 0.3 & 0.03 & 0.02 & 0.32 & 0.2 & 0.1 & 0.2 & 0.01 & 0.00 & 0.15 \\
gLV & 0.1 & 1.5 & 0.1 & 0.18 & 0.21 & 0.67 & 0.1 & 0.7 & 0.1 & 0.05 & 0.05 & 0.37 \\
NM & 7.8 & 0.4 & 1.3 & 0.52 & 0.72 & 0.45 & 7.9 & 0.3 & 1.8 & 0.56 & 0.68 & 0.57 \\
EUT1 & 4.5 & 1.5 & 0.6 & 0.47 & 0.17 & 0.78 & 4.1 & 3.1 & 0.8 & 0.50 & 0.33 & 0.90 \\
EUT2 & 0.3 & 0.4 & 0.4 & 0.04 & 0.04 & 0.17 & 0.3 & 0.4 & 0.4 & 0.04 & 0.04 & 0.18 \\
VEG1 & 13.3 & 10.4 & 0.7 & 0.45 & 0.96 & 0.47 & 10.6 & 12.7 & 5.8 & 0.41 & 0.97 & 5.86 \\
VEG2 & 0.2 & 0.5 & 1.2 & 0.02 & 0.05 & 0.15 & 0.2 & 0.6 & 1.4 & 0.02 & 0.06 & 0.19 \\
LDW & 14.3 & 34.6 & 38.0 & 0.42 & 3.33 & 0.79 & 11.3 & 21.6 & 42.9 & 0.32 & 1.99 & 0.77 \\
\bottomrule
\end{tabular}
\end{table}

\subsection{Across-dynamics reconstruction}

We estimate the range mapping and the parameters $(a,b)$ from $41$
calibration control-parameter values of the target dynamics and evaluate
the reconstruction on the remaining $10$ values. Thus, the target-side
evaluation is out of sample in the same sense as the source-side evaluation.
We select the $41$ target calibration values using the same three protocols
as for the source training values. Under the interspersed calibration used
in the main text, we select $41$ of the $M=51$ values uniformly at random,
such that they generally bracket the $10$ evaluation values. Under the
first-$41$ calibration, the first $41$ values of the target sweep are used
for calibration, and the $10$ evaluation values lie beyond the upper end of
the calibration range. Under the last-$41$ calibration, the last $41$
values are used for calibration, and the evaluation values lie beyond the
lower end. In the two contiguous cases, the range mapping and the parameters
$(a,b)$ must therefore be extrapolated to the evaluation values.

The source-side and target-side choices of the calibration values are
independent. We therefore evaluate all nine combinations of the three
source-side and three target-side splits for each of the four combinations
of decoder architecture and sentinel-selection method.
Table~\ref{tab:grid} reports the cross-transfer performance, evaluated in
each case on the $10$ held-out control-parameter values; thus, both the
source and target sides are out of sample throughout. The source-side split
has little effect on performance, indicating that cross-dynamics transfer
does not require the decoder to have encountered the portion of the source
sweep on which it is evaluated. By contrast, the target-side calibration
split matters. For example, the median n-RMSE over the
adjacency-to-adjacency pairs is $6.7$--$8.3\%$ with interspersed
calibration, increases to $8.0$--$9.1\%$ with first-$41$ calibration, and
is $4.0$--$7.5\%$ with last-$41$ calibration, where each range covers the
four combinations of decoder architecture and sentinel-selection method.
Similarly, the median $r(\mathrm{S}\to\mathrm{T})$ over these pairs is
$0.41$--$0.64$ with interspersed calibration, remains similar at
$0.39$--$0.59$ with first-$41$ calibration, and increases to
$1.15$--$1.71$ with last-$41$ calibration. The fraction of pairs satisfying
$r(\mathrm{S}\to\mathrm{T})<1$ decreases from $63$--$99\%$ under
interspersed calibration to $62$--$90\%$ under first-$41$ calibration and
$25$--$45\%$ under last-$41$ calibration.

The two contiguous calibration protocols impair performance for different
reasons. Under first-$41$ calibration, the $10$ evaluation values lie above
the entire calibration block, and the decoder must predict node states in a
range of $x_i^*$ not observed during calibration. This is a genuinely
difficult extrapolation task. Under last-$41$ calibration, by contrast, the
decoder is evaluated in a region where the node states have changed little.
For many dynamics, the first control-parameter values, which form the
evaluation set under last-$41$ calibration, correspond to a trivial state
in which the $x_i^*$ values are nearly identical and often close to zero,
such as the disease-free state of epidemic models. The mean-prediction
baseline assigns every node the mean of the sentinel observations at the
same control-parameter value and is therefore highly accurate when the
nodes differ little from one another. As a result, although the cross-transfer decoder also has a small
n-RMSE in this regime, the baseline has similarly low error, causing
$r(\mathrm{S}\to\mathrm{T})$ to exceed one.

While we have discussed pairs of adjacency-coupled dynamics 
in Table~\ref{tab:grid} so far, the same ordering holds when we consider all source--target pairs with
$\mathrm{S}\neq\mathrm{T}$. For example, the fraction
satisfying $r(\mathrm{S}\to\mathrm{T})<1$ decreases from $41$--$59\%$ under
interspersed calibration to $35$--$53\%$ under first-$41$ calibration and
$11$--$26\%$ under last-$41$ calibration. In every row of
Table~\ref{tab:grid}, and for each of the four reported performance
measures, performance over all source--target pairs is worse than that over
the adjacency-to-adjacency pairs. This result is expected because the full
set also includes diffusively coupled dynamics, which the main text shows
do not belong to the transferable class; transfer into or out of these
dynamics is poorer regardless of the calibration protocol.

\begin{table}[b]
\centering
\caption{Cross-transfer performance under different splits of the
equilibrium data for calibration and evaluation. ``intersp.'' denotes the
interspersed split. Each case is evaluated on the $10$ control-parameter
values outside its calibration set. We normalize the RMSE by the target
dynamics' full range over all $51$ control-parameter values to obtain the
n-RMSE. We use the full range rather than the range spanned by the $10$
evaluation values because the latter depends strongly on the location of the
evaluation window. For example, the $10$ values at one end of a sweep, as
used under the first-$41$ and last-$41$ protocols, may span only a small
fraction of the target dynamics' full range. Normalizing by this smaller
range would substantially inflate the n-RMSE because of the location of the
evaluation window rather than because of poorer reconstruction. Using the
full range places all nine combinations on a common scale. We include only
pairs of different dynamics, i.e., $\mathrm{S}\neq\mathrm{T}$. For each
source--target pair, we first average the n-RMSE and the baseline n-RMSE
over the $14$ networks. Quantity $r(\mathrm{S}\to\mathrm{T})$ is the ratio of
these two averages. We then compute the reported median and mean over the
$272$ pairs with $\mathrm{S}\neq\mathrm{T}$ or over the $110$ adjacency-to-adjacency pairs. The columns labeled ``$r<1$~\%'' show the
percentage of these pairs with $r(\mathrm{S}\to\mathrm{T})<1$. For
$r(\mathrm{S}\to\mathrm{T})$, we report only the median because the
baseline n-RMSE is close to zero for a few source--target pairs, yielding
large ratios that substantially inflate the mean.}
\label{tab:grid}
\scriptsize\setlength{\tabcolsep}{2.6pt}
\begin{tabular}{ll*{8}{>{\centering\arraybackslash}p{0.073\textwidth}}}
\toprule
& & \multicolumn{4}{c}{all pairs, $\mathrm{S}\neq\mathrm{T}$}
& \multicolumn{4}{c}{adjacency $\to$ adjacency, $\mathrm{S}\neq\mathrm{T}$}\\
\cmidrule(lr){3-6}\cmidrule(lr){7-10}
decoder, sentinels & source/target & \shortstack{median\\ n-RMSE} & \shortstack{mean\\ n-RMSE} & \shortstack{median\\ $r(\mathrm{S}\!\to\!\mathrm{T})$} & \shortstack{$r<1$\\ \%} & \shortstack{median\\ n-RMSE} & \shortstack{mean\\ n-RMSE} & \shortstack{median\\ $r(\mathrm{S}\!\to\!\mathrm{T})$} & \shortstack{$r<1$\\ \%}\\
\midrule
\multirow{9}{*}{NN, MinCorr} & intersp./intersp. & 12.5 & 15.0 & 1.03 & 41 & 8.1 & 10.0 & 0.64 & 63 \\
 & intersp./first-$41$ & 13.4 & 18.5 & 1.08 & 35 & 8.6 & 13.6 & 0.59 & 62 \\
 & intersp./last-$41$ & 7.3 & 15.9 & 2.20 & 11 & 5.0 & 9.9 & 1.44 & 25 \\
 & first-$41$/intersp. & 12.3 & 15.6 & 1.03 & 39 & 8.2 & 10.2 & 0.62 & 65 \\
 & first-$41$/first-$41$ & 13.4 & 19.2 & 1.14 & 39 & 9.1 & 14.0 & 0.59 & 65 \\
 & first-$41$/last-$41$ & 7.4 & 16.8 & 2.17 & 13 & 5.3 & 10.3 & 1.39 & 31 \\
 & last-$41$/intersp. & 11.5 & 15.1 & 1.03 & 39 & 7.8 & 10.0 & 0.66 & 66 \\
 & last-$41$/first-$41$ & 12.3 & 18.6 & 1.08 & 34 & 8.3 & 13.6 & 0.57 & 65 \\
 & last-$41$/last-$41$ & 7.6 & 16.2 & 2.30 & 11 & 4.7 & 10.0 & 1.63 & 25 \\
\addlinespace
\multirow{9}{*}{NN, QR} & intersp./intersp. & 11.7 & 16.1 & 0.99 & 58 & 8.3 & 11.2 & 0.48 & 98 \\
 & intersp./first-$41$ & 12.5 & 17.8 & 1.00 & 52 & 9.1 & 13.7 & 0.44 & 90 \\
 & intersp./last-$41$ & 8.3 & 18.5 & 1.94 & 11 & 7.5 & 15.7 & 1.71 & 25 \\
 & first-$41$/intersp. & 12.5 & 16.5 & 0.99 & 55 & 8.4 & 11.5 & 0.46 & 98 \\
 & first-$41$/first-$41$ & 12.8 & 18.2 & 1.00 & 48 & 9.4 & 13.9 & 0.45 & 89 \\
 & first-$41$/last-$41$ & 8.9 & 19.1 & 2.06 & 12 & 8.2 & 16.1 & 1.79 & 30 \\
 & last-$41$/intersp. & 12.2 & 16.3 & 1.00 & 50 & 8.7 & 11.5 & 0.51 & 92 \\
 & last-$41$/first-$41$ & 13.2 & 18.4 & 1.02 & 42 & 9.2 & 14.0 & 0.46 & 79 \\
 & last-$41$/last-$41$ & 8.2 & 18.4 & 2.00 & 8 & 7.8 & 14.9 & 1.73 & 19 \\
\addlinespace
\multirow{9}{*}{linear, MinCorr} & intersp./intersp. & 12.1 & 15.3 & 1.05 & 43 & 7.2 & 9.5 & 0.59 & 66 \\
 & intersp./first-$41$ & 12.6 & 17.7 & 1.13 & 43 & 8.3 & 13.3 & 0.56 & 65 \\
 & intersp./last-$41$ & 4.6 & 14.5 & 1.37 & 18 & 4.0 & 8.5 & 1.25 & 35 \\
 & first-$41$/intersp. & 12.1 & 15.7 & 1.03 & 46 & 7.3 & 9.7 & 0.55 & 66 \\
 & first-$41$/first-$41$ & 12.5 & 18.3 & 1.09 & 47 & 8.5 & 13.6 & 0.56 & 65 \\
 & first-$41$/last-$41$ & 4.8 & 14.9 & 1.30 & 21 & 4.2 & 8.6 & 1.23 & 35 \\
 & last-$41$/intersp. & 11.6 & 15.1 & 1.04 & 38 & 7.3 & 9.3 & 0.59 & 67 \\
 & last-$41$/first-$41$ & 12.3 & 17.5 & 1.08 & 38 & 7.8 & 13.0 & 0.55 & 65 \\
 & last-$41$/last-$41$ & 4.2 & 14.3 & 1.27 & 21 & 3.6 & 8.2 & 1.23 & 39 \\
\addlinespace
\multirow{9}{*}{linear, QR} & intersp./intersp. & 11.5 & 16.2 & 0.97 & 59 & 6.7 & 10.1 & 0.41 & 99 \\
 & intersp./first-$41$ & 11.7 & 16.6 & 0.99 & 53 & 8.0 & 13.7 & 0.39 & 87 \\
 & intersp./last-$41$ & 4.4 & 15.3 & 1.20 & 26 & 4.4 & 10.8 & 1.15 & 45 \\
 & first-$41$/intersp. & 12.4 & 16.7 & 0.98 & 55 & 7.1 & 10.1 & 0.40 & 98 \\
 & first-$41$/first-$41$ & 11.6 & 17.1 & 0.99 & 53 & 8.5 & 13.8 & 0.39 & 88 \\
 & first-$41$/last-$41$ & 4.9 & 15.7 & 1.24 & 26 & 5.2 & 11.0 & 1.20 & 46 \\
 & last-$41$/intersp. & 12.4 & 16.2 & 0.99 & 55 & 7.0 & 10.2 & 0.45 & 92 \\
 & last-$41$/first-$41$ & 12.1 & 16.7 & 1.00 & 50 & 8.2 & 13.8 & 0.40 & 77 \\
 & last-$41$/last-$41$ & 4.5 & 15.2 & 1.20 & 21 & 4.7 & 10.6 & 1.30 & 36 \\
\bottomrule
\end{tabular}
\end{table}

Finally, we examine how cross-transfer performance depends on the
number of target-dynamics control-parameter values used for calibration
rather than evaluation. Because Table~\ref{tab:grid} indicates that the
first-$41$ and last-$41$ target-side calibrations perform poorly, we
restrict this analysis to interspersed target-side calibration. Let
$p\in\{50\%,60\%,70\%,80\%,90\%\}$ denote the fraction of the $M=51$
target control-parameter values used for calibration. The corresponding
calibration sets contain $26$, $31$, $36$, $41$, and $46$ values,
respectively. The default used elsewhere is $p=80\%$, corresponding to $41$
calibration values and $10$ evaluation values. On the source side, we retain
the three splits used in Table~\ref{tab:grid}, namely, interspersed,
first-$41$, and last-$41$, each with $41$ training control-parameter values.
Table~\ref{tab:tfrac} reports the cross-transfer performance. Performance
depends only weakly on $p$; in particular, we observe no appreciable
difference between $p=50\%$ and $p=80\%$. These results indicate that
cross-dynamics transfer is robust to a reduction in the density of the
target-dynamics control-parameter sweep used for calibration.

\begin{table}[t]
\centering
\caption{Cross-transfer performance as a function of the number of
target-dynamics control-parameter values used for calibration. We select an
interspersed $50\%$, $60\%$, $70\%$, $80\%$, or $90\%$ of the $M=51$
target-dynamics control-parameter values, corresponding to $26$, $31$, $36$,
$41$, or $46$ calibration values, respectively. The source-side split is one
of the three protocols used in Table~\ref{tab:grid}, each with $41$ training
control-parameter values. All other conventions are the same as in
Table~\ref{tab:grid}. We evaluate each case on the target-dynamics
control-parameter values outside its calibration set. The rows for
``$80\%$'' are identical to the corresponding rows of
Table~\ref{tab:grid} with an interspersed target-side split because we used
the same random-number-generator seed to select the interspersed calibration
indices in both analyses. The table continues on the next page with the
results for the linear decoder.}
\label{tab:tfrac}
\scriptsize\setlength{\tabcolsep}{2.6pt}
\begin{tabular}{ll*{8}{>{\centering\arraybackslash}p{0.073\textwidth}}}
\toprule
& & \multicolumn{4}{c}{all pairs, $\mathrm{S}\neq\mathrm{T}$}
& \multicolumn{4}{c}{adjacency $\to$ adjacency, $\mathrm{S}\neq\mathrm{T}$}\\
\cmidrule(lr){3-6}\cmidrule(lr){7-10}
decoder, sentinels & source/target & \shortstack{median\\ n-RMSE} & \shortstack{mean\\ n-RMSE} & \shortstack{median\\ $r(\mathrm{S}\!\to\!\mathrm{T})$} & \shortstack{$r<1$\\ \%} & \shortstack{median\\ n-RMSE} & \shortstack{mean\\ n-RMSE} & \shortstack{median\\ $r(\mathrm{S}\!\to\!\mathrm{T})$} & \shortstack{$r<1$\\ \%}\\
\midrule
\multirow{17}{*}{NN, MinCorr} & intersp./$50\%$ & 12.7 & 15.4 & 1.04 & 40 & 8.1 & 10.1 & 0.64 & 62 \\
 & intersp./$60\%$ & 12.5 & 15.4 & 1.04 & 40 & 7.8 & 10.0 & 0.65 & 62 \\
 & intersp./$70\%$ & 12.7 & 15.0 & 1.03 & 41 & 7.8 & 9.9 & 0.64 & 63 \\
 & intersp./$80\%$ & 12.5 & 15.0 & 1.03 & 41 & 8.1 & 10.0 & 0.64 & 63 \\
 & intersp./$90\%$ & 11.7 & 14.4 & 1.02 & 41 & 7.7 & 9.5 & 0.66 & 64 \\
\addlinespace[3pt]
 & first-$41$/$50\%$ & 13.0 & 16.0 & 1.03 & 39 & 7.9 & 10.4 & 0.62 & 65 \\
 & first-$41$/$60\%$ & 12.8 & 16.1 & 1.03 & 39 & 7.8 & 10.3 & 0.63 & 65 \\
 & first-$41$/$70\%$ & 12.5 & 15.6 & 1.02 & 40 & 8.2 & 10.1 & 0.64 & 66 \\
 & first-$41$/$80\%$ & 12.3 & 15.6 & 1.03 & 39 & 8.2 & 10.2 & 0.62 & 65 \\
 & first-$41$/$90\%$ & 12.0 & 15.0 & 1.02 & 40 & 8.2 & 9.7 & 0.67 & 66 \\
\addlinespace[3pt]
 & last-$41$/$50\%$ & 11.5 & 15.5 & 1.04 & 38 & 7.7 & 10.1 & 0.66 & 66 \\
 & last-$41$/$60\%$ & 11.0 & 15.6 & 1.04 & 38 & 7.5 & 10.1 & 0.67 & 65 \\
 & last-$41$/$70\%$ & 11.6 & 15.1 & 1.03 & 39 & 7.9 & 9.9 & 0.66 & 66 \\
 & last-$41$/$80\%$ & 11.5 & 15.1 & 1.03 & 39 & 7.8 & 10.0 & 0.66 & 66 \\
 & last-$41$/$90\%$ & 11.2 & 14.5 & 1.03 & 41 & 8.0 & 9.5 & 0.69 & 67 \\
\addlinespace[8pt]
\multirow{17}{*}{NN, QR} & intersp./$50\%$ & 11.6 & 16.7 & 0.99 & 57 & 8.5 & 11.8 & 0.50 & 95 \\
 & intersp./$60\%$ & 11.4 & 16.8 & 0.99 & 55 & 8.4 & 11.9 & 0.51 & 92 \\
 & intersp./$70\%$ & 11.9 & 16.3 & 0.99 & 57 & 8.3 & 11.2 & 0.48 & 98 \\
 & intersp./$80\%$ & 11.7 & 16.1 & 0.99 & 58 & 8.3 & 11.2 & 0.48 & 98 \\
 & intersp./$90\%$ & 11.2 & 15.8 & 0.99 & 56 & 8.1 & 10.9 & 0.49 & 97 \\
\addlinespace[3pt]
 & first-$41$/$50\%$ & 13.2 & 17.1 & 0.99 & 53 & 8.5 & 12.0 & 0.49 & 95 \\
 & first-$41$/$60\%$ & 12.9 & 17.2 & 1.00 & 50 & 8.5 & 12.2 & 0.50 & 89 \\
 & first-$41$/$70\%$ & 12.6 & 16.7 & 0.99 & 55 & 8.3 & 11.5 & 0.47 & 98 \\
 & first-$41$/$80\%$ & 12.5 & 16.5 & 0.99 & 55 & 8.4 & 11.5 & 0.46 & 98 \\
 & first-$41$/$90\%$ & 11.5 & 16.2 & 0.99 & 56 & 8.1 & 11.1 & 0.48 & 97 \\
\addlinespace[3pt]
 & last-$41$/$50\%$ & 13.2 & 16.9 & 1.00 & 47 & 8.7 & 11.9 & 0.52 & 86 \\
 & last-$41$/$60\%$ & 12.8 & 16.9 & 1.01 & 45 & 8.6 & 12.0 & 0.53 & 82 \\
 & last-$41$/$70\%$ & 12.8 & 16.4 & 1.00 & 50 & 8.7 & 11.4 & 0.51 & 91 \\
 & last-$41$/$80\%$ & 12.2 & 16.3 & 1.00 & 50 & 8.7 & 11.5 & 0.51 & 92 \\
 & last-$41$/$90\%$ & 11.9 & 15.9 & 1.00 & 51 & 8.2 & 11.1 & 0.54 & 92 \\
\bottomrule
\end{tabular}
\end{table}

\begin{table}[t]\ContinuedFloat
\centering
\caption{(Continued.)}
\scriptsize\setlength{\tabcolsep}{2.6pt}
\begin{tabular}{ll*{8}{>{\centering\arraybackslash}p{0.073\textwidth}}}
\toprule
& & \multicolumn{4}{c}{all pairs, $\mathrm{S}\neq\mathrm{T}$}
& \multicolumn{4}{c}{adjacency $\to$ adjacency, $\mathrm{S}\neq\mathrm{T}$}\\
\cmidrule(lr){3-6}\cmidrule(lr){7-10}
decoder, sentinels & source/target & \shortstack{median\\ n-RMSE} & \shortstack{mean\\ n-RMSE} & \shortstack{median\\ $r(\mathrm{S}\!\to\!\mathrm{T})$} & \shortstack{$r<1$\\ \%} & \shortstack{median\\ n-RMSE} & \shortstack{mean\\ n-RMSE} & \shortstack{median\\ $r(\mathrm{S}\!\to\!\mathrm{T})$} & \shortstack{$r<1$\\ \%}\\
\midrule
\multirow{17}{*}{linear, MinCorr} & intersp./$50\%$ & 12.5 & 15.7 & 1.05 & 43 & 7.4 & 9.7 & 0.59 & 65 \\
 & intersp./$60\%$ & 12.1 & 15.7 & 1.06 & 43 & 7.1 & 9.5 & 0.59 & 65 \\
 & intersp./$70\%$ & 12.3 & 15.4 & 1.04 & 43 & 7.3 & 9.4 & 0.59 & 66 \\
 & intersp./$80\%$ & 12.1 & 15.3 & 1.05 & 43 & 7.2 & 9.5 & 0.59 & 66 \\
 & intersp./$90\%$ & 11.3 & 14.8 & 1.04 & 44 & 7.3 & 9.0 & 0.61 & 67 \\
\addlinespace[3pt]
 & first-$41$/$50\%$ & 12.4 & 16.1 & 1.03 & 46 & 7.2 & 9.8 & 0.55 & 66 \\
 & first-$41$/$60\%$ & 12.3 & 16.1 & 1.03 & 46 & 7.1 & 9.6 & 0.56 & 66 \\
 & first-$41$/$70\%$ & 12.3 & 15.8 & 1.02 & 46 & 7.3 & 9.6 & 0.57 & 66 \\
 & first-$41$/$80\%$ & 12.1 & 15.7 & 1.03 & 46 & 7.3 & 9.7 & 0.55 & 66 \\
 & first-$41$/$90\%$ & 11.4 & 15.1 & 1.01 & 47 & 7.4 & 9.1 & 0.59 & 69 \\
\addlinespace[3pt]
 & last-$41$/$50\%$ & 11.9 & 15.5 & 1.04 & 39 & 7.0 & 9.4 & 0.58 & 66 \\
 & last-$41$/$60\%$ & 11.9 & 15.5 & 1.04 & 39 & 6.9 & 9.3 & 0.59 & 66 \\
 & last-$41$/$70\%$ & 11.9 & 15.2 & 1.03 & 39 & 7.2 & 9.2 & 0.59 & 70 \\
 & last-$41$/$80\%$ & 11.6 & 15.1 & 1.04 & 38 & 7.3 & 9.3 & 0.59 & 67 \\
 & last-$41$/$90\%$ & 10.8 & 14.6 & 1.03 & 40 & 7.0 & 8.8 & 0.62 & 70 \\
\addlinespace[8pt]
\multirow{17}{*}{linear, QR} & intersp./$50\%$ & 11.2 & 16.5 & 0.97 & 58 & 6.7 & 10.3 & 0.42 & 99 \\
 & intersp./$60\%$ & 11.0 & 16.5 & 0.98 & 59 & 6.5 & 10.1 & 0.44 & 99 \\
 & intersp./$70\%$ & 11.5 & 16.3 & 0.97 & 58 & 6.6 & 10.0 & 0.42 & 98 \\
 & intersp./$80\%$ & 11.5 & 16.2 & 0.97 & 59 & 6.7 & 10.1 & 0.41 & 99 \\
 & intersp./$90\%$ & 10.7 & 15.8 & 0.98 & 58 & 6.4 & 9.4 & 0.42 & 98 \\
\addlinespace[3pt]
 & first-$41$/$50\%$ & 12.4 & 17.0 & 0.98 & 54 & 6.9 & 10.3 & 0.40 & 98 \\
 & first-$41$/$60\%$ & 12.2 & 17.0 & 0.98 & 55 & 6.9 & 10.2 & 0.41 & 98 \\
 & first-$41$/$70\%$ & 12.3 & 16.8 & 0.98 & 55 & 7.0 & 10.1 & 0.40 & 98 \\
 & first-$41$/$80\%$ & 12.4 & 16.7 & 0.98 & 55 & 7.1 & 10.1 & 0.40 & 98 \\
 & first-$41$/$90\%$ & 11.4 & 16.3 & 0.99 & 54 & 6.5 & 9.4 & 0.41 & 98 \\
\addlinespace[3pt]
 & last-$41$/$50\%$ & 12.1 & 16.5 & 0.99 & 55 & 6.9 & 10.3 & 0.45 & 91 \\
 & last-$41$/$60\%$ & 11.9 & 16.5 & 0.99 & 55 & 6.8 & 10.1 & 0.46 & 91 \\
 & last-$41$/$70\%$ & 12.3 & 16.3 & 0.99 & 55 & 7.0 & 10.1 & 0.46 & 92 \\
 & last-$41$/$80\%$ & 12.4 & 16.2 & 0.99 & 55 & 7.0 & 10.2 & 0.45 & 92 \\
 & last-$41$/$90\%$ & 10.8 & 15.8 & 0.99 & 56 & 6.8 & 9.5 & 0.47 & 92 \\
\bottomrule
\end{tabular}
\end{table}

\newpage
\clearpage

\section{Coupled double-well dynamics initialized in the upper state}\label{si:dw}

\subsection{Model and numerical methods}

The double-well dynamics of an isolated node is bistable. We therefore
expect the equilibrium of the coupled double-well dynamics (DW) to depend on
the initial condition. In the numerical simulations reported in the main
text, we initialize every node in the lower state. In this section, we
instead initialize every node in the upper state, i.e.,
$x_i(0)=r_3=5$ for all $i$, and refer to the resulting dynamics as DW-up.
We verified that DW and DW-up converge to different equilibrium branches for
many nodes over a wide range of control-parameter values. In particular,
under DW-up, all $N$ nodes remain in the upper state at every value of $D$
considered, whereas under DW, nodes undergo a cascade in which they
gradually transition from the lower to the upper state as $D$ increases.

As in Figs.~\ref{fig:heat-combined} and~\ref{fig:ratio} of the main
text, we compute the n-RMSE and $r(\mathrm{S}\to\mathrm{T})$ for cases in
which DW-up is either the source or the target dynamics. All other aspects
of the evaluation follow the protocol used for those figures. In
particular, we use the interspersed $41/10$ split of the control-parameter
values on both the source- and target-dynamics sides.

\subsection{Results}

\begin{figure}[t]
\centering
\includegraphics[width=\textwidth]{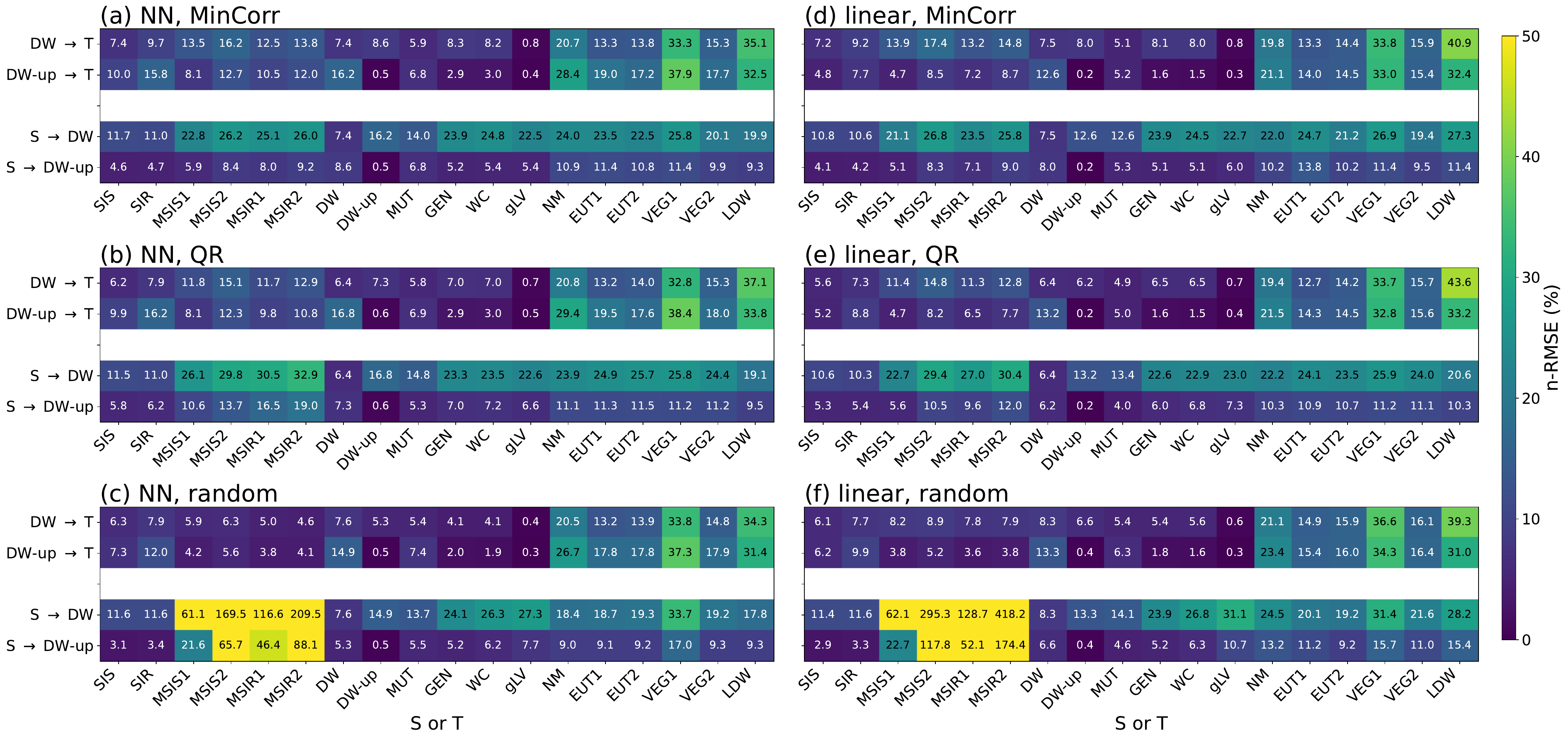}
\caption{n-RMSE of within-dynamics reconstruction and cross-dynamics
transfer when DW or DW-up is used as the source or target dynamics. As in
Fig.~\ref{fig:heat-combined} of the main text, each cell shows the average
over the $14$ networks.}
\label{fig:dwheat}
\end{figure}

\begin{figure}
\centering
\includegraphics[width=\textwidth]{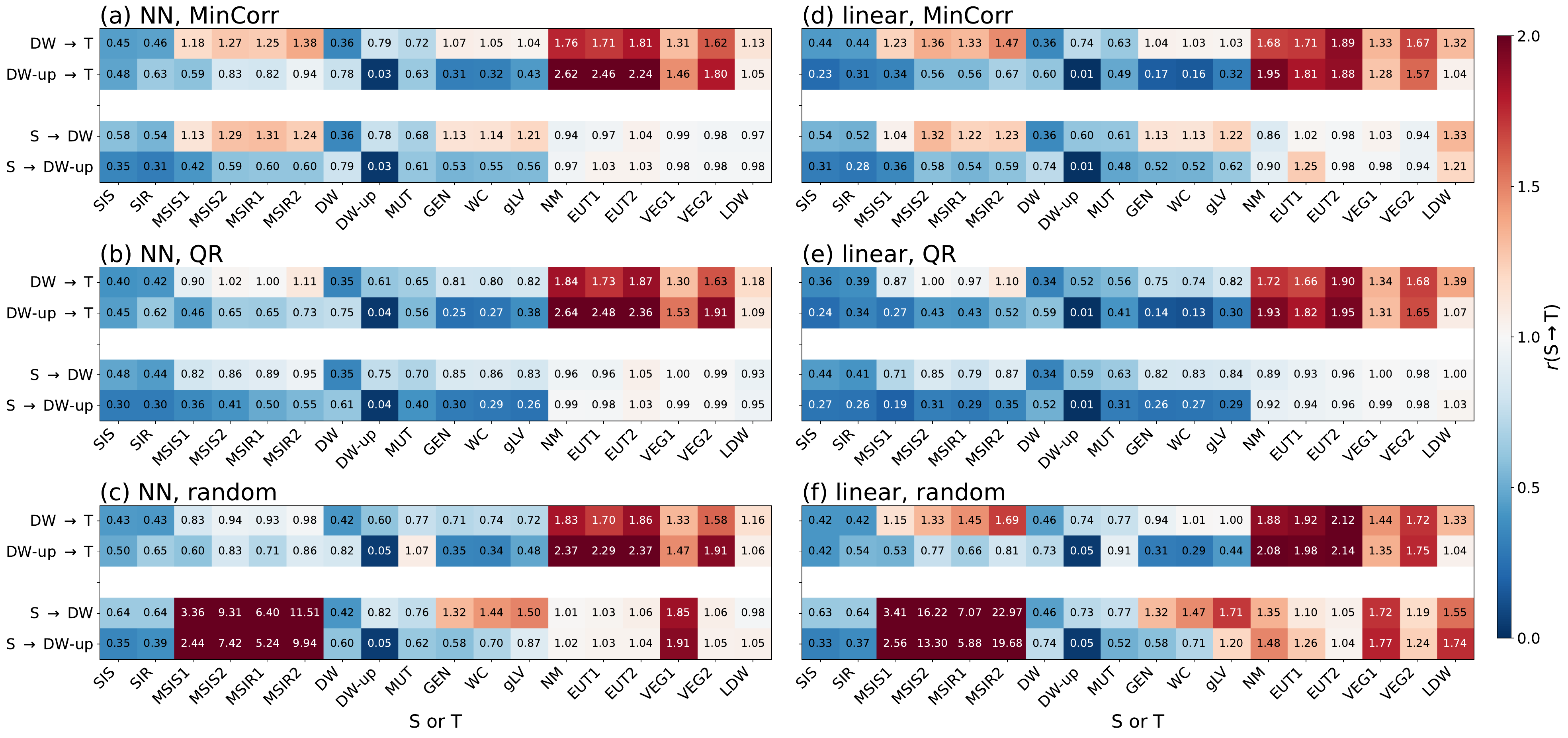}
\caption{Ratio $r(\mathrm{S}\to\mathrm{T})$ for within-dynamics
reconstruction and cross-dynamics transfer when DW or DW-up is used as the
source or target dynamics. As in Fig.~\ref{fig:ratio} of the main text,
each cell shows the n-RMSE averaged over the $14$ networks divided by the
mean-baseline n-RMSE averaged over the $14$ networks.}
\label{fig:dwratio}
\end{figure}

Figures~\ref{fig:dwheat} and~\ref{fig:dwratio} show the resulting
n-RMSE and $r(\mathrm{S}\to\mathrm{T})$, respectively. We show only the
source--target pairs involving DW or DW-up. In each panel, the first two
rows show the results when DW or DW-up is used as the source dynamics and
evaluated on each target dynamics. The last two rows show the results for
each source decoder evaluated on DW or DW-up as the target dynamics.

For within-dynamics reconstruction, both equilibrium branches are
reconstructed accurately. Averaged over the four combinations of decoder
architecture (NN or linear) and non-random sentinel-selection method
(MinCorr or QR-pivoted), the n-RMSE is $6.9\%$ for DW and $0.4\%$ for
DW-up. We exclude the random-sentinel results from these and the following
summary statistics because random sentinels perform extremely poorly for
some source dynamics, as shown, for example, in
Fig.~\ref{fig:heat-combined}; the same behavior occurs in
Figs.~\ref{fig:dwheat} and~\ref{fig:dwratio}.

For cross-dynamics transfer, we average over the same four combinations of
decoder architecture and sentinel-selection method and over all other
dynamics paired with DW or DW-up. Thus, for example, when DW is the source
dynamics, we average over all other target dynamics. As a source dynamics,
the average n-RMSE over the other $17$ target dynamics is $13.7\%$ for DW
and $13.1\%$ for DW-up. When the targets are restricted to the $11$
adjacency-coupled dynamics, these values are $8.9\%$ and $7.3\%$,
respectively. As a target dynamics, DW-up is substantially easier to
reconstruct: the average n-RMSE over the other $17$ source dynamics is
$21.6\%$ for DW and $8.6\%$ for DW-up. When the sources are restricted to
the adjacency-coupled dynamics, these values are $20.6\%$ and $7.3\%$,
respectively.

The ratio $r(\mathrm{S}\to\mathrm{T})$ yields the same ordering. For
within-dynamics reconstruction, the average ratio is $0.35$ for DW and
$0.02$ for DW-up. As a source dynamics, the average ratio is $1.12$ for DW
and $0.94$ for DW-up over the other $17$ target dynamics, and $0.86$ and
$0.47$, respectively, when the targets are restricted to the
adjacency-coupled dynamics. As a target dynamics, the corresponding values
are $0.91$ for DW and $0.63$ for DW-up over the other $17$ source dynamics,
and $0.86$ and $0.43$, respectively, when the sources are restricted to the
adjacency-coupled dynamics. These results indicate that DW-up is both a
better source dynamics for constructing decoders and an easier target
dynamics to reconstruct than DW. More generally, different equilibrium
branches of multistable network dynamics can serve as viable source and
target dynamics in our framework.

\newpage
\clearpage

\section{Two-parameter families of dynamics}\label{si:2d}

In this section, we examine the reconstruction of network dynamics
governed by two control parameters, including cross-dynamics transfer from
and to the one-parameter families considered in the main text. We construct
two two-parameter families from models already used in this study.

\subsection{Models and numerical methods}

The first family, which we call MSIS2P, is the metapopulation SIS model
described in the main text, with the diffusion rate of infected individuals,
$D_{\rm I}$, introduced as a second control parameter; the infection rate,
$\beta$, remains the first control parameter. We retain the other parameter
values and the initial condition used for MSIS1 and MSIS2 in the main text.
MSMS1 and MSIS2 use $D_{\rm I}=1$ and $D_{\rm I}=0.2$, respectively. Here, we
vary $D_{\rm I}$ from $0.2$ to $1$ using eight equally spaced values. We
also use eight equally spaced values of $\beta$ in
$[0,\beta_{\max}]$. We determine $\beta_{\max}$ at $D_{\rm I}=1$ using the
procedure described in the main text. As for MSIS1 and MSIS2, the observable
is the density of infectious individuals.

The second family, which we call MUT2P, is based on the mutualistic
model (MUT) described in the main text, with the immigration rate, $B$,
introduced as a second control parameter; the coupling strength, $D$,
remains the first control parameter. We retain the other parameter values
and the initial condition used for MUT in the main text. We use eight
equally spaced values of $D$ in $[0,D_{\max}]$, where
$D_{\max}\langle k\rangle=10.26$, following the rule used in the main text.
We also use eight equally spaced values of $B$ in $[0.08,0.15]$, an interval
that contains the value $B=0.1$ used for MUT in the main text. This interval
is relatively narrow because it is approximately the widest one for which
the large regime shift observed as $D$ varies, suggestive of a saddle-node
bifurcation, remains within $[0,D_{\max}]$.

We simulate MSIS2P and MUT2P at $M=8\times8=64$ parameter pairs,
i.e., $(\beta,D_{\rm I})$ for MSIS2P and $(D,B)$ for MUT2P, on each of the
$14$ networks. For MSIS2P, at each fixed value of $D_{\rm I}$, we sweep
$\beta$ downward from $\beta_{\max}$, following the protocol used for MSIS2
in the main text. For MUT2P, at each fixed value of $B$, we sweep $D$
upward using a forward warm start.

Following the out-of-sample evaluation protocol used throughout this
study, we randomly hold out $13$ of the $64$ parameter pairs
($\approx20\%$) for evaluation and use the remaining $51$ pairs for
training or calibration, on both the source- and target-dynamics sides. As
in Figs.~\ref{fig:heat-combined} and~\ref{fig:ratio} of the main text and
Section~\ref{si:dw}, we compute the n-RMSE and
$r(\mathrm{S}\to\mathrm{T})$ for cases in which MSIS2P or MUT2P is used as
either the source or target dynamics. All other aspects of the analysis
follow the protocols used for those figures.

\subsection{Results}

\begin{figure}[t]
\centering
\includegraphics[width=\textwidth]{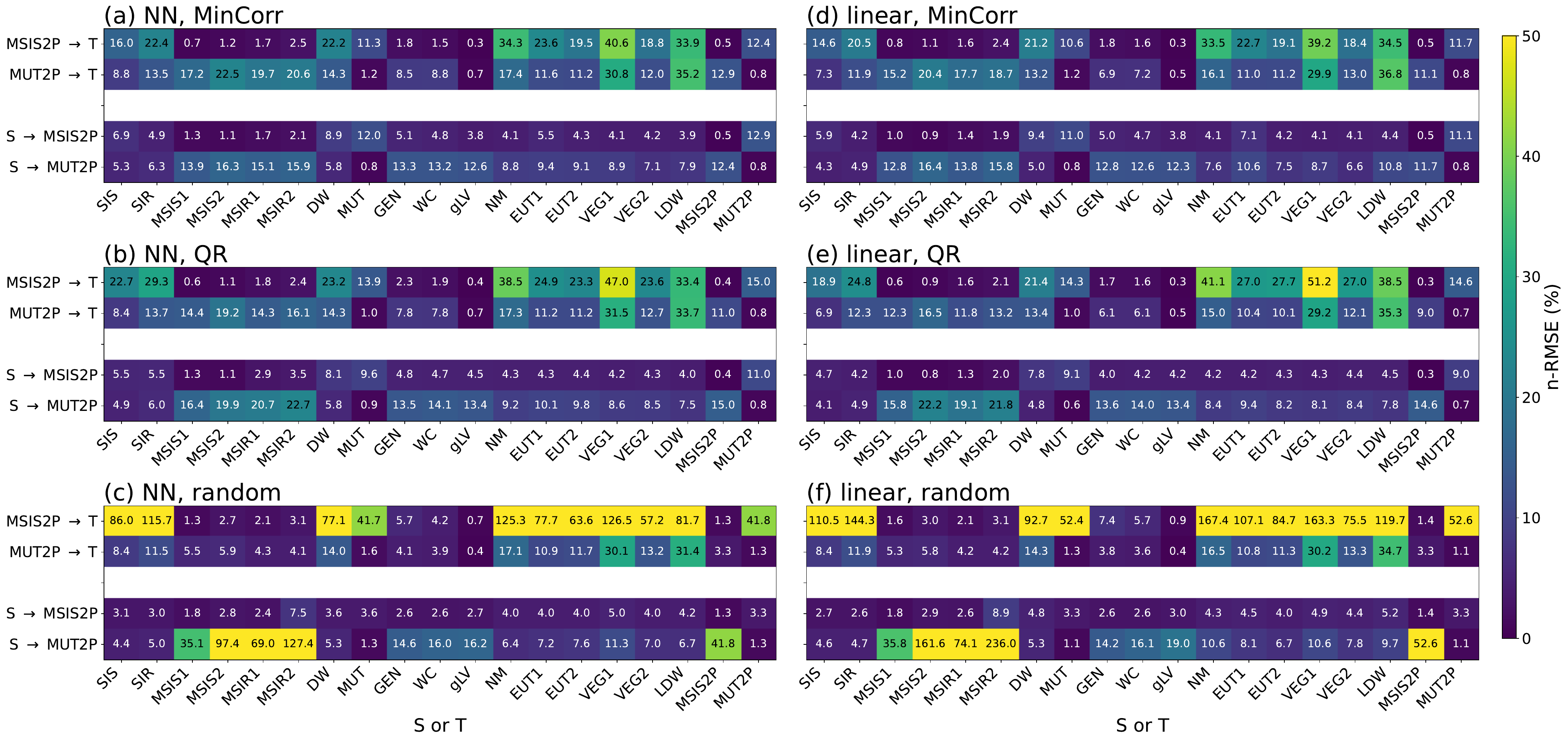}
\caption{n-RMSE of within-dynamics reconstruction and cross-dynamics
transfer when MSIS2P or MUT2P is used as the source or target dynamics. As
in Fig.~\ref{fig:heat-combined} of the main text, each cell shows the
average over the $14$ networks.}
\label{fig:twodheat}
\end{figure}

\begin{figure}[t]
\centering
\includegraphics[width=\textwidth]{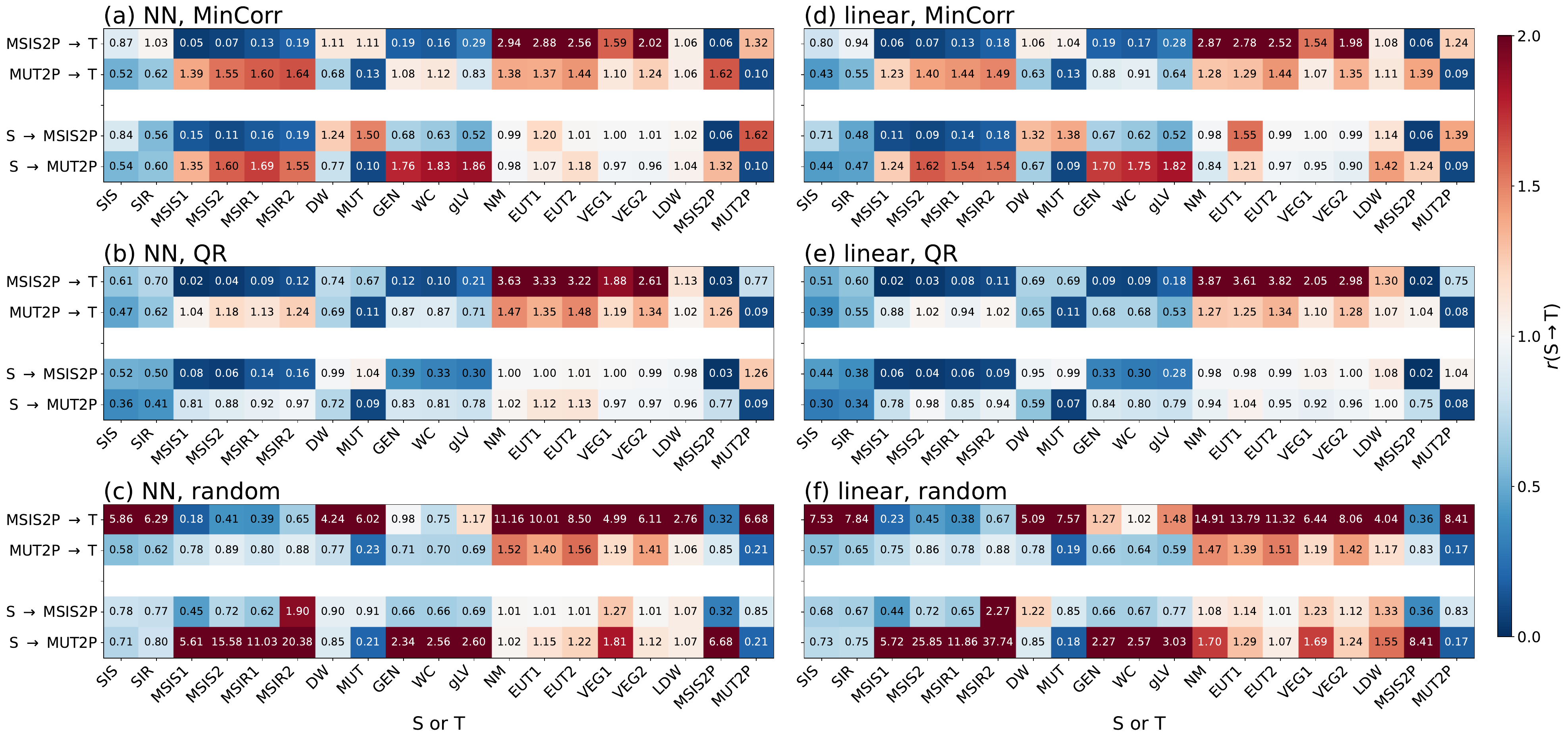}
\caption{Ratio $r(\mathrm{S}\to\mathrm{T})$ for within-dynamics
reconstruction and cross-dynamics transfer when MSIS2P or MUT2P is used as
the source or target dynamics. As in Fig.~\ref{fig:ratio} of the main text,
each cell shows the n-RMSE averaged over the $14$ networks divided by the
mean-baseline n-RMSE averaged over the $14$ networks.}
\label{fig:twodratio}
\end{figure}

Figures~\ref{fig:twodheat} and~\ref{fig:twodratio} show the n-RMSE and
$r(\mathrm{S}\to\mathrm{T})$, respectively. As in
Figs.~\ref{fig:dwheat} and~\ref{fig:dwratio}, we show only the
source--target dynamics pairs involving MSIS2P or MUT2P.

The two-parameter families behave similarly to the one-parameter
dynamics. To quantify this similarity, we compute the same averages as in
Section~\ref{si:dw}: we average over the four combinations of decoder
architecture and non-random sentinel-selection method and, for
cross-dynamics transfer, over all other dynamics paired with the family.
For within-dynamics reconstruction, the two two-parameter families are
reconstructed more accurately than a typical one-parameter dynamics. Their
n-RMSE values are $0.4\%$ for MSIS2P and $0.8\%$ for MUT2P, compared with
the reference value of $2.8\%$, obtained by averaging over the
one-parameter dynamics considered in the main text. As source dynamics, the
average n-RMSE over the other $18$ target dynamics is $15.8\%$ for MSIS2P
and $13.7\%$ for MUT2P, compared with the all-pair reference value of
$15.7\%$. When the targets are restricted to the $12$ adjacency-coupled
dynamics, the corresponding values are $8.3\%$ for MSIS2P, $10.8\%$ for
MUT2P, and $10.2\%$ for the reference, where the reference is averaged over
pairs of adjacency-coupled dynamics. As target dynamics, the average n-RMSE
over the other $18$ source dynamics is $4.8\%$ for MSIS2P and $10.6\%$ for
MUT2P, compared with the all-pair reference value of $15.7\%$. When the
sources are restricted to the adjacency-coupled dynamics, the corresponding
values are $5.0\%$, $11.6\%$, and $10.2\%$, respectively.

One caveat is that the n-RMSE is normalized by the target dynamics'
full range. For example, the full range of MSIS2P is larger than that of
MSIS1 because the two-dimensional parameter grid spans a broader set of
conditions, which may reduce the n-RMSE of MSIS2P. The ratio
$r(\mathrm{S}\to\mathrm{T})$ does not share this normalization issue. For
within-dynamics reconstruction, the average ratio is $0.04$ for MSIS2P and
$0.09$ for MUT2P, compared with the reference value of $0.15$. As source
dynamics, the average ratio over the other $18$ target dynamics is $1.11$
for MSIS2P and $1.01$ for MUT2P, compared with the all-pair reference value
of $1.15$. When the targets are restricted to the adjacency-coupled
dynamics, the corresponding values are $0.43$, $0.89$, and $0.60$. As
target dynamics, the average ratio over the other $18$ source dynamics is
$0.71$ for MSIS2P and $0.97$ for MUT2P, compared with the all-pair reference
value of $1.15$. When the sources are restricted to the adjacency-coupled
dynamics, the corresponding values are $0.55$, $0.95$, and $0.60$. For
MSIS2P, the results based on $r(\mathrm{S}\to\mathrm{T})$ are consistent
with those based on the n-RMSE, with performance at least as good as the
reference in every comparison. For MUT2P, the two measures agree when all
dynamics pairs are considered. When the partner dynamics is restricted to
the adjacency-coupled class, however, MUT2P has a larger (i.e., worse)
ratio than the reference. Nevertheless, its average ratio remains below
one, indicating better performance than the mean-prediction baseline.
Furthermore, Figs.~\ref{fig:twodratio}(a), (b), (d), and (e) suggest that
QR-pivoted sentinels substantially outperform MinCorr sentinels for both
MSIS2P and MUT2P, suggesting a direction for future investigation. Overall,
these results suggest that our framework generalizes to two-parameter
families of nonlinear network dynamics without substantial difficulty.

\newpage
\clearpage

\section{Variability in reconstruction performance across networks and
random factors}\label{si:var}

In Figs.~\ref{fig:heat-combined} and~\ref{fig:ratio} of the main text,
we reported the n-RMSE and $r(\mathrm{S}\to\mathrm{T})$, respectively,
averaged over the $14$ networks. In this section, we examine the variability
of these quantities across networks and across the random factors in our
simulation and evaluation protocols.

\subsection{Random factors and their variation}

We vary four random factors in this section. First, we vary the random
seed used to initialize the shallow NN decoder. Second, under the
interspersed splitting protocol, we select $41$ of the $51$
control-parameter values uniformly at random and use the same selection on
the source- and target-dynamics sides, as in the main text. Third, for random sentinel selection, we select
$n$ sentinel nodes uniformly at random from the $N$ nodes. Fourth, five
dynamical systems have heterogeneous node-level parameters: $r_i$ in NM,
$b_i$ in EUT1 and EUT2, and $h_i$ in VEG1 and VEG2. For each of these
dynamics, we draw the relevant parameter independently and uniformly at
random for every node.

For each of the six combinations of decoder architecture and
sentinel-selection method and each source--target dynamics pair, we first
compute the standard deviation of the n-RMSE across the $14$ networks. We
then compute the standard deviation while varying both the network and one
of the four random factors. For each network, we compute the n-RMSE using
$R$ different random-number seeds for the selected factor, yielding
$14R$ samples. We compare the standard deviation over these samples with
the reference standard deviation obtained when only the network varies. A
substantially larger standard deviation would indicate that the additional
random factor contributes appreciably to the variability of the mean values
shown in Fig.~\ref{fig:heat-combined} of the main text. We set $R=20$. For
the heterogeneous node parameters of the five relevant dynamics, we use
only the $11$ smallest networks because of the computational cost. We
compute the corresponding network-only reference over the same $11$
networks to ensure comparability.

\subsection{Results}

Tables~\ref{tab:std-nrmse} and~\ref{tab:std-ratio} report, for three
subsets of source--target cells, the median per-cell standard deviation.
These results indicate that most of the variability in the n-RMSE and
$r(\mathrm{S}\to\mathrm{T})$ arises from differences among the $14$
networks rather than from the other random factors. For example, the
standard deviation of either measure over the $280$ samples in which both
the network and NN initialization vary is similar to that computed over the
$14$ networks with a single NN initialization. The exceptions are the
choice of random sentinels and the choice of random node parameters when the
decoder is combined with random sentinel selection. Because
Figs.~\ref{fig:heat-combined} and~\ref{fig:ratio} of the main text show that
random sentinels perform worse than MinCorr and QR-pivoted sentinels, and
because we do not use random sentinels in the subsequent analyses, these
exceptions do not affect our main conclusions.

\begin{table}[tp]
\centering
\caption{Median per-cell standard deviation of the n-RMSE (\%) across
selected source--target dynamics pairs. Each combination of decoder
architecture and sentinel-selection method occupies three rows, which
correspond to the cell subsets specified in the ``cells'' column. By ``all'', we denote
all $17^2=289$ source--target dynamics pairs. By ``node param.'', we denote
the $145$ pairs in which the source or target is one of the five dynamics
with heterogeneous node parameters. This subset enables comparison between
variability due to the network alone and that due jointly to the network and
node-parameter realizations. By ``adjacency'', we denote the $11^2=121$ pairs in
which both the source and target dynamics are adjacency coupled; these
dynamics constitute the transfer class examined in
Section~\ref{sec:wccnet} of the main text. The ``network only'' column is
the reference condition, in which only the network varies. In each remaining
column, both the network and one additional random factor vary. ``$+$ NN
init.'' additionally varies the initial weights of the shallow NN decoder.
``$+$ splits'' additionally varies the interspersed division of the
control-parameter values into training and held-out sets, using the same
division on the source- and target-dynamics sides. ``$+$ sentinels'' additionally varies
the uniformly random sentinel set. ``$+$ node param.'' additionally varies
the heterogeneous node parameters of NM, EUT1, EUT2, VEG1, and VEG2. A dash
indicates that the combination does not apply.}
\label{tab:std-nrmse}
\small
\begin{tabular}{ll*{5}{>{\centering\arraybackslash}p{0.098\textwidth}}}
\toprule
decoder, sentinels & cells & \shortstack{network\\ only}
& \shortstack{$+$ NN\\ init.} & \shortstack{$+$\\ splits}
& \shortstack{$+$\\ sentinels} & \shortstack{$+$ node\\ param.}\\
\midrule
NN, MinCorr & all & 5.14 & 4.93 & 5.74 & -- & -- \\
 & node param. & 5.10 & -- & -- & -- & 4.78 \\
 & adjacency & 3.99 & 3.90 & 4.02 & -- & -- \\
\addlinespace
NN, QR & all & 5.14 & 4.95 & 5.64 & -- & -- \\
 & node param. & 5.33 & -- & -- & -- & 5.00 \\
 & adjacency & 3.37 & 3.53 & 3.44 & -- & -- \\
\addlinespace
NN, random & all & 5.51 & 5.43 & 5.77 & 7.53 & -- \\
 & node param. & 6.19 & -- & -- & -- & 7.08 \\
 & adjacency & 3.56 & 3.47 & 3.99 & 3.99 & -- \\
\addlinespace
linear, MinCorr & all & 5.42 & -- & 5.98 & -- & -- \\
 & node param. & 5.46 & -- & -- & -- & 5.11 \\
 & adjacency & 4.03 & -- & 4.00 & -- & -- \\
\addlinespace
linear, QR & all & 5.43 & -- & 5.78 & -- & -- \\
 & node param. & 5.65 & -- & -- & -- & 5.35 \\
 & adjacency & 3.48 & -- & 3.54 & -- & -- \\
\addlinespace
linear, random & all & 7.34 & -- & 8.30 & 8.80 & -- \\
 & node param. & 7.52 & -- & -- & -- & 8.23 \\
 & adjacency & 3.75 & -- & 4.14 & 4.34 & --\\
\bottomrule
\end{tabular}
\end{table}

\begin{table}[t]
\centering
\caption{Median per-cell standard deviation of
$r(\mathrm{S}\to\mathrm{T})$ across selected source--target dynamics pairs.
The rows, columns, and all other conventions are the same as in
Table~\ref{tab:std-nrmse}.}
\label{tab:std-ratio}
\small
\begin{tabular}{ll*{5}{>{\centering\arraybackslash}p{0.098\textwidth}}}
\toprule
decoder, sentinels & cells & \shortstack{network\\ only}
& \shortstack{$+$ NN\\ init.} & \shortstack{$+$\\ splits}
& \shortstack{$+$\\ sentinels} & \shortstack{$+$ node\\ param.}\\
\midrule
NN, MinCorr & all & 0.261 & 0.259 & 0.301 & -- & -- \\
 & node param. & 0.199 & -- & -- & -- & 0.172 \\
 & adjacency & 0.321 & 0.318 & 0.331 & -- & -- \\
\addlinespace
NN, QR & all & 0.207 & 0.198 & 0.216 & -- & -- \\
 & node param. & 0.157 & -- & -- & -- & 0.152 \\
 & adjacency & 0.217 & 0.211 & 0.225 & -- & -- \\
\addlinespace
NN, random & all & 0.229 & 0.219 & 0.260 & 0.467 & -- \\
 & node param. & 0.250 & -- & -- & -- & 0.413 \\
 & adjacency & 0.226 & 0.215 & 0.229 & 0.289 & -- \\
\addlinespace
linear, MinCorr & all & 0.280 & -- & 0.305 & -- & -- \\
 & node param. & 0.275 & -- & -- & -- & 0.241 \\
 & adjacency & 0.305 & -- & 0.305 & -- & -- \\
\addlinespace
linear, QR & all & 0.221 & -- & 0.220 & -- & -- \\
 & node param. & 0.206 & -- & -- & -- & 0.201 \\
 & adjacency & 0.193 & -- & 0.208 & -- & -- \\
\addlinespace
linear, random & all & 0.457 & -- & 0.485 & 0.806 & -- \\
 & node param. & 0.319 & -- & -- & -- & 0.474 \\
 & adjacency & 0.346 & -- & 0.339 & 0.347 & --\\
\bottomrule
\end{tabular}
\end{table}

\begin{figure}[htbp]
\centering
\includegraphics[width=\textwidth]{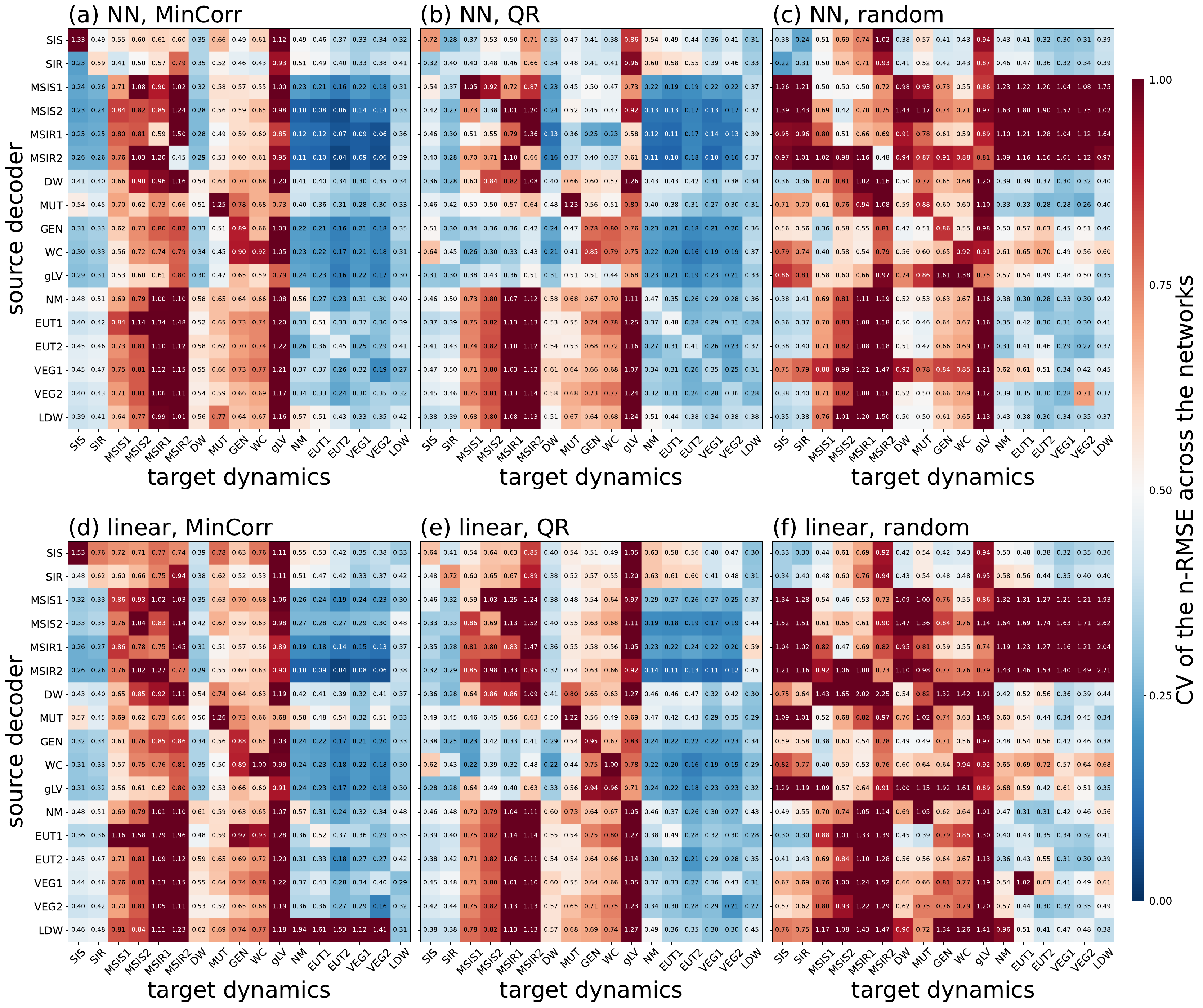}
\caption{Coefficient of variation (CV) of the n-RMSE across the $14$
networks for each source--target dynamics pair. Each cell shows the
standard deviation of the n-RMSE across the networks divided by its mean
across the networks.}
\label{fig:cv}
\end{figure}

Based on these results, we next focus on variability in reconstruction
accuracy across networks. To quantify this variability for each
source--target dynamics pair, Fig.~\ref{fig:cv} shows the coefficient of
variation (CV) of the n-RMSE, defined as its standard deviation across the
$14$ networks divided by its mean across the same networks. The denominator
is therefore the value shown in Fig.~\ref{fig:heat-combined} of the main
text. Averaged over the four combinations with non-random sentinels shown in
Figs.~\ref{fig:cv}(a), (b), (d), and (e), the mean and standard deviation
of the CV across the cells are $0.55$ and $0.29$, respectively. Thus, for a
given source--target dynamics pair, reconstruction performance varies
appreciably across networks. Nevertheless, the CV values are generally
moderate, suggesting that few, if any, networks yield catastrophically
poorer performance than the others. The CV tends to be small when the
target dynamics is Laplacian-coupled, corresponding to the final six
columns of each panel in Fig.~\ref{fig:cv}. As reported in the main text,
reconstruction of Laplacian-coupled target dynamics is generally poor;
the present results indicate that this poor performance varies relatively
little across networks.

\begin{figure}[htbp]
\centering
\includegraphics[width=\textwidth]{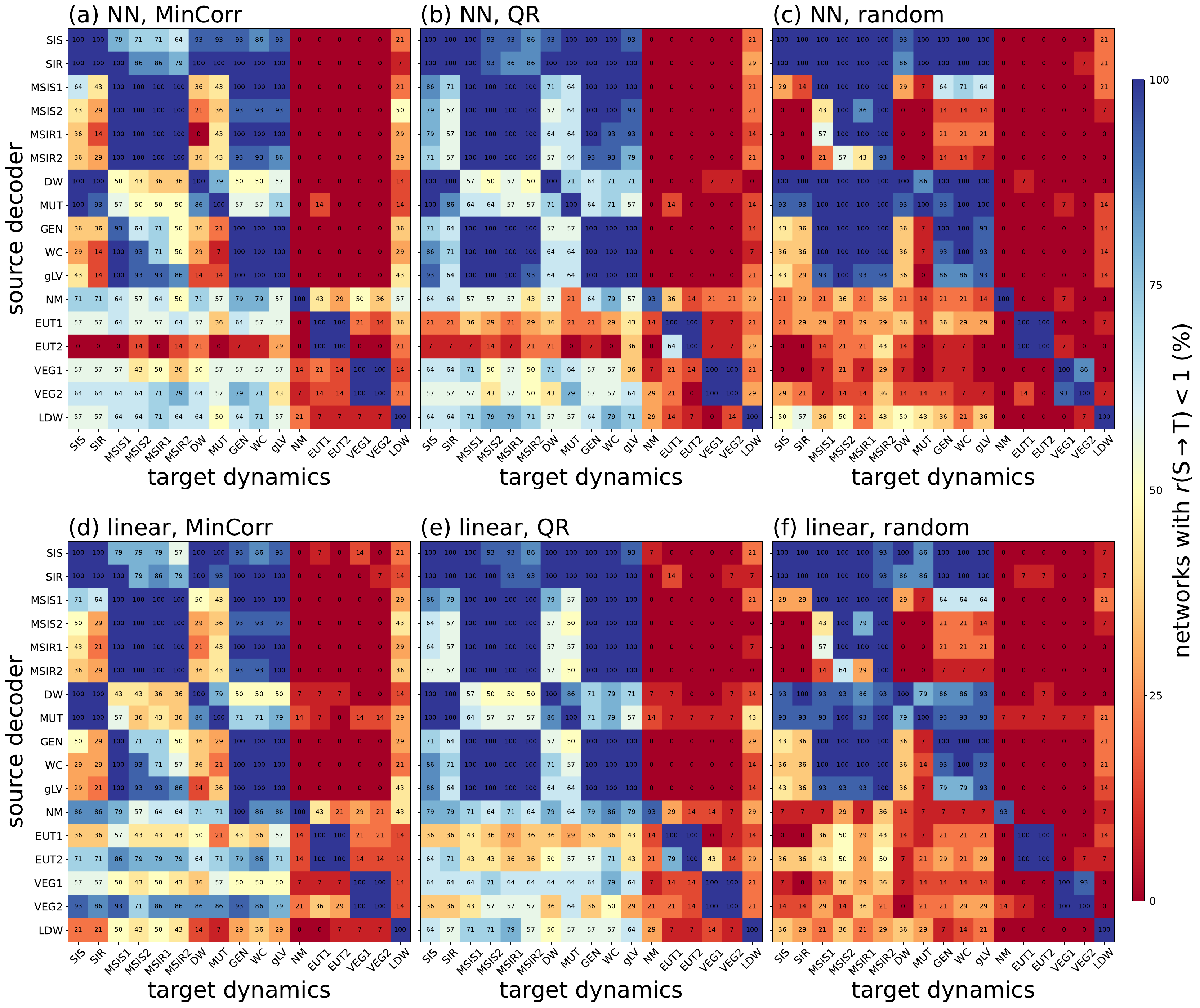}
\caption{Percentage of the $14$ networks for which the decoder
outperforms the mean-prediction baseline, i.e.,
$r(\mathrm{S}\to\mathrm{T})<1$, for each source--target dynamics pair.}
\label{fig:frac}
\end{figure}

Figure~\ref{fig:frac} shows, for each source--target dynamics pair, the
percentage of the $14$ networks for which
$r(\mathrm{S}\to\mathrm{T})<1$, i.e., the decoder outperforms the
mean-prediction baseline. These results are consistent with the values of
$r(\mathrm{S}\to\mathrm{T})$ averaged over the $14$ networks, which are shown in
Fig.~\ref{fig:ratio} of the main text. When the target dynamics are
restricted to the adjacency-coupled class, as in the later analyses of the
main text, some source dynamics, particularly SIS and SIR, yield
$r(\mathrm{S}\to\mathrm{T})<1$ for a large majority of the networks. This
result is consistent with Section~\ref{sec:decoder-ranking}, in which SIS
and SIR rank relatively highly as source dynamics and their decoders
reconstruct other adjacency-coupled target dynamics reasonably accurately.

We conclude that the quantities averaged over the $14$ networks in
Figs.~\ref{fig:heat-combined} and~\ref{fig:ratio} of the main text do not
fully represent the results for individual networks, although they are not
strongly affected by the other random factors examined here. Despite the
variation across networks, carefully selected source decoders reconstruct
adjacency-coupled target dynamics more accurately than the mean-prediction
baseline for a large majority of the networks. This finding motivates the
subsequent analyses of transfer classes in the main text.

\newpage
\clearpage

\section{Reconstruction under measurement noise}\label{si:noise}

All other analyses in this study use equilibrium values without
measurement error. In this section, we repeat the four cases examined in
Section~\ref{si:nodewise}, comprising two within-dynamics reconstructions
and two cross-dynamics transfers, after adding Gaussian noise to every
recorded equilibrium value.

For a dynamics $\mathrm{d}$, let $\mathrm{range}_i(\mathrm{d})$ denote
the dynamic range of node $i$ across the $M=51$ control-parameter values,
and let $R(\mathrm{d})$ denote the median of
$\mathrm{range}_i(\mathrm{d})$ across the $N$ nodes. At relative noise level
$\epsilon$, we add an independent Gaussian random variable with mean zero
and standard deviation $\epsilon R(\mathrm{d})$ to each of the $MN$ entries
of the equilibrium matrix for that dynamics. Thus, the noise standard
deviation is the same for all nodes and control-parameter values within a
given dynamics--network pair.

Using the resulting noisy equilibrium data, we select sentinel nodes
with the MinCorr method and train the decoders. This combination of
sentinel-selection method and decoder architecture is the same as that used
in Section~\ref{si:nodewise}. All other settings are the same as in the
main text: we use the shallow NN decoder without bias terms, the
interspersed $41/10$ split of the control-parameter values, and evaluation
on the $10$ held-out values. The noisy target data are used in every step
that requires target data, i.e., the input and output maps and the
mean-prediction baseline. However, we compare the reconstructed state with
the noise-free target equilibria and normalize the RMSE by the full range of
the noise-free target equilibria. Therefore, the n-RMSE quantifies how
accurately the true state is recovered from noisy observations. We generate $20$ independent noise
realizations at each value of $\epsilon$.

\begin{figure}[H]
\centering
\includegraphics[width=\textwidth]{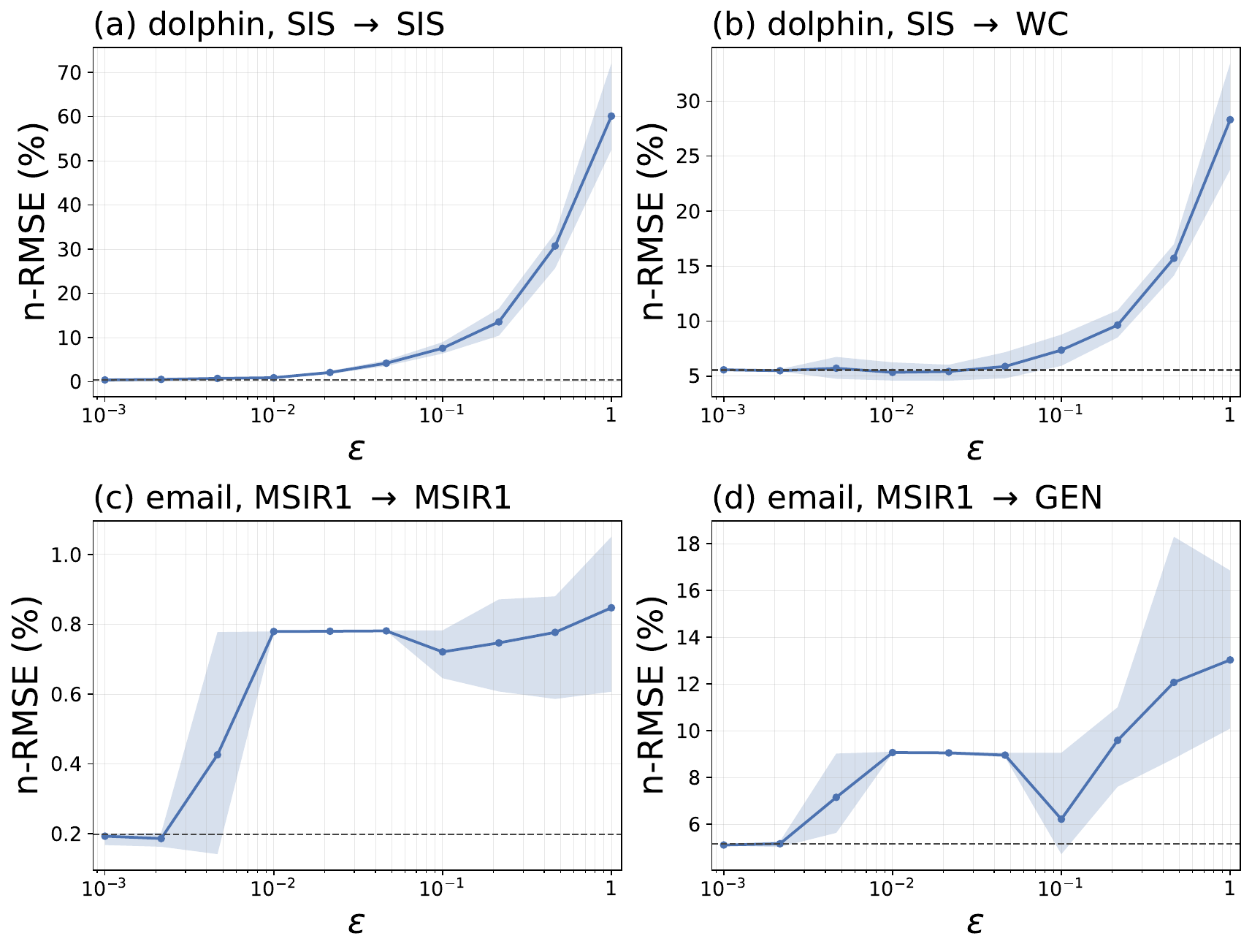}
\caption{n-RMSE as a function of the relative noise level,
$\epsilon$, for the four cases examined in Section~\ref{si:nodewise}.
(a) Dolphin network, with the SIS decoder evaluated on SIS. (b) Dolphin
network, with the SIS decoder evaluated on WC. (c) Email network, with the
MSIR1 decoder evaluated on MSIR1. (d) Email network, with the MSIR1 decoder
evaluated on GEN. The solid line shows the mean over $20$ independent noise
realizations, and the shaded region shows the $10$th--$90$th percentile
range. The dashed horizontal line indicates the noise-free value.}
\label{fig:noise-nrmse}
\end{figure}

\begin{figure}[H]
\centering
\includegraphics[width=\textwidth]{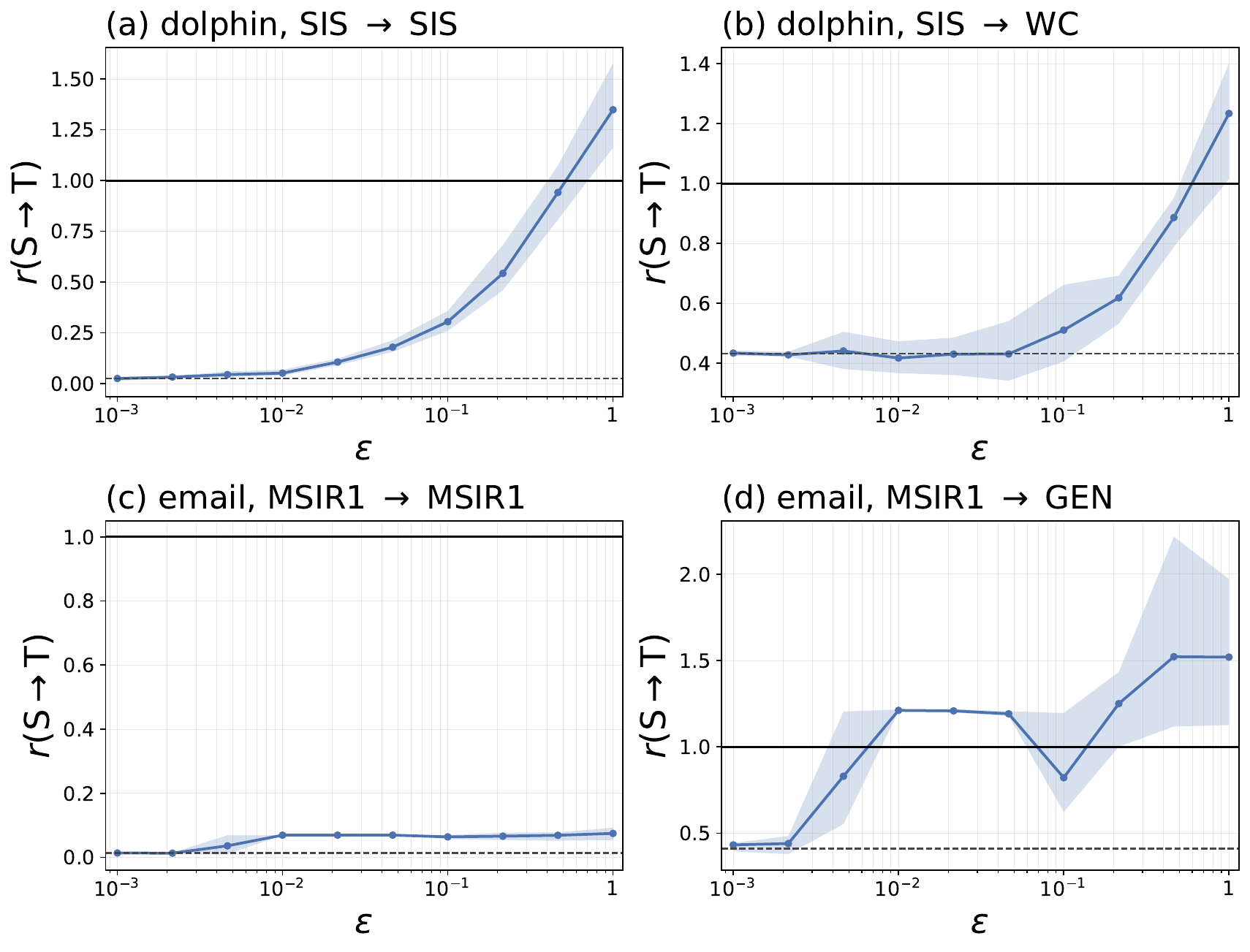}
\caption{Ratio $r(\mathrm{S}\to\mathrm{T})$ as a function of the
relative noise level, $\epsilon$. The four panels and graphical conventions
are the same as in Fig.~\ref{fig:noise-nrmse}. The solid horizontal line
indicates $r(\mathrm{S}\to\mathrm{T})=1$; above this value, the
mean-prediction baseline is more accurate than the decoder.}
\label{fig:noise-ratio}
\end{figure}

Figures~\ref{fig:noise-nrmse} and~\ref{fig:noise-ratio} show the
n-RMSE and $r(\mathrm{S}\to\mathrm{T})$, respectively, as functions of
$\epsilon$. We consider noise levels up to $\epsilon=1$, at which the noise
standard deviation equals the full dynamic range of the median node. The
reconstruction performance degrades gradually as the noise amplitude
increases and remains relatively close to its noise-free value up to
approximately $\epsilon=0.1$, except in Fig.~\ref{fig:noise-ratio}(d).
Although this analysis is limited to four cases, the results suggest that
our decoding framework is reasonably robust to measurement noise.

\newpage
\clearpage

\section{Robustness to the number of sentinel nodes}
\label{sec:c-sweep}

Our default choice for the number of sentinel nodes is
$n=\lfloor \ln N \rfloor$. In this section, we assess whether the main results depend
on this choice by carrying out complementary analyses.

\subsection{Effective rank of the equilibrium data across network
sizes}\label{si:c-rank}

We first ask why a number of sentinels of order $\ln N$ is sufficient.
Section~\ref{si:rank} showed that the equilibrium data for a network have
low effective ranks in most cases. If these effective ranks grow no faster
than $\ln N$ as $N$ increases, then choosing
$n=\lfloor\ln N\rfloor$ sentinels may be sufficient to capture the
variation in the equilibrium data that the decoder must reconstruct.

We use the four effective-rank measures for the centered equilibrium
matrix computed in Section~\ref{si:rank}. For each of the $14$ networks and
each effective-rank measure, we average the effective rank over the $11$
adjacency-coupled dynamics, which constitute the transferable class of
primary interest. The network sizes range from $N=62$ to $N=11558$.

\begin{figure}[H]
\centering
\includegraphics[width=\textwidth]{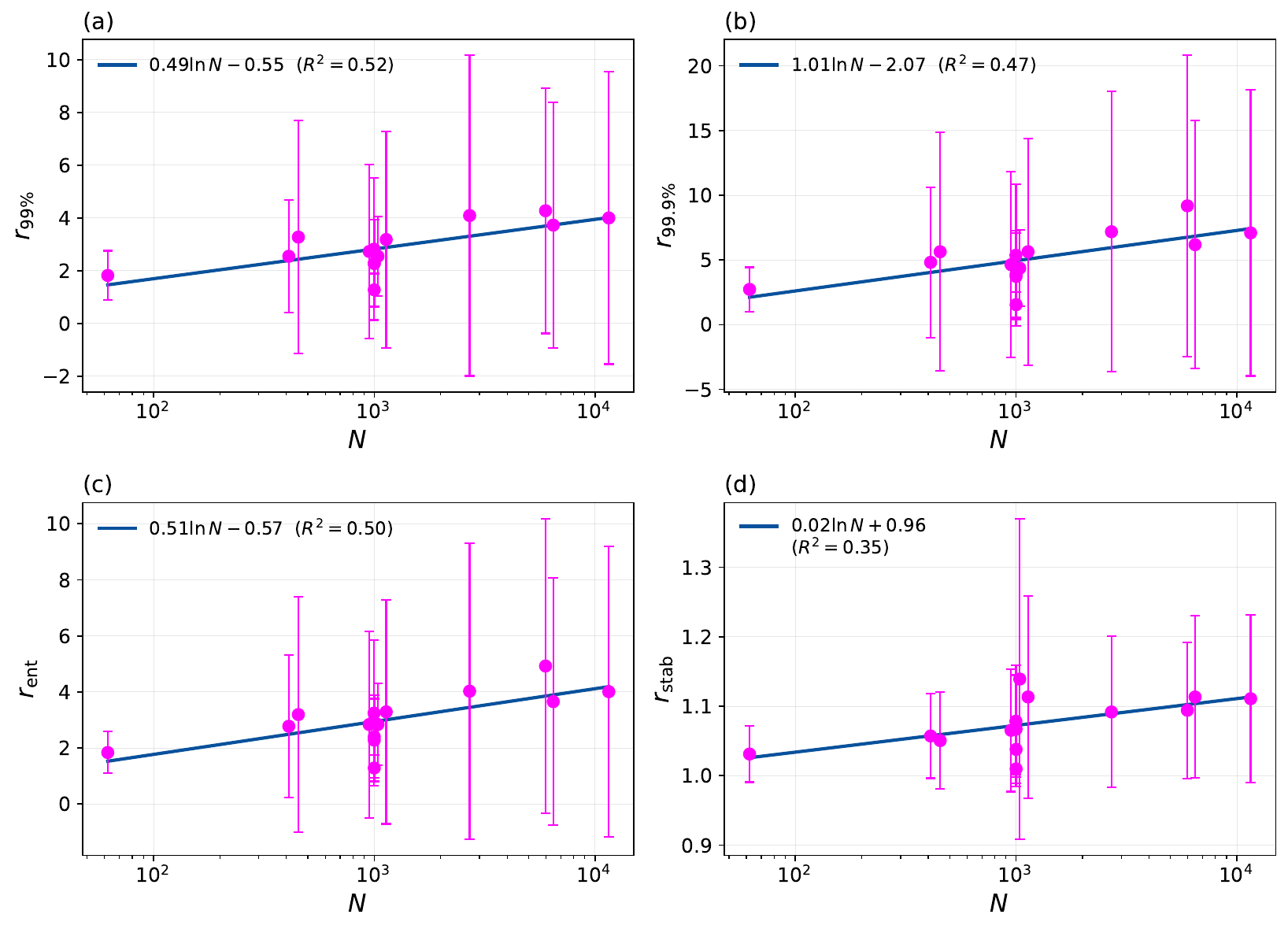}
\caption{Effective ranks of the centered equilibrium matrix for the
$14$ networks. (a) Energy-based effective rank $r_{99\%}$.
(b) Energy-based effective rank $r_{99.9\%}$. (c) Entropy effective rank.
(d) Stable rank. Each filled circle represents one network and shows
the mean over the $11$ adjacency-coupled dynamics; the error bars indicate
one standard deviation across these dynamics. The solid line shows the
least-squares fit as a linear function of $\ln N$. Each legend reports the
fitted relation and its coefficient of determination, $R^2$.}
\label{fig:si-ranks}
\end{figure}

Figure~\ref{fig:si-ranks} shows the results. All four effective-rank
measures increase with network size. A least-squares fit as a linear
function of $\ln N$ provides a moderate description of each measure, with
the coefficient of determination $R^2$ ranging from $0.35$ to $0.52$. The
fastest growth occurs for $r_{99.9\%}$ (Fig.~\ref{fig:si-ranks}(b)), for
which the fitted relation is
$r_{99.9\%}\approx1.01\ln N-2.07$ ($R^2=0.47$). Its fitted slope is close
to one, corresponding to growth at approximately the same rate as $\ln N$.
The energy-based rank $r_{99\%}$ and entropy effective rank increase at
approximately half this rate (Figs.~\ref{fig:si-ranks}(a) and (c)),
whereas the stable rank is nearly constant as $N$ increases
(Fig.~\ref{fig:si-ranks}(d)). These results provide no evidence that any of
the effective-rank measures grows faster than $\ln N$. Therefore, they
support $n=\lfloor\ln N\rfloor$ as a reasonable choice.

\subsection{Varying the number of sentinel nodes and network size
together}\label{si:c-ba}

The $14$ networks differ not only in the number of nodes but also in other structural
properties. Therefore, the analysis in Section~\ref{si:c-rank} does not
isolate the effect of network size. To do so, we generate
Barab\'asi--Albert networks~\cite{BarabasiAlbert1999Science-SI} of six sizes:
$N=32$ ($\approx10^{1.5}$), $100$, $316$ ($\approx10^{2.5}$), $1000$,
$3162$ ($\approx10^{3.5}$), and $10^4$. We initialize network growth from
a triangle, i.e., the complete graph on three nodes, and connect each
new node to $m=3$ existing nodes. Thus, the mean degree is
$\langle k\rangle\approx6$ regardless of $N$. We compute the equilibria of
each dynamics on each network and vary the number of sentinel nodes, $n$,
from $2$ to $10$ for every network size.

\begin{figure}[H]
\centering
\includegraphics[width=\textwidth]{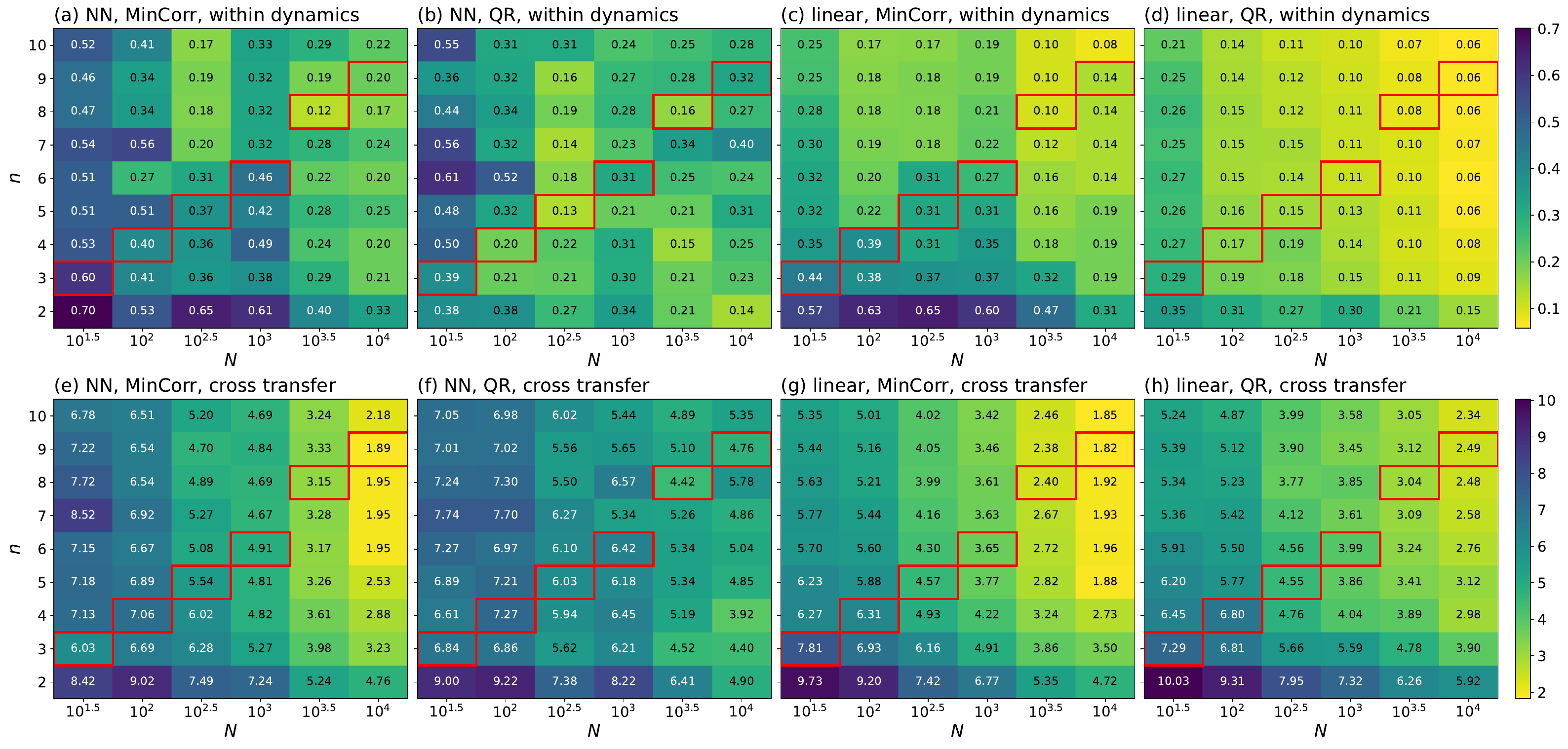}
\caption{Median n-RMSE over the $11$ adjacency-coupled dynamics on
Barab\'asi--Albert networks as a function of the number of nodes, $N$, and
the number of sentinel nodes, $n$. (a)--(d) Within-dynamics reconstruction,
with each cell showing the median over the $11$ dynamics. (e)--(h)
Cross-dynamics transfer, with each cell showing the median over the $110$
ordered pairs of distinct source and target dynamics among the $11$
dynamics. Panels (a)--(d) share one color scale, and panels (e)--(h) share
another. The cells with red outlines indicate $n=\lfloor\ln N\rfloor$.}
\label{fig:si-scale-nrmse}
\end{figure}

\begin{figure}[H]
\centering
\includegraphics[width=\textwidth]{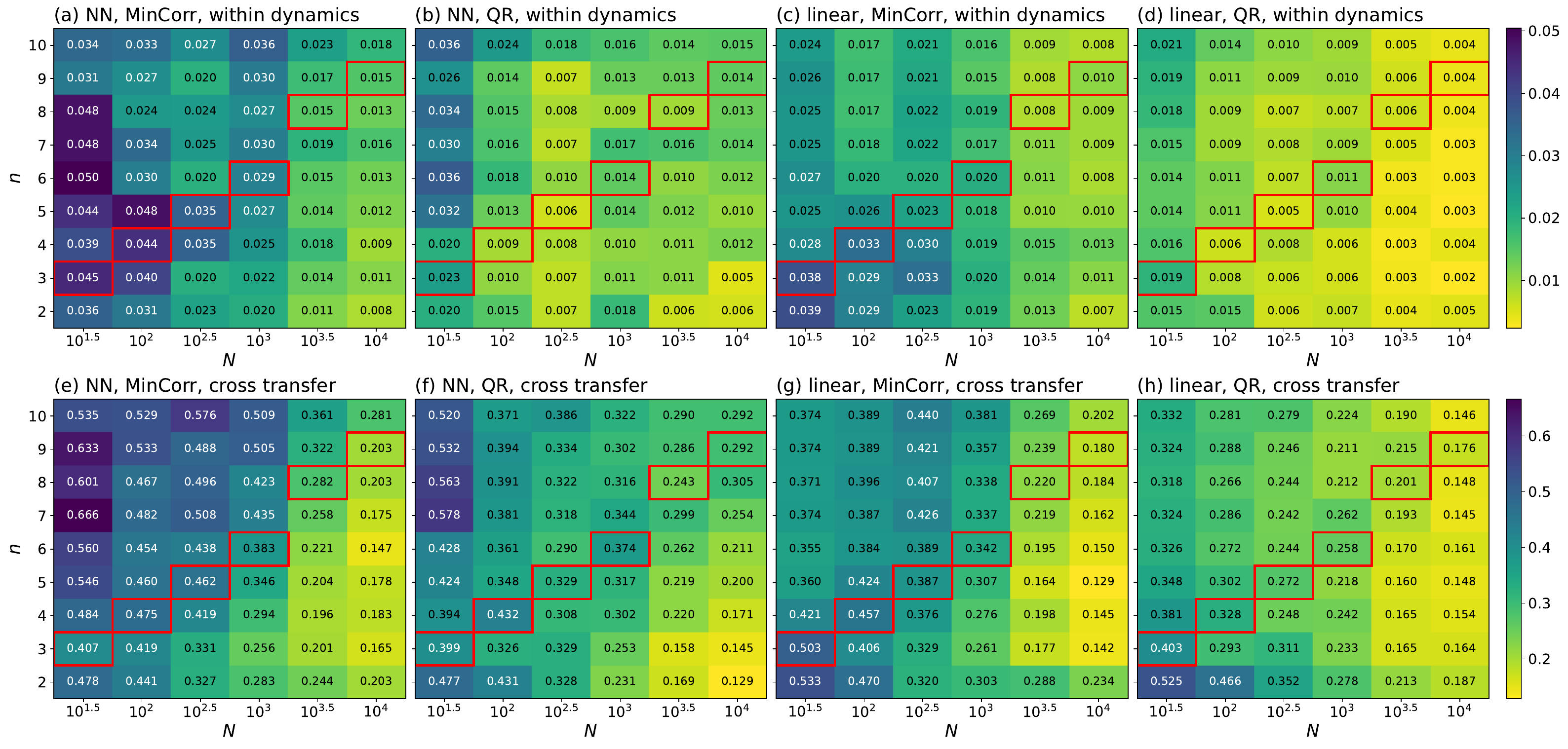}
\caption{Median $r(\mathrm{S}\to\mathrm{T})$ over the $11$
adjacency-coupled dynamics on Barab\'asi--Albert networks as a function of
the number of nodes, $N$, and the number of sentinel nodes, $n$. The
conventions are the same as in Fig.~\ref{fig:si-scale-nrmse}.}
\label{fig:si-scale-ratio}
\end{figure}

Figures~\ref{fig:si-scale-nrmse} and~\ref{fig:si-scale-ratio} show the
median n-RMSE and median $r(\mathrm{S}\to\mathrm{T})$, respectively, for
each combination of decoder architecture, non-random sentinel-selection
method, and reconstruction setting (within dynamics or cross dynamics).
Each panel shows the performance measure as a function of $N$ and $n$. For
within-dynamics reconstruction
(Figs.~\ref{fig:si-scale-nrmse}(a)--(d) and
\ref{fig:si-scale-ratio}(a)--(d)), each cell is the median over the $11$
adjacency-coupled dynamics. For cross-dynamics transfer
(Figs.~\ref{fig:si-scale-nrmse}(e)--(h) and
\ref{fig:si-scale-ratio}(e)--(h)), each cell is the median over the $110$
ordered pairs of distinct adjacency-coupled dynamics. We focus on these
$11$ dynamics because they form the principal transferable class. We
consider only non-random sentinel-selection methods because uniformly
random sentinels generally perform worse. The cells outlined in red
correspond to $n=\lfloor\ln N\rfloor$.

\begin{figure}[t]
\centering
\includegraphics[width=\textwidth]{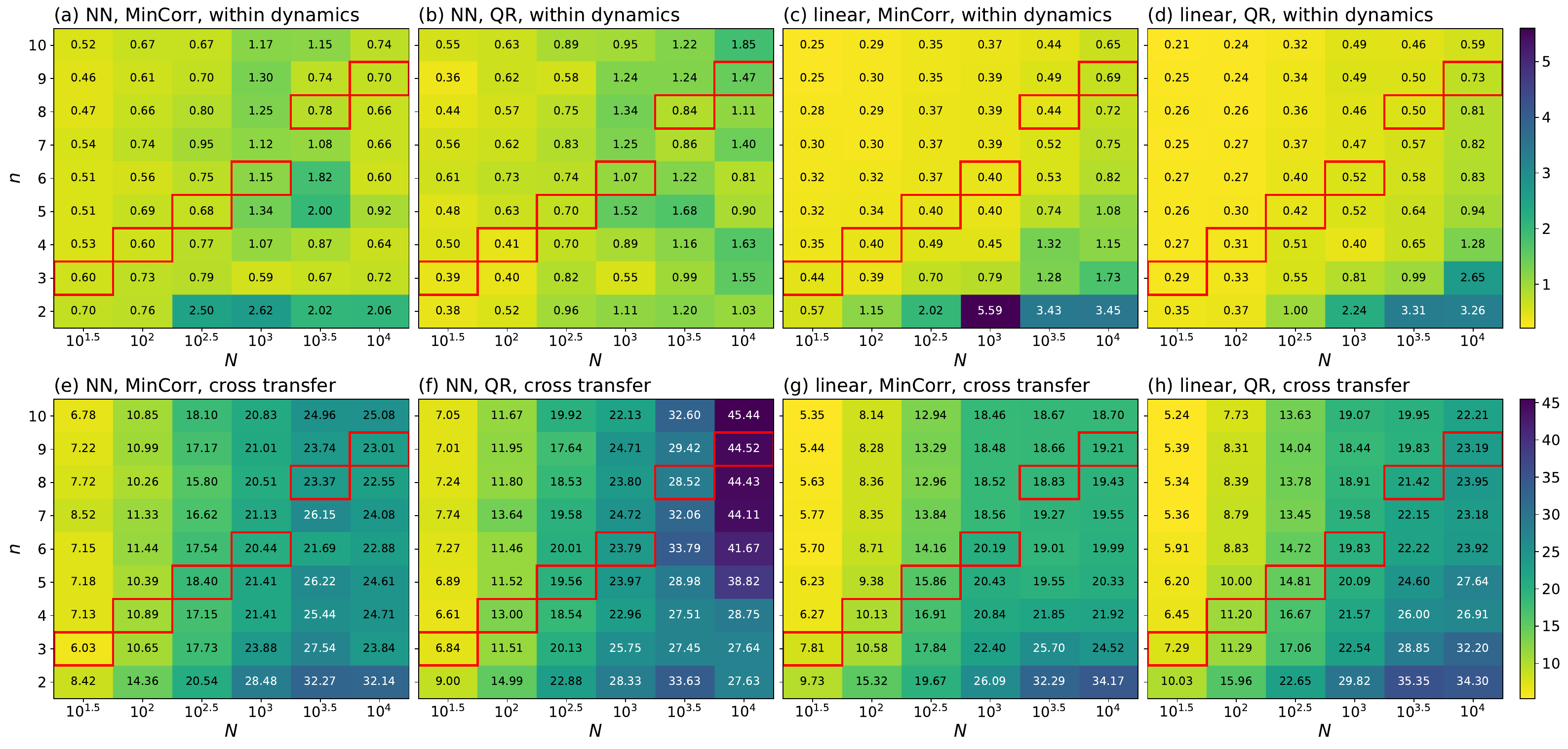}
\caption{RMSE normalized by the range of the target dynamics on the
smallest network, $N=32$, rather than by the target dynamics' range on each
network. All other conventions are the same as in
Fig.~\ref{fig:si-scale-nrmse}.}
\label{fig:si-scale-abs}
\end{figure}

The dominant pattern in both figures is horizontal rather than
diagonal: for a fixed number of sentinels, the n-RMSE decreases as $N$
increases. This result is counterintuitive because reconstructing
equilibria on a larger network using the same number of sentinels would
seem to be more difficult. The pattern arises mainly from the normalization
used in the n-RMSE. By definition, the n-RMSE divides the RMSE by the global
range of the target dynamics. On a Barab\'asi--Albert network, this range
is determined by the largest-degree nodes and, for some dynamics, increases
algebraically with $N$. For dynamics whose equilibrium values increase with
degree, the denominator therefore grows with network size, reducing the
n-RMSE. If the RMSE is instead normalized by the target-dynamics range on
the smallest network, $N=32$, the resulting measure increases with $N$ for
each fixed $n$ (Fig.~\ref{fig:si-scale-abs}).

\begin{figure}[H]
\centering
\includegraphics[width=\textwidth]{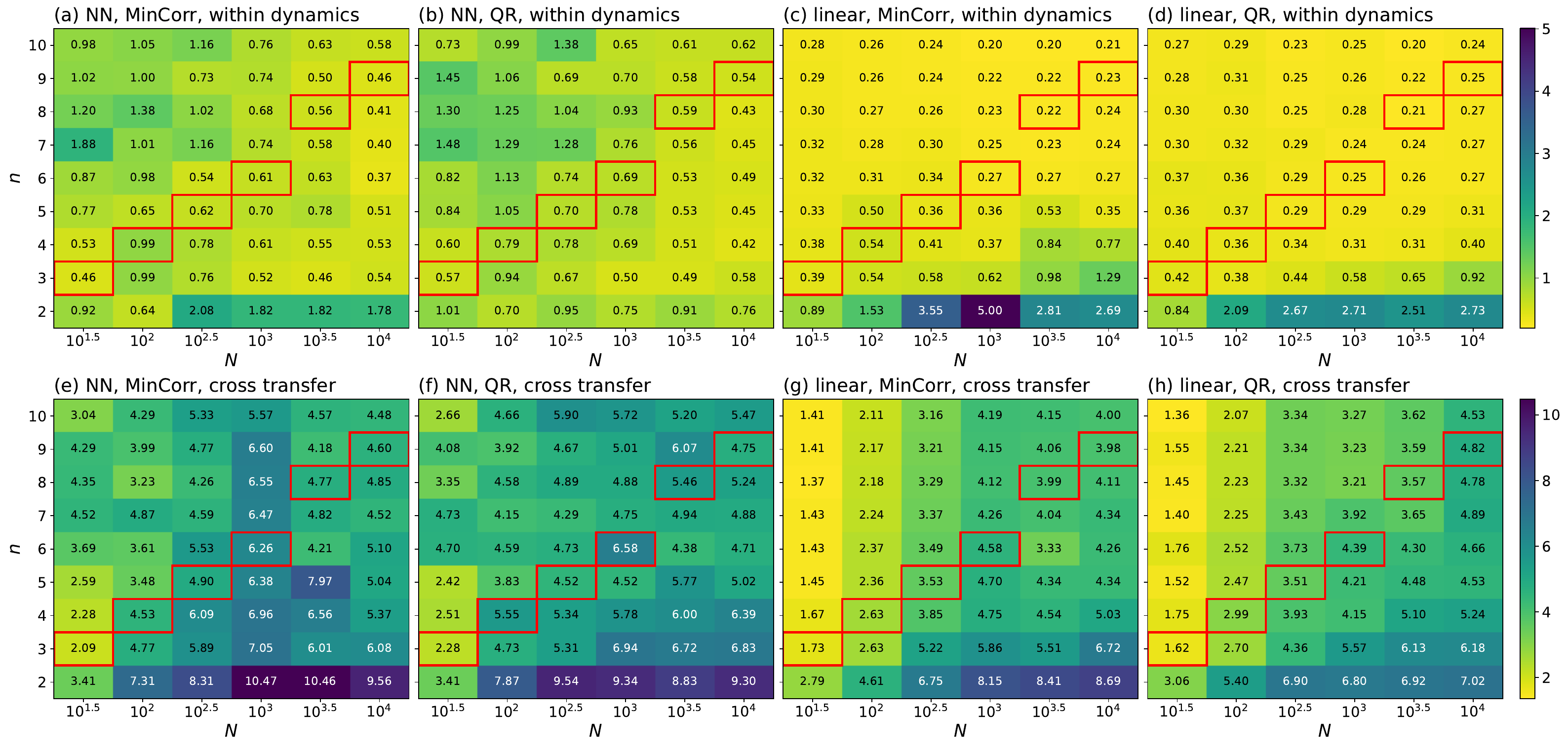}
\caption{Median n-RMSE over the four dynamics whose global range
remains approximately bounded as the network size increases, which are SIS, SIR, DW,
and MUT. All other conventions are the same as in
Fig.~\ref{fig:si-scale-nrmse}.}
\label{fig:si-scale-nrmse-sub}
\end{figure}

\begin{figure}[H]
\centering
\includegraphics[width=\textwidth]{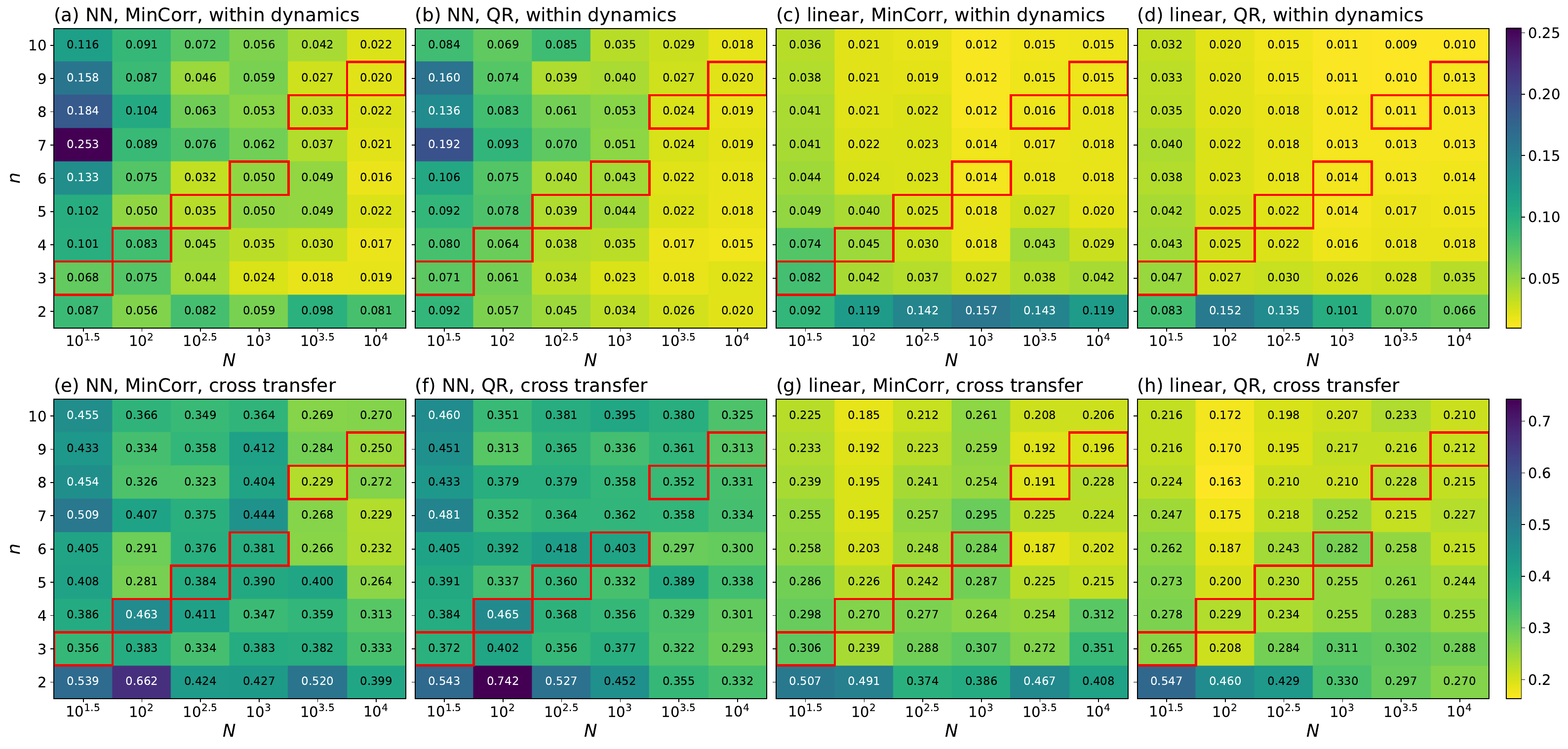}
\caption{Median $r(\mathrm{S}\to\mathrm{T})$ over the same four
bounded-range dynamics as in Fig.~\ref{fig:si-scale-nrmse-sub}. All other
conventions are the same as in Fig.~\ref{fig:si-scale-nrmse}.}
\label{fig:si-scale-ratio-sub}
\end{figure}

Given this normalization effect, we repeat the analyses in
Figs.~\ref{fig:si-scale-nrmse} and~\ref{fig:si-scale-ratio} using only
dynamics whose global range remains approximately bounded as the network
grows. We define this condition as an increase of less than a factor of
three over the entire range of $N$, which selects SIS, SIR, DW,
and MUT. Figures~\ref{fig:si-scale-nrmse-sub} and
\ref{fig:si-scale-ratio-sub} show the resulting n-RMSE and
$r(\mathrm{S}\to\mathrm{T})$, respectively. With $n$ fixed, the
cross-dynamics n-RMSE now increases with $N$
(Fig.~\ref{fig:si-scale-nrmse-sub}(e)--(h)). By contrast, the
within-dynamics n-RMSE
(Fig.~\ref{fig:si-scale-nrmse-sub}(a)--(d)), within-dynamics
$r(\mathrm{S}\to\mathrm{T})$
(Fig.~\ref{fig:si-scale-ratio-sub}(a)--(d)), and cross-dynamics
$r(\mathrm{S}\to\mathrm{T})$
(Fig.~\ref{fig:si-scale-ratio-sub}(e)--(h)) still decrease with $N$ at
fixed $n$. Thus, a genuine improvement with network size remains for these
three measures, although it is substantially weaker than in
Figs.~\ref{fig:si-scale-nrmse} and~\ref{fig:si-scale-ratio}.

Except for the cross-dynamics n-RMSE, the performance measures in the
red-outlined cells of Figs.~\ref{fig:si-scale-nrmse-sub} and
\ref{fig:si-scale-ratio-sub} do not systematically worsen as $N$ increases.
For the cross-dynamics n-RMSE
(Fig.~\ref{fig:si-scale-nrmse-sub}(e)--(h)), the unusually small values at
$N=10^{1.5}$ and $N=10^2$ create an apparent increase along
$n=\lfloor\ln N\rfloor$. Nevertheless, over
$N\in[10^{2.5},10^4]$, the n-RMSE along
$n=\lfloor\ln N\rfloor$ is approximately constant. Although this analysis
is limited to a particular network model with approximately constant mean
degree, the results support the sufficiency of
$n=\lfloor\ln N\rfloor$ sentinels for this family of networks.

\newpage
\clearpage

\subsection{Varying the number of sentinel nodes on the $14$
networks}\label{si:c-fixed}

Finally, we varied the number of sentinels around $n=\lfloor \ln N \rfloor$ 
and measured reconstruction performance for the leading source
decoders. Specifically, for each network we set
\begin{equation}
n = \max\left\{1,\left\lfloor c\lfloor \ln N \rfloor\right\rfloor\right\},
\end{equation}
where $c \in [0.25,2]$. The lower endpoint, $c=0.25$, is the smallest
value for which the dolphin network, the smallest network in our data
set ($N=62$, $\lfloor \ln N \rfloor=4$), has at least one sentinel
node.

\begin{figure}[H]
\centering
\includegraphics[width=\textwidth]{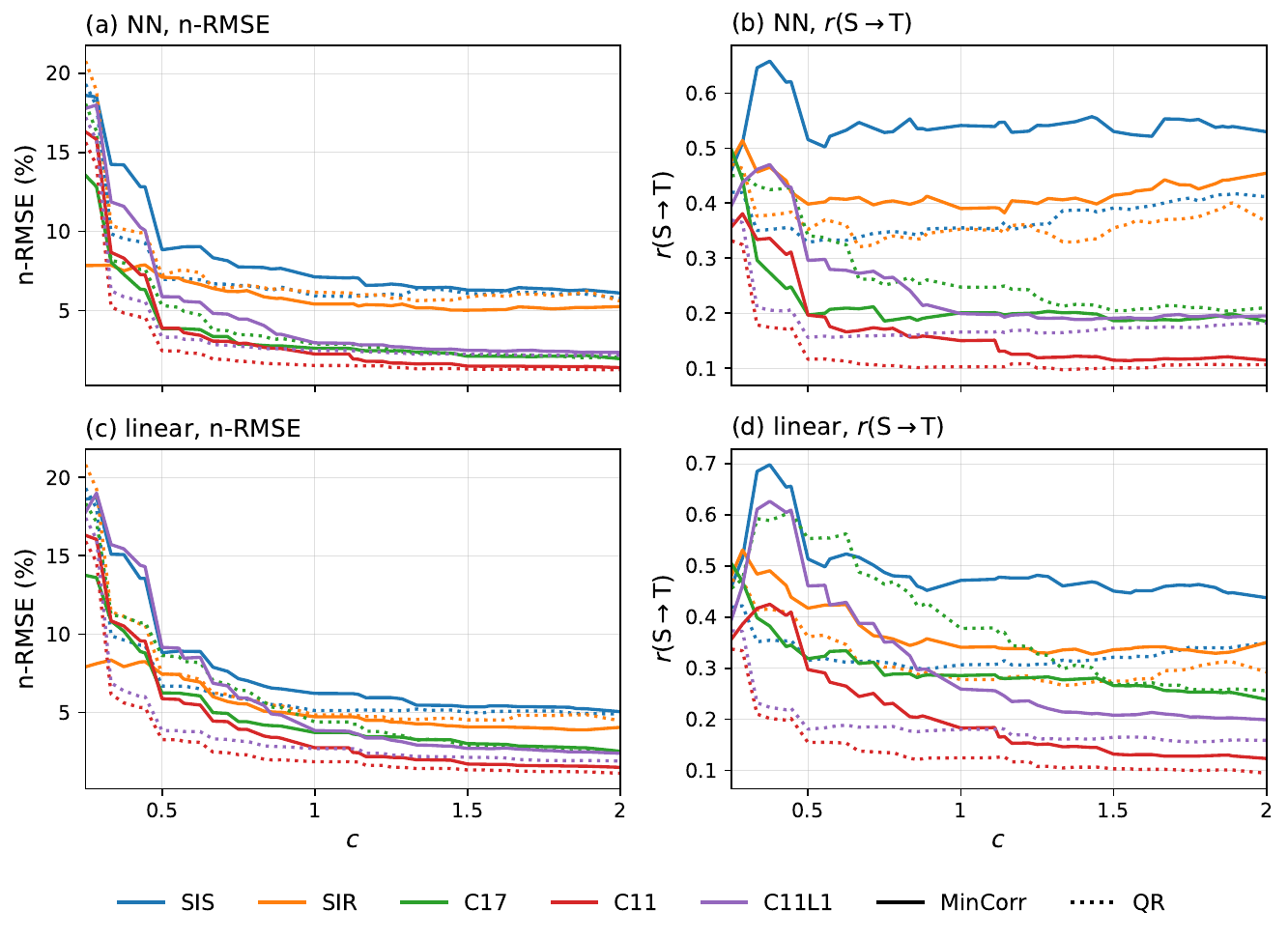}
\caption{Reconstruction performance as the number of sentinel nodes
varies. (a)~NN decoder, n-RMSE. (b)~NN decoder,
$r(\mathrm{S}\to\mathrm{T})$. (c)~Linear decoder, n-RMSE.
(d)~Linear decoder, $r(\mathrm{S}\to\mathrm{T})$. Each panel shows
performance for combinations of source decoder (SIS, SIR, C17, C11,
or C11L1), non-random sentinel-selection method (MinCorr or QR), and
sentinel-count multiplier $c$. The
default choice $n=\lfloor \ln N \rfloor$ corresponds to $c=1$.}
\label{fig:c-sweep}
\end{figure}

Figure~\ref{fig:c-sweep} shows the n-RMSE and
$r(\mathrm{S}\to\mathrm{T})$ as functions of $c$ for each combination
of source decoder, decoder architecture, and non-random
sentinel-selection method. We include the leading source decoders
SIS, SIR, C17, C11, and C11L1, and the two non-random
sentinel-selection methods, MinCorr and QR. We omit random sentinels
from this analysis because their performance is generally worse than
that of MinCorr and QR.

The results are stable across the tested range of sentinel counts.
The n-RMSE generally decreases as $c$ increases, but the gain beyond
the default choice $c=1$, i.e., $n=\lfloor \ln N \rfloor$, is modest.
The decrease in the n-RMSE as $c$ increases is modest in $c \in [0.5, 1]$ as well.
The ratio $r(\mathrm{S}\to\mathrm{T})$ improves gradually and modestly as
$c$ increases in the range $c \ge 0.5$ for the pooled decoders C17, C11, and C11L1, whereas it
remains approximately constant for the per-dynamics SIS and SIR
decoders. The relative ranking of the source decoders is preserved
for approximately $c \ge 0.8$, which
includes the default choice.
Specifically, C11 consistently outperforms C17,
C11L1 lies between C11 and the per-dynamics SIS/SIR sources, and the
pooled decoders remain substantially better than the per-dynamics
sources. We therefore conclude that
$n=\lfloor \ln N \rfloor$ is an arbitrary but practically reasonable
default choice.

\newpage
\clearpage

\section{Robustness of the cross-transfer network structure}
\label{app:wcc-robust}

\subsection{Component structure across decoder architectures and
sentinel-selection methods}

\begin{figure}[t]
\centering
\includegraphics[width=\textwidth]{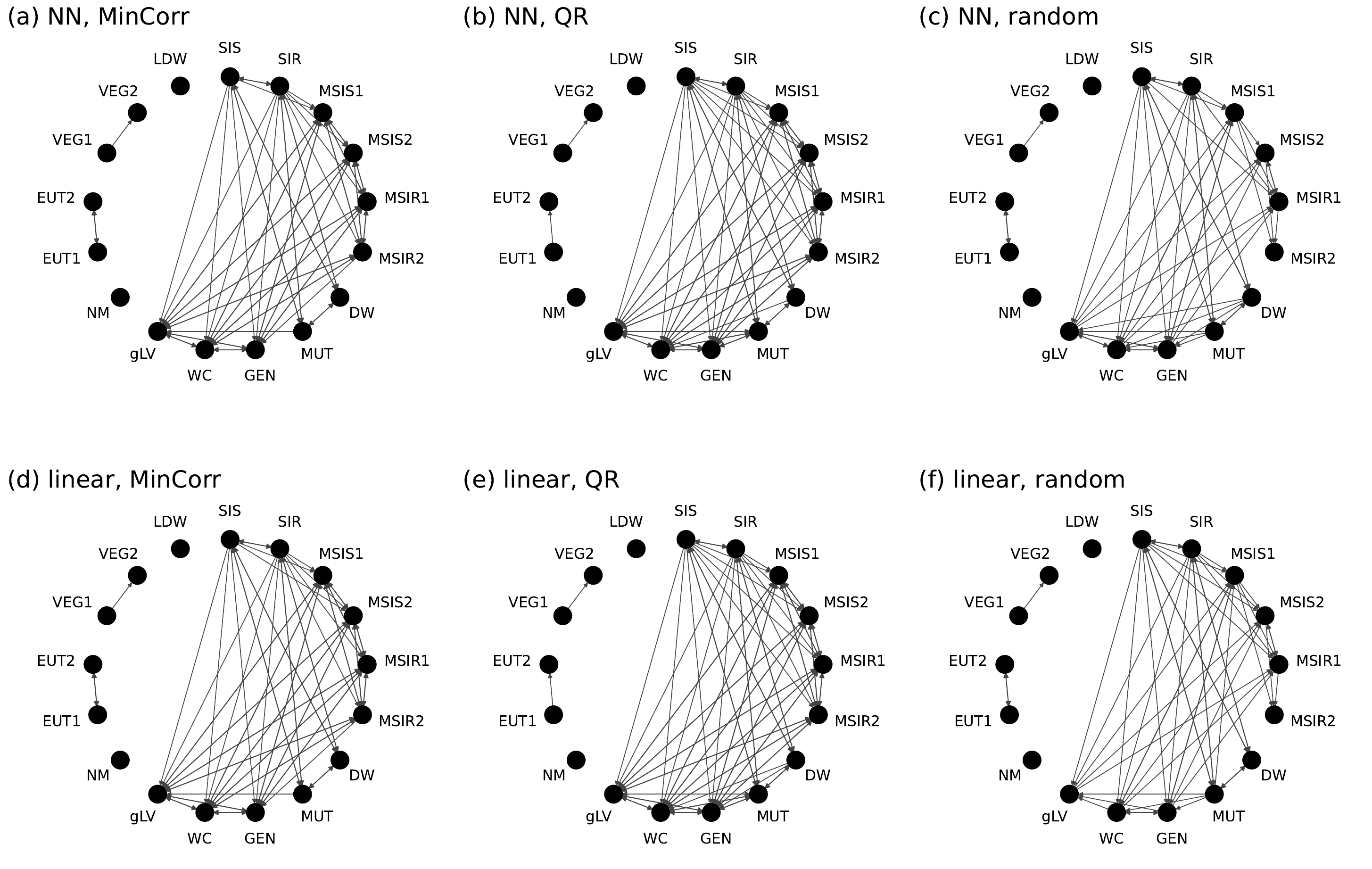}
\caption{Cross-transfer networks for the six combinations of decoder
architecture and sentinel-selection method. Each node represents one of the
$17$ dynamics. A directed edge is drawn from source dynamics $\mathrm{S}$
to target dynamics $\mathrm{T}$ if
$\mathrm{n\text{-}RMSE}(\mathrm{S}\to\mathrm{T})<0.15$ and
$r(\mathrm{S}\to\mathrm{T})<0.8$. Panel (a) reproduces
Fig.~\ref{fig:wccnet} of the main text.}
\label{fig:si-nets}
\end{figure}

Figure~\ref{fig:si-nets} shows the cross-transfer network at the
reference thresholds, $\mathrm{n\text{-}RMSE}=0.15$ and
$r(\mathrm{S}\to\mathrm{T})=0.8$, for all six combinations of decoder
architecture and sentinel-selection method.
Figure~\ref{fig:si-nets}(a) reproduces Fig.~\ref{fig:wccnet} of the main
text. For every combination, the network contains one densely interconnected
largest weakly connected component (LWCC) comprising $11$ dynamics, along
with four smaller components of sizes $2$ (EUT1 and EUT2), $2$ (VEG1 and
VEG2), $1$, and $1$. The members of the LWCC are exactly the $11$
adjacency-coupled dynamics.

Table~\ref{tab:si-wcc} summarizes the structure of the cross-transfer
networks and their LWCCs. Cross-transfer networks obtained with random
sentinels are sparser, exhibit lower reciprocity, and have LWCCs divided
into more strongly connected components (four, compared with one or two)
than those obtained with the
non-random sentinel-selection methods, MinCorr and QR. Because
Figs.~\ref{fig:heat-combined} and~\ref{fig:ratio} of the main text and the
other analyses show that decoders with random sentinels perform worse than
those with non-random sentinels, we focus on MinCorr and QR. For these two
methods, the LWCC is substantially denser than the full network, with edge
density at least $0.591$. The full cross-transfer networks are also highly
reciprocal, with $\rho\ge0.676$, while the reciprocity of their LWCCs
satisfies $\rho\ge0.319$; see the caption of
Table~\ref{tab:si-wcc} for the definition of $\rho$. The LWCC is itself strongly connected for QR sentinels and is
divided into two strongly connected components (SCCs) for MinCorr
sentinels. In the latter case, one SCC comprises MSIS1, MSIS2, MSIR1,
MSIR2, GEN, WC, and gLV, whereas the other comprises SIS, SIR, DW, and MUT.
Thus, with non-random sentinel-selection methods, the LWCC formed by the
$11$ adjacency-coupled dynamics is a densely connected and highly
reciprocal component, even when it is not itself an SCC. We conclude that
the principal cross-transfer component is not an artifact of defining
transfer classes using weak connectivity.

\begin{table}[ht]
\centering
\caption{Structural properties of the cross-transfer networks at the
reference thresholds, $\mathrm{n\text{-}RMSE}=0.15$ and
$r(\mathrm{S}\to\mathrm{T})=0.8$. The edge density is the number of
directed edges divided by the number of ordered pairs of distinct nodes.
Reciprocity (abbreviated as ``recip.'') is defined as
$\rho=(r-a)/(1-a)$, where $r$ is the fraction of directed edges for which
the reverse edge is also present and $a$ is the edge
density~\cite{Garlaschelli2004PRLrecip-SI}. A value of $\rho=0$ indicates that
reciprocal edges occur at the frequency expected by chance at the given
density; $\rho>0$ and $\rho<0$ indicate overrepresentation and
underrepresentation, respectively. The final two columns report the number
of strongly connected components (SCCs) within the largest weakly connected
component (LWCC) and the size of the largest SCC.}
\label{tab:si-wcc}
\small
\begin{tabular}{lccccccc}
\toprule
 & \multicolumn{2}{c}{entire network} &
\multicolumn{5}{c}{largest weakly connected component}\\
\cmidrule(lr){2-3}\cmidrule(lr){4-8}
decoder, sentinels & density & recip. & size & density & recip.
& \#SCC & largest SCC\\
\midrule
NN, MinCorr & 0.250 & 0.725 & 11 & 0.591 & 0.511 & 2 & 7 \\
NN, QR & 0.283 & 0.692 & 11 & 0.682 & 0.371 & 1 & 11 \\
NN, random & 0.202 & 0.339 & 11 & 0.473 & $-0.021$ & 4 & 4 \\
linear, MinCorr & 0.254 & 0.709 & 11 & 0.600 & 0.470 & 2 & 7 \\
linear, QR & 0.287 & 0.676 & 11 & 0.691 & 0.319 & 1 & 11 \\
linear, random & 0.195 & 0.274 & 11 & 0.455 & $-0.100$ & 4 & 4\\
\bottomrule
\end{tabular}
\end{table}

\subsection{Robustness with respect to the threshold values}

In the main text, we constructed a cross-transfer network
using the reference thresholds
$\mathrm{n\text{-}RMSE}=0.15$ and
$r(\mathrm{S}\to\mathrm{T})=0.8$. At these thresholds, the network has
five weakly connected components (WCCs), with sizes
$\{11,2,2,1,1\}$, and the component membership is identical across all
six combinations of decoder architecture and sentinel-selection
method. In this section, we quantify how far the two thresholds can be
varied before this WCC pattern changes.

We first fix the n-RMSE threshold at $0.15$ and lower the
$r(\mathrm{S}\to\mathrm{T})$ threshold until the WCC partition first
differs from the reference partition. We do not increase the
$r(\mathrm{S}\to\mathrm{T})$ threshold because we regard a
$20\%$ improvement over the mean-prediction baseline as a minimum
criterion for a successful cross transfer. The second column of
Table~\ref{tab:wcc-robust} reports the smallest
$r(\mathrm{S}\to\mathrm{T})$ threshold for which the reference WCC
partition is preserved. Across all decoder and sentinel combinations,
the $\{11,2,2,1,1\}$ WCC structure remains unchanged down to
$r(\mathrm{S}\to\mathrm{T})=0.596$.

We next fix the $r(\mathrm{S}\to\mathrm{T})$ threshold and lower the n-RMSE
threshold. The third and fourth columns of Table~\ref{tab:wcc-robust}
show the smallest n-RMSE thresholds that preserve the reference WCC
partition when the ratio threshold is fixed at $0.8$ or $0.65$,
respectively. With $r(\mathrm{S}\to\mathrm{T})=0.8$, the WCC
structure remains unchanged for n-RMSE thresholds down to $0.060$
across all six decoder/sentinel combinations. Similar results are
obtained when the ratio threshold is fixed at $0.65$.

Increasing the n-RMSE threshold above the reference value $0.15$ does
not change the WCC partition for any decoder/sentinel combination. In
our data, every source--target pair with
$\mathrm{n\text{-}RMSE}\geq 0.15$ also has
$r(\mathrm{S}\to\mathrm{T})\geq 0.8$, so increasing only the n-RMSE
threshold does not add any edge satisfying both criteria.

These results show that the five-component WCC structure is not an
artifact of the particular reference thresholds. It is robust to
substantial changes in both the n-RMSE and $r(\mathrm{S}\to\mathrm{T})$ thresholds,
as well as to the choice of decoder architecture and
sentinel-selection method.

\begin{table}[t]
\centering\small
\caption{Robustness of the weakly connected component (WCC) pattern of
the cross-transfer network. The reference thresholds are
$\mathrm{n\text{-}RMSE}=0.15$ and
$r(\mathrm{S}\to\mathrm{T})=0.8$, which yield five WCCs with sizes
$11$, $2$, $2$,  $1$, and $1$. In the second column, we fix
$\mathrm{n\text{-}RMSE}=0.15$ and report the smallest
$r(\mathrm{S}\to\mathrm{T})$ threshold for which this WCC partition is
preserved. In the third and fourth columns, we fix
$r(\mathrm{S}\to\mathrm{T})=0.8$ or $0.65$, respectively, and report
the smallest n-RMSE threshold for which the same WCC partition is
preserved.}
\label{tab:wcc-robust}
\begin{tabular}{l *{3}{>{\centering\arraybackslash}p{3.0cm}}}
\toprule
fixed threshold & $\mathrm{n\text{-}RMSE}=0.15$ & $r(\mathrm{S}\to\mathrm{T})=0.8$ & $r(\mathrm{S}\to\mathrm{T})=0.65$ \\
\midrule
smallest threshold for & $r(\mathrm{S}\to\mathrm{T})$ & n-RMSE & n-RMSE \\
\midrule
NN, MinCorr & 0.494 & 0.060 & 0.075 \\
NN, QR & 0.515 & 0.058 & 0.063 \\
NN, random & 0.596 & 0.056 & 0.064 \\
linear, MinCorr & 0.436 & 0.052 & 0.052 \\
linear, QR & 0.360 & 0.050 & 0.050 \\
linear, random & 0.584 & 0.054 & 0.062\\
\bottomrule
\end{tabular}
\end{table}

To further examine the robustness of the clustering of the $17$
dynamics to the choice of threshold, we perform hierarchical clustering,
which does not require thresholding the n-RMSE. We retain the condition
$r(\mathrm{S}\to\mathrm{T})<0.8$ to determine whether an ordered pair of
dynamics constitutes an edge, but we use the n-RMSE itself as the edge
weight rather than thresholding it; a smaller n-RMSE therefore indicates a
stronger link. We symmetrize the weights by averaging the n-RMSE values in
the two directions. This averaging requires assigning a value to a
direction with no edge, i.e., one for which
$r(\mathrm{S}\to\mathrm{T})\ge0.8$, when the reverse direction has an edge.
We assign such a direction the largest n-RMSE observed for the same
combination of decoder architecture and sentinel-selection method. Thus, a
failed transfer is treated as the poorest transfer observed rather than
being omitted. We then apply average-linkage hierarchical clustering to the
resulting symmetric distance matrix.

\begin{figure}[ht]
\centering
\includegraphics[width=\textwidth]{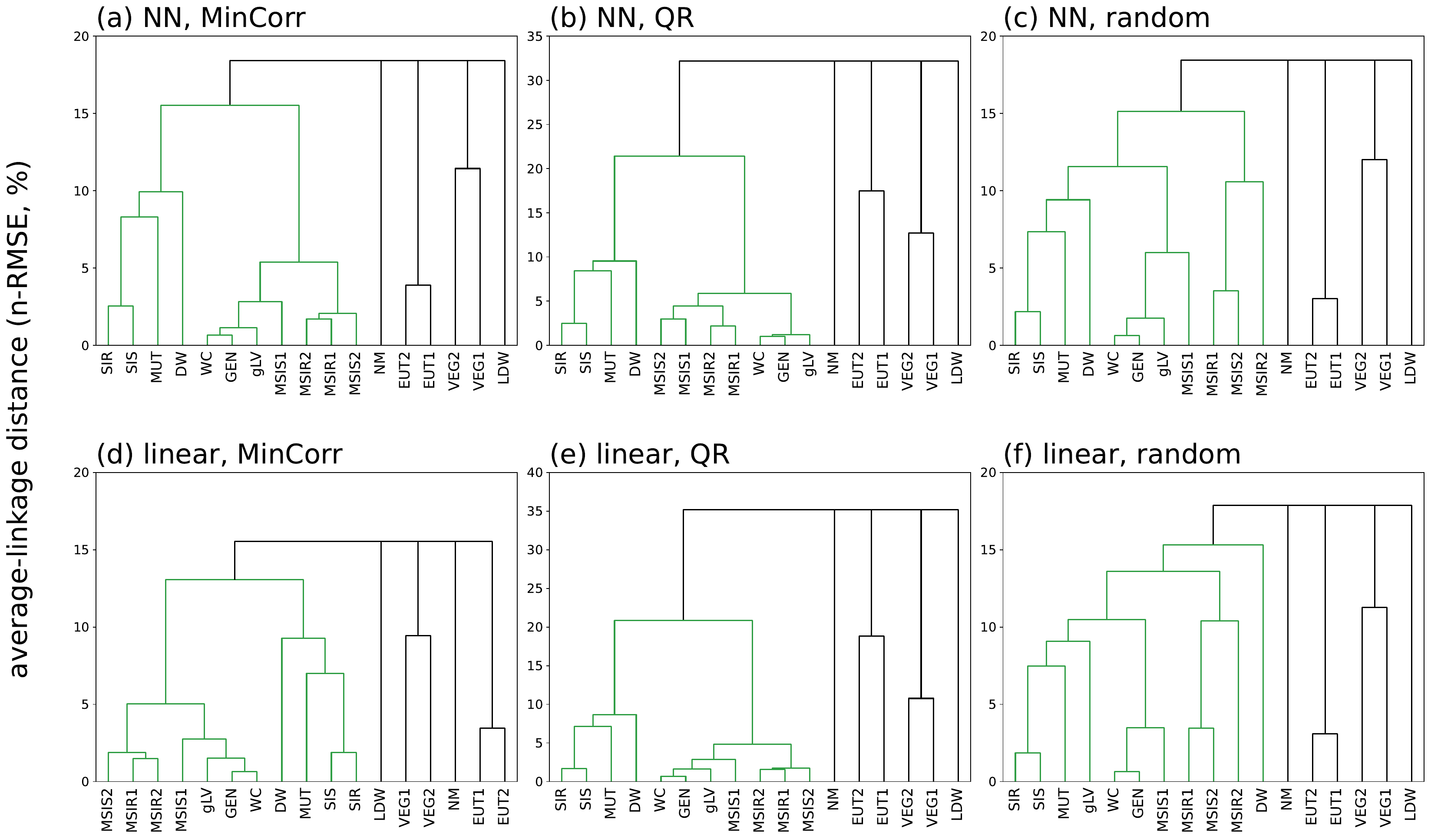}
\caption{Hierarchical clustering of the dynamical systems based on
their cross-transfer n-RMSE values. Each panel corresponds to one
combination of decoder architecture and sentinel-selection method. In every
panel, the $11$ adjacency-coupled dynamics form a single branch, shown in
green, that is distinct from the six Laplacian-coupled dynamics.}
\label{fig:si-dendro}
\end{figure}

Figure~\ref{fig:si-dendro} shows the dendrogram for each combination
of decoder architecture and sentinel-selection method. In all six
dendrograms, the $11$ adjacency-coupled dynamics merge with one another at
substantially smaller distances, as shown in green, than the distance at
which any of them merges with a Laplacian-coupled dynamics. This result
provides further evidence that the cross-transfer component comprising the
$11$ adjacency-coupled dynamics is robust to the threshold used for
clustering.

\subsection{Individual networks}

The preceding analyses use n-RMSE values averaged over the $14$
networks. To examine whether the cross-transfer network among dynamical
systems is also meaningful for individual networks, we apply the same
edge-creation rule separately to each of the $14$ networks, using the
n-RMSE and $r(\mathrm{S}\to\mathrm{T})$ obtained for that network.

Table~\ref{tab:si-pernet} reports the size of the LWCC of the
cross-transfer network separately for each network and each combination of
decoder architecture and sentinel-selection method. With the non-random
sentinel-selection methods, MinCorr and QR, the LWCC generally comprises
exactly the $11$ adjacency-coupled dynamics or these $11$ dynamics plus one
additional dynamics. The main exception is the linear decoder with MinCorr
sentinels on the proximity network, for which the LWCC contains four
additional dynamics. We verified that whenever the LWCC contains more than
$11$ dynamics, it includes all $11$ adjacency-coupled dynamics. In $42$ of
the $56$ combinations of network, decoder architecture, and non-random
sentinel-selection method ($14\times2\times2$), the LWCC consists of exactly
these $11$ dynamics. With uniformly random sentinels, by contrast, the LWCC
often omits some adjacency-coupled dynamics. We conclude that the
cross-transfer network and its component structure are not artifacts of
averaging over the $14$ networks, provided that we use non-random sentinels. This result also reinforces the importance of selecting sentinel
nodes carefully.

\begin{table}[ht]
\centering
\caption{Largest weakly connected component (LWCC) of the
cross-transfer network constructed separately for each of the $14$
networks, using that network's own n-RMSE and
$r(\mathrm{S}\to\mathrm{T})$. Each column corresponds to a combination of
decoder architecture and sentinel-selection method, and each entry gives
the number of dynamics in the LWCC. Whenever the LWCC contains at least
$11$ dynamics, it includes all $11$ adjacency-coupled dynamics.}
\label{tab:si-pernet}
\small
\begin{tabular}{lcccccc}
\toprule
\multirow{2}{*}{network} & NN, & NN, & NN, & linear, & linear, & linear,\\
 & MinCorr & QR & random & MinCorr & QR & random\\
\midrule
Dolphin & 11 & 11 & 11 & 11 & 11 & 10 \\
Proximity & 11 & 12 & 11 & 15 & 14 & 10 \\
Metabolic & 12 & 12 & 11 & 12 & 12 & 11 \\
GKK & 11 & 11 & 11 & 12 & 12 & 11 \\
LFR & 11 & 11 & 8 & 11 & 11 & 8 \\
ER & 11 & 11 & 11 & 11 & 11 & 11 \\
BA & 11 & 11 & 11 & 11 & 11 & 11 \\
HK & 11 & 11 & 7 & 11 & 11 & 7 \\
Euroroad & 11 & 11 & 11 & 11 & 11 & 11 \\
Email & 11 & 13 & 11 & 11 & 13 & 10 \\
Flybi & 11 & 11 & 8 & 11 & 11 & 8 \\
Reactome & 11 & 11 & 10 & 11 & 11 & 8 \\
Route views & 11 & 11 & 8 & 11 & 11 & 7 \\
Spanish & 12 & 11 & 7 & 12 & 12 & 7\\
\bottomrule
\end{tabular}
\end{table}

\newpage
\clearpage

\section{Cross-transfer classes with four additional dynamics}\label{si:fourdyn}

In this section, we examine the cross-transfer networks obtained after
adding four dynamical systems to the $17$ dynamics considered in the main
text. We construct each additional dynamics from a model in the main text
by replacing adjacency-matrix coupling with Laplacian coupling, or vice
versa. These four dynamics are not intended as models of real-world
phenomena. Rather, they serve as benchmarks for testing whether transfer
classes are determined by the form of the coupling. For all four additional
dynamics, we retain the reaction (i.e., single-node dynamics) term, node-level random parameters, and
the number $M=51$ of control-parameter values used in the corresponding
original model.

\subsection{Models}

\textbf{LMUT:} The mutualistic dynamics (MUT) in the main text uses
adjacency-matrix coupling through the saturating kernel
$g(a,b)=ab/(\tilde{D}+Ea+Hb)$. Replacing this coupling with its diffusive
counterpart gives the Laplacian-coupled mutualistic dynamics (LMUT):
\begin{equation}
\dot{x}_i = B + x_i\Bigl(1-\frac{x_i}{K}\Bigr)\Bigl(\frac{x_i}{C}-1\Bigr)
 + D\sum_{j=1}^{N}A_{ij}\bigl[g(x_i,x_j)-g(x_i,x_i)\bigr],
\label{eq:lmut}
\end{equation}
where $B=0.1$, $K=5$, $C=1$, $\tilde{D}=5$, $E=0.9$, and $H=0.1$, as in
MUT, and $D$ is the control parameter. The coupling term vanishes for any
spatially uniform state. Therefore, the uniform equilibria
$x_i=0.116$ and $x_i=5.025$ are stable, whereas the uniform equilibrium
$x_i=0.860$ is unstable. This property does not preclude the existence of
other stable equilibria.

\textbf{ANM:} Replacing the Laplacian coupling of the Noy--Meir (NM)
model in the main text with adjacency-matrix coupling gives the
adjacency-coupled NM model (ANM):
\begin{equation}
\dot{x}_i = r_i x_i\left(1-\frac{x_i}{K}\right)
           - c\,\frac{x_i^2}{x_i^2+1}
           + D\sum_{j=1}^{N}A_{ij}x_j .
\label{eq:anm}
\end{equation}
As in NM, we set $K=10$, use the same node-specific growth rates $r_i$, and
vary the grazing pressure $c$ as the control parameter.

\textbf{AEUT1:} Replacing the Laplacian coupling of EUT1 with
adjacency-matrix coupling gives its adjacency-coupled variant (AEUT1):
\begin{equation}
\dot{x}_i = \alpha - b_i x_i + \frac{x_i^8}{x_i^8+1}
           + D\sum_{j=1}^{N}A_{ij}x_j .
\label{eq:aeut}
\end{equation}
We use the same node-specific loss rates $b_i$ as in EUT1 and vary $\alpha$
as the control parameter.

\textbf{AVEG1:} Replacing the Laplacian coupling of VEG1 with
adjacency-matrix coupling gives its adjacency-coupled variant (AVEG1):
\begin{align}
\dot{x}_i &= r_V x_i\Bigl[1-\frac{x_i(h_{E,i}^4+E_i^4)}{h_{E,i}^4}\Bigr]
 + D\sum_{j=1}^{N}A_{ij}x_j , \label{eq:aveg}\\
E_i &= \frac{E_{\rm o} h_V}{h_V+x_i}.
\end{align}
We use the same $r_V=0.5$, $h_V=0.2$, and node-specific half-saturation
parameters $h_{E,i}$ as in VEG1 and vary the background turbidity
$E_{\rm o}$ as the control parameter.

For ANM, AEUT1, and AVEG1, the node-specific random parameters $r_i$,
$b_i$, and $h_{E,i}$ are numerically identical to those used for NM, EUT1,
and VEG1, respectively, on the same network. Thus, each corresponding pair,
such as VEG1 and AVEG1, differs only in its coupling term. Likewise, MUT and
LMUT differ only in their coupling term.

\subsection{Numerical methods}

For these four dynamics, we use the settling-aware integration
procedure described for LDW in the Methods section of the main text.
Specifically, we initially integrate until $t_{\max}=200$. If
$\max_i|\dot{x}_i|\ge\varepsilon_{\rm res}$ at $t=t_{\max}$, we repeat the
integration after increasing $t_{\max}$ by a factor of ten. We continue
until $\max_i|\dot{x}_i|<\varepsilon_{\rm res}$. We set
$\varepsilon_{\rm res}=10^{-6}$ for LMUT, AEUT1, and AVEG1. We set
$\varepsilon_{\rm res}=10^{-5}$ for ANM because some nodes relax relatively
slowly near the grazing transition.

For ANM, AEUT1, and AVEG1, we use the same uniform initial conditions
as for the corresponding dynamics in the main text:
$x_i(0)=8$, $0.5$, and $0.6$, respectively. We cannot use a uniform initial
condition for LMUT for the same reason as for LDW: the two uniform states
are exact equilibria, and dynamics initialized at either state would never
leave it. We therefore draw each $x_i(0)$ independently from the uniform
distribution on $[0.116,5.025]$ once per network and reuse the resulting
initial condition for every control-parameter value.

For ANM, AEUT1, and AVEG1, the coupling strength must depend on the
network because, under adjacency-matrix coupling, the input to node $i$ is
approximately $Dk_i\bar{x}$ and therefore scales with degree. For ANM and
AVEG1, we follow the convention used for DW and WC in the main text and
hold $D\langle k\rangle$ fixed at
$0.05\langle k\rangle_{\rm Dolphin}=0.257$, where
$\langle k\rangle_{\rm Dolphin}$ is the mean degree of the dolphin network.
For AEUT1, the
large-$x$ behavior is governed by the linear term
$\mathbf{J}_\infty=-\mathrm{diag}(b)+D\bA$. A negative largest eigenvalue
of $\mathbf{J}_\infty$ provides a sufficient condition for the dynamics to
remain bounded. Following the convention used for GEN in the main text, we
hold $D\alpha_{\max}$ fixed at
$0.05\alpha_{\max,\mathrm{Dolphin}}
=0.05\times7.194=0.360$, where
$\alpha_{\max,\mathrm{Dolphin}}$ is the largest eigenvalue of the adjacency
matrix of the dolphin network.

For AEUT1 and AVEG1, we retain the control-parameter ranges used for
EUT1 and VEG1, namely, $\alpha\in[0.1,0.9]$ and
$E_{\rm o}\in[2,12]$, respectively, because the transition points remain
within these ranges after replacing Laplacian coupling with
adjacency-matrix coupling. The exceptions are the RouteViews and
Spanish-language networks, for which $28\%$ and $38\%$ of the nodes,
respectively, remain vegetated at $E_{\rm o}=12$ under AVEG1. Nevertheless,
we retain $E_{\rm o}\in[2,12]$ to enable direct comparison between VEG1 and
AVEG1. For LMUT and ANM, the transition points depend substantially on the
network, so we determine the control-parameter ranges separately for each
network. For LMUT, we identify $D_{\rm c}$, the smallest value of $D$ at
which the equilibrium becomes spatially uniform, defined by
$\max_i x_i-\min_i x_i<10^{-6}$, and sweep
$D\in[0,D_{\rm c}/0.9]$. For ANM, we identify $c_{\rm c}$, the smallest
value of $c$ at which no node remains above $x=2$, and sweep
$c\in[1,c_{\rm c}/0.95]$. We locate $D_{\rm c}$ and $c_{\rm c}$ by
bisection.

\subsection{Transfer classes}

Adding the four dynamics increases the total number of dynamical
systems to $21$, of which $14$ are adjacency coupled and seven are
Laplacian coupled. We repeat the analysis of
Section~\ref{app:wcc-robust} for these $21$ dynamics.

\begin{figure}
\centering
\includegraphics[width=\textwidth]{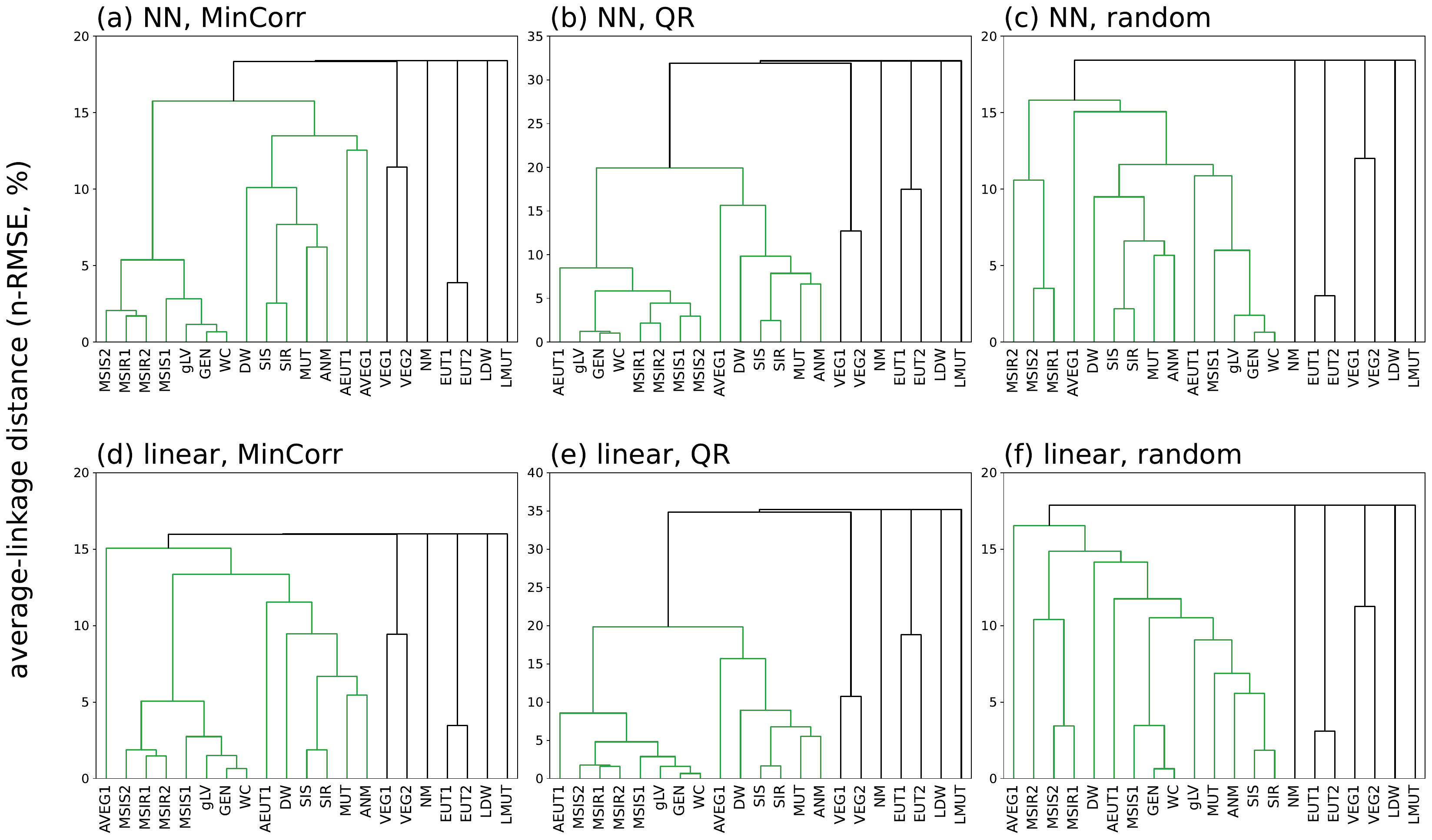}
\caption{Average-linkage hierarchical clustering of the $21$ dynamics
based on their cross-transfer n-RMSE values. In each panel, corresponding to
one combination of decoder architecture and sentinel-selection method, the
$14$ adjacency-coupled dynamics form a single branch, shown in green. All
other conventions are the same as in Fig.~\ref{fig:si-dendro}.}
\label{fig:si-dendro21}
\end{figure}

Table~\ref{tab:si-21} summarizes the cross-transfer networks, focusing
on their LWCCs, using the same edge-creation rule as before:
$\mathrm{n\text{-}RMSE}(\mathrm{S}\to\mathrm{T})<0.15$ and
$r(\mathrm{S}\to\mathrm{T})<0.8$. For both decoder architectures with
QR-pivoted or uniformly random sentinels, the LWCC consists of exactly the
$14$ adjacency-coupled dynamics. With MinCorr sentinels, the LWCC contains
these $14$ dynamics together with VEG1 and VEG2. Thus, the three newly
introduced adjacency-coupled dynamics join the LWCC, whereas the
Laplacian-coupled dynamics from which they were constructed (i.e., NM, EUT1, and
VEG1) generally remain outside it. In addition, LDW and LMUT remain outside the
LWCC, whereas DW and MUT, their adjacency-coupled counterparts, belong to the LWCC.
For the non-random sentinel-selection methods, the LWCC is again reasonably
dense and has positively reciprocal edges. The hierarchical-clustering
results in Fig.~\ref{fig:si-dendro21} support the same conclusion: the $14$
adjacency-coupled dynamics form a single subtree, shown in green, for every
combination of decoder architecture and sentinel-selection method.

The results in Table~\ref{tab:si-21} and
Fig.~\ref{fig:si-dendro21} strongly support the conclusion that equilibrium
reconstruction transfers successfully between source and target dynamics
when both are adjacency coupled. Importantly, the Laplacian-coupled
dynamics do not form a comparably cohesive cluster, indicating that sharing
Laplacian coupling alone is insufficient for successful cross-dynamics
transfer.

\begin{table}[t]
\centering
\caption{Structural properties of the cross-transfer networks for the
$21$ dynamics, comprising $14$ adjacency-coupled and seven
Laplacian-coupled dynamics, at the reference thresholds
$\mathrm{n\text{-}RMSE}=0.15$ and
$r(\mathrm{S}\to\mathrm{T})=0.8$. All conventions are the same as in
Table~\ref{tab:si-wcc}.}
\label{tab:si-21}
\small
\begin{tabular}{lccccccc}
\toprule
 & \multicolumn{2}{c}{entire network} &
\multicolumn{5}{c}{largest weakly connected component}\\
\cmidrule(lr){2-3}\cmidrule(lr){4-8}
decoder, sentinels & density & recip. & size & density & recip.
& \#SCC & largest SCC\\
\midrule
NN, MinCorr & 0.226 & 0.606 & 16 & 0.388 & 0.491 & 4 & 7 \\
NN, QR & 0.286 & 0.627 & 14 & 0.648 & 0.277 & 1 & 14 \\
NN, random & 0.200 & 0.315 & 14 & 0.445 & $-0.001$ & 5 & 6 \\
linear, MinCorr & 0.224 & 0.616 & 16 & 0.383 & 0.506 & 4 & 13 \\
linear, QR & 0.283 & 0.637 & 14 & 0.643 & 0.306 & 2 & 13 \\
linear, random & 0.186 & 0.307 & 14 & 0.412 & 0.025 & 5 & 6\\
\bottomrule
\end{tabular}
\end{table}

\newpage
\clearpage

\section{Why the transferable dynamics transfer}\label{si:why}

To clarify why adjacency-coupled dynamics are largely transferable
whereas Laplacian-coupled dynamics are not, we hypothesize that the
equilibrium state vectors $(x_1^*,\ldots,x_N^*)$ are similar across
adjacency-coupled dynamics, with high-degree nodes tending to have large
$x_i^*$. To test this hypothesis, we compute the Pearson correlation
coefficient between the equilibrium state vectors for every pair of
dynamics. For each dynamics--network pair, we first select the
control-parameter value that maximizes the sample variance of
$(x_1^*,\ldots,x_N^*)$ across nodes. To the resulting $17$ state vectors,
we add the degree vector $(k_1,\ldots,k_N)$ and log-degree vector
$(\ln k_1,\ldots,\ln k_N)$ as references. We then compute the Pearson
correlation for every pair of these $19$ vectors and average each entry of
the resulting $19\times19$ correlation matrix over the $14$ networks.

\begin{figure}[t]
\centering
\includegraphics[width=\textwidth]{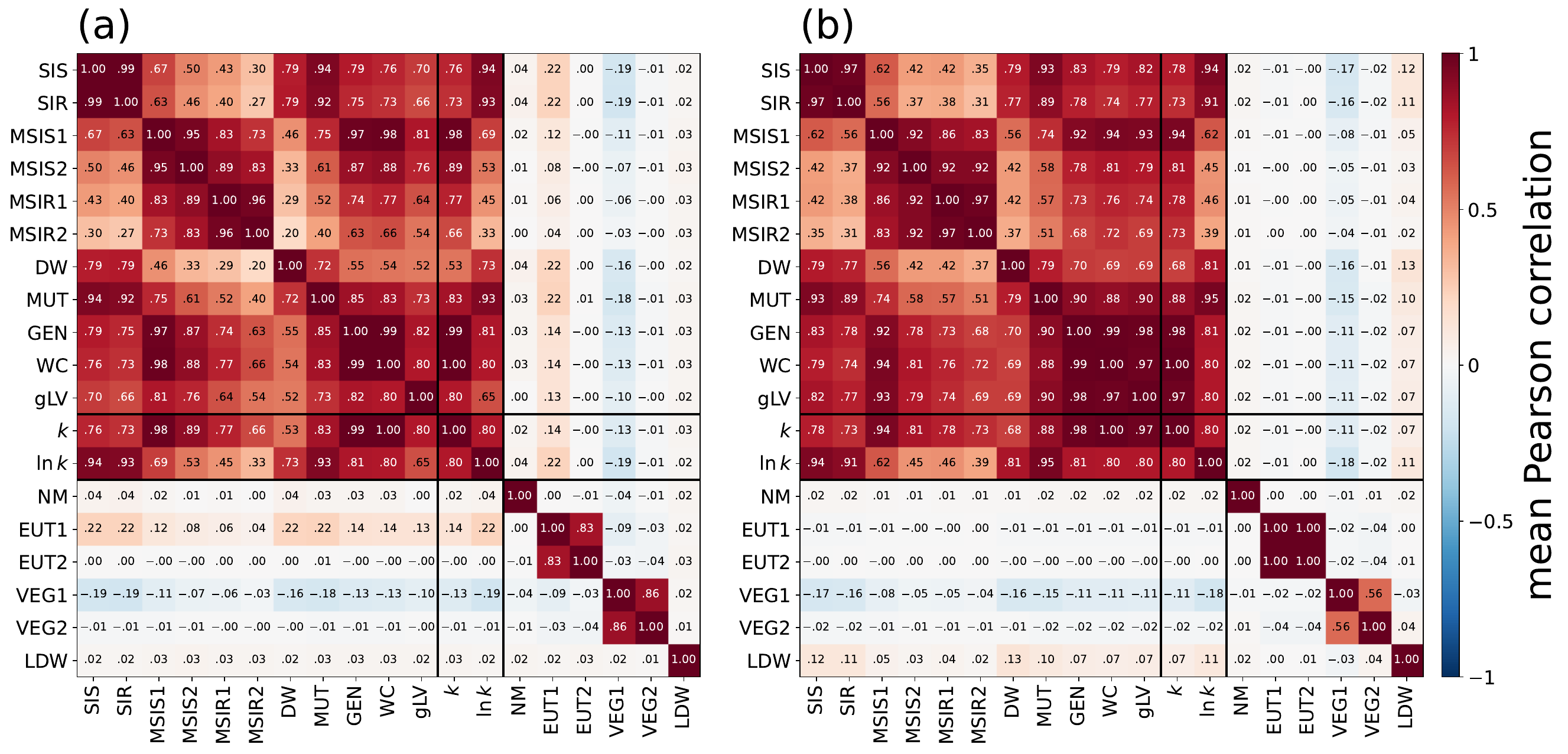}
\caption{Pearson correlations between equilibrium state vectors,
averaged over the $14$ networks, together with correlations involving the
degree vector (denoted by $k$ in the figure) and log-degree vector (denoted by $\ln k$). (a) For each
dynamics--network pair, the state vector is evaluated at the
control-parameter value that maximizes the sample variance of
$(x_1^*,\ldots,x_N^*)$ across nodes. (b) The state vector is evaluated at
the control-parameter value whose sample variance across nodes is the
median among the $M=51$ values. The thick solid lines separate the $11$
adjacency-coupled dynamics, the two degree-based reference vectors, and the
six Laplacian-coupled dynamics.}
\label{fig:si-why}
\end{figure}

Figure~\ref{fig:si-why}(a) shows the resulting correlation matrix. The
$11$ adjacency-coupled dynamics and the two degree-based reference vectors
form a positively correlated block. The mean off-diagonal correlation
within the $11\times11$ block of adjacency-coupled dynamics is $0.688$.
The six Laplacian-coupled dynamics do not form a corresponding block,
although the correlations are large for EUT1--EUT2 and VEG1--VEG2, as
expected. The degree-based vectors are essentially uncorrelated with the
state vectors of the Laplacian-coupled dynamics, and the mean correlation
between an adjacency-coupled and a Laplacian-coupled dynamics is $0.011$.
To quantify the separation between the two classes, we compare the $110$
off-diagonal entries of the $11\times11$ adjacency-to-adjacency block with
the $132$ entries in the two cross-class blocks, namely, the
$11\times6$ adjacency-to-Laplacian and $6\times11$
Laplacian-to-adjacency blocks. We exclude the diagonal entries, which equal
one by definition. We summarize the separation using the area under the
receiver operating characteristic curve (AUC), for which perfect
separation yields $\mathrm{AUC}=1$ and no separation yields
$\mathrm{AUC}=0.5$. We obtain $\mathrm{AUC}=0.999$.

We repeat this analysis using, for each dynamics--network pair, the
equilibrium vector at the control-parameter value whose sample variance
across nodes is the median among the $M=51$ values. The results, shown in
Fig.~\ref{fig:si-why}(b), are similar to those in
Fig.~\ref{fig:si-why}(a). The mean correlation within the $11\times11$
adjacency-coupled block is $0.729$, that within the $6\times6$
Laplacian-coupled block is $0.099$, and that between the two classes is
$-0.007$. We obtain $\mathrm{AUC}=1.000$.

If the equilibrium state vectors of different dynamical systems are
approximately linearly related, reconstruction from sentinel observations
is likely to transfer accurately between them. To examine this relationship,
we restrict attention to the $110$ ordered pairs of distinct
adjacency-coupled dynamics and compare the equilibrium-state correlations
shown in Fig.~\ref{fig:si-why} with the corresponding
$\mathrm{n\text{-}RMSE}$ and $r(\mathrm{S}\to\mathrm{T})$ values.
Table~\ref{tab:si-why} reports the Pearson correlation, across these $110$
ordered pairs, between the equilibrium-state correlation and each
cross-transfer performance measure. The equilibrium-state correlation is
strongly negatively associated with both $\mathrm{n\text{-}RMSE}$ and
$r(\mathrm{S}\to\mathrm{T})$, particularly for the non-random
sentinel-selection methods, MinCorr and QR. Thus, greater similarity between
the equilibria of two dynamical systems is associated with more accurate
cross-dynamics reconstruction.

Taken together, these results suggest a mechanism underlying the
transfer classes identified in the main text. The adjacency-coupled
dynamics organize node states in similar ways, approximately according to a
monotonic function of node degree, which facilitates successful transfer
among them. The Laplacian-coupled dynamics lack this common organization.

\begin{table}[htbp]
\centering
\caption{Pearson correlations, across the $110$ ordered pairs of
distinct adjacency-coupled dynamics, between the equilibrium-state
correlations shown in Fig.~\ref{fig:si-why} and the two cross-transfer
performance measures.}
\label{tab:si-why}
\small
\begin{tabular}{lcccc}
\toprule
& \multicolumn{2}{c}{against $\mathrm{n\text{-}RMSE}$}
& \multicolumn{2}{c}{against $r(\mathrm{S}\to\mathrm{T})$}\\
\cmidrule(lr){2-3}\cmidrule(lr){4-5}
decoder, sentinels & Fig.~\ref{fig:si-why}(a) & Fig.~\ref{fig:si-why}(b)
& Fig.~\ref{fig:si-why}(a) & Fig.~\ref{fig:si-why}(b)\\
\midrule
NN, MinCorr & $-0.603$ & $-0.645$ & $-0.656$ & $-0.624$\\
NN, QR & $-0.583$ & $-0.640$ & $-0.750$ & $-0.711$\\
NN, random & $-0.519$ & $-0.580$ & $-0.494$ & $-0.541$\\
linear, MinCorr & $-0.609$ & $-0.637$ & $-0.660$ & $-0.612$\\
linear, QR & $-0.579$ & $-0.619$ & $-0.704$ & $-0.649$\\
linear, random & $-0.496$ & $-0.545$ & $-0.485$ & $-0.524$\\
\bottomrule
\end{tabular}
\end{table}

\newpage
\clearpage

\section{Source-decoder rankings}
\label{sec:decoder-ranking}

\subsection{$11$ adjacency-coupled target dynamics}\label{sec:rank11}

In this section, we compare source decoders by their average
reconstruction performance over different sets of target dynamics. We
first focus on the $11$ transferable dynamics: SIS, SIR, MSIS1,
MSIS2, MSIR1, MSIR2, DW, MUT, GEN, WC, and gLV. The candidate source
decoders are the $11$ per-dynamics decoders trained separately on
these dynamics, the C17 decoder trained on pooled data from all
$17$ dynamics, the C11 decoder trained on pooled data from the
$11$ transferable dynamics, and C11L1, the leave-one-dynamics-out
pooled decoder over the $11$ transferable dynamics. Thus, there are
$14$ candidate source decoders in this comparison.

Table~\ref{tab:rank-rmse} ranks these $14$ source decoders by mean
n-RMSE for each combination of decoder architecture and
sentinel-selection method. The values are averaged over the
$11$ transferable target dynamics and over all $14$ networks. For
C11L1, each target-specific n-RMSE is obtained from the
leave-one-dynamics-out decoder whose training pool excludes the
target dynamics. Table~\ref{tab:rank-ratio} gives the corresponding
rankings by the mean $r(\mathrm{S}\to\mathrm{T})$.

The rankings show that pooled training gives the best performance.
C11 is ranked first in each of the $12$ ranked lists obtained from the six
combinations of decoder architecture and sentinel-selection method and the
two performance measures. C11L1 ranks second in six of the $12$ lists,
whereas C17 ranks second in five lists; neither ranks second in the remaining
list. Notably, C11L1 has never been trained on the target dynamics, whereas
C17 has. Thus, restricting the training pool to the $11$ transferable
dynamics is beneficial. Among the per-dynamics decoders, SIS and SIR are
the best sources, occupying the two highest
per-dynamics ranks in eight of the $12$ lists; in the remaining four,
MSIS1 displaces one of them.

In the preceding comparison, C11 and C11L1 use more training data than
each per-dynamics decoder. A per-dynamics decoder is trained on the $41$
control-parameter values of a single dynamics, whereas C11 is trained on
$11\times41=451$ values across dynamics and C11L1 on
$10\times41=410$ values. To distinguish the benefit of pooling multiple
dynamics from that of simply using more training data, we construct two
budget-matched decoders, C11B and C11L1B, each trained on exactly $41$
control-parameter values in total. Because $41=8\times4+3\times3$, eight of
the $11$ adjacency-coupled dynamics contribute four control-parameter values each to the C11B
training pool, and the remaining three dynamics contribute three values each. We
select uniformly at random the eight dynamics that contribute four values
and select the three or four values from each dynamics uniformly at random,
following the interspersed protocol. For C11L1B, the training pool contains
$10$ dynamics. Because $41=1\times5+9\times4$, one dynamics contributes
five control-parameter values, and the remaining nine dynamics contribute four values each. For C11B and
C11L1B, we normalize the equilibrium data from each dynamics to $[0,1]$
using only the control-parameter values contributed by that dynamics,
because these are the only data available to the decoder. All other aspects
of the procedure remain unchanged.

Matching the training-data budget in this manner does not qualitatively
change the results. Across the six combinations of decoder architecture and
sentinel-selection method, the mean n-RMSE ranges from $2.02\%$ to $3.21\%$
for C11B, compared with $1.52\%$ to $3.01\%$ for C11. If C11B and C11L1B
were both added to Table~\ref{tab:rank-rmse}, C11B would rank second, behind only C11 and above
every per-dynamics decoder, in all six lists. The same is true in
Table~\ref{tab:rank-ratio}. The mean n-RMSE ranges from $2.82\%$ to
$3.79\%$ for C11L1B, compared with $2.47\%$ to $3.84\%$ for C11L1.
C11L1B would rank between third and fifth in all six lists in
Table~\ref{tab:rank-rmse}. Thus, pooled training outperforms
per-dynamics training even when the total amount of training data is held
fixed.

\subsection{All $17$ target dynamics}\label{sec:rank17}

We next repeat the ranking over all $17$ target dynamics. The
candidate source decoders are the $17$ per-dynamics decoders, C17, and
C17L1, the leave-one-dynamics-out pooled decoder over all $17$
dynamics. Thus, there are $19$ candidate source decoders in this
comparison. Tables~\ref{tab:rank16-rmse} and~\ref{tab:rank16-ratio}
rank these sources by mean n-RMSE and mean
$r(\mathrm{S}\to\mathrm{T})$, respectively, averaged over all
$17$ target dynamics and over all $14$ networks. C17 is ranked first
in all $12$ ranked lists, and C17L1 is ranked second in all
$12$ lists. These results confirm that pooled training is the most
effective strategy when performance is averaged over all target
dynamics, although the main text shows that this broader pooling comes
at a cost relative to restricting the pool to the $11$ transferable
dynamics.

We also examined budget-matched variants of the two pooled decoders,
constructed by adapting the procedure in Section~\ref{sec:rank11}. Across
the six combinations of decoder architecture and sentinel-selection method,
the mean n-RMSE ranges from $5.53\%$ to $7.52\%$ for C17B, compared with
$4.12\%$ to $6.34\%$ for C17. The mean n-RMSE ranges from $7.50\%$ to
$9.44\%$ for C17L1B, compared with $7.37\%$ to $8.73\%$ for C17L1. If
C17B and C17L1B were both added to Tables~\ref{tab:rank16-rmse}
and~\ref{tab:rank16-ratio}, C17B
would rank second and C17L1B fourth in every column. Thus, they would rank
behind C17 and C17L1, respectively, but above every per-dynamics decoder in
all $12$ ranked lists. Therefore, the conclusion that pooled training is
effective across all target dynamics does not depend on the pooled decoders
having more training data.

\begin{table}[H]
\centering\footnotesize
\setlength{\tabcolsep}{3pt}
\caption{Ranking by mean n-RMSE (in percent), averaged over the
$11$ transferable target dynamics and over all $14$ networks. The $14$ candidate sources are the
$11$ transferable per-dynamics decoders, C17, C11, and C11L1. Lower
values are better.}
\label{tab:rank-rmse}
\begin{tabular}{r|l@{\hspace{3pt}}c|l@{\hspace{3pt}}c|l@{\hspace{3pt}}c||l@{\hspace{3pt}}c|l@{\hspace{3pt}}c|l@{\hspace{3pt}}c}
\toprule
 & \multicolumn{2}{c|}{NN, MinCorr} & \multicolumn{2}{c|}{NN, QR} & \multicolumn{2}{c|}{NN, random} & \multicolumn{2}{c|}{linear, MinCorr} & \multicolumn{2}{c|}{linear, QR} & \multicolumn{2}{c}{linear, random} \\
\cmidrule{2-3} \cmidrule{4-5} \cmidrule{6-7} \cmidrule{8-9} \cmidrule{10-11} \cmidrule{12-13}
rank & source & value & source & value & source & value & source & value & source & value & source & value \\
\midrule
1 & C11 & 2.24 & C11 & 1.52 & C11 & 2.70 & C11 & 2.74 & C11 & 1.84 & C11 & 3.01 \\
2 & C17 & 2.61 & C11L1 & 2.47 & C17 & 3.18 & C17 & 3.71 & C11L1 & 2.68 & SIS & 3.74 \\
3 & C11L1 & 2.95 & C17 & 2.87 & C11L1 & 3.60 & C11L1 & 3.84 & C17 & 4.38 & C11L1 & 3.74 \\
4 & SIR & 5.40 & SIS & 5.93 & SIS & 3.98 & SIR & 4.72 & SIR & 4.84 & SIR & 3.75 \\
5 & SIS & 7.12 & SIR & 6.15 & SIR & 3.99 & SIS & 6.21 & SIS & 5.12 & C17 & 4.03 \\
6 & MSIS1 & 8.30 & DW & 8.41 & DW & 5.24 & MSIS1 & 7.45 & DW & 8.02 & MUT & 5.83 \\
7 & MSIR1 & 9.28 & MUT & 9.85 & MUT & 6.01 & MSIR1 & 8.48 & MSIS1 & 8.49 & DW & 6.54 \\
8 & gLV & 9.34 & gLV & 10.41 & GEN & 9.17 & MSIS2 & 9.54 & MUT & 9.16 & GEN & 9.13 \\
9 & DW & 9.42 & MSIS1 & 10.54 & WC & 10.34 & DW & 9.58 & GEN & 9.94 & WC & 10.46 \\
10 & MSIS2 & 9.52 & GEN & 10.70 & gLV & 10.71 & gLV & 9.61 & gLV & 10.46 & gLV & 12.94 \\
11 & MSIR2 & 9.92 & WC & 11.22 & MSIS1 & 24.94 & MSIR2 & 9.74 & WC & 10.68 & MSIS1 & 25.65 \\
12 & WC & 10.36 & MSIS2 & 12.15 & MSIR1 & 47.86 & WC & 10.19 & MSIR1 & 10.84 & MSIR1 & 52.60 \\
13 & GEN & 10.40 & MSIR1 & 13.24 & MSIS2 & 70.31 & GEN & 10.29 & MSIS2 & 11.69 & MSIS2 & 121.69 \\
14 & MUT & 11.65 & MSIR2 & 14.81 & MSIR2 & 85.73 & MUT & 10.80 & MSIR2 & 12.59 & MSIR2 & 171.80 \\
\bottomrule
\end{tabular}
\end{table}

\begin{table}[H]
\centering\footnotesize
\setlength{\tabcolsep}{3pt}
\caption{Ranking by mean $r(\mathrm{S}\to\mathrm{T})$,
averaged over the $11$ transferable target dynamics and over all
$14$ networks, for the same $14$ candidate sources as
in Table~\ref{tab:rank-rmse}. Lower values are better.}
\label{tab:rank-ratio}
\begin{tabular}{r|l@{\hspace{3pt}}c|l@{\hspace{3pt}}c|l@{\hspace{3pt}}c||l@{\hspace{3pt}}c|l@{\hspace{3pt}}c|l@{\hspace{3pt}}c}
\toprule
 & \multicolumn{2}{c|}{NN, MinCorr} & \multicolumn{2}{c|}{NN, QR} & \multicolumn{2}{c|}{NN, random} & \multicolumn{2}{c|}{linear, MinCorr} & \multicolumn{2}{c|}{linear, QR} & \multicolumn{2}{c}{linear, random} \\
\cmidrule{2-3} \cmidrule{4-5} \cmidrule{6-7} \cmidrule{8-9} \cmidrule{10-11} \cmidrule{12-13}
rank & source & value & source & value & source & value & source & value & source & value & source & value \\
\midrule
1 & C11 & 0.14 & C11 & 0.10 & C11 & 0.35 & C11 & 0.17 & C11 & 0.12 & C11 & 0.38 \\
2 & C17 & 0.19 & C11L1 & 0.16 & C17 & 0.40 & C11L1 & 0.26 & C11L1 & 0.17 & C11L1 & 0.44 \\
3 & C11L1 & 0.20 & C17 & 0.23 & C11L1 & 0.43 & C17 & 0.27 & MSIS1 & 0.29 & C17 & 0.49 \\
4 & SIR & 0.42 & MSIS1 & 0.36 & SIS & 0.54 & SIR & 0.36 & SIR & 0.30 & SIS & 0.51 \\
5 & MSIS1 & 0.52 & MSIS2 & 0.38 & SIR & 0.55 & MSIS1 & 0.47 & SIS & 0.33 & SIR & 0.51 \\
6 & SIS & 0.57 & SIS & 0.38 & DW & 0.72 & SIS & 0.49 & MSIR1 & 0.34 & MUT & 0.69 \\
7 & MSIS2 & 0.60 & SIR & 0.39 & MUT & 0.74 & MSIR1 & 0.57 & C17 & 0.35 & GEN & 0.83 \\
8 & MSIR2 & 0.62 & gLV & 0.41 & GEN & 0.84 & MSIS2 & 0.60 & MSIS2 & 0.36 & WC & 0.93 \\
9 & MSIR1 & 0.63 & MSIR1 & 0.42 & WC & 0.92 & MSIR2 & 0.60 & MSIR2 & 0.39 & DW & 0.97 \\
10 & gLV & 0.71 & WC & 0.45 & gLV & 1.00 & WC & 0.71 & gLV & 0.41 & gLV & 1.21 \\
11 & WC & 0.72 & MSIR2 & 0.46 & MSIS1 & 2.01 & GEN & 0.72 & GEN & 0.42 & MSIS1 & 2.06 \\
12 & GEN & 0.73 & GEN & 0.46 & MSIR1 & 3.98 & gLV & 0.74 & WC & 0.43 & MSIR1 & 4.32 \\
13 & DW & 0.93 & MUT & 0.71 & MSIS2 & 5.90 & MUT & 0.87 & MUT & 0.65 & MSIS2 & 9.50 \\
14 & MUT & 0.95 & DW & 0.75 & MSIR2 & 7.73 & DW & 0.94 & DW & 0.72 & MSIR2 & 14.03 \\
\bottomrule
\end{tabular}
\end{table}

\begin{table}[H]
\centering\footnotesize
\setlength{\tabcolsep}{3pt}
\caption{Ranking by mean n-RMSE (in percent), averaged over all
$17$ target dynamics and over all $14$ networks. The $19$ candidate sources
are the $17$ per-dynamics decoders, C17, and C17L1, the
leave-one-dynamics-out pooled decoder over all $17$ dynamics.
}
\label{tab:rank16-rmse}
\begin{tabular}{r|l@{\hspace{3pt}}c|l@{\hspace{3pt}}c|l@{\hspace{3pt}}c||l@{\hspace{3pt}}c|l@{\hspace{3pt}}c|l@{\hspace{3pt}}c}
\toprule
 & \multicolumn{2}{c|}{NN, MinCorr} & \multicolumn{2}{c|}{NN, QR} & \multicolumn{2}{c|}{NN, random} & \multicolumn{2}{c|}{linear, MinCorr} & \multicolumn{2}{c|}{linear, QR} & \multicolumn{2}{c}{linear, random} \\
\cmidrule{2-3} \cmidrule{4-5} \cmidrule{6-7} \cmidrule{8-9} \cmidrule{10-11} \cmidrule{12-13}
rank & source & value & source & value & source & value & source & value & source & value & source & value \\
\midrule
1 & C17 & 4.16 & C17 & 4.12 & C17 & 4.90 & C17 & 6.34 & C17 & 6.30 & C17 & 6.26 \\
2 & C17L1 & 7.44 & C17L1 & 7.37 & C17L1 & 7.70 & C17L1 & 8.44 & C17L1 & 8.73 & C17L1 & 8.51 \\
3 & VEG2 & 10.86 & LDW & 11.04 & SIR & 9.93 & VEG2 & 10.46 & SIR & 11.17 & SIS & 9.75 \\
4 & SIR & 11.04 & SIR & 11.93 & SIS & 10.13 & SIR & 10.64 & SIS & 11.19 & SIR & 10.05 \\
5 & LDW & 11.43 & SIS & 12.11 & MUT & 10.39 & SIS & 11.05 & VEG2 & 11.93 & MUT & 10.30 \\
6 & EUT2 & 12.24 & VEG2 & 12.13 & LDW & 10.54 & EUT2 & 11.46 & LDW & 11.95 & EUT2 & 10.88 \\
7 & SIS & 12.26 & VEG1 & 12.14 & VEG2 & 10.70 & NM & 12.74 & VEG1 & 12.22 & VEG2 & 11.69 \\
8 & VEG1 & 12.40 & NM & 12.96 & EUT1 & 11.04 & VEG1 & 12.81 & EUT2 & 12.32 & EUT1 & 12.21 \\
9 & NM & 12.84 & EUT2 & 13.22 & NM & 11.06 & MUT & 14.25 & MUT & 12.75 & DW & 12.70 \\
10 & EUT1 & 13.42 & DW & 13.27 & DW & 11.07 & DW & 14.33 & NM & 12.85 & NM & 14.08 \\
11 & DW & 13.84 & EUT1 & 13.29 & EUT2 & 11.07 & EUT1 & 14.80 & EUT1 & 13.37 & LDW & 15.20 \\
12 & MUT & 14.87 & MUT & 13.71 & VEG1 & 15.14 & MSIS1 & 15.07 & DW & 13.38 & VEG1 & 15.38 \\
13 & MSIS1 & 15.89 & gLV & 17.43 & GEN & 18.09 & MSIR1 & 15.88 & gLV & 17.42 & GEN & 17.82 \\
14 & gLV & 16.17 & GEN & 17.62 & gLV & 19.06 & gLV & 16.24 & GEN & 17.68 & WC & 20.07 \\
15 & MSIR1 & 16.63 & WC & 18.12 & WC & 19.62 & WC & 16.66 & WC & 17.87 & gLV & 20.80 \\
16 & WC & 16.99 & MSIS1 & 18.33 & MSIS1 & 43.03 & GEN & 16.82 & MSIS1 & 18.70 & MSIS1 & 45.08 \\
17 & GEN & 17.08 & MSIR1 & 20.57 & MSIR1 & 77.29 & MSIR2 & 17.20 & MSIR1 & 21.18 & MSIR1 & 85.64 \\
18 & MSIS2 & 17.08 & MSIS2 & 20.70 & MSIS2 & 105.41 & MSIS2 & 17.75 & MSIS2 & 23.16 & MSIS2 & 210.63 \\
19 & MSIR2 & 17.26 & MSIR2 & 21.79 & MSIR2 & 122.87 & LDW & 20.37 & MSIR2 & 23.25 & MSIR2 & 285.77 \\
\bottomrule
\end{tabular}
\end{table}

\begin{table}[H]
\centering\footnotesize
\setlength{\tabcolsep}{3pt}
\caption{Ranking by mean $r(\mathrm{S}\to\mathrm{T})$,
averaged over all $17$ target dynamics and over all $14$ networks, for
the same $19$ candidate sources as in
Table~\ref{tab:rank16-rmse}.}
\label{tab:rank16-ratio}
\begin{tabular}{r|l@{\hspace{3pt}}c|l@{\hspace{3pt}}c|l@{\hspace{3pt}}c||l@{\hspace{3pt}}c|l@{\hspace{3pt}}c|l@{\hspace{3pt}}c}
\toprule
 & \multicolumn{2}{c|}{NN, MinCorr} & \multicolumn{2}{c|}{NN, QR} & \multicolumn{2}{c|}{NN, random} & \multicolumn{2}{c|}{linear, MinCorr} & \multicolumn{2}{c|}{linear, QR} & \multicolumn{2}{c}{linear, random} \\
\cmidrule{2-3} \cmidrule{4-5} \cmidrule{6-7} \cmidrule{8-9} \cmidrule{10-11} \cmidrule{12-13}
rank & source & value & source & value & source & value & source & value & source & value & source & value \\
\midrule
1 & C17 & 0.28 & C17 & 0.30 & C17 & 0.45 & C17 & 0.43 & C17 & 0.45 & C17 & 0.57 \\
2 & C17L1 & 0.47 & C17L1 & 0.48 & C17L1 & 0.61 & C17L1 & 0.55 & C17L1 & 0.58 & C17L1 & 0.69 \\
3 & SIR & 0.81 & SIR & 0.84 & SIR & 0.89 & SIR & 0.78 & SIR & 0.78 & SIS & 0.86 \\
4 & SIS & 0.92 & SIS & 0.87 & SIS & 0.91 & SIS & 0.82 & SIS & 0.79 & SIR & 0.88 \\
5 & VEG2 & 0.93 & VEG2 & 0.94 & MUT & 0.94 & VEG2 & 0.90 & MUT & 0.87 & MUT & 0.90 \\
6 & VEG1 & 0.95 & VEG1 & 0.96 & EUT1 & 0.98 & EUT2 & 0.93 & EUT2 & 0.92 & EUT2 & 0.96 \\
7 & NM & 0.96 & MUT & 0.96 & EUT2 & 0.99 & NM & 0.96 & VEG2 & 0.93 & VEG2 & 1.09 \\
8 & EUT2 & 0.99 & EUT1 & 0.97 & VEG2 & 1.01 & VEG1 & 0.97 & EUT1 & 0.96 & EUT1 & 1.10 \\
9 & EUT1 & 1.03 & NM & 0.99 & NM & 1.01 & MUT & 1.05 & VEG1 & 0.96 & DW & 1.24 \\
10 & LDW & 1.04 & EUT2 & 0.99 & DW & 1.02 & MSIS1 & 1.10 & NM & 0.98 & NM & 1.25 \\
11 & MUT & 1.12 & LDW & 1.00 & LDW & 1.05 & DW & 1.17 & DW & 1.03 & VEG1 & 1.44 \\
12 & DW & 1.15 & DW & 1.05 & VEG1 & 1.45 & MSIR1 & 1.19 & LDW & 1.08 & GEN & 1.50 \\
13 & MSIS1 & 1.16 & gLV & 1.13 & GEN & 1.52 & EUT1 & 1.20 & gLV & 1.13 & LDW & 1.52 \\
14 & gLV & 1.22 & WC & 1.17 & gLV & 1.62 & WC & 1.21 & WC & 1.17 & WC & 1.67 \\
15 & GEN & 1.23 & GEN & 1.17 & WC & 1.64 & GEN & 1.21 & GEN & 1.20 & gLV & 1.79 \\
16 & WC & 1.24 & MSIS1 & 1.19 & MSIS1 & 3.50 & gLV & 1.24 & MSIS1 & 1.28 & MSIS1 & 3.66 \\
17 & MSIS2 & 1.25 & MSIR1 & 1.29 & MSIR1 & 6.44 & MSIR2 & 1.27 & MSIR1 & 1.41 & MSIR1 & 7.07 \\
18 & MSIR1 & 1.26 & MSIS2 & 1.32 & MSIS2 & 9.03 & MSIS2 & 1.30 & MSIR2 & 1.53 & MSIS2 & 17.06 \\
19 & MSIR2 & 1.28 & MSIR2 & 1.33 & MSIR2 & 10.76 & LDW & 1.78 & MSIS2 & 1.54 & MSIR2 & 23.50 \\
\bottomrule
\end{tabular}
\end{table}

\newpage
\clearpage

\section{Robustness to NN bias terms and ridge strength}
\label{app:baseline}

In this section, we assess the robustness of the main results to two
decoder specifications: the presence or absence of bias terms in the
NN decoder, and the ridge regularization strength in the linear
decoder.

The bias-restored NN decoder is
\begin{equation}
\hat{\tilde{\bx}}_{\rm bias}(\tilde{\bs})
=
\mathbf{W}_2
\mathrm{ReLU}\!\left(\mathbf{W}_1\tilde{\bs}+\mathbf{b}_1\right)
+\mathbf{b}_2,
\label{eq:nn-decoder-baseline}
\end{equation}
where $\mathbf{b}_1$ and $\mathbf{b}_2$ are bias terms in the hidden
and output layers, respectively. This is the natural default shallow
decoder architecture used in previous work~\cite{Erichson2020PRSA-SI,wzk2024-SI}. In the main text, we use the bias-free version in
Eq.~\eqref{eq:nn-decoder-nobias} because it maps zero input to zero
output, which is desirable for dynamics with a physical zero state.

\begin{figure}
\centering
\includegraphics[width=\textwidth]{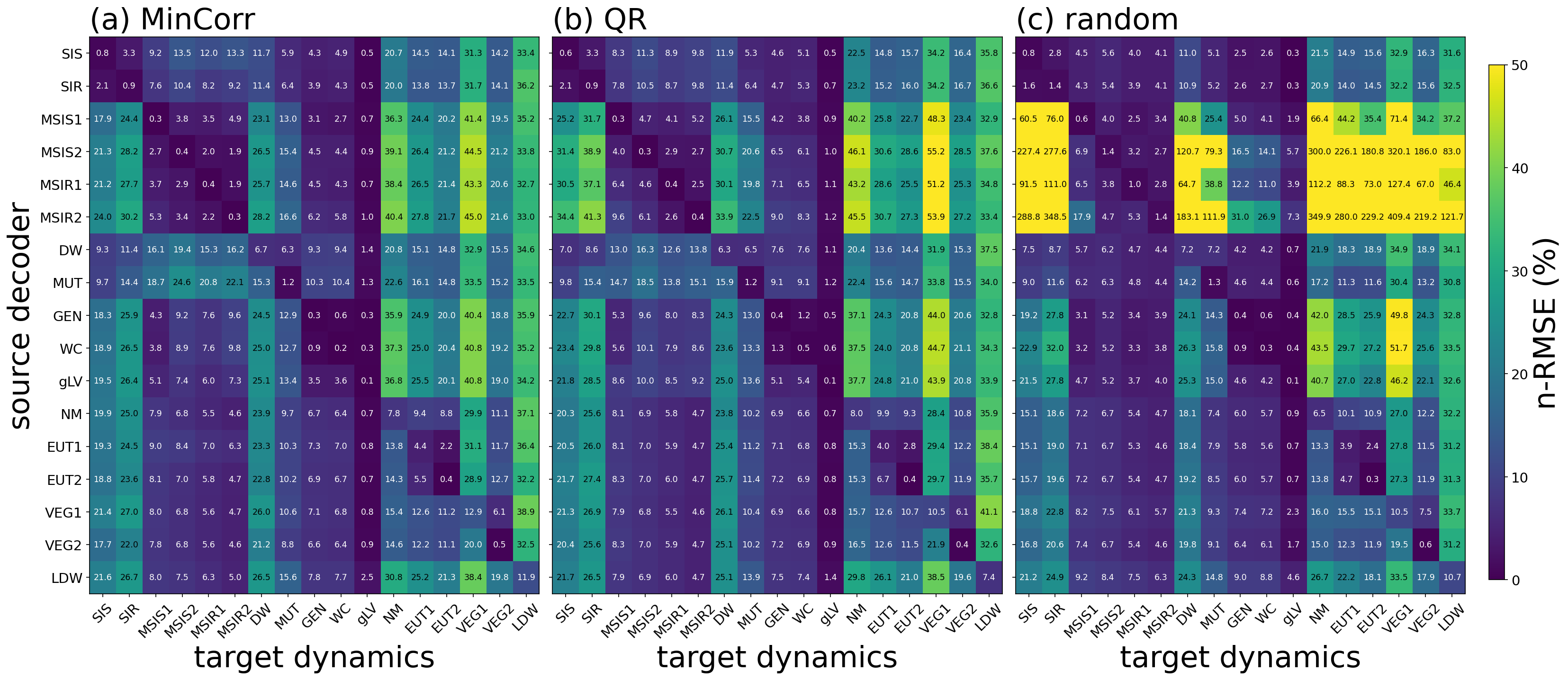}
\caption{n-RMSE under the NN decoder with bias terms restored. As in
Fig.~\ref{fig:heat-combined}, each cell shows the n-RMSE averaged over
the $14$ networks for a source decoder evaluated on a target dynamics.
The sentinel-selection methods are (a)~MinCorr, (b)~QR, and
(c)~uniformly random.}
\label{fig:robustness-heat-nn}
\end{figure}

First, we evaluated this bias-restored NN decoder instead of the
bias-free architecture used in the main text.
Figure~\ref{fig:robustness-heat-nn}
shows the cross-transfer n-RMSE for the bias-restored NN decoder under
the three sentinel-selection methods. The results are qualitatively
similar to those for the bias-free NN decoder shown in
Fig.~\ref{fig:heat-combined}(a)--(c) in the main text. The diagonal cells remain small,
the block of large cross-transfer errors for the last six target
dynamics (i.e., NM, EUT1, EUT2, VEG1, VEG2, and LDW) is preserved, and random
sentinel selection amplifies errors when metapopulation dynamics
(i.e., MSIS and MSIR) are used as source dynamics.

Quantitatively, for the two non-random sentinel methods, MinCorr and
QR, the cell-by-cell n-RMSE shift from the corresponding bias-free
results in Fig.~\ref{fig:heat-combined}(a)--(b) is within $1$ percentage point
for approximately $74\%$ of the cells, and within
$2$ percentage points for approximately $90\%$ of the cells. The
random sentinels (see Fig.~\ref{fig:robustness-heat-nn}(c)) yield
larger shifts, driven by a few large changes in the metapopulation
source dynamics rows.

\begin{figure}
\centering
\includegraphics[width=\textwidth]{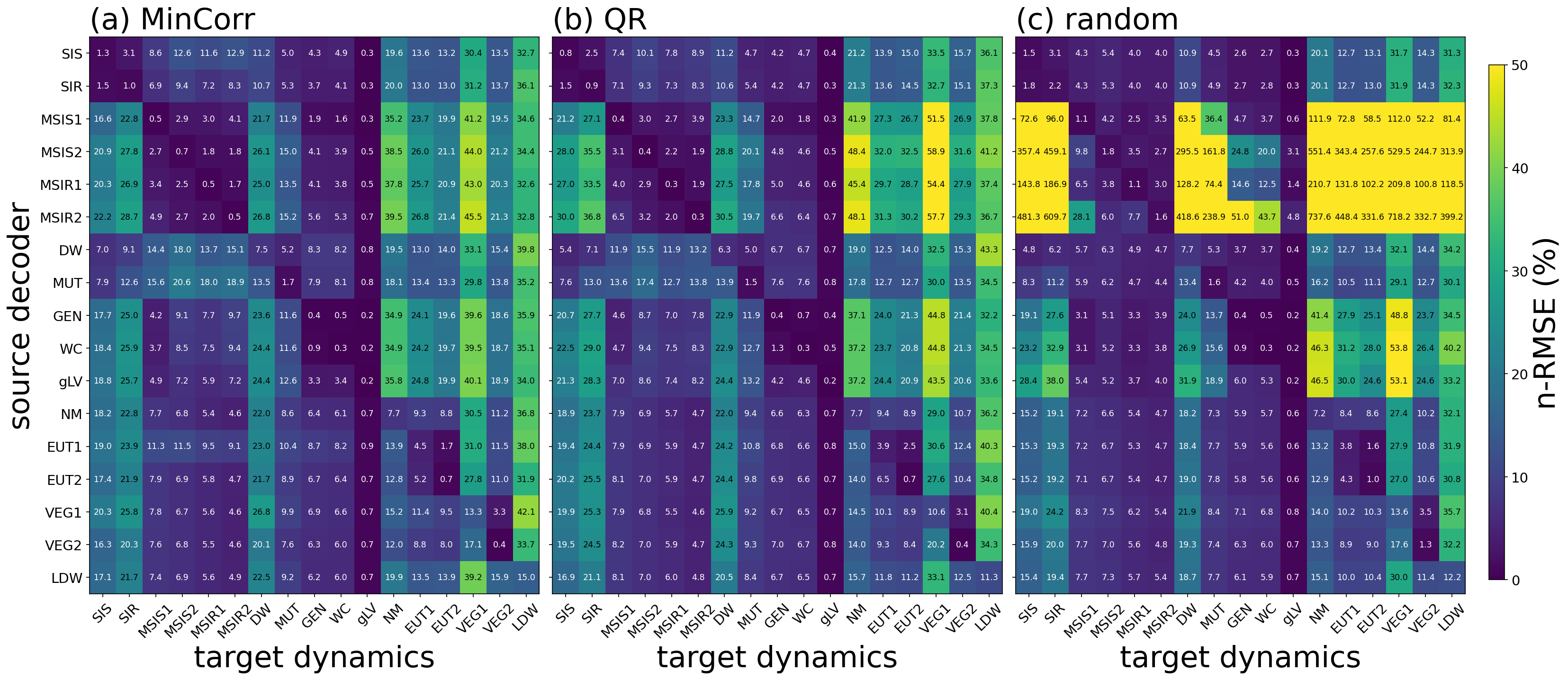}
\caption{n-RMSE under the linear ridge decoder with regularization
strength $\alpha=10^{-1}$. The sentinel-selection methods are
(a)~MinCorr, (b)~QR, and (c)~uniformly random.}
\label{fig:robustness-heat-lin}
\end{figure}

Second, we increased the ridge regularization strength of the linear
decoder from $\alpha=10^{-2}$, used in the main text, to
$\alpha=10^{-1}$. Figure~\ref{fig:robustness-heat-lin} shows the
cross-transfer n-RMSE for this stronger regularization. The results
are similar to those obtained with $\alpha=10^{-2}$ shown in
Fig.~\ref{fig:heat-combined}(d)--(f) in the main text. Increasing $\alpha$ by one order
of magnitude shifts the diagonal cells upward by about
$0.18$ percentage points on average and the off-diagonal cells
downward by about
$0.24$ percentage points on average for the two non-random sentinel
methods. For random sentinels, the off-diagonal mean shift is again
dominated by a few large outliers in the metapopulation source dynamics rows,
similar to the behavior observed for the bias-restored NN decoder.

Tables~\ref{tab:robustness-rmse} and~\ref{tab:robustness-ratio}
summarize the effects of these two decoder modifications on
reconstruction across the $11$ transferable target dynamics.
Specifically, we report the mean n-RMSE and the mean baseline ratio
$r(\mathrm{S}\to\mathrm{T})$, respectively, for $15$ candidate source
decoders: the $11$ transferable per-dynamics source decoders, the C11
pooled decoder, the C11L1 leave-one-dynamics-out pooled decoder,
and the budget-matched pooled decoders C11B and C11L1B.
The columns labeled ``bias-free'' and ``$\alpha=10^{-2}$'' correspond
to the decoder specifications used in the main text.

The budget-matched decoders C11B and C11L1B, introduced in
Section~\ref{sec:rank11}, respond similarly to both decoder modifications.
For the two non-random sentinel-selection methods, their mean n-RMSE changes
by at most $0.34$ percentage points, and both continue to outperform every
per-dynamics source decoder. The only exceptions involve uniformly random
sentinels: C11L1B performs slightly worse than the best per-dynamics source,
with n-RMSE values of $4.04\%$ versus $3.86\%$ for the bias-restored NN
decoder and $3.98\%$ versus $3.93\%$ for the linear decoder with
$\alpha=10^{-1}$. These equal-training-budget comparisons reinforce our
conclusion that pooled training outperforms per-dynamics training, at least
for the non-random sentinel-selection methods on which we primarily focus
in the main text.

Both modifications change reconstruction errors, and the changes
are the largest for random sentinels and metapopulation source dynamics.
Nevertheless, the main qualitative conclusions are unchanged. The
pooled decoders remain the best-performing sources, random sentinels
remain substantially worse than MinCorr and QR sentinels, and the
broad cross-transfer structure is preserved.

\begin{table}
\centering\small\setlength{\tabcolsep}{5pt}
\caption{Mean n-RMSE (in percent), averaged over the $11$
transferable target dynamics and over all $14$ networks, for
$15$ candidate source decoders (i.e., the $11$ transferable per-dynamics
decoders, C11, C11B, C11L1, and C11L1B). The top half compares the bias-free NN
decoder used in the main text with the bias-restored NN decoder. The
bottom half compares the linear ridge decoder with $\alpha=10^{-2}$,
used in the main text, with the same decoder at $\alpha=10^{-1}$.
In the column headers, ``bias rest.'' denotes the bias-restored NN
decoder.}
\label{tab:robustness-rmse}
\begin{tabular}{>{\raggedright\arraybackslash}p{1.3cm}*{6}{>{\centering\arraybackslash}p{1.95cm}}}
\toprule
 & \multicolumn{2}{c}{NN, MinCorr} & \multicolumn{2}{c}{NN, QR} & \multicolumn{2}{c}{NN, random} \\
\cmidrule{2-3} \cmidrule{4-5} \cmidrule{6-7}
source & \multicolumn{1}{c}{bias-free} & \multicolumn{1}{c}{bias rest.} & \multicolumn{1}{c}{bias-free} & \multicolumn{1}{c}{bias rest.} & \multicolumn{1}{c}{bias-free} & \multicolumn{1}{c}{bias rest.} \\
\midrule
SIS & 7.12 & 7.23 & 5.93 & 6.33 & 3.98 & 3.93 \\
SIR & 5.40 & 5.91 & 6.15 & 6.20 & 3.99 & 3.86 \\
MSIS1 & 8.30 & 8.85 & 10.54 & 11.06 & 24.94 & 20.38 \\
MSIS2 & 9.52 & 9.82 & 12.15 & 13.19 & 70.31 & 68.69 \\
MSIR1 & 9.28 & 9.78 & 13.24 & 13.28 & 47.86 & 31.57 \\
MSIR2 & 9.92 & 11.21 & 14.81 & 15.41 & 85.73 & 93.34 \\
DW & 9.42 & 10.98 & 8.41 & 9.12 & 5.24 & 5.52 \\
MUT & 11.65 & 13.53 & 9.85 & 11.25 & 6.01 & 6.13 \\
GEN & 10.40 & 10.32 & 10.70 & 11.22 & 9.17 & 9.30 \\
WC & 10.36 & 10.42 & 11.22 & 11.33 & 10.34 & 10.37 \\
gLV & 9.34 & 10.68 & 10.41 & 12.34 & 10.71 & 10.54 \\
C11 & 2.24 & 2.35 & 1.52 & 1.70 & 2.70 & 3.18 \\
C11B & 2.48 & 2.59 & 2.02 & 2.36 & 3.06 & 3.36 \\
C11L1 & 2.95 & 3.27 & 2.47 & 2.69 & 3.60 & 3.76 \\
C11L1B & 3.41 & 3.75 & 2.82 & 3.13 & 3.79 & 4.04 \\
\bottomrule
\end{tabular}
\vspace{0.8em}
\begin{tabular}{>{\raggedright\arraybackslash}p{1.3cm}*{6}{>{\centering\arraybackslash}p{1.95cm}}}
\toprule
 & \multicolumn{2}{c}{linear, MinCorr} & \multicolumn{2}{c}{linear, QR} & \multicolumn{2}{c}{linear, random} \\
\cmidrule{2-3} \cmidrule{4-5} \cmidrule{6-7}
source & \multicolumn{1}{c}{$\alpha{=}10^{-2}$} & \multicolumn{1}{c}{$\alpha{=}10^{-1}$} & \multicolumn{1}{c}{$\alpha{=}10^{-2}$} & \multicolumn{1}{c}{$\alpha{=}10^{-1}$} & \multicolumn{1}{c}{$\alpha{=}10^{-2}$} & \multicolumn{1}{c}{$\alpha{=}10^{-1}$} \\
\midrule
SIS & 6.21 & 6.88 & 5.12 & 5.70 & 3.74 & 3.94 \\
SIR & 4.72 & 5.31 & 4.84 & 5.42 & 3.75 & 3.93 \\
MSIS1 & 7.45 & 7.95 & 8.49 & 9.14 & 25.65 & 26.24 \\
MSIS2 & 9.54 & 9.58 & 11.69 & 11.79 & 121.69 & 121.76 \\
MSIR1 & 8.48 & 9.30 & 10.84 & 11.37 & 52.60 & 52.39 \\
MSIR2 & 9.74 & 10.40 & 12.59 & 12.98 & 171.80 & 171.95 \\
DW & 9.58 & 9.77 & 8.02 & 8.23 & 6.54 & 4.85 \\
MUT & 10.80 & 11.41 & 9.16 & 9.96 & 5.83 & 5.85 \\
GEN & 10.29 & 9.99 & 9.94 & 10.24 & 9.13 & 9.17 \\
WC & 10.19 & 10.07 & 10.68 & 10.84 & 10.46 & 10.48 \\
gLV & 9.61 & 10.31 & 10.46 & 11.59 & 12.94 & 13.36 \\
C11 & 2.74 & 2.83 & 1.84 & 1.86 & 3.01 & 2.99 \\
C11B & 3.03 & 2.69 & 2.39 & 2.39 & 3.21 & 3.36 \\
C11L1 & 3.84 & 3.55 & 2.68 & 2.71 & 3.74 & 3.60 \\
C11L1B & 3.75 & 4.01 & 3.09 & 3.42 & 3.63 & 3.98 \\
\bottomrule
\end{tabular}
\end{table}

\begin{table}
\centering\small\setlength{\tabcolsep}{5pt}
\caption{Mean baseline ratio $r(\mathrm{S}\to\mathrm{T})$, averaged
over the $11$ transferable target dynamics and over all $14$ networks,
for the same $15$ candidate source decoders and the same six
decoder/sentinel pairs as in Table~\ref{tab:robustness-rmse}.
Conventions are the same as in Table~\ref{tab:robustness-rmse};
in particular, ``bias rest.'' denotes the bias-restored NN decoder.}
\label{tab:robustness-ratio}
\begin{tabular}{>{\raggedright\arraybackslash}p{1.3cm}*{6}{>{\centering\arraybackslash}p{1.95cm}}}
\toprule
 & \multicolumn{2}{c}{NN, MinCorr} & \multicolumn{2}{c}{NN, QR} & \multicolumn{2}{c}{NN, random} \\
\cmidrule{2-3} \cmidrule{4-5} \cmidrule{6-7}
source & \multicolumn{1}{c}{bias-free} & \multicolumn{1}{c}{bias rest.} & \multicolumn{1}{c}{bias-free} & \multicolumn{1}{c}{bias rest.} & \multicolumn{1}{c}{bias-free} & \multicolumn{1}{c}{bias rest.} \\
\midrule
SIS & 0.57 & 0.59 & 0.38 & 0.42 & 0.54 & 0.54 \\
SIR & 0.42 & 0.47 & 0.39 & 0.41 & 0.55 & 0.54 \\
MSIS1 & 0.52 & 0.60 & 0.36 & 0.41 & 2.01 & 1.88 \\
MSIS2 & 0.60 & 0.66 & 0.38 & 0.43 & 5.90 & 5.97 \\
MSIR1 & 0.63 & 0.68 & 0.42 & 0.44 & 3.98 & 3.10 \\
MSIR2 & 0.62 & 0.73 & 0.46 & 0.49 & 7.73 & 8.30 \\
DW & 0.93 & 1.13 & 0.75 & 0.85 & 0.72 & 0.79 \\
MUT & 0.95 & 1.17 & 0.71 & 0.84 & 0.74 & 0.76 \\
GEN & 0.73 & 0.74 & 0.46 & 0.50 & 0.84 & 0.88 \\
WC & 0.72 & 0.73 & 0.45 & 0.47 & 0.92 & 0.95 \\
gLV & 0.71 & 0.82 & 0.41 & 0.49 & 1.00 & 1.02 \\
C11 & 0.14 & 0.16 & 0.10 & 0.12 & 0.35 & 0.46 \\
C11B & 0.15 & 0.18 & 0.12 & 0.17 & 0.37 & 0.48 \\
C11L1 & 0.20 & 0.24 & 0.16 & 0.19 & 0.43 & 0.52 \\
C11L1B & 0.22 & 0.26 & 0.16 & 0.20 & 0.45 & 0.55 \\
\bottomrule
\end{tabular}
\vspace{0.8em}
\begin{tabular}{>{\raggedright\arraybackslash}p{1.3cm}*{6}{>{\centering\arraybackslash}p{1.95cm}}}
\toprule
 & \multicolumn{2}{c}{linear, MinCorr} & \multicolumn{2}{c}{linear, QR} & \multicolumn{2}{c}{linear, random} \\
\cmidrule{2-3} \cmidrule{4-5} \cmidrule{6-7}
source & \multicolumn{1}{c}{$\alpha{=}10^{-2}$} & \multicolumn{1}{c}{$\alpha{=}10^{-1}$} & \multicolumn{1}{c}{$\alpha{=}10^{-2}$} & \multicolumn{1}{c}{$\alpha{=}10^{-1}$} & \multicolumn{1}{c}{$\alpha{=}10^{-2}$} & \multicolumn{1}{c}{$\alpha{=}10^{-1}$} \\
\midrule
SIS & 0.49 & 0.55 & 0.33 & 0.37 & 0.51 & 0.54 \\
SIR & 0.36 & 0.42 & 0.30 & 0.34 & 0.51 & 0.54 \\
MSIS1 & 0.47 & 0.50 & 0.29 & 0.31 & 2.06 & 2.11 \\
MSIS2 & 0.60 & 0.61 & 0.36 & 0.37 & 9.50 & 9.51 \\
MSIR1 & 0.57 & 0.63 & 0.34 & 0.36 & 4.32 & 4.29 \\
MSIR2 & 0.60 & 0.65 & 0.39 & 0.40 & 14.03 & 14.05 \\
DW & 0.94 & 0.96 & 0.72 & 0.74 & 0.97 & 0.68 \\
MUT & 0.87 & 0.94 & 0.65 & 0.73 & 0.69 & 0.72 \\
GEN & 0.72 & 0.70 & 0.42 & 0.44 & 0.83 & 0.84 \\
WC & 0.71 & 0.70 & 0.43 & 0.45 & 0.93 & 0.94 \\
gLV & 0.74 & 0.81 & 0.41 & 0.46 & 1.21 & 1.26 \\
C11 & 0.17 & 0.18 & 0.12 & 0.12 & 0.38 & 0.39 \\
C11B & 0.18 & 0.16 & 0.13 & 0.14 & 0.40 & 0.43 \\
C11L1 & 0.26 & 0.23 & 0.17 & 0.17 & 0.44 & 0.44 \\
C11L1B & 0.24 & 0.24 & 0.17 & 0.19 & 0.44 & 0.49 \\
\bottomrule
\end{tabular}
\end{table}

\newpage
\clearpage

\section{Sentinel selection based only on network structure}\label{si:sentsel}

We have shown in the main text that the MinCorr and QR heuristics
outperform uniformly random sentinel selection. In this section, we examine
which properties distinguish MinCorr and QR sentinels from other nodes and
whether these properties enable effective sentinel selection based only on
network structure. We restrict the analysis to the $11$ adjacency-coupled
dynamics because their equilibria are strongly correlated with node degree
(Section~\ref{si:why}) and are therefore likely to be more readily
characterized in terms of network structure than those of the
Laplacian-coupled dynamics.

We first ask whether the node sets selected by MinCorr and QR differ
systematically from uniformly random sets of the same size. For each
dynamics--network pair, we obtain the $n=\lfloor\ln N\rfloor$ MinCorr
sentinels and the $n$ QR sentinels. For each sentinel set, we compute the
mean degree, mean betweenness centrality, standard deviation of degree
within the set, and mean shortest-path distance between node pairs within
the set. We also compute the number of nodes shared by the MinCorr and QR
sets. We convert each statistic to a $z$-score relative to a null
distribution generated from $10^3$ uniformly random sets of $n$ nodes. For
the overlap statistic, the null distribution is generated from the number
of nodes shared between the same heuristic set, either MinCorr or QR, and a
uniformly random set.

Table~\ref{tab:sentsel-z} summarizes these $z$-scores across the
$11\times14=154$ dynamics--network pairs. For example, the reported mean
$z$-score is the average over these $154$ pairs. Both MinCorr and QR select
nodes at strongly non-random locations in the network, and the two selected
sets tend to overlap. Both methods favor sets with higher mean degree,
greater variation in degree, and higher mean betweenness centrality than
uniformly random sets. By contrast, their mean pairwise distances do not
consistently differ from the random-set null. Thus, network structure
contains information about which nodes MinCorr and QR tend to select.

However, this information alone is insufficient to construct an
effective sentinel set. To demonstrate this, we define two structural
sentinel-selection rules. The degree rule selects the $n$ nodes with the
largest degrees, whereas the betweenness rule selects the $n$ nodes with the
largest betweenness centralities. In both cases, we break ties uniformly at
random. These rules use no equilibrium data but require knowledge of the
network structure. We evaluate the resulting sentinel sets using the
n-RMSE averaged over the $14$ networks. We report both the median and mean
of this network-averaged n-RMSE across source--target dynamics pairs because
the difference between these summaries is informative.

Table~\ref{tab:sentsel-struct} shows the results. For within-dynamics
reconstruction, represented by the diagonal entries, the two structural
rules are comparable to MinCorr and QR in terms of the median n-RMSE but
worse in terms of the mean. For cross-dynamics transfer, represented by the
off-diagonal entries, both structural rules perform clearly worse than
MinCorr and QR in terms of both the median and mean. Moreover, the
structural rules do not consistently outperform uniformly random sentinels
across the four summary measures. Random sentinels yield the smallest
off-diagonal median but by far the largest off-diagonal mean among the five
methods. Thus, random sentinels are often adequate but occasionally
catastrophic, whereas MinCorr and QR are more reliable.

A likely reason why the degree and betweenness rules perform worse than
MinCorr and QR is redundancy among the selected nodes. For example, the
highest-degree nodes in a network may behave similarly, such that observing
$n$ of them provides less information than observing a more diverse,
carefully selected set of $n$ nodes. MinCorr penalizes this redundancy by
construction, whereas QR avoids it through pivoting. Requiring a sentinel
set to have a broad degree distribution, as suggested by
Table~\ref{tab:sentsel-z}, may reduce this redundancy by diversifying the
properties of the selected nodes. However, it is unclear how to balance a
large mean degree, which Table~\ref{tab:sentsel-z} also favors, against
large degree dispersion, which generally requires including some
low-degree nodes.

In summary, although network structure contains information about the
nodes selected by MinCorr and QR, it does not by itself provide a
sufficient design principle for constructing an effective sentinel set.
These results support the use of data-driven selection methods such as
MinCorr and QR and of equilibrium data, or potentially other dynamical
observations, to construct reliable sentinel sets. This conclusion is
consistent with our previous study, in which optimized sentinel sets for
reconstructing the scalar $\sum_{i=1}^{N}x_i^*/N$ required an algorithmic
selection procedure and could not be reproduced using network structure
alone~\cite{maclaren2025-SI}.

\begin{table}[htbp]
\centering
\caption{$z$-scores of five network-structural statistics for the
sentinel sets relative to null distributions generated from $10^3$
uniformly random node sets of the same size, $n$. The reported statistics
are computed across the $154$ combinations of the $14$ networks and $11$
adjacency-coupled dynamics. For the overlap statistic, the MinCorr--QR
overlap is compared with the overlap between the corresponding heuristic
set, either MinCorr or QR, and a uniformly random set. The column
$|z|>2$ gives the percentage of the $154$ dynamics--network pairs for which
the absolute $z$-score exceeds $2$.}
\label{tab:sentsel-z}
\small
\begin{tabular}{lcccccc}
\toprule
& \multicolumn{3}{c}{MinCorr} & \multicolumn{3}{c}{QR}\\
\cmidrule(lr){2-4}\cmidrule(lr){5-7}
statistic & mean $z$ & median $z$ & $|z|>2$ & mean $z$ & median $z$ & $|z|>2$\\
\midrule
mean degree & 7.19 & 3.68 & 69\% & 16.09 & 9.27 & 90\%\\
mean betweenness & 9.07 & 4.66 & 77\% & 18.15 & 11.01 & 87\%\\
standard deviation of the degrees & 8.16 & 5.65 & 93\% & 7.44 & 5.02 & 84\%\\
mean pairwise distance & 1.10 & 0.56 & 37\% & $-1.76$ & $-1.56$ & 50\%\\
MinCorr--QR overlap & 8.24 & 5.96 & 84\% & 8.50 & 5.68 & 84\%\\
\bottomrule
\end{tabular}
\end{table}

\begin{table}[htbp]
\centering
\caption{Median and mean n-RMSE (\%) across source--target pairs of the
$11$ adjacency-coupled dynamics. We compute these summaries separately for
within-dynamics reconstruction (the diagonal entries) and cross-dynamics
transfer (the off-diagonal entries) of the $11\times11$ n-RMSE matrix. Each
matrix entry is the n-RMSE averaged over the $14$ networks. The degree and
betweenness rules select the $n$ nodes with the largest degree and
betweenness centrality, respectively.}
\label{tab:sentsel-struct}
\small
\begin{tabular}{llcccc}
\toprule
& & \multicolumn{2}{c}{median} & \multicolumn{2}{c}{mean}\\
\cmidrule(lr){3-4}\cmidrule(lr){5-6}
decoder & sentinels & diagonal & off-diagonal & diagonal & off-diagonal\\
\midrule
\multirow{5}{*}{shallow NN decoder} & MinCorr & 0.40 & 8.09 & 1.14 & 9.96\\
 & QR & 0.36 & 8.34 & 0.99 & 11.24\\
 & random & 0.97 & 5.10 & 1.52 & 27.67\\
 & degree & 0.50 & 16.07 & 3.22 & 17.40\\
 & betweenness & 0.44 & 13.50 & 2.76 & 14.89\\
\addlinespace
\multirow{5}{*}{linear ridge decoder} & MinCorr & 0.39 & 7.23 & 1.13 & 9.55\\
 & QR & 0.17 & 6.75 & 0.83 & 10.10\\
 & random & 0.98 & 5.39 & 1.61 & 42.25\\
 & degree & 0.59 & 13.90 & 3.17 & 18.49\\
 & betweenness & 0.35 & 12.02 & 2.74 & 16.45\\
\bottomrule
\end{tabular}
\end{table}

\newpage
\clearpage

\section{Positivity and stability of the gLV coexistence
equilibrium}\label{si:glv}

For the gLV dynamics, the coexistence equilibrium is positive at every
node and locally asymptotically stable. To establish these properties,
recall that $c=\alpha_{\max}+1$, $\bW=D\bA-c\bI$, and that the coexistence
equilibrium is
\begin{equation}
\bx^*
=
-\overline{\lambda}\bW^{-1}\bone
=
\overline{\lambda}(c\bI-D\bA)^{-1}\bone.
\label{eq:glv-positive-equilibrium}
\end{equation}
Equation~\eqref{eq:glv-positive-equilibrium} is equivalent to
Eq.~\eqref{eq:glv-equilibrium} in the main text. Every value of $D$ in our
sweep satisfies
\begin{equation}
0\leq D<D_{\rm c}=\frac{c}{\alpha_{\max}}.
\end{equation}
Therefore,
\begin{equation}
\rho\!\left(\frac{D}{c}\bA\right)
=
\frac{D\alpha_{\max}}{c}
<1,
\end{equation}
where $\rho(\cdot)$ denotes the spectral radius. The inverse in
Eq.~\eqref{eq:glv-positive-equilibrium} consequently has the convergent
Neumann-series representation
\begin{equation}
(c\bI-D\bA)^{-1}
=
\frac{1}{c}
\left(\bI-\frac{D}{c}\bA\right)^{-1}
=
\frac{1}{c}
\sum_{\ell=0}^{\infty}
\left(\frac{D}{c}\bA\right)^\ell.
\label{eq:glv-neumann-series}
\end{equation}
Because $\bA$ is entrywise nonnegative, every term in
Eq.~\eqref{eq:glv-neumann-series} is entrywise nonnegative. Hence,
\begin{equation}
\bx^*
=
\frac{\overline{\lambda}}{c}
\sum_{\ell=0}^{\infty}
\left(\frac{D}{c}\bA\right)^\ell\bone
\geq
\frac{\overline{\lambda}}{c}\bone
>
\mathbf{0}.
\label{eq:glv-strict-positivity}
\end{equation}
Thus, every component of the coexistence equilibrium is strictly positive.

To establish local asymptotic stability, we examine the Jacobian at
$\bx^*$. From the original ODE
(Eq.~\eqref{eq:glv-componentwise} in the main text), the $(i,k)$ entry of
the Jacobian is
\begin{equation}
J_{ik} =
\delta_{ik}
\left(\overline{\lambda}
+\sum_{j=1}^{N}W_{ij}x_j\right)
+x_iW_{ik},
\label{eq:glv-jacobian-entry-general}
\end{equation}
where $\delta_{ik}$ is the Kronecker delta. At a coexistence equilibrium
$\bx^*>\mathbf{0}$, the equilibrium condition $\dot{x}_i=0$ implies
\begin{equation}
\overline{\lambda}
+\sum_{j=1}^{N}W_{ij}x_j^*
=
0
\quad\text{for all }i,
\label{eq:glv-equilibrium-condition}
\end{equation}
because $x_i^*>0$. Substituting
Eq.~\eqref{eq:glv-equilibrium-condition} into
Eq.~\eqref{eq:glv-jacobian-entry-general} yields
\begin{equation}
J_{ik}(\bx^*)=x_i^*W_{ik}.
\end{equation}
Therefore, in matrix form,
\begin{equation}
\bJ(\bx^*)
=
\operatorname{diag}(x_1^*,\ldots,x_N^*)\bW
=
\operatorname{diag}(\bx^*)\bW.
\label{eq:glv-jacobian}
\end{equation}

Because $\bA$ is symmetric, $\bW$ is also symmetric. Furthermore, its
largest eigenvalue satisfies
\begin{equation}
\lambda_{\max}(\bW)
=
D\alpha_{\max}-c
<0,
\end{equation}
so $\bW$ is negative definite. Define
\begin{equation}
\bP
=
\operatorname{diag}
\left(\sqrt{x_1^*},\ldots,\sqrt{x_N^*}\right).
\end{equation}
The strict positivity of $\bx^*$ implies that $\bP$ is invertible, and
\begin{equation}
\bP^{-1}\bJ(\bx^*)\bP
=
\bP\bW\bP.
\label{eq:glv-jacobian-similarity}
\end{equation}
The matrix on the right-hand side is symmetric and negative definite.
Therefore, all eigenvalues of $\bJ(\bx^*)$ are real and negative, proving
that $\bx^*$ is locally asymptotically stable.

\newpage
\clearpage

\section{Networks}
\label{si:networks}

Table~\ref{tab:networks} lists the $14$ undirected and unweighted
networks used in this study, together with their basic statistics. The
networks are drawn from the data set used in our previous
study~\cite{maclaren2025-SI}.

We excluded the six largest networks from the network collection used
in~\cite{maclaren2025-SI} because the computational cost of simulating
all dynamics and training all decoders on those networks was not
feasible on our hardware. The retained $14$ networks nevertheless span
more than two orders of magnitude in the number of nodes and nearly
three orders of magnitude in maximum degree.

\begin{table}[H]
\centering
\caption{The $14$ networks used in this study. $N$ is the number of
nodes in the largest connected component; $|E|$ is the number of
edges; $\langle k \rangle = 2|E|/N$ is the mean degree; $k_{\max}$ is
the maximum degree; and $n=\lfloor \ln N \rfloor$ is the number of
sentinel nodes.}
\label{tab:networks}
\small
\begin{tabular}{l r r r r r l}
\toprule
network & $N$ & $|E|$ & $\langle k \rangle$ & $k_{\max}$ & $n$ & ref. \\
\midrule
Dolphin social network                       & 62      & 159     & 5.13  & 12      & 4 & \cite{Lusseau2003BES-SI} \\
Human face-to-face proximity                 & 410     & 2{,}765 & 13.49 & 50      & 6 & \cite{Isella2011JTheorBiol-SI} \\
\emph{C.\,elegans} metabolic network         & 453     & 2{,}025 & 8.94  & 237     & 6 & \cite{Jeong2000Nature-SI} \\
Goh--Kahng--Kim model                        & 949     & 2{,}496 & 5.26  & 26      & 6 & \cite{Goh2001PRL-SI} \\
Lancichinetti--Fortunato--Radicchi benchmark & 998     & 1{,}988 & 3.98  & 36      & 6 & \cite{Lancichinetti2008PRE-SI} \\
Erd\H{o}s--R\'enyi random graph              & 1{,}000 & 25{,}132 & 50.26 & 76      & 6 & \cite{Erdos1960Publ-SI} \\
Barab\'asi--Albert model                     & 1{,}000 & 1{,}996 & 3.99  & 81      & 6 & \cite{BarabasiAlbert1999Science-SI} \\
Holme--Kim model                             & 1{,}000 & 1{,}996 & 3.99  & 102     & 6 & \cite{HolmeKim2002PRE-SI} \\
European road network                        & 1{,}039 & 1{,}305 & 2.51  & 10      & 6 & \cite{Subelj2011EurPhysJB-SI} \\
University email                             & 1{,}133 & 5{,}451 & 9.62  & 71      & 7 & \cite{Guimera2003PRE-SI} \\
\emph{Drosophila} protein--protein binary    & 2{,}705 & 8{,}458 & 6.25  & 129     & 7 & \cite{Tang2023NatCommun-SI} \\
Reactome biochemical-reactions network       & 5{,}973 & 145{,}778 & 48.81 & 855     & 8 & \cite{JoshiTope2005NAR-SI} \\
Internet AS-level (RouteViews)               & 6{,}474 & 12{,}572 & 3.88  & 1{,}458 & 8 & \cite{Leskovec2005KDD-SI} \\
Spanish-language word-association network    & 11{,}558 & 43{,}050 & 7.45 & 2{,}986 & 9 & \cite{Batagelj2002LangTech-SI} \\
\bottomrule
\end{tabular}
\end{table}

\end{document}